%% file: Thesis.tex
\documentclass[a4paper,11pt,chapterprefix=off]{scrreprt}
\usepackage[T1]{fontenc}
\usepackage[utf8]{inputenc}
\usepackage{adjustbox}
\usepackage{amsfonts}
\usepackage{amsmath}
\usepackage{amsthm}
\usepackage[toc,page]{appendix}
\usepackage{array} 
\usepackage{cancel}
\usepackage{caption}
\usepackage{comment} 
\usepackage{csquotes}
\usepackage{etoolbox}
\usepackage{fancybox}
\usepackage[margin=2.5cm]{geometry}
\usepackage{graphics}
\usepackage{humanist}
\usepackage{hyperref}
\usepackage{imakeidx}
\usepackage{mathrsfs}
\usepackage{mathtools}
\usepackage{subcaption}
\usepackage{subfiles}
\usepackage{tikz}
\usepackage{varioref}
\usepackage{xcolor}

\setkomafont{chapter}{\LARGE}
\setkomafont{section}{\Large}
\setkomafont{subsection}{\large}
\setkomafont{subsubsection}{\normalsize}
\setkomafont{paragraph}{\normalsize}
\setkomafont{subparagraph}{\normalsize}

\usetikzlibrary {arrows.meta, positioning, quotes, decorations.pathreplacing, shapes}

\newcommand{\equals}{=}
\newcommand{\cee}{\ell}
\newcommand{\cor}{\hat{~=~}}
\newcommand{\cupdot}{\hspace*{2.5pt}\dot\cup\hspace*{2.5pt}}
\newcommand{\ind}{f_\#}
\newcommand{\mc}{\mathcal}
\newcommand{\mf}{\mathfrak}
\newcommand{\komma}{,}

\newcommand{\xarrow}{-{Stealth[length=2mm]}}
\newcommand{\xor}{\oplus}

\newcommand{\CQ}{\text{\textsc{Cliq}} }
\newcommand{\DHC}{\text{\textsc{DHC}} }
\newcommand{\DHP}{\text{\textsc{DHP}} }
\newcommand{\DM}{\text{\textsc{\small{3}DeM}} }
\newcommand{\DMu}{\text{\textsc{\tiny{3}\scriptsize{DeM}}} }
\newcommand{\ODM}{\text{\textsc{\small{1+3}DM}} }
\newcommand{\ODMu}{\text{\textsc{\tiny{1+3}\scriptsize{DM}}} }
\newcommand{\DS}{\text{\textsc{DS}} }
\newcommand{\EC}{\text{\textsc{EC}} }
\newcommand{\FAS}{\text{\textsc{Fas}} }
\newcommand{\FVS}{\text{\textsc{Fvs}} }
\newcommand{\HS}{\text{\textsc{HS}} }
\newcommand{\IP}{\text{\textsc{IP}} }
\newcommand{\KS}{\text{\textsc{KS}} }
\newcommand{\MDS}{\text{\textsc{Mds}} }
\newcommand{\MIS}{\text{\textsc{Mis}} }
\newcommand{\MVC}{\text{\textsc{Mvc}} }

\renewcommand{\P}{\text{\textsc{Part}} }
\newcommand{\PCen}{\text{\textsc{pCen}} }
\newcommand{\PMed}{\text{\textsc{pMed}} }
\newcommand{\SAT}{\text{\textsc{Sat}} }
\newcommand{\TSAT}{\text{\textsc{\small{3}Sat}} }
\newcommand{\TSATu}{\text{\textsc{\tiny{3}\scriptsize{Sat}}} }
\newcommand{\ESAT}{\text{\textsc{E\small{3}Sat}} }
\newcommand{\ESATu}{\text{\textsc{\scriptsize{E}\tiny{3}\scriptsize{Sat}}} }
\newcommand{\OSAT}{\text{\textsc{\small{1}in\small{3}Sat}} }
\newcommand{\OSATu}{\text{\textsc{\tiny{1}\scriptsize{in}\tiny{3}\scriptsize{Sat}}} }

\renewcommand{\SS}{\text{\textsc{SubSum}} }
\newcommand{\SC}{\text{\textsc{SC}} }
\newcommand{\SP}{\text{\textsc{SP}} }
\newcommand{\STT}{\text{\textsc{StT}} }
\newcommand{\TMS}{\text{\textsc{TMS}} }
\newcommand{\TSP}{\text{\textsc{TSP}} }
\newcommand{\UHC}{\text{\textsc{UHC}} }
\newcommand{\UHP}{\text{\textsc{UHP}} }
\newcommand{\UFL}{\text{\textsc{UFL}} }
\newcommand{\VC}{\text{\textsc{VC}} }
\newcommand{\VCV}{\text{\textsc{VCV}} }

\newcommand{\SPR}{\leq_{\text{\tiny SPR}}}
\newcommand{\SSP}{\leq_{\text{\tiny SSP}}}
\newcommand{\SWPR}{\leq_{\text{\tiny WPR}}^{\text{\tiny SSP}}}
\newcommand{\SSPR}{\leq_{\text{\tiny SPR}}^{\text{\tiny SSP}}}
\newcommand{\WPR}{\leq_{\text{\tiny WPR}}}

\renewcommand{\epsilon}{\varepsilon}
\renewcommand{\implies}{~\Rightarrow~}
\renewcommand{\ne}{\overline}
\renewcommand{\mod}{\text{ mod }}

\let\oldsmall\small
\renewcommand{\small}[1]{\oldsmall{#1}\normalsize}

\renewcommand{\S}[1]{S\in\mc S_{#1}}
\newcommand{\Sr}[1]{S'\in\mc S_{#1}}
\newcommand{\Srep}{\mc S'_{\mf R}}
\newcommand{\srep}{R_{S'}}
\newcommand{\Sall}{\mathcal S'_\infty}
\newcommand{\Slink}{\mathcal S'_{\mathfrak L}}
\newcommand{\slink}{\mf L(f(S))}
\newcommand{\slinkr}{\mf L(\srep)}
\newcommand{\Snev}{\mathcal S'_\emptyset}

\newcommand{\SSPP}{\normalfont{SSP-\#P}}

\newtheorem{definition}{Definition}
\newtheorem{lemma}{Lemma}
\newtheorem{theorem}{Theorem}

\newcommand{\lemref}[1]{\autoref{#1}}

\newcolumntype{L}{>{$}l<{$}}
\newcolumntype{R}{>{$}r<{$}}

\AtBeginEnvironment{align}{\setcounter{equation}{0}}
\labelformat{equation}{(#1)}

\newcommand{\emptypage}{\newpage\null\thispagestyle{empty}\newpage}

\let\oldsection\section
\let\oldsubsection\subsection
\let\oldsubsubsection\subsubsection

\renewcommand{\section}{\chapter}
\renewcommand{\subsection}{\oldsection}
\renewcommand{\subsubsection}{\oldsubsection}
\renewcommand{\paragraph}{\oldsubsubsection*}

\title{A Compendium of SSP and Parsimonious Reductions}
\author{Celina Janet Bartlett}

\makeindex

\begin{document}
\def\biblio{}

\subfile{Sections/Titlepage/Titlepage.tex}
\emptypage
\subfile{Sections/Abstract/Abstract.tex}
\emptypage

\tableofcontents

\subfile{Sections/Introduction/Introduction.tex}

\subfile{Sections/General_Framework/General_Framework.tex}

\subfile{Sections/Overview/Overview.tex}

\section{Reductions}
\label{chap:4}
This chapter contains all of the reductions we examined in order to create this compendium. Most of these are based off of reductions found 
in \cite{AB09}, \cite{ssp}, \cite{garey}, \cite{karp} and \cite{Sip97}. For these, and some other, reductions we included a detailed description, a proof and 
an example reduction. \\
However we also worked closely with the author of the \cite{Femke} paper to create this compendium, which already examines certain 
reductions in search of the SSP property and includes proofs and examples in the same style as ours. Therefore we do not include these 
for their reductions, except for those cases, where we found the Parsimonious property to be violated.\\
As we have a lot of reductions betweens \textsc{Satisfiability} problems, we abuse some notations as shorthand for increased legibility.
For a literal set $L$ we define the two sets $L^+$ and $L^-$ with $L^+\cupdot L^- = L$, where $L^+$ contains all the positive 
literals $\ell\in L$ and $L^-$ all the negated ones $\ne \ell\in L$ and we define the set $L^2\subseteq L^+ \times L^-$, 
containing all the pairs of positive and negated literals $\ell,\ne\ell$ 
as tuples i.e. $\forall \ell \in L^+:(\ell,\ne\ell)\in L^2$. This also means we can denote the number of distinct literals as $|L^2|$.

\subfile{Reductions/Sat/Sat}

\subfile{Reductions/3-Sat/3-Sat}

\subfile{Reductions/Exact_3-Sat/Exact_3-Sat}

\subfile{Reductions/1-in-3-Sat/1-in-3-Sat}

\subfile{Reductions/Vertex_Cover/Vertex_Cover}

\subfile{Reductions/Independent_Set/Independent_Set.tex}

\subfile{Reductions/Clique/Clique.tex}

\subfile{Reductions/Set_Packing/Set_Packing.tex}

\subfile{Reductions/Subset_Sum/Subset_Sum.tex}

\subfile{Reductions/Partition/Partition.tex}

\subfile{Reductions/Exact_Cover/Exact_Cover.tex}

\subfile{Reductions/3-Dimensional_Matching/3-Dimensional_Matching.tex}

\subfile{Reductions/Directed_Hamiltonian_Cycle/Directed_Hamiltonian_Cycle.tex}

\subfile{Reductions/UHC/UHC.tex}

\subfile{Reductions/Directed_Hamiltonian_Path/Directed_Hamiltonian_Path.tex}

\subfile{Reductions/UHP/UHP.tex}

\subfile{Sections/Conclusion/Conclusion.tex}

\subfile{Sections/Appendix/Appendix.tex}

\newpage

\bibliographystyle{alpha}
\bibliography{ref}

\end{document}

%% file: Sections/Titlepage/Titlepage.tex
\begin{titlepage}
    \begin{center}
        \vspace*{1cm}
 
        \Huge
        \textbf{A Compendium of Subset Search Problems and Reductions relating to the Parsimonious Property}
 
        \vspace{0.5cm}
        \Large
        Bachelor's Thesis
             
        \vspace{1.5cm}
 
        \textbf{Celina Janet Bartlett}\\
        \today

        \vspace{3cm}
        \normalsize
        Examiners: \\
        Prof. Dr. Peter Rossmanith\\ 
        Prof. Dr. Martin Hoefer \\~\\
        Supervisor: \\
        Christoph Grüne
 
        \vfill

        Department of Computer Science I1\\
        RWTH Aachen University\\
        Germany\\
        Winter Semester 2024/25
             
    \end{center}
 \end{titlepage}

\ifSubfilesClassLoaded{
  \bibliography{../../ref}
}{}
\end{document}

%% file: Sections/Abstract/Abstract.tex
\begin{abstract}
    \begin{centering}
        \textbf{Abstract}\\
    \end{centering}
        This thesis centers around the concept of Subset Search Problems (SSP), a type of computational problem
        introduced by Grüne and Wulf to analyze the complexity of more intricate optimization problems. 
        These problems are given an input set, a so-called universe, and their solution lies within their own universe, e.g.
        the shortest path between two point is a subset of all possible paths. \\
        Due to this, reductions upholding the SSP property require an injective embedding from the universe of the first problem
        into that of the second. This, however, appears inherently similar to the concept of a Parsimonious reduction,
        a reduction type requiring a bijective function between the solution spaces of the two problems. Parsimonious
        reductions are mainly used within the complexity 
        class \#P, as this class of problems concerns itself with the number of possible solutions in a given problem.\\
        These two concepts, SSP and Parsimonious reductions, are inherently similar but, crucially,
        not equivalent. \\
        We therefore explore the interplay between reductions upholding the SSP and Parsimonious properties, 
        highlighting both the similarities and differences by providing a comprehensive theorem delineating the properties 
        required for reductions to uphold both attributes. \\
        We also compile and evaluate 46 reductions between 30 subset search variants of computational problems, 
        including those of classic NP-complete problems such as \textsc{Satisfiability}, \textsc{Vertex Cover}, 
        \textsc{Hamiltonian Cycle}, the \textsc{Traveling Salesman Problem} and \textsc{Subset Sum},
        providing reduction proofs, illustrative examples and insights as to where the SSP and Parsimonious properties 
        coexist or diverge. 
        \\
        With this compendium we contribute to the understanding of the computational complexity of bilevel and robust 
        optimization problems, by contributing a vast collection of proven SSP- and \#P-complete problems. 
        We also provide a basis for enhancing the compendium website “The Reduction Network”
        by reworking many of their given reductions into the Parsimonious framework and expanding the existing 
        collection of SSP-NP-complete problems. 
\end{abstract}

\ifSubfilesClassLoaded{
  \bibliography{../../ref}
}{}
\end{document}

%% file: Sections/Introduction/Introduction.tex
\section{Introduction}
\label{chap:1}
In the recent decades, the concept of optimization has been a core research topic in fields like 
theoretical computer science, 
applied mathematics and operations research, to name a few. The question of \enquote{How can 
we do this better?} has been bugging\footnote{Pun intended} researchers in all fields for ages. \\
Among the more advanced types of optimization, robust and bilevel optimization have been crucial for
understanding and improving many aspects of every day life.\\
Robust optimization addresses complex real-world challenges in uncertain environments, as the real world does 
not behave consistently. Unforeseen challenges and unreliable expectations are part and parcel of every day life.\\
Take, for instance, a delivery company sending out its goods via trucks. They can plan the route, check the weather
and traffic forecasts and preform regular maintenance on the trucks to keep them in perfect condition. However, these aspects
are all uncertain and cannot be controlled; Accidents can render planned roads impassable, weather and traffic 
conditions can change on a moment's notice and even a well-maintained truck can break down. This makes exact planning 
impossible and so any of the company's planned deliveries would have to work well even if the conditions deviate
from their expectations. This is in essence what robust optimization strives to do.\\
Bilevel optimization, on the other hand, is concerned with optimizing problems with two agents,
each having differing levels of control on the situation. These are so-called hierarchical decision-making problems, i.e. problems
where one party makes a decision and the other then reacts accordingly. \\
Say our delivery company operates in a state in which the government is trying to impose tolls on certain highways.
The government now initiates the hierarchical decision-making process by placing toll booths and setting the toll prices.
In reaction to this, the company now has to decide if the money lost on the tolls is worth the loss of time a detour would 
entail. If there is a simple alternative, they might decide to send their trucks on the longer, yet cheaper path. However, 
if the detour becomes to long they would send the truck through the toll booths. This of course depends 
on the prices of the tolls, so the government might then try to either reduce or increase the price at certain points in order 
to encourage more people to pay the tolls instead of circumnavigating them or to earn more money on the points that get used 
regularly, respectively. This has created a bilevel optimization problem where the company and the government have differing goals.\\
Now, to solve these problems and therefore maximize its profits, a company would have to create models of these 
real-world problems and then try to solve them as efficiently as possible. For this reason it is pertinent to know 
how hard these problems are to solve, as this dictates the amount of resources needed to find
viable solutions. In more technical terms, we need to know on which 
level of the polynomial hierarchy these problems are situated.\\
To understand the polynomial hierarchy, we first need the complexity class \textbf{NP}, i.e. the class of 
problems solvable in polynomial time on a nondeterministic machine. With this we define the polynomial 
hierarchy consisting of the classes $\Sigma_k^p$, for every $k\in \mathbb N$. The first level, 
$\Sigma_1^p$, consists of all NP-problems with only one existential quantifier. This level is defined by
questions like \enquote{Does a solution exist?}
The next level, $\Sigma_2^p$, involves two alternating quantifiers, existential followed 
by universal. Here the questions are of the form \enquote{Does a solution exist, so that for all ...?}
Under the widespread belief that the polynomial 
hierarchy does not collapse to a level below $\Sigma^p_2$, a $\Sigma^p_2$-complete problem would thus be 
harder to solve than NP-complete problems. \\
To more adequately describe and define the concept of robust and bilevel optimization  complexity classes,
Grüne and Wulf introduced the concept of \textbf{SSP-NP}-completeness \cite[pp. 12-13]{ssp} over 
\textit{Subset Search Problems}, for which the solution is a subset of the instance. For instance,
if we return to our delivery company, the route a certain truck takes can be seen as a subset of all
the roads that lie between the start and destination points, and an example problem would be to find the shortest 
possible route.\\
Now, a reduction upholding the Subset Search Problems property, a so-called SSP reduction, would, very 
basically, require an injective embedding from 
the elements in the first problem to those of the second, so that a solution of the first problem 
can be mapped to a subset of 
the elements in the corresponding solutions of the second problem.\\
This reduction property sounds inherently similar to a better-known reduction type, which upholds the 
exact number of possible solutions by creating a bijective 
mapping between these solution spaces. This reduction property is called \enquote{Parsimonious} and is 
used in reductions proving \textbf{\#P}-completeness \cite[pp. 189-201]{sharp1}.
The complexity class \#P concerns itself with so-called \enquote{counting problems}, where 
the number of possible solutions becomes relevant and is defined by questions like \enquote{How many solutions 
exist?}\\
It is therefore easy to see how the two complexity classes SSP-NP and \#P can be confused and 
how their respective reduction properties can be seen as equal.
Our paper concerns itself with the inherent similarities and differences between a SSP reduction and 
a Parsimonious one.

\subsection{Related Work}
\label{sec:related}
As we provide a clear line between SSP-completeness and \#P-completeness, works concerning the two topics 
are closely related to this paper. The concept of SSP-completeness was first introduced in 
a paper by Grüne and Wulf \cite{ssp} to better describe the complexity 
of bilevel and robust optimization problems. As for \#P-completeness, we point to Valiant's 1979 article \cite{sharp1} in which he 
introduces the concept of counting problems and the complexity class \#P, and to Barak's paper \cite{sharp3}
in which the definition of \#P is broadened to 
its current meaning, a complexity class over problems counting the number of solutions. \\
The aforementioned paper was also used in Aurora and Barak 2009 book \cite{AB09}, which serves as a detailed compendium of computational 
complexity theory, covering essential topics such as NP-completeness, probabilistic algorithms and cryptography, and is 
a comprehensive resource for understanding the intricacies of computational problems and their inherent difficulty.\\
Similarly important compendiums concerning the topic of P- and NP-completeness can be found in various books and articles.
One of the most famous is Karp's \enquote{Reducibility Among Combinatorial Problems} \cite[1972]{karp}, a paper in which he stipulates that
the notorious 21 problems are NP-complete. 
In 1979, Garey and Johnson provided a 
foundational overview of NP-completeness in their book \cite{garey}, offering a detailed exploration of NP-hard and NP-complete problems and their 
implications for computational theory and problem solving. \\
Later, Greenlaw, Hoover and Ruzzo explore the concept of P-completeness in their 1995 book \cite{comp4}, 
examining the inherent difficulty of parallelizing certain problems within the class P and the limitations this 
imposes on parallel computation. 
For a rigorous introduction to the fundamental principles of computational theory, (covering topics such as automata theory, 
formal languages, and Turing machines) we point to Sipser's 1996 book \cite{Sip97}.
It also explores important concepts in complexity theory, decidability and the limits of computation, 
with a focus on formal proofs and mathematical foundations.
Ausiello, Crescenzi, Gambosi, Kann, Marchetti-Spaccamela and Protasi's 1999 book \cite{comp2} 
presents another extensive collection of NP-hard optimization problems, outlining their approximation possibilities. 
Their findings on NP-completeness are also presented on a compendium website \cite{www1}. \\
Umans and Schaefer's 2002 article \cite{comp1} on the
higher levels of the polynomial-time hierarchy offers a comprehensive collection of problems 
and techniques related to the higher complexity classes, such as the $\Sigma_2^p$ and $\Sigma_3^p$ classes, also relating
to SSP-NP-completeness. \\
Further works relating to $\Sigma^p_2$ are Woeginger's article \cite{comp5}, which
discusses the challenges posed by the second quantifier while providing a comprehensive survey of the results, and 
Pfaue's paper \cite{Femke}, which provides a collection of SSP-NP-complete problems. \\
This latter paper also provides the basis for a compendium website \cite{TRN},
containing a multitude of NP-complete problems and reductions with the SSP property. Outside of this site and \cite{www1},
further compendium websites include the \enquote{Complexity Zoo} \cite{www2}, which contains a comprehensive guide to the various
complexity classes, and \enquote{The House of Graphs} \cite{www3}, which is a repository and database of well-known and 
interesting graphs in graph theory and provides detailed information about various types of graphs.\\
Finally, for further reading on the topics of robust and bilevel optimization we suggest
Gabrel, Murat and Thiele's compendium of robust optimization research from 2007 to 2014 \cite{GMT14}
and Colson, Marcotte and Savard's \enquote{An overview of bilevel optimization} \cite[2007]{CMS07}. Both works included
detailed introduction into their respective fields as well as a comprehensive analysis of the various key techniques.

\subsection{Contribution}
SSP-completeness was introduced by Grüne and Wolf \cite{ssp} in their paper \enquote{Completeness in the 
Polynomial Hierarchy for many natural Problems in Bilevel and Robust Optimization}, to find a large 
collection of $\Sigma^p_2$ and $\Sigma^p_3$-complete problems in one go. On an initial viewing, the concept
of an SSP reduction, as presented in this paper, seems indistinguishable to the concept of a Parsimonious reduction
proving \#P-completeness.\\
In this paper we clearly outline the difference between these two 
complexity classes by providing a comprehensive theorem, which states the properties needed in
a reduction for it to be able to prove both SSP-NP- and \#P-completeness.  
We then use this theorem to prove the SSP-NP- and \#P-hardness of 
the following thirty problems:
\begin{quote}
  \textsc{
    Satisfiability, 3-Satisfiability, Exact 3-Satisfiability, 1-in-3-Satisfiability, Clique,
    Dominating Set, Minimum Cardinality Dominating Set, Maximum Cardinality Independent Set, Steiner Tree,
    Traveling Salesman Problem, Undirected Hamiltonian Cycle, Directed Hamiltonian Cycle, 
    Undirected Hamiltonian Path, Directed Hamiltonian Path, Vertex Cover,
    Minimum Cardinality Vertex Cover, 
    Feedback Arc Set, Feedback Vertex Set, 0-1-Integer Programming, 3-Dimensional Exact Matching,
    Exact Cover, Hitting Set, Knapsack, 
    Partition, p-Center, p-Median, Set Cover, Set Packing, Subset Sum,
    Two-Machine-Scheduling, Uncapacitated Facility Location
    }
\end{quote}
Most of these reductions are based on existing reductions presented in papers relating to 
NP-completeness, such as \enquote{Computers and Intractability: A Guide to the Theory of NP-Completeness},
by Garey and Johnson \cite{garey}, Aurora and Barak's \enquote{Computational Complexity: A Modern Approach} \cite{AB09},
Sipser's \enquote{Introduction to the Theory of Computation} \cite{Sip97} and, of course, Karp's 
\enquote{Reducibility Among Combinatorial Problems} \cite{karp}.
We modified some of these reductions to fit the context of 
Subset Search Problems and to adhere to the reduction properties we favour in our paper. \\
As Pfaue's \enquote{Exploring the Reductions Between SSP-NP-complete Problems and Developing a 
Compendium Website Displaying the Results} \cite{Femke}, already contains a multitude of reductions
fitted to the constraints for Subset Search Problems, we took these reductions and tried to apply our theorem 
checking for \SSPP-completeness. Our compendium also provides further contents for the website 
\enquote{The Reduction Network: A compendium of reductions for various complexity classes} \cite{TRN}, 
originally written by Pfaue. For this reason, reductions already included in their paper are left unproven here,
whereas for any other reduction we also provide a proof, a minimal example and information on the SSP and Parsimonious
properties inherent in the reductions. 
\\
Furthermore, we prove that Cook's Theorem
\cite[pp. 151-158]{Coo71} is applicable to the complexity class \SSPP, in order to show that
the \textsc{Satisfiability} problem is SSP-NP- and \#P-complete. We later use this to prove the existence of
a transitive SSP and Parsimonious reduction for shown reductions lacking one or both properties.

\subsection{Structure}
This paper is divided up into five distinct chapters. 
\autoref{chap:2} introduces the necessary theoretical framework consisting of Subset Search problems and reductions
as well as the concept of Parsimonious reductions. We then provide a theorem as to when exactly these two 
types of reductions intersect.
In \autoref{chap:3} we provide an overview of the reductions contained in this compendium and some general
observations we gleaned from them. 
\autoref{chap:4} contains our compendium of SSP-NP- and \#P-hard problems, each reduction providing
a proof of concept and an illustrating example, as well as insights, as to which of the properties hold for each 
specific reduction. All of this is summarized in \autoref{chap:5}, in which we condense the essence
of our findings into a conclusion.
Finally, in \autoref{appendix}, we list all of the computational problems presented in 
this paper.

\ifSubfilesClassLoaded{
  \bibliography{../../ref}
}{}
\end{document}

%% file: Sections/General_Framework/General_Framework.tex
\section{Theoretical Framework}
\label{chap:2}
This chapter focuses mainly on the theoretical groundwork we use as a basis for our paper and on further framework we
build specifically for this context. Here we introduce the concept of \SSPP-completeness and provide a theorem on which 
properties a reduction between problems of this class needs to uphold.

\subsection{Preliminaries}
In our paper we use certain notations to ease our work with sets and functions. We denote the set of numbers $\{1,...,n\}\subseteq \mathbb{N}$ as $[n]$ and 
the i$^{th}$ element of a set $S$ as $S[i]$. For the power set of a set $S$ we use the notation $2^S$.\\
We define a function $f$ over a set $X$ as $f(X)=\{f(x)\mid x\in X\}$.
For a bijective function $f$ we use the notation $f^{-1}$ to denote the inverse function, where $f(f^{-1}(x))=x$. As we use the index of an element
in some of our reductions, let $\ind:X\rightarrow \mathbb N$ define the index function, mapping an element $x$ of a set $X$ to its index. Likewise
let $x_{n}$ for a $n\in\mathbb N$ denote the element $x\in X$ at index $n$. Groups of functions $(f_I)_{I\in \mc I}$ shall be denoted by $f$, when it 
is clear over which set they are defined.

\subsection{Problem and Reduction Framework}
This work is a compendium of reductions between Subset Search Problems, a variant of the general computational problem, 
for which the solution is a subset of the given problems instance. As such, we adapt the general 
computational problem to this premise, so that it now consists of instances and a solution set. 
This solution set $\mc S$ of a problem $\Pi$ is denoted by $\mc S_\Pi$ and is necessary, as our later reductions make
statements about a problems solutions.
\begin{definition}[Computational Problem]
    A computational problem $\Pi$ is a tuple $(\mc I, \mc S_\Pi)$, such that 
    \begin{itemize}
    \setlength\itemsep{0cm}
        \item $\mc I\subseteq\{0,1\}^*$ is a language. We call $\mc I$ the set of instances of $\Pi$.
        \item For each instance $I \in \mc I$, there exists a (potentially empty) set $\mc S_\Pi(I) \subseteq \{0,1\}^*$, which we call
        the solution set associated to the instance $I$.
        \item A solution $\S\Pi(I)$ can be verified in polynomial time using a deterministic turing machine. 
    \end{itemize}
\end{definition}
\subsubsection{Subset Search Problems}
As mentioned previously, we are analyzing the properties of reductions between \textit{Subset Search Problems}, SSP for short. We take this idea directly from \cite{ssp},
where the concepts of Subset Search Problems and SSP reductions is introduced as follows.
\begin{definition}[Subset Search Problem]
    A subset search problem $\Pi$ is a tuple $(\mc I,\mc U, \mc S_\Pi)$, such that 
    \begin{itemize}
        \setlength\itemsep{0cm}
        \item $\mc I\subseteq\{0,1\}^*$ is a language. We call $\mc I$ the set of instances of $\Pi$.
        \item To each instance $I \in \mc I$, there is some set $\mc U(I)$ which we call the universe associated to the instance $I$.
        \item To each instance $I \in \mc I$, there is some (potentially empty) set $\mc S_\Pi(I) \subseteq 2^{\mc U(I)}$ which we call
        the solution set associated to the instance $I$. An instance $I\in \mc I$ is called a Yes-instance, if $\mc S_\Pi(I)\neq \emptyset$.
        \item The decision problem associated to the SSP problem $\Pi$ is the language $\{I\in \mc I:\mc S_\Pi(I)\neq \emptyset\}$ of all Yes-instances.
    \end{itemize}
\end{definition}
We also use the concept of the \textit{SSP property}, where \enquote{there is an injective embedding 
of one SSP problem into another, such that the solution sets of the two SSP problems correspond one-to-one to each other in a
strict fashion. An SSP reduction is a usual many-one reduction which additionally has the
SSP property.} \cite[p. 12]{ssp}. With this we can introduce \textit{SSP reductions} to 
adhere to the SSP property.
\begin{definition}[SSP Reduction] Let $\Pi = (\mc I, \mc U, \mc S_\Pi)$ and $\Pi' = (\mc I', \mc U', \mc S_{\Pi'})$ be two SSP
problems. We say that there is an SSP reduction from $\Pi$ to $\Pi'$, and write $\Pi \SSP \Pi'$, if
\begin{itemize}
    \setlength\itemsep{0cm}
    \item There exists a function $g : \{0,1\}^* \rightarrow \{0,1\}^*$ computable in polynomial time in the input size $|I|$,
    such that $I$ is a Yes-instance iff $g(I)$ is a Yes-instance 
    $$\text{(i.e. }\mc S_\Pi(I) \neq \emptyset \Leftrightarrow \mc S_{\Pi'}(g(I)) \neq \emptyset\text{).}$$
    \item There exist functions $(f_I)_{I\in \mc I}$ computable in polynomial time in $|I|$ such that for all
    instances $I \in \mc I$, we have that $f_I : \mc U(I) \rightarrow \mc U'(g(I))$ is an injective function mapping
    from the universe of the instance $I$ to the universe of the instance $g(I)$ such that
    $$\{f_I(S) : S \in \mc S_\Pi(I)\} = \{S' \cap f_I(\mc U(I)) : S' \in \mc S_{\Pi'}(g(I))\}.$$
\end{itemize}
\end{definition}
\noindent Note that for a given instance $I\in \mc I$ of an SSP problem $\Pi:= (\mc I,\mc U, \mc S)$ we write $\mc U$ instead of $\mc U(I)$ and 
$\mc S$ instead of $\mc S(I)$ and we denote a specific instance $I\in \mc I$ of a problem $\Pi$ as $\Pi:=[I]$.\\
For many of our reductions we also use the transitive properties of an SSP reduction, which are proven in \cite[p. 13]{ssp}.
\begin{lemma}
    SSP reductions are transitive, i.e. for SSP problems $\Pi_1, \Pi_2$ and $\Pi_3$ with $\Pi_1 \SSP \Pi_2$
    and $\Pi_2 \SSP \Pi_3$, it holds that $\Pi_1 \SSP \Pi_3$.
\end{lemma}
\noindent Finally, we use the class SSP-NP, which is defined the analogously to the class NP restricted to SSP problems, as defined in \cite[p. 13]{ssp}.
\begin{definition}[SSP-NP] The class SSP-NP consists out of all the SSP problems which are
    polynomial-time verifiable. Formally, an SSP problem $\Pi=(\mc I,\mc U,\mc S_\Pi)$ belongs to SSP-NP, if
    $|\mc U(I)| =\text{poly}(|I|)$ and if there is an algorithm receiving tuples of an instance $I\in\mc I$ and a
    subset $S\subseteq \mc U(I)$ as input and decides in time polynomial in $|I|$, whether $S\in \mc S_\Pi(I)$.
\end{definition} 
Now, a Subset Search Problem $\Pi$ is SSP-NP-hard, iff for all problems $\Pi'$ in SSP-NP, 
    there exists an SSP reduction, so that $\Pi'\SSP\Pi$, and therefore SSP-NP-complete, iff 
    it is both in SSP-NP, as well as SSP-NP-hard.

\subsubsection{The Parsimonious property}
In a given SSP reduction, we look for the existence of a second reduction property, the so-called Parsimonious property.
The word \enquote{parsimonious} means \enquote{to be very unwilling to use resources}. In the case of 
Parsimonious reductions it is used to denote a direct one-to-one correspondence between the two solution sets of the problems in a reduction $\Pi\leq\Pi'$. 
This means that not only does $\Pi'$ have a solution if and only if $\Pi$ has a solution, but there also exist the exact same amount of solutions in the two 
solution sets $\mc S_{\Pi}$ and $\mc S_{\Pi'}$.
\begin{definition}[Weakly Parsimonious Reduction] Let $\Pi = (\mc I, \mc S_{\Pi})$ and $\Pi' = (\mc I', \mc S_{\Pi'})$ be two computational problems. 
    We say that there is a Weakly Parsimonious reduction from $\Pi$ to $\Pi'$ ($\Pi \WPR \Pi'$) if
    \begin{itemize}
        \setlength\itemsep{0cm}
        \item There exists a function $g : \{0,1\}^* \rightarrow \{0,1\}^*$ computable in polynomial time in the input size $|I|$,
        such that $I$ is a Yes-instance iff $g(I)$ is a Yes-instance.
        \item The number of solutions $\big|\mc S_\Pi(I)\big|$ equals the number of solutions $\big|\mc S_{\Pi'}(I')\big|$.
    \end{itemize}
\end{definition}    
\noindent This concept can be taken further, as we can try to find a bijective function between the two solution spaces.
\begin{definition}[Strongly Parsimonious Reduction] Let $\Pi = (\mc I, \mc S)$ and $\Pi' = (\mc I', \mc S')$ be two computational problems. 
    We say that there is an Strongly Parsimonious reduction from $\Pi$ to $\Pi'$ ($\Pi \SPR \Pi'$) if
    \begin{itemize}
        \setlength\itemsep{0cm}
        \item There exists a function $g : \{0,1\}^* \rightarrow \{0,1\}^*$ computable in polynomial time in the input size $|I|$,
        such that $I$ is a Yes-instance iff $g(I)$ is a Yes-instance.
        \item There exist functions $(p_I)_{I\in \mc I}$ computable in polynomial time in $|I|$ such that for all
        instances $I \in \mc I$, we have that $p_I : \mc S(I) \rightarrow \mc S'(g(I))$ is an bijective function mapping
        from the solution set of the instance $I$ to the solution set of the instance $g(I)$ such that
        $$\{p_I(S) : S \in \mc S_\Pi(I)\} = \{S'\in \mc S_{\Pi'}(g(I))\}.$$
    \end{itemize}
\end{definition}

\noindent Parsimonious reductions are mainly used for the complexity class \#P, concerning itself with counting problems. We use the 
definition used by Barak \cite[p. 104]{sharp3}.
\begin{definition}[\#P]
    \#P is the set of all functions $g : \{0, 1\}^* \rightarrow\mathbb N$ such
    that there is a polynomial time nondeterministic turing machine $M$ such that for all $x \in \{0, 1\}^*$,
    $g (x)$ equals the number of accepting branches in $M $'s computation graph on $x$.
\end{definition}
It follows that a computational problem $\Pi$ is \#P-hard, iff for every problem $\Pi'$ in \#P, 
    there exists a Parsimonious reduction, so that $\Pi'\SPR\Pi$, and therefore 
\#P-complete, iff it is both in \#P, as well as \#P-hard.\\

Similarly to an SSP reduction, these reduction types are also transitive, a property which we heavily rely on. The following
shows this for both weakly and strongly Parsimonious reductions.
\begin{lemma}
    Weakly Parsimonious reductions are transitive, i.e. for computational problems $\Pi_1, \Pi_2$ and $\Pi_3$ with $\Pi_1 \WPR \Pi_2$
    and $\Pi_2 \WPR \Pi_3$, it holds that $\Pi_1 \WPR \Pi_3$.
\end{lemma}
\begin{proof}
    Let $\Pi_1 := (\mc I_1, \mc S_{\Pi_1}), \Pi_2 := (\mc I_1, \mc S_{\Pi_2})$ and $\Pi_3 := (\mc I_3, \mc S_{\Pi_3})$ where 
    $\Pi_1 \WPR \Pi_2$ and $\Pi_2 \WPR \Pi_3$ hold. Therefore there exist reduction functions $g_1$ and $g_2$ computable in polynomial 
    time for which the concatenation $g_2\circ g_1$ is also computable in polynomial time. Also, as these reductions are weakly Parsimonious
    both $|\mc S_{\Pi_1}| = |\mc S_{\Pi_2}|$ and $|\mc S_{\Pi_2}| = |\mc S_{\Pi_3}|$ hold.
    Thus it follows that with the function $g:=g_2\circ g_1$ the reduction $\Pi_1 \leq \Pi_3$ is weakly Parsimonious as, due to the 
    transitive nature of equality, $|\mc S_{\Pi_1}| = |\mc S_{\Pi_3}|$ also holds.
\end{proof}

\begin{lemma}
    Strongly Parsimonious reductions are transitive, i.e. for computational problems $\Pi_1, \Pi_2$ and $\Pi_3$ with $\Pi_1 \SPR \Pi_2$
    and $\Pi_2 \SPR \Pi_3$, it holds that $\Pi_1 \SPR \Pi_3$.
\end{lemma}
\begin{proof}
    Let $\Pi_1 := (\mc I_1, \mc S_{\Pi_1}), \Pi_2 := (\mc I_1, \mc S_{\Pi_2})$ and $\Pi_3 := (\mc I_3, \mc S_{\Pi_3})$ where $\Pi_1 \SPR \Pi_2$
    and $\Pi_2 \SPR \Pi_3$ hold. It follows that there exist bijective functions $(p_I)_{I\in \mc I_1}$ and $(p_{I'})_{I'\in \mc I_2}$
    henceforth refered to as $p_1$ and $p_2$, as well as functions $g_1$ and $g_2$ for which the following statements hold. 
     $$\{p_1(S) : S \in \mc S_{\Pi_1}(I)\} = \{S'\in \mc S_{\Pi_2}(g_1(I))\} \text{ and } 
     \{p_2(S') : S' \in \mc S_{\Pi_2}(I')\} = \{S''\in \mc S_{\Pi_3}(g_2(I'))\}$$
    As both $p_1$ and $p_2$ consist of bijective functions, $p_2\circ p_1$ is also bijective. Therefore we get
    \begin{flalign*}
        \{p_2(p_1(S)) : S \in \mc S_{\Pi_1}(I)\} &= \{p_2(S'):S'\in \mc S_{\Pi_2}(g_1(I))\}\\
        &= \{S'' \in \mc S_{\Pi_3}(g_2(g_1(I)))\}
    \end{flalign*}
    And since concatenations of functions computable in polynomial time are also computable in polynomial time, we get that 
    $$\Pi_1 \SPR \Pi_2 ~\land~ \Pi_2 \SPR \Pi_3 \implies \Pi_1 \SPR \Pi_3$$
    holds and hence strongly Parsimonious reductions are transitive.
\end{proof}
 Finally, it is sufficient to show that there is a strongly Parsimonious reduction between two problems, as this property
directly implies the weakly Parsimonious property.
\begin{lemma}
    Every strongly Parsimonious reduction is also weakly Parsimonious, i.e. for computational problems $\Pi_1$ and $\Pi_2$ if 
    $\Pi_1 \SPR \Pi_2$ holds then $\Pi_1 \WPR \Pi_2$ also holds.
\end{lemma}
\begin{proof}
    Let $\Pi_1 := (\mc I_1, \mc S_{\Pi_1})$ and $\Pi_2 := (\mc I_2,  \mc S_{\Pi_2})$ where $\Pi_1 \SPR \Pi_2$ holds. It follows that there 
    exists a bijective function between the solution spaces $\mc S_{\Pi_1}$ and $\mc S_{\Pi_2}$. However, for this function to be bijective,
    both solution spaces have to be of equal size and therefore 
    $$\Pi_1 \SPR \Pi_2 \implies |\mc S_{\Pi_1}|=|\mc S_{\Pi_2}| \implies \Pi_1 \WPR \Pi_2.$$
\end{proof}
\noindent We denote a reduction having a certain property as
$\SSP$, $\WPR$, $\SPR$ or as $\SWPR$ and $\SSPR$ if it has multiple properties.

\subsubsection[Relation between SSP and SPR]{The relation between Subset Search and Parsimonious Reductions}
\label{sec:SSP and SPR}
\begin{figure}[h]
    \begin{center}
        \begin{tikzpicture}[vertex/.style = {draw, circle, fill, inner sep = 1.5}, node distance = 1.5cm]
            \node[](G){$G:$};
            \node[vertex, below right = 0 and .75 of G, label = $u$](u){};
            \node[vertex, below left = 1 and .66 of u, label = below:$v$](v){};
            \node[vertex, below right= 1 and .66 of u, label = below:$w$](w){};
            \node[below = .5 of v](k){$k=3$};
  
            \node[above right = .33 and 1.5cm of w](to){$\mapsto$};
  
            \node[right = 4.5 of G](G'){$\ne G:$};
            \node[vertex, below right = 0 and .75 of G', label = $u$](u'){};
            \node[vertex, below left = 1 and .66 of u', label = below:$v$](v'){};
            \node[vertex, below right= 1 and .66 of u', label = below:$w$](w'){};
            \node[below = .5 of v'](k){$k=0$};
  
            \draw (u)--(w)
                  (u)--(v)
                  (w)--(v)
                  
            ;
        \end{tikzpicture} 
    \caption{\footnotesize{Example \textsc{Clique} to \textsc{Vertex Cover} is not SSP}}
    \label{fig: CQ to VC}
    \end{center}
  \end{figure}
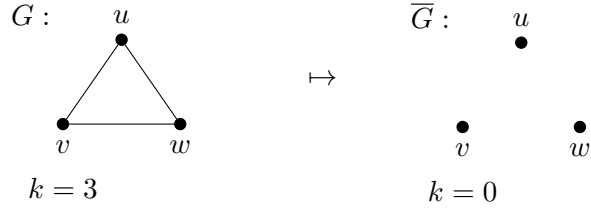

Intuitively, there should be a correlation between SSP and Parsimonious reductions. However they are not the same. There exist both SSP reductions that 
are not Parsimonious, as well as Parsimonious reductions which are not SSP. Take, for instance, the well known reduction from \textsc{Satisfiability}
to \textsc{3-Satisfiability} (\ref{prob:SAT}) as seen in \cite{ssp}, which is a modified version of Karp's original reduction, where every clause $\{\cee_1,\cee_2,\cee_3, ...,\cee_n\}$ 
with more than three literals gets recursively replaced by two new clauses $\{\cee_1,\cee_2,h_1\},\{\ne h_1,\cee_3, ...,\cee_n\}$ until no clause has more 
than three literals.\\
The problem with this reduction, is that, while it has the SSP properties, it is not Parsimonious in any way. \\
By defining the functions 
$f_I:L\rightarrow L', ~f(\cee)=\cee'$, where $\cee$ and $\cee'$ correspond to the same literal in the two instances respectively, we can see in the following
that the SSP properties hold. 
\begin{flalign*}
    \{f(S):S\in \mathcal{S}_{\text{\textsc{Sat}}}\} &= \big\{f\big(\{\cee \in S\}\big) : S\in\mc S_\SAT\big\}\\
    &= \big\{\{\cee'\in S'\}\backslash\{h \in S'\} : S'\in\mc S_\TSATu\}\\
    &= \{S'\cap f(L):S'\in\mathcal{S}_{\text{\textsc{3-Sat}}}\}
\end{flalign*}
However, as we see in the following example, this reduction is not Parsimonious. Let 
$$\SAT:=\big[L=\{\ell_1,...,\ell_4,\ne\ell_1,...,\ne\ell_4\},~\mc C=\big\{\{\ell_1,\ell_2,\ell_3,\ell_4\}\big\}\big],$$
then our reduction would produce
$$\TSAT:=\big[L'=\{\ell_1,...,\ell_4,\ne\ell_1,...,\ne\ell_4,h_1,\ne h_1\},~\mc C'=\{\{\ell_1,\ell_2,h_1\},\{\ne h_1,\ell_3,\ell_4\}\}\big]$$
as the corresponding \textsc{3-Satisfiability} instance. \\
Now, take the solution $S=\{\ell_1,\ne\ell_2,\ell_3,\ne\ell_4\}\in\mc S_\SAT$. As this solution already satisfies both new clauses in the 
\TSAT instance either $h_1$ or $\ne h_1$ can be included in the solution, giving $S$ two corresponding solutions in $\mc S_\TSATu$, 
those being $\{\ell_1,\ne\ell_2,\ne\ell_3,\ne\ell_4,h_1\}$ and $\{\ell_1,\ne\ell_2,\ne\ell_3,\ne\ell_4,\ne h_1\}$. It follows that 
$|\mathcal{S}_{\text{\textsc{Sat}}}|<|\mathcal{S}_{\text{\textsc{3-Sat}}}|$ and therefore that this reduction is not even weakly Parsimonious and 
definitely not strongly Parsimonious, as there cannot be a bijective function between two sets of different sizes. 
Thus, an SSP reduction is not necessarily also Parsimonious.\\~\\
The obvious next step is to see if Parsimonious reductions are always SSP. This is however also not the case as we can see in
the reduction \textsc{Clique} $\leq$ \textsc{Vertex Cover} (\autoref{prob:CQ}), which works by mapping the graph $G$ to its 
complement $\overline{G}$. The vertices not included in the clique found in $G$ then form a vertex cover in $\ne G$, making this 
a strongly Parsimonious reduction, as we can easily define a bijective function mapping every solution set to its complement.
Despite this we cannot define an injective function mapping vertices in a clique to vertices in a vertex cover for every instance. Take
the example shown in \autoref*{fig: CQ to VC}.
In $G$ there is exactly one clique of size $3$, that being $\{u,v,w\}$. This can be mapped to the vertex cover $\emptyset$ in 
$\ne G$, yet we cannot injectively map the elements in $\{u,v,w\}$ to the elements in $\emptyset$, as there are none. 
Thus, a strongly Parsimonious reduction also is not necessarily SSP.

Knowing that the reductions are not necessarily equivalent, i.e. having one characteristic does not directly imply the other, we can take a closer look
at the actual relation between these reduction constraints. In the following we introduce a set of properties that, if met by a reduction, are necessary
for this reduction to be both a Subset Search-, as well as a Parsimonious reduction.
\begin{theorem}
    \label{theorem 1}
    A reduction, defined by the function $g$, between two SSP Problems $\Pi:= (\mc I, \mc U, \mc S_{\Pi})$ and 
    $\Pi':= (\mc I', \mc U', \mc S_{\Pi'})$, is both SSP and strongly Parsimonious 
    $(\Pi \SSPR \Pi')$ if, and only if, there exists a partition of the universe $\mc U'(g(I))$ of $\Pi'$ into the three sets $\Snev\big(g(I)\big)$,
    $\Srep\big(g(I)\big)$ and $\Slink\big(g(I)\big)$, where, for ever instance $I\in\mc I$,
    \begin{itemize}
        \setlength\itemsep{0cm}
        \item $\Snev$ is a fixed set that is \textbf{never} a subset of any $\Sr{\Pi'}$, i.e. $\forall \Sr{\Pi'}:\Snev \cap S' =\emptyset$, 
        \item $\Srep$ is a set of \textbf{representatives}, where for every $\Sr{\Pi'}$, there exists a subset $\srep\subseteq \Srep$, 
            which uniquely defines $S'$ via the corresponding solution $\S\Pi$. This means that 
            there exist bijective functions $(r_I)_{I\in \mc I},~r_I:\Srep\rightarrow \mc U$ so that 
            \begin{flalign*}
                \bigl\{r(S'\cap\Srep):\Sr{\Pi'} \bigr\} = \bigl\{\S\Pi\bigr\}
            \end{flalign*}
        \item $\Slink$ is a set of \textbf{linked elements}, with $\Slink\subseteq 2^{\hspace*{1pt}\mc U'}$,
            where for every $\Sr{\Pi'}$, the subset $\srep\in\Srep$ is linked to an element $\slinkr\in\Slink$,
            so that $S'=\srep \cupdot \slinkr$ i.e. 
            for each instance $I$ there exists an injective function $\mathfrak L_I$ so that
            $$\mathfrak L_I: 2^{\hspace*{1pt}\Srep}\rightarrow 2^{\hspace*{1pt}\mc U'\backslash (\Srep\cupdot\Snev)},~
                \slinkr = S'\backslash\Srep$$
            and therefore $\Slink$ is defined as
            \begin{flalign*}
                \Slink :=& \bigcup_{\Sr{\Pi'}} \mf L\big(S'\cap \Srep\big).
            \end{flalign*}
    \end{itemize}
\end{theorem}
\begin{proof}
    Let $\Pi:= (\mc I, \mc U, \mc S_{\Pi})$ and $\Pi':= (\mc I', \mc U', \mc S_{\Pi'})$ be two SSP Problems and let 
    $g$ define the reduction $\Pi\leq\Pi'$.
    \begin{enumerate}
        \item \textit{Let the reduction $\Pi\leq\Pi'$ have the above partition of $\mc U'(g(I))$.}\\
            First, to see that the reduction has the SSP property, we look at the functions $(r_I)_{I\in \mc I}$. As these are bijective, there exist
            functions 
            $$r^{-1}_I:\mc S_\Pi(I)\rightarrow \Srep\big(g(I)\big),~ r^{-1}(S) = \srep$$
            Now we let the SSP functions $(f_I)_{I\in \mc I}$ be defined as 
            $$f_I:\mc U(I)\rightarrow \mc U'\big(g(I)\big),~f_I(x)=\begin{cases}
                r_I^{-1}(x),&x\in S,~S\in\mc S_\Pi \\
                x,&\text{otherwise}
            \end{cases}$$
            and we see that
            \begin{flalign*}
                \{f_I(S):\S\Pi(I)\} &= \{r^{-1}_I(S):\S\Pi(I)\} \\
                &= \big\{\srep:\Sr{\Pi'}\big(g(I)\big)\big\}\\
                &= \big\{S'\cap f_I\big(\mc U(I)\big):\Sr{\Pi'}\big(g(I)\big)\big\}
            \end{flalign*}
            The functions $(f_I)_{I\in \mc I}$ are injective as
            \begin{flalign*}
                \mc U'(g(I)) &=  \Srep\cupdot\Slink\cupdot\Snev \\
                &= f_I\big(\mc U(I)\big)\cupdot \Slink\cupdot\Snev
            \end{flalign*}
            Therefore is follows, that a reduction with the above properties is SSP.\\
            ~\\
            Now, to see that the reduction is also strongly Parsimonious, we again look at the functions $(r_I)_{I\in \mc I}$ and $(r^{-1}_I)_{I\in \mc I}$.
            We can define our Parsimonious functions $(p_I)_{I\in \mc I}$ as 
            \begin{flalign*}
                p_I&:\mc S_\Pi\rightarrow \mc S_{\Pi'}, p_I(S)=r_I^{-1}(S)\cupdot \mf L_I\big(r_I(S)\big)\\
                p_I^{-1}&:S_{\Pi'} \rightarrow S_{\Pi},~ p^{-1}_I(S')=r_I\big(S'\backslash \Slink\big)
            \end{flalign*} 
            We can 
            now use these functions to prove that $p_I\big(\mc S_\Pi(I)\big)=\mc S_{\Pi'}(g(I))$
                \begin{flalign}
                    \{p_I(S):\S\Pi(I)\} &= \big\{r_I^{-1}(S)\cupdot \mf L_I\big(r_I(S)^{-1}\big): S \in \mc S_{\Pi}(I)\big\}\\
                    &= \big\{\srep\cupdot \mf L_I\big(\srep\big) : S \in \mc S_{\Pi}(I)\big\}\\
                    &= \{S' : S' \in \mc S_{\Pi'}(g(I))\}\label{eq:1}\\
                    &= \mc S_{\Pi'}(g(I))
                \end{flalign} 
            \autoref*{eq:1} holds, as the representatives imply a unique solution $S'$ and that solution can then only contain representatives, 
            and elements linked to a certain solution. Elements that are never in any solution cannot affect the ammount of possible solutions.
            The other direction also holds, showing that both directions of this function are surjective and therefore the function itself is bijective.
                \begin{flalign}
                    \{p^{-1}_I(S'):\Sr{\Pi'}(g(I))\}&= \big\{r_I\big(S'\backslash (\Sall\cup\Slink)\big) : \Sr{\Pi'}(g(I))\big\}\\
                    &= \big\{r_I(\srep) : \Sr{\Pi'}(g(I))\big\}\label{eq:2}\\
                    &= \{S:\S\Pi(I)\}\label{eq:3}\\
                    &= \mc S_{\Pi}(I)
                \end{flalign}
            Here \autoref*{eq:2} holds, as $S'\backslash \Slink$ has to be the set of representatives $\srep$, as otherwise the solution would not be
            clearly defined by the set of representatives and \autoref*{eq:3} holds as per our definition. 
            Therefore $(p_I)_{I\in \mc I}$ are bijective functions between the solution spaces. $\Pi\SPR\Pi'$ holds. \\
            As the reduction is both SSP and SPR, we get that $\Pi\SSPR\Pi'$ holds for a reduction $\Pi\leq\Pi'$ with the above properties.
        
        \item \textit{Let the reduction $\Pi\leq\Pi'$ not have the above partition of $\mc U'(g(I))$.}\\
            Now, if no set of representatives $\Srep$ exists, then there cannot be an injective function mapping the elements in 
            $\mc U(I)$ to elements in $\mc U'(g(I))$. Therefore the reduction cannot have the SSP property. \\
            However, if a set of representatives $\Srep\subseteq \mc U'(g(I))$ exists, but the elements in $\mc U'(g(I))\backslash \Srep$ cannot be
            linked to specific solutions $\S\Pi$, then the number of solutions $\Sr{\Pi'}$ differs from $|\mc S_\Pi|$. It follows, that there cannot
            be a bijective function between the two solution spaces as they are of different sizes.\\
            Therefore a reduction $\Pi\leq\Pi'$, which does not have the above partition of $\mc U'(g(I))$, cannot have both the Subset Search and the Parsimonious
            properties.
    \end{enumerate}
    It follows, that a reduction $\Pi\leq\Pi'$ is both a Subset Search- and a Parsimonious reduction if, and only if, the above partition of $\mc U'(g(I))$ exists.
\end{proof}
As we now know exactly when both properties apply to a reduction, we can define a new complexity class, \SSPP, to combine both properties.
\begin{definition}[\SSPP]
    The complexity class \SSPP~\textit{includes all counting problems, where the solution is a subset of the instance.}
\end{definition}
Therefore a subset search problem $\Pi$ is \SSPP-hard, iff for all problems $\Pi'$ in \SSPP{}
    there exists a reduction, where \autoref{theorem 1} applies and therefore $\Pi'\SSPR\Pi$ holds, and it is
    \SSPP-complete, iff it is both in \SSPP{} and \SSPP-hard.\\

Many of the problems we observe in our compendium stem from either \cite{ssp} or \cite{Femke}, 
both of which already prove the SSP property of a number of reductions.
Therefore, for these cases, we use a modified version of the above theorem, where proof of the SSP property already is provided. 

\begin{theorem}
    \label{lem:SSP and SPR}
    An SSP reduction, defined by the functions $g$ and $(f_I)_{I\in\mc I}$, between two SSP Problems $\Pi:= (\mc I, \mc U, \mc S_{\Pi})$ and 
    $\Pi':= (\mc I', \mc U', \mc S_{\Pi'})$, is also strongly Parsimonious 
    $(\Pi \SSPR \Pi')$ if there exists a partition of the universe $\mc U'(g(I))$ of $\Pi'$ into the four sets 
    $\Srep\big(g(I)\big)$, $\Sall\big(g(I)\big)$, $\Snev\big(g(I)\big)$ and $\Slink\big(g(I)\big)$, where, for ever instance $I\in\mc I$,
    \begin{itemize}
        \setlength\itemsep{0cm}
        \item $\Srep = f_I\big(\mc U(I)\big)$ is the set that the universe of the given instance is mapped to,
        \item $\Sall$ is a fixed subset of \textbf{every} $\Sr{\Pi'}$, i.e. $\forall \Sr{\Pi'}:\Sall\subseteq S'$,
        \item $\Snev$ is a fixed set that is \textbf{never} a subset of any $\Sr{\Pi'}$, i.e. $\forall \Sr{\Pi'}:\Snev \cap S' =\emptyset$ and 
        \item $\Slink$ is a set of \textbf{linked elements}, with $\Slink\subseteq 2^{\hspace*{1pt}\mc U'}$,
        where for every $\Sr{\Pi'}$  each element $ \mf L(x)\in \Slink$ is linked to an element $x\in \srep$ i.e. 
        for each instance $I$ there exists an injective function $\mathfrak L_I$ so that
        $$\mathfrak L_I: \Srep\rightarrow 2^{\hspace*{1pt}\mc U'\backslash (\Srep\cup\hspace*{1pt}\Sall)},~
        x\in\srep\Leftrightarrow \mathfrak L(x)\subseteq \big(S'\backslash\Srep\big)$$
        and therefore $\Slink$ is defined as
        \begin{flalign*}
            \Slink :=& \bigcup_{\Sr{\Pi'}} \mf L\big(S'\cap \Srep\big).
        \end{flalign*}
    \end{itemize}
\end{theorem}
    \begin{proof}
        Let $\Pi:= (\mc I, \mc U, \mc S_{\Pi})$ and $\Pi':= (\mc I', \mc U', \mc S_{\Pi'})$ be two SSP Problems and let 
        $g$ and $(f_I)_{I\in\mc I}$ define the reduction $\Pi\SSP\Pi'$.\\
        The functions $(f_I)_{I\in\mc I}$ are injective functions mapping from $\mc U(I)$ to $\mc U'(g(I))$. However, if the above
        partition exists, then the functions are bijective when confined to 
        $$f_I:\mc U(I)\rightarrow \mc U'(g(I))\backslash(\Sall\cupdot\Snev\cupdot\Slink).$$
        As the reduction is SSP, we also know that
        $\{f(S):\S\Pi\}=\{S'\cap f(\mc U):\Sr{\Pi'}\}$
        and, with the above, it follows that 
        $$\{f(S):\S\Pi\}=\{S'\cap f(\mc U):\Sr{\Pi'}\} = \{\srep :\Sr{\Pi'}\}$$
        Now, $\mc U'(g(I))\backslash(\Sall\cupdot\Snev\cupdot\Slink) = \Srep$ and therefore we can define the functions $(r_I)_{I\in\mc I}$ as 
        $$r_I:\Srep \rightarrow \mc U(I), r_I(\srep) = f_I^{-1}(\srep)$$
        Finally, with 
        $$\mc S_{\mf L}^{''}:=\bigcup_{\Sr{\Pi'}} \mf L\big(S'\cap \Srep\big)\cup \Sall$$
        then we have a partition of $\mc U'(g(I))$ into the three sets $\Snev$, $\Srep$ and $\mc S_{\mf L}^{''}$, where the representatives 
        $\srep\subseteq \Srep$ correspond to solutions $\S\Pi$ via the bijective funtions $(r_I)_{I\in\mc I}$. Therefore this reduction is both 
        SSP and SPR, according to \autoref{theorem 1}.
    \end{proof}

Finally, we also introduce the following more restricted, less applicable, version of the theorem, for cases in which the SSP property is already
proven, and the functions $(f_I)_{I\in \mc I}$ are already bijective.
    \begin{theorem}
        \label{lem:SSP and SPR 2}
        An SSP reduction $(g, (f_I)_{I\in \mc I})$ is also strongly Parsimonious, if the functions $(f_I)_{I\in \mc I}$ are bijective, i.e a reduction between two SSP Problems 
        $\Pi:= (\mc I, \mc U, \mc S_{\Pi})$ and $\Pi':= (\mc I', \mc U', \mc S_{\Pi'})$ is both SSP and strongly Parsimonious 
        $(\Pi \SSPR \Pi')$ if there exist functions $(f_I)_{I\in \mc I}$, computable in polynomial time in $|I|$, such that for all
                instances $I \in \mc I$, we have that $f_I : \mc U(I) \rightarrow \mc U'(g(I))$ is a \textbf{bijective} function mapping
                from the universe of the instance $I$ to the universe of the instance $g(I)$ such that
                $$\{f_I(S): S \in \mc S_\Pi(I)\} = \{S' \cap f_I(\mc U(I)): S' \in \mc S_{\Pi'}(g(I))\}= \{S' \in \mc S_{\Pi'}(g(I))\}.$$
    \end{theorem}
    \begin{proof}
        Let $\Pi:= (\mc I, \mc U, \mc S_{\Pi})$ and $\Pi':= (\mc I', \mc U', \mc S_{\Pi'})$ be two SSP Problems. For this proof we use Lemma 5, as 
        the functions $(f_I)_{I\in \mc I}$ are bijective and therefore also bijective when restricted as follows
        $$f_I' : \mc U(I) \rightarrow \big(\mc U'(g(I))\backslash (\Sall\cup \Snev\cup \Slink)\big),$$
        with $\Sall = \Snev = \Slink =\emptyset$. Thus the functions $(p_I)_{I\in\mc I}$ can be defined as follows
        \begin{flalign*}
            p_I&: S_{\Pi} \rightarrow S_{\Pi'},~ p_I(S)=f_I(S)\\
            p_I^{-1}&:S_{\Pi'} \rightarrow S_{\Pi},~ p^{-1}_I(S')=f_I^{-1}(S')
        \end{flalign*} 
        and we get 
        \begin{flalign*}
            \{p_I(S):\S\Pi\} &=\{f_I(S): S \in \mc S_\Pi(I)\}\\
            & =\{S' \cap f_I(\mc U(I)): S' \in \mc S_{\Pi'}(g(I))\} \\
            &= \{S' \in \mc S_{\Pi'}(g(I))\}\\
            &= \mc S_{\Pi'}(g(I))\\
            \text{and }\{p^{-1}_I(S'):\Sr{\Pi'}(g(I))\}&=\{f_I^{-1}(S'): S' \in \mc S_{\Pi'}(g(I))\}\\
            &=\big\{f_I^{-1}\big(S' \cap f_I(\mc U(I))\big): S' \in \mc S_{\Pi'}(g(I))\big\}\\
            &= \{S:\S\Pi(I)\}\\
            &= \mc S_{\Pi}(I)
        \end{flalign*}
        Therefore $(p_I)_{I\in \mc I}$ are bijective functions between the solution spaces. $\Pi\SSPR\Pi'$ holds.
\end{proof} 

\subsubsection{Cook's Theorem}
Cook's theorem, or the Cook-Levin theorem states, that the \textsc{Satisfiability} problem is NP-complete, as any problem in this
complexity class can be reduced to it. Therefore it is pertinent to our thesis, to examine whether or not there exists a similar
reduction for problems in \SSPP, which is both SSP and Parsimonious. Thankfully, Grüne and Wulf \cite[pp. 13-14]{ssp} already outline a proof that
$\Pi \SSP \SAT$ holds, for any SSP Problem $\Pi =(\mc I, \mc U, \mc S_{\Pi})$. We use their proof to show that the reduction is also
Parsimonious. 
\begin{theorem}
    \label{sec:Cook}
    Every SSP Problem $\Pi =(\mc I, \mc U, \mc S_{\Pi})$ can be reduced to \textsc{Satisfiability} with a 
    reduction that is both SSP and strongly Parsimonious.
\end{theorem}
\begin{proof}
    Let $\Pi =(\mc I, \mc U, \mc S_{\Pi})$ be an arbitrary SSP problem in NP and let \textsc{Satisfiability} be defined as in
    \autoref{sec:Sat Problems}. As $\Pi$ is in NP, there exists a verifying deterministic Turing machine $M^D_\Pi$, which is given a 
    binary encoded tuple $(I,S)$, consisting of an instance $I\in \mc I$ and a possible solution $S\subseteq \mc U(I)$, that can  
    now verify a.k.a. decide, whether $\S\Pi$ or not. This is done in polynomially many steps, depending on the input length. \\
    Now, let $M_\Pi$ be a nondeterministic Turing machine solving $\Pi$. Then this can be split into two parts. The first part is the 
    nondeterministic part, which constructs solutions for $\Pi$. The second part then verifies these solutions. For our purposes we split 
    $M_\Pi$ into these two part, one being the nondeterministic machine $M_\Pi^N$ and one the deterministic one $M_\Pi^D$. We now modify 
    $M_\Pi^N$, so that its output is the binary encoding of $(I,S)$, with the solution $S\in \{0,1\}^{\mc U(I)}$ starting at the 
    $n^{th}$ bit of the output. The $\mc U(I)$ can then be mapped to this solution, more explicitly to the bits in 
    $\big[n;\big(|S|+n-1\big)\big]$. \\
    $M_\Pi^D$ then gets $bin\big((I,S)\big)$ as input and verifies this solution. As $M_\Pi^D$ is deterministic, there is a 
    unique sequence of states in the Turing machine for every possible input. And since Cook's reduction maps each configuration for every step of the 
    machine to variables in the \SAT formula, the sequence of configurations visited for any accepted input (where $M_\Pi^D$ accepts) is mapped 
    to a unique solution in the \SAT instance. \\
    This means every input $bin\big((I,S)\big)$  for $M_\Pi^D$, with $\S\Pi$, can be mapped to exactly one solution $\Sr\SAT$. Therefore there exists a 
    bijective function between $\mc S_\Pi$ and $\mc S_\SAT$ and thus \textsc{Satisfiability} is \SSPP-complete. 
    $\Pi \SSPR \SAT$ holds for every arbitrary computational problem 
    $\Pi =(\mc I, \mc U, \mc S_{\Pi})$ in NP.
\end{proof}

\subsubsection{Minimum and Maximum Cardinality Problems}
In this paper we use the minimum or maximum cardinality version of problems that allow for finding a solution of size \textit{at most}
$k$. As shown below, some reductions between the general versions of these problems are only Parsimonious if an optimal
$k$ is chosen. Take for instance the two problems \textsc{Vertex Cover} and \textsc{Dominating Set}, as defined in \autoref{prob:VC}.\\
The reduction \textsc{Vertex Cover} $\leq$ \textsc{Dominating Set} works similarly to the reduction presented in \autoref{sec:MVC to MDS},
where we add extra vertices into the \textsc{Dominating Set} instance for every edge in the \textsc{Vertex Cover} instance, to ensure that one or both
of the two endpoints of the edge is in the resulting dominating set. \\
For this example it suffices to consider a graph with no isolated nodes. 
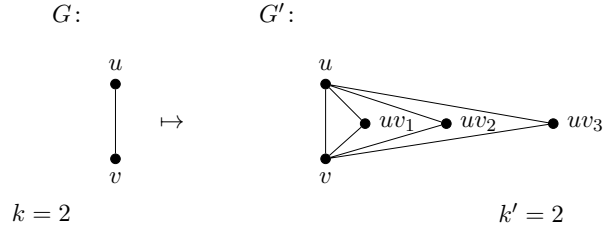
\begin{figure}[h]
    \begin{center}
    \scalebox{.85}{
        \begin{tikzpicture}[vertex/.style = {draw, circle, fill, inner sep = 1.5}, node distance = 1.5cm]
            \node[label = above:$G\colon$](G){};
            \node[vertex, below right = .5 and .5 of G, label = $u$](u){};
            \node[vertex, below = 1 of u, label = below:$v$](v){};
            \node[below left = .5 and .5 of v](){$k\equals 2$};
            
            \node[below right = .35 and .5cm of u](to){$\mapsto$};

            \node[right = 3 of G, label = above:$G'\colon$](G'){};
            \node[vertex, below right = .5 and .5 of G', label = $u$](u'){};
            \node[vertex, below = 1 of u', label = below:$v$](v'){};
            \node[vertex, below right = .5 and .5 of u', label = right:$uv_1$](1){};
            \node[vertex, right = 1.1 of 1, label = right:$uv_2$](2){};
            \node[vertex, right = 1.5 of 2, label = right:$uv_3$](3){};
            \node[below right = .5 and 2.5 of v'](){$k'\equals 2$};

            \draw   (u) -- (v)
                    (u') -- (v')
                    (u') -- (1)
                    (u') -- (2)
                    (u') -- (3)
                    (v') -- (1)
                    (v') -- (2)
                    (v') -- (3)
            ;
        \end{tikzpicture} 
    }
    \caption{\footnotesize{ Example \textsc{Vertex Cover} to \textsc{Dominating Set}} }
    \end{center}
\end{figure}\\
Here the solution set for the \textsc{Vertex Cover} instance is $\mc S_\VC = \big\{\{u\}, \{v\}, \{u,v\}\big\}$, but the  
solution set for the \textsc{Dominating set} instance can be defined as 
$$\mc S_\DS =  \mc S_\VC \cup \big\{\{u, uv_i\}, \{v, uv_i\}\mid i\in [3]\big\}$$
and therefore the solution sets have different sizes as $|\mc S_\VC| = 3 < 9 = |\mc S_\DS| $. It follows, that this reduction not Parsimonious 
for any non-minimal $k$, as there cannot be
a bijective function between sets of different sizes.\\
However, for minimal $k$ it is easy to see that the reduction is indeed Parsimonious as the solution set for \textsc{Dominating Set} is equivalent
to the \textsc{Vertex Cover} solution set.\\
Note that for all the reductions presented in this paper using the minimum or maximum version of a problem, the SSP reduction properties are upheld
even in the normal versions. It is only the Parsimonious properties which then do not hold anymore.

\ifSubfilesClassLoaded{
  \bibliography{../../ref}
}{}
\end{document}

%% file: Sections/Overview/Overview.tex
\section{Overview}
\label{chap:3}
In this paper we examine a total of \textit{46} reductions between \textit{30} SSP problems, not counting Cook's reduction we use in
in \autoref{sec:Cook}. This compendium is just a fragment of the vast quantity of possible reductions, but it suffices to examine certain 
similarities between the SSP and Parsimonious properties. \autoref*{Reduction Map} depicts the problems and reductions regarded in this paper, with 
the arrows depicting the properties existing in the reductions between two problems. A black arrow depicts a reduction that has both the Subset Search and the
strongly Parsimonious property, whereas a dotted or dashed arrow shows a reduction with only the SSP property 
or the strongly Parsimonious property
respectively.

\subsection{Reduction Map}
\subfile{Reduction_Map/Reduction_Map.tex}

\subsection{Observations}
As we can see above in \autoref{Reduction Map}, not every reduction we examine
is both SSP and SPR. This is especially prevalent with reductions concerning the \textsc{Exact 3-Satisfiability} and 
\textsc{Maximum Cardinality Vertex Cover} problems. It appears that the nature of these properties is not well suited for these two problems
in particular. However, in regards the the \textsc{Exact 3-Satisfiability} problem, we can see that the \textsc{1-in-3-Satisfiability} variant seems 
better suited for these reduction types.\\
That being said, we also see that with the help of \autoref*{sec:Cook} we can find an alternative, transitive reduction,
upholding both properties, for every single one of the reductions studied here.\\
Another interesting observation, is that none of the reductions we examine have only the weakly Parsimonious property. It seems that, 
if the two solution sets of the problems in a given reduction are of equal size, we can always find a bijective function mapping from
one set to the other.

\ifSubfilesClassLoaded{
  \bibliography{../../ref}
}{}
\end{document}

%% file: Sections/Overview/Reduction_Map/Reduction_Map.tex
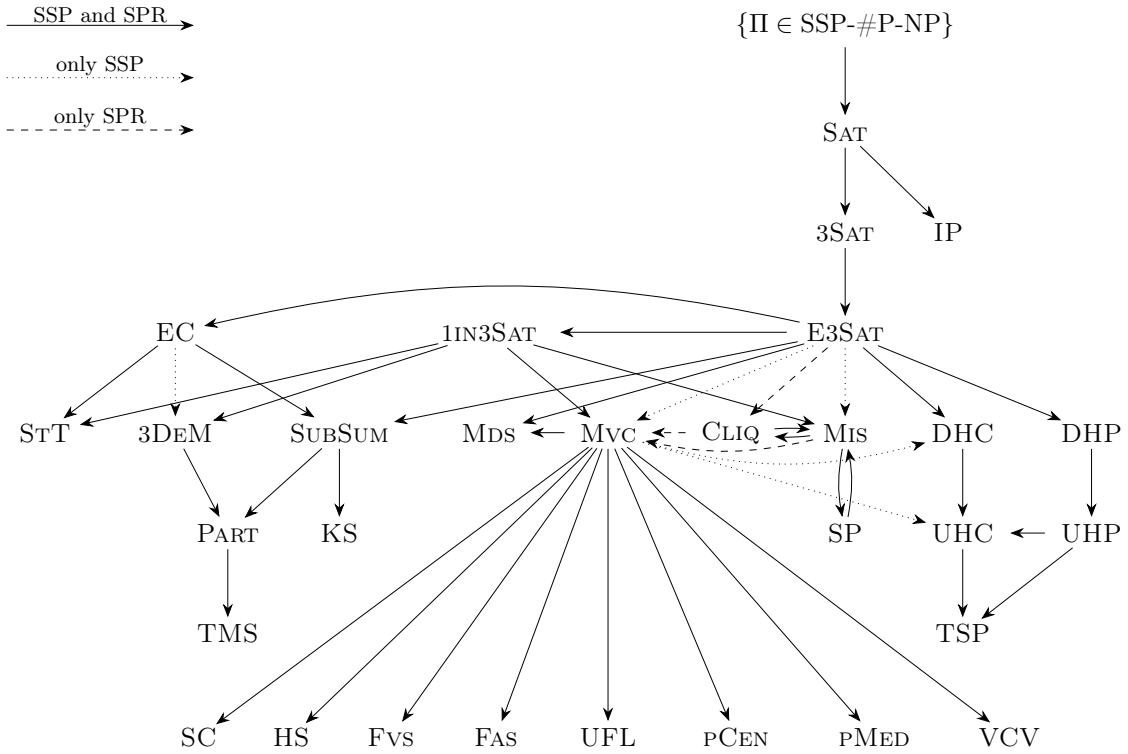
\begin{figure}[h]
    \begin{center}
    \scalebox{.9}{
        \begin{tikzpicture}[vertex/.style = {ellipse, inner sep = 0, outer sep = 0.05cm}, node distance = 1.5cm, 
                            every edge quotes/.style = {font=\footnotesize}, tight/.style={inner sep=1pt}]
            \node[vertex](SAT){\SAT};
            \node[vertex, above = 1cm of SAT](Pi){$\{\Pi\in \text{\SSPP-NP}\}$};
            \node[vertex, below = 1cm of SAT](3SAT){\TSAT};
            \node[vertex, right = .5 of 3SAT](IP){\IP};
            \node[vertex, below = 1cm of 3SAT](E3SAT){\ESAT};

            \node[vertex, below = 1cm of E3SAT](MIS){\MIS};
            \node[vertex, left = .5cm of MIS](CLIQ){\CQ};
            \node[vertex, left = .5cm of CLIQ](MVC){\MVC};
            \node[vertex, left = .5cm of MVC](MDS){\MDS};
            \node[vertex, left = .5cm of MDS](SS){\SS};
            \node[vertex, right = .5cm of MIS](DHC){\DHC};
            \node[vertex, right = .5cm of DHC](DHP){\DHP};
            \node[vertex, below = 1cm of DHC](UHC){\UHC};
            \node[vertex, below = 1cm of UHC](TSP){\TSP};
            
            \node[vertex, below = 1cm of MIS](SP){\SP};

            \node[vertex, below = 4 of MVC](UFL){\UFL};
            \node[vertex, right = .5cm of UFL](PCEN){\PCen};
            \node[vertex, right = .5cm of PCEN](PMED){\PMed};
            \node[vertex, right = .5cm of PMED](VCV){\VCV};
            \node[vertex, left = .5 of UFL](FAS){\FAS};
            \node[vertex, left = .5cm of FAS](FVS){\FVS};
            \node[vertex, left = .5cm of FVS](HS){\HS};
            \node[vertex, left = .5cm of HS](SC){\SC};

            \node[vertex, below = 1cm of SS](KS){\KS};
            \node[vertex, left = .5cm of KS](PART){\P};
            \node[vertex, below = 1cm of PART](TMS){\TMS};

            \node[vertex, left = .5cm of SS](DM){\DM};
            \node[vertex, left = .5cm of DM](STT){\STT};
            \node[vertex, above = 1cm of MDS](OSAT){\OSAT};
            \node[vertex, above = 1 of DM](EC){\EC};

            \node[vertex, below = 1cm of DHP](UHP){\UHP}; 

            \node[left = 10 of Pi](1){};
            \node[below = .5cm of 1](2){};
            \node[right = 2.75cm of 1](3){};
            \node[below = .5cm  of 3](4){};
            \node[below = .5cm of 2](5){};
            \node[below = .5cm of 4](6){};

            \draw   (1) edge[\xarrow, "SSP and SPR" tight, above] (3)
            (2) edge[\xarrow, "only SSP" tight, above, dotted] (4)
            (5) edge[\xarrow, "only SPR" tight, above, dashed] (6)
            ; 

            \draw   (Pi) edge[\xarrow] (SAT)
                    (SAT) edge[\xarrow] (3SAT)
                    (SAT) edge[\xarrow] (IP)
                    (3SAT) edge[\xarrow] (E3SAT)
                    
                    (E3SAT) edge[\xarrow] (OSAT)
                    (E3SAT) edge[\xarrow, dotted] (MIS)
                    (E3SAT) edge[\xarrow, dashed] (CLIQ)
                    (E3SAT) edge[\xarrow, dotted] (MVC)
                    (E3SAT) edge[\xarrow] (MDS)
                    (E3SAT) edge[\xarrow] (SS)
                    (E3SAT) edge[\xarrow] (DHP)
                    (E3SAT) edge[\xarrow] (DHC)
                    (E3SAT) edge[\xarrow, bend right = 12.5] (EC)
                    
                    (DHC) edge[\xarrow] (UHC)
                    (UHC) edge[\xarrow] (TSP)

                    (MVC) edge[\xarrow] (SC)
                    (MVC) edge[\xarrow] (HS)
                    (MVC) edge[\xarrow] (FVS)
                    (MVC) edge[\xarrow] (FAS)
                    (MVC) edge[\xarrow] (UFL)
                    (MVC) edge[\xarrow] (PCEN)
                    (MVC) edge[\xarrow] (PMED)
                    (MVC) edge[\xarrow] (MDS)
                    (MVC) edge[\xarrow, bend right = 15, dotted] (DHC)
                    (MVC) edge[\xarrow, dotted] (UHC)
                    (MVC) edge[\xarrow] (VCV)

                    (MIS) edge[\xarrow, bend left = 5] (CLIQ)
                    (MIS) edge[\xarrow, bend left = 12.5, dashed] (MVC)

                    (CLIQ) edge[bend left = 5, \xarrow] (MIS)
                    (CLIQ) edge[\xarrow, dashed] (MVC)
                    
                    (SS) edge[\xarrow] (KS)
                    (SS) edge[\xarrow] (PART)
                    (PART) edge[\xarrow] (TMS)

                    (OSAT) edge[\xarrow] (STT)
                    (OSAT) edge[\xarrow] (MVC)
                    (OSAT) edge[\xarrow] (MIS)
                    (OSAT) edge[\xarrow] (DM)
                    
                    (MIS) edge[\xarrow, bend right = 10] (SP)
                    (SP) edge[\xarrow, bend right = 10] (MIS)

                    (EC) edge[\xarrow] (STT)
                    (EC) edge[\xarrow, dotted] (DM)
                    (EC) edge[\xarrow] (SS)
                    
                    (DHP) edge[\xarrow] (UHP)
                    (UHP) edge[\xarrow] (TSP)
                    (UHP) edge[\xarrow] (UHC)

                    (DM) edge[\xarrow] (PART)
            ;
        \end{tikzpicture} 
    }
    \caption{Reduction Map}
    \label{Reduction Map}
    \end{center}
\end{figure}

\ifSubfilesClassLoaded{
  \bibliography{../../ref}
}{}
\end{document}

%% file: Reductions/Sat/Sat.tex
\subsection{Reductions from Satisfiability}
\subfile{../../Problems/SAT.tex}
\subsubsection[... to 3-Satisfiability]{Satisfiability to 3-Satisfiability}
\label{sec:SAT to 3SAT}
\paragraph{Problem Definition}
\subfile{../../Problems/TSAT}

\paragraph{Reduction}
As we show in \autoref{sec:SSP and SPR} the simplest reduction $\SAT\leq\TSAT$ is not Parsimonious. Because of this, we look at a modification presented by Karp 
\cite{karp}. The reduction now looks as follows. Each clause $C_j\in \mc C$ in the \SAT instance with $|C_j|>3$ 
gets replaced by two new clauses $C_{j}^i$ and $C_{j}^{i'}$, the first one containing the first two literals of the split clause and a new literal 
$h_j^i\notin L$ and the second clause containing $\ne h_{j}^{i} \notin L$ and the remaining literals. In addition to this we also create clauses
$\{\ne\cee_i, h_j^i\}$ for every literal $\cee_i$ in the new clause $C_{j}^{i'}$. We do this recursively, until no clause has more than three literals. 
The literal $h_{j}^{i}$ denotes the number of times $i$ the clause $C_j$ has 
been split.\\
More formally each clause $C_j\in \mc C$ gets split as follows.
\begin{flalign*}
    C_j = \{\cee_1,\cee_2,\cee_3,\cee_4,...,\cee_n\} \mapsto ~& \{\cee_1,\cee_2,h_{j}^{1}\},\{\ne h_{j}^{1},\cee_3,\cee_4,...,\cee_n\},\{h_j^1,\cee_3\},...,\{h_j^1,\cee_n\}\\
    C_{j}^{1'} =\{\ne h_{j}^{1},\cee_3,\cee_4,...,\cee_n\} \mapsto ~& \{\ne h_{j}^{1},\cee_3,h_{j}^{2}\},\{\ne h_{j}^{2},\cee_4, ...,\cee_n\},\{h_j^2,\cee_4\},...,\{h_j^2,\cee_n\}\\
    ... & \hspace*{7cm} (\text{if } n > 4)\\
    C_{j}^{(n-4)'} =\{\ne h_{j}^{(n-4)},\cee_{n-2},\cee_{n-1},\cee_n\} \mapsto ~&\{\ne h_{j}^{(n-4)},\cee_{n-2},h_{j}^{(n-3)}\},\{\ne h_{j}^{(n-3)}\cee_{n-1},\cee_n\},\\ 
    &\{h_j^{(n-3)},\cee_{n-1}\},\{h_j^{(n-3)},\cee_n\}
\end{flalign*}
Also $L':=L\cup \big\{h_j^i,\ne h_j^i \mid j\in\big[|\mc C|\big], i\in\big[|L^2|-3\big]\big\}$. \autoref*{fig:Sat to 3Sat} shows an example reduction.
\begin{figure}[h]
    \begin{center}
        \begin{flalign*}
            \SAT:=&[L,\mc C],\text{ with}\\
            L&:=\{\ell_1,...,\ell_5,\ne\ell_1,...,\ne\ell_5\},\\
            \mc C&:=\big\{\{\ell_1,\ne\ell_2,\ne\ell_3, \ell_4, \ne \ell_5\}\big\}\\
            \mapsto\TSAT:=&[L',\mc C'],\text{ with}\\
            L'&:=\{\ell_1,...,\ell_5,\ne\ell_1,...,\ne\ell_5, h_1^1,h_1^2,\ne h_1^1,\ne h_1^2\},\\
            \mc C'&:=\big\{\{\ell_1,\ne\ell_2,h_1^1\},\{\ne h_1^1,\ne\ell_3,h_1^2\},\{\ne h_1^2,\ell_4,\ne\ell_5\},\\
            &\hspace*{.85cm}\{\ell_3, h_1^1\},\{\ne\ell_4, h_1^1\},\{\ne\ell_5, h_1^1\},\{\ne\ell_4, h_1^2\},\{\ne\ell_5, h_1^2\}
                \big\}
        \end{flalign*}
        \caption{\footnotesize{Example \textsc{Satisfiability} to \textsc{3-Satisfiability}}}
        \label{fig:Sat to 3Sat}
    \end{center}
\end{figure}

\begin{proof} Let $S$ be a satifying assignment for \SAT. It suffices to look at a single clause, as every clause needs to be individually satisfied. Therefore we leave out the indices $j$
    denoting which clause was split in both $C_j$ and $h_j^i$, which will now be denoted as $C$ and $h_i$.\\
    Let $\mathcal C_k:=C$, be the set of clauses with $k:=|C|$ and the solution set $\mc S_\SAT^k$. 
    We state without loss of generality that all literals $\ell_i\in C$ differ from each other and only one literal per pair $(\ell,\ne \ell)\in L^2$ is in the clause.\\
    We now use the prinicple of complete induction over the clause length to prove that this reduction holds. Let $\SAT:=[L,\mc C_k]$ and $\TSAT:=[L',\mc C'_k]$.
    \begin{itemize}
        \item[\textbf{IB}]
        For the base lengths $k\leq 4$ we get that $\mc C_k$ = $\mc C'_k$, as clauses of at most three literals do not get split and thus that the solutions are also
        the same. Now let $k=4$, then the clause $C=\{\cee_1,\cee_2,\cee_3,\cee_4\}$ gets mapped to $\{\cee_1,\cee_2,h_1\},\{\ne h_1,\cee_3,\cee_4\},\{\ne \cee_3,h_1\}$
        and $\{\ne \cee_4,h_1\}$. \\
        C has $2^4$ possible solutions. In the following we show how each solution 
            $S\in \mc S_\SAT^4$ is equivalent to exactly one solution in $S'\in \mc S_\TSATu^4$. Let $C_i'$ denote the clauses in $\mc C'_4$. If
            \begin{itemize}
                \item[...] $S = \emptyset $ then $ S' = \emptyset$.
                \item[...] $\{\ne \cee_1, \ne \cee_2\}\subseteq S $ then $ S' = S\cup \{h_1\}$ as the clause $C_1'$ needs to be satisfied.
                \item[...] $\{\ne \cee_3, \ne \cee_4\}\subseteq S $ then $ S' = S\cup \{\ne h_1\}$ as the clause $C_2'$ needs to be satisfied.
                \item[...] $\{\cee_3\} \subseteq S $ then $ S'=S\cup\{h_1\}$ as the clause $C_3'$ needs to be satisfied.
                \item[...] $\{\cee_4\} \subseteq S $ then $ S'=S\cup\{h_1\}$ as the clause $C_4'$ needs to be satisfied.
            \end{itemize}
        As every possible solution of $\mc S_\SAT^4$ falls into (at least) one of the five cases stated above, each $S\in \mc S_\SAT^4$ is equivalent to exactly one 
        $S'\in \mc S_\TSATu^4$.
        \item[\textbf{IS}]
            We denote the $i^{th}$ split by the new variable $h_i$ we create for it. Now let $k\mapsto k+1$. We get
            \begin{flalign*}
                \mathcal{C}_{k+1}=&\big\{\{\cee_1,\cee_2,...,\cee_{k+1}\}\big\}\\
                \mapsto \mathcal{C}_{k+1}'= &\big\{\{\cee_1,\cee_2,h_1\}\big\}\cup \bigcup_{i=3}^{(k+1)-2}\big\{\{\ne h_{i-2},\cee_i,h_{i-1}\}\big\}\cup
                \big\{\{\ne h_{(k+1)-3},\cee_{(k+1)-1},\cee_{k+1}\}\big\}\\
                &\cup \bigcup_{j=1}^{(k+1)-3} \Big\{\bigcup_{i=j+2}^{k+1}\big\{\{\ne \cee_i, h_j\}\big\}\Big\}
                \\
                =& \big\{\{\cee_1,\cee_2,h_1\}\big\} \cup \bigcup_{i=3}^{k-2}\big\{\{\ne h_{i-2},\cee_i,h_{i-1}\}\big\}\cup\big\{\{\ne h_{k-3}, \cee_{k-1}, h_{k-2}\}
                ,\{\ne h_{k-2},\cee_{k},\cee_{k+1}\}\big\}\\
                &\cup \bigcup_{j=1}^{k-3} \Big\{\bigcup_{i=j+2}^{k}\big\{\{\ne \cee_i, h_j\}\big\}\Big\} 
                \cup\bigcup_{i=1}^{k-2}\big\{\{\ne \cee_{k+1}, h_i\}\big\}\cup \big\{\{\ne \cee_{k}, h_{k-2}\}\big\}
                \\
                =& \big\{\{\cee_1,\cee_2,h_1\}\big\}\cup \bigcup_{i=3}^{k-2}\big\{\{\ne h_{i-2},\cee_i,h_{i-1}\}\big\} 
                \cup \bigcup_{j=1}^{k-3} \Big\{\bigcup_{i=j+2}^k\big\{\{\ne \cee_i, h_j\}\big\}\Big\} \\
                &\cup \big\{\{\ne h_{k-3}, \cee_{k-1}, h_{k-2}\},\{\ne h_{k-2},\cee_{k},\cee_{k+1}\}\big\} \cup\bigcup_{i=1}^{k-2}\big\{\{\ne \cee_{k+1}, h_i\}\big\}
                \cup \big\{\{\ne \cee_{k}, h_{k-2}\}\big\}
                \\
                \stackrel{\text{IH}}{=}&~ \mc C_k \backslash \{\ne h_{k-3},\cee_{k-1},\cee_k\} \\
                &\cup\big\{\{\ne h_{k-3}, \cee_{k-1}, h_{k-2}\},\{\ne h_{k-2},\cee_{k},\cee_{k+1}\},\{\ne \cee_{k}, h_{k-2}\}\big\}
                \cup\bigcup_{i=1}^{k-2}\big\{\{\ne \cee_{k+1}, h_i\}\big\}
            \end{flalign*}
            We see that for $k+1$ we are at the $(k-2)^{nd}$ split and therefore split the clause \\
            $$\{\ne h_{k-3},\cee_{k-1},\cee_k,\cee_{k+1}\} \mapsto \{\ne h_{k-3}, \cee_{k-1}, h_{k-2}\}, \{\ne h_{k-2},\cee_{k},\cee_{k+1}\}.$$ 
            We also add the clause $\{\ne \cee_{k+1}, h_i\}$ for every existing split $i\leq k-3$, as well as the two clauses
            $\{\ne \cee_k, h_{k-2}\}$ and $\{\ne \cee_{k+1}, h_{k-2}\}$ for the $k-2^{nd}$ split.\\
            Consider the literal $\cee_{k+1}$ and its possible assignments in $S'\in\mc S_\TSATu^{k+1}$. For $\cee_{k+1}\in S'$ the clause $\{\ne \cee_{k+1}, h_{k+1}\}$
            directly implies that $h_{k-2}$ is also in $S'$. If $\ne \cee_{k+1}\in S'$ we can conclude from the clauses $\{\ne h_{k-2},\cee_{k},\cee_{k+1}\}, \{\ne \cee_k, h_{k-2}\}$
            that $\cee_k\in S' \implies h_{k-2}\in S'$ and $\ne \cee_k\in S' \implies \ne h_{k-2}\in S'$. With this we get
            \begin{flalign*}
                \cee_{k+1} \in S' ~\Leftrightarrow ~& h_{k-2}\in S'\\
                \ne \cee_{k+1}\in S' ~\Leftrightarrow ~ & (h_{k-2}\in S' \Leftrightarrow \cee_k \in S)
            \end{flalign*}
    \end{itemize}
        As both $\ell_{k+1}\in S$ and $\ne\ell_{k+1}\in S$ for $S\in \mc S_\SAT^k$ (and therefore in $S'\in \mc S_\TSATu^k$) directly imply an assignment of $h_{k-2}$ in $S'\in \mc S_\TSATu^k$
        we can state that for every $\S\SAT$ we have an equivalent $\Sr\TSATu$.\\
        Now let $S'$ be a solution for $\TSAT$. As the split clauses still contain the same literals that the original $\SAT$ clause contained, a satisfying assignment
        would also satisfy the \SAT clauses, as \TSAT insatnce only adds restraints.
\end{proof}
\noindent The reduction function $g$ is computable in polynomial time and is dependent on the input size of the \SAT instance $\big|[L,\mc C]\big|$,
as for each clause in $\mc C$ we create a fixed number of clauses and literals in the new instance. Therefore $\SAT\leq \TSAT$ holds.

\paragraph{SSP \& Parsimonious}
As seen above, there is a one-to-one correspondence between the literals $\ell\in S$ and the corresponding literals $\ell'\in S'$, as each \TSAT solution 
$S'$ consists of the same literals as $S$ in addition to the literals $h_j^i$. We can now use \autoref{theorem 1} to prove that this reduction is both
SSP and SPR.
\begin{proof}
    Let the set of representatives $\srep$ for each solution $\Sr\TSATu$ be defined as 
    $$\srep:=\{r(\ell)\mid\ell\in S\},\text{ with } r:L\rightarrow L',~r(\ell)=\ell'$$
    where $\cee$ and $\cee'$ correspond to the same literal in the two instances respectively. Therefore the representatives are the 
    literals corresponding to the literals in the original \SAT instance. \\
    As we have shown above, every solution $\S\SAT$ directly implies a fixed set of $h\in L'$ to be in the corresponding solution $\Sr\TSATu$. 
    It follows that every $\srep$ is linked to a set of $h\in L'$. Therefore our linked set $\Slink$ is defined by
    $$\Slink:=\bigcup_{\Sr\TSATu}\slinkr =\big\{h_j^i\mid j\in \big[|\mc C|], i\in\big[|L^2|-3\big]\big\}$$
    Now $\mc U'(g(I))=L'$ can be partitioned into $\Srep\cupdot \Slink=L'$ 
    and therefore according to \autoref{theorem 1} $\SAT\SSPR\TSAT$ holds.
\end{proof}

\subsubsection[... to 0-1-Integer Programming]{Satisfiability to 0-1-Integer Programming}
\label{sec:SAT to IP}
\paragraph{Problem Definition}
\subfile{../../Problems/IP}

\paragraph{Reduction}
The reduction presented in \cite[p. 15-17]{Femke} is already both SSP and SPR and, as that thesis includes the whole reduction, 
proof and an example, we only briefly introduce the idea behind the reduction here.\\
Let $\SAT:=[L,\mc C]$, then the corresponding $\IP$ instance is created by mapping every literal pair $(\ell_i,\ne \ell_i)\in L^2$
to two columns of the matrix $m_{2i-1}$ and $m_{2i}$ respectively. These columns contain the value '$-$1' for each of the occurrences of the 
literals each of the clauses, as well as entries ensuring that only one of the two columns can be in the solution.

\paragraph{SSP \& Parsimonious}
As seen above, there exists a direct one-to-one correspondence between the literals satisfying the \textsc{Satisfiability} instance
and the columns of $M$ chosen for the \textsc{0-1-Integer Programming} solution. We can therefore apply \lemref{lem:SSP and SPR 2} 
as follows.
\begin{proof}
    Let $\Sall=\Slink=\Snev=\emptyset$ and let the functions $f_I$ be defined as 
    $$f_I:L\rightarrow m_1,...,m_n;~f(\cee_i) = \begin{cases}
        m_{2i-1} &\mid \cee_i\in L^+ \\
        m_{2i} &\mid \cee_i\in L^- 
    \end{cases}$$
    We see that $f_I$ are bijective functions between the solution spaces. With these we get
    \begin{align}
        \{f(S):\S\SAT\} &=\big\{\{f(\cee)\mid \ell\in S\}:\S\SAT\big\}\\
        &=\big\{\{m\in S'\}:\Sr\IP\big\}\\\label{eq:SAT1}
        &=\{\Sr\IP\}\\ \label{eq:SAT2}
        &=\{S'\cap f(L):\Sr\IP\}
    \end{align}
    where \ref{eq:SAT1} and \ref{eq:SAT2} hold, as the functions $f_I$ are bijective and therefore all elements in $L$ are mapped to exactly one
    corresponding element in $\{m_1,...,m_n\}$. It follows that $\SAT\SSPR\IP$ holds according to \autoref*{lem:SSP and SPR 2}.
\end{proof}

\ifSubfilesClassLoaded{
  \bibliography{../../ref}
}{}
\end{document}

%% file: Problems/SAT.tex
\noindent\fbox{
    \parbox{\textwidth}{
        \textsc{Satisfiability ($\SAT$)}\\
        \begin{tabular}{cl}
        ~ & \textbf{Instances:} Literal Set $L = \{\ell_1, . . . , \ell_n\} \cup \{\ne\ell_1, . . . , \ne\ell_n\},$ Clauses $\mc C \subseteq 2^L$.\\
        ~ & \textbf{Universe:} $\mathcal{U}:=L$.\\
        ~ & \textbf{Solution set:} The set of all sets $S\subseteq L$ such that for all $i \in\{1,...,n\}$ we have \\
        ~ &$|S\cap\{\ell_i,\overline{\ell_i}\}|=1$, and such that $|S\cap C|\geq 1$ for all $C\in\mc C$.
        \end{tabular}
    }
}

\ifSubfilesClassLoaded{
  \bibliography{../ref}
}{}
\end{document}

%% file: Problems/TSAT.tex
\noindent\fbox{
    \parbox{\textwidth}{
        \textsc{3-Satisfiability ($\TSAT$)}\\
        \begin{tabular}{cl}
        ~ & \textbf{Instances:} Literal Set $L = \{\ell_1, . . . , \ell_n\} \cup \{\ne\ell_1, . . . , \ne\ell_n\},$ Clauses $\mc C \subseteq 2^L$ s.t. \\
        ~ & $\forall C\in \mc C:|C|\leq 3$. \\
        ~ & \textbf{Universe:} $\mathcal{U}:=L$.\\
        ~ & \textbf{Solution set:} The set of all sets $S\subseteq L$ such that for all $i \in\{1,...,n\}$ we have \\
        ~ &$|S\cap\{\ell_i,\ne\ell_i\}|=1$, and such that $|S\cap C|\geq 1$ for all $C\in \mc C$.
        \end{tabular}
    }
}

\ifSubfilesClassLoaded{
  \bibliography{../ref}
}{}
\end{document}

%% file: Problems/IP.tex
\noindent\fbox{
    \parbox{\textwidth}{
        \textsc{0-1-Integer Programming ($\IP$)}\\
        \begin{tabular}{cl}
        ~ & \textbf{Instances:} Matrix $M\in \mathbb Z^{m\times n}$, vector $b\in \mathbb Z^{m\times 1}$.\\
        ~ & \textbf{Universe:} Matrix columns $\mathcal{U}:=m_1,...,m_n$.\\
        ~ & \textbf{Solution set:} The set $S\subseteq m_1,...,m_n$, such that $M\cdot x \leq b$, with $x$ being a vector\\
        ~ & over $\{0,1\}^n$, with $x_i=1 \Leftrightarrow m_i\in S$.
        \end{tabular}
    }
}

\ifSubfilesClassLoaded{
  \bibliography{../ref}
}{}
\end{document}

%% file: Reductions/3-Sat/3-Sat.tex
\subsection{Reductions from 3-Satisfiability}
\paragraph{Problem definition}

\subfile{../../Problems/TSAT.tex}
\subsubsection[... to Exact 3-Satisfiability]{3-Satisfiability to Exact 3-Satisfiability}
\label{sec:3SAT to ESAT}
\paragraph{Problem Definition}
\subfile{../../Problems/ESAT.tex}

\paragraph{Reduction}
Let $(L,\mc C)$ be a \textsc{3-Satisfiability} instance, then $(L',\mc C')$ is a \textsc{Exact 3-Satisfiability} instance with the following reduction
\TSAT $\SSPR$ \ESAT. To begin, we create 6 new literals $h_i,\ne h_i \notin L$ for $i\in [3]$ and then unite them with $L$. Next we take every clause 
$C\in \mc C$ with $|C|<3$ and create two new clauses $C_{1} = C\cup \{h_1\}$ and  $C_{1'} = C\cup \{\ne h_1\}$. We do this recursively, adding $h_2$ and 
$h_3$ respectively, until every clause contains exactly three literals. Note how this also covers the case $\{\}\in\mc C$.
Next we create 7 further new clauses, which are designed to force the assignment $\{h_1,h_2,h_3\}\in S'$ for all $\Sr\ESATu$. We call the set of new clauses $H$.
\begin{flalign*}
    H:=\big\{&\{h_1,h_2,h_3\},\{h_1,\ne h_2,h_3\},\{h_1,h_2,\ne h_3\},\{h_1,\ne h_2,\ne h_3\},\\
    &\{\ne h_1,h_2,h_3\},\{\ne h_1,h_2,\ne h_3\},\{\ne h_1,\ne h_2, h_3\}\big\}
\end{flalign*}
Finally, let $L':= L\cup \{h_1,h_2,h_3,\ne h_1,\ne h_2,\ne h_3\}$ and $\mc C' :=\mc C\cup H$ and then do the following for all $1\leq i\leq 3$ until every clause in $\mc C'$ 
has exactly 3 literals
$$\forall C\in \mc C', |C|<3: \mc C' = (\mc C'\backslash C)\cup \big\{C \cup \{\ne h_i\}\big\}$$
\autoref*{fig:3Sat to E3S} shows an example reduction.
\begin{figure}[h]
    \begin{center}
        \begin{flalign*}
            \TSAT:=&[L,\mc C],\text{ with}\\
            L&:=\{\ell_1, \ell_2, \ell_3, \ne \ell_1, \ne \ell_2, \ne \ell_3\},\\
            \mc C&:=\big\{\{\ell_1\},\{\ell_2,\ell_3\}\big\}\\
            \mapsto\ESAT:=&[L',\mc C'],\text{ with}\\
            L'&:=\{\ell_1,\ell_2,\ell_3,\ne\ell_1,\ell_2,\ne\ell_3, h_1,h_2,h_3,\ne h_1,\ne h_2,\ne h_3\},\\
            \mc C'&:=\big\{\{\ell_1, \ne h_1, \ne h_2\},\{\ell_2, \ell_3,\ne h_1\}\big\} \cup H
        \end{flalign*}
        \caption{\footnotesize{Example \textsc{3-Satisfiability} to \textsc{Exact 3-Satisfiability}}}
        \label{fig:3Sat to E3S}
    \end{center}
\end{figure}
\begin{proof}
    Any solution $S\in \mc S_\TSATu$ is now a solution $S'$ in $\mc S_\ESATu$ when combined with $\{h_1,h_2,h_3\}$:\\
    Let $C \in \mc C$ be a \TSAT clause satisfied by a solution $S$. If $|C|=3$ then there exists an equivalent clause $C = C'\in \mc C'$ in \ESAT which is satisfied
    as well. If $|C|<3$ there exits a clause $C'$ that contains the literals in $C$ and up to three literals $\ne h_i$. As $S'$ contains $\{h_1,h_2,h_3\}$, this clause 
    $C'$ can only be satisfied if $C$ is satisfied. All other new clauses contain a literal $h_i$ and are therefore satisfied. \\
    Let $S' \in \mc S_\ESATu$ be a \ESAT solution. The clauses in $H$ can only be satisfied by the partial assignment $\{h_1,h_2,h_3\}$. Now $S'\backslash \{h_1,h_2,h_3\}$
    is a \TSAT solution, as any clause $C'$ in the \ESAT instance not containing literals $h_i$ or $\ne h_i$ is equivalent to a \TSAT clause.\\
    And for every satisfied clause $C' \in \mc C'$ containing $h_i$ we also have the same clause, but with $\ne h_i$ in place of $h_i$. As this clause is also satisfied 
    by the solution $S'$, there is a clause $C'\backslash \{\ne h_i\}$ that is also satisfied. 
    These clauses are equivalent to satisfied clauses $C$ in $\mc C$ as well.
\end{proof}
\noindent The reduction function $g$ is computable in polynomial time and is dependent on the input size of the \TSAT instance $\big|[L,\mc C]\big|$,
as for each clause in $\mc C$ we create a fixed number of clauses and literals in the new instance. Therefore $\TSAT\leq \ESAT$ holds.

\paragraph{SSP \& Parsimonious}
As any solution $S\in \mc S_\TSATu$ is equivalent to the solution $S'=S\cup \{h_1,h_2,h_3\},~ S'\in \mc S_\ESATu$
we can use Lemma \lemref{lem:SSP and SPR} with the following bijective functions $f_I$ and fixed sets $\Sall$ and $\Snev$
to show that the reduction is both SSP and strongly Parsimonious. 
\begin{proof}
Let $\Sall:=\{h_1,h_2,h_3\}$, $\Snev:=\{\ne h_1,\ne h_2,\ne h_3\}$ and $\Slink:=\emptyset$ and let
$$f_I:L\rightarrow L', ~ f(\cee)=\cee',$$
where $\cee$ and $\cee'$ are the same literal in the \TSAT and \ESAT instances respectively. Now 
$$f_I':L\rightarrow L'\backslash (\Sall\cup\Snev)$$ 
is bijective, as the reduction only adds the 
literals $h_i$ and $\ne h_i$
With these we get 
\begin{align}
    \{f(S):S\in\mc S_\TSATu\} & = \big\{\{f(\ell)\mid\cee \in S\}:S\in\mc S_\TSATu\big\}\\
    & = \{S'\backslash \{h_1,h_2,h_3\} :S'\in\mc S_\ESATu\} \label{eq:TSAT1}\\
    & = \{S'\cap f(L) :S'\in\mc S_\ESATu\}
\end{align}
where \ref{eq:TSAT1} holds, as $\{h_1,h_2,h_3\}$ is part of every solution and otherwise $S$ and $S'$ are equivalent.
Therefore, as this shows that the reduction is SSP, \TSAT $\SSPR$ \ESAT holds according to \lemref{lem:SSP and SPR}.
\end{proof}

\ifSubfilesClassLoaded{
  \bibliography{../../ref}
}{}
\end{document}

%% file: Problems/ESAT.tex
\noindent\fbox{
    \parbox{\textwidth}{
        \textsc{Exact 3-Satisfiability ($\ESAT$)}\\
        \begin{tabular}{cl}
        ~ & \textbf{Instances:} Literal Set $L = \{\ell_1, . . . , \ell_n\} \cup \{\ne\ell_1, . . . , \ne\ell_n\},$ Clauses $\mc C \subseteq 2^L$ s.t. \\
        ~ & $\forall C\in \mc C:|C| = 3$. \\
        ~ & \textbf{Universe:} $\mathcal{U}:=L$.\\
        ~ & \textbf{Solution set:} The set of all sets $S\subseteq L$ such that for all $i \in\{1,...,n\}$ we have \\
        ~ &$|S\cap\{\ell_i,\ne\ell_i\}|=1$, and such that $|S\cap C|\geq 1$ for all $C\in \mc C$.
        \end{tabular}
    }
}

\ifSubfilesClassLoaded{
  \bibliography{../ref}
}{}
\end{document}

%% file: Reductions/Exact_3-Sat/Exact_3-Sat.tex
\subsection{Reductions from Exact 3-Satisfiability}

\subfile{../../Problems/ESAT.tex}
\subsubsection[... to 1-in-3-Satisfiability]{Exact 3-Satisfiability to 1-in-3-Satisfiability}
\label{sec:ESAT to OSAT}
\paragraph{Problem Definition}
\subfile{../../Problems/OSAT.tex}

\paragraph{Reduction}
This reduction is based on the reduction presented by Arora and Barak \cite{AB09}, but is modified to be Parsimonious. The idea remains the same, 
we introduce new helper variables for each clause and map each \ESAT clause to a set of clauses, with each literal in the clause now being respresented
by its own clause. However we need to add more clauses than originally to avoid symmetrical solutions. \\
More formally, the reduction looks as follows. Let $\ESAT:=[L,\mc C]$ and $\OSAT:=[L',\mc C']$, then we map each clause $C_j\in \mc C$
to seven new clauses, using nine new literals, as follows. 
\begin{flalign*}
    C_j:=\{\cee_j^1, \cee_j^2,\cee_j^3\} \mapsto C_j^1:=&\{\ne \cee^1_j, z_j^1, h_j^1\},\\
    C_j^2:=&\{\ne \cee^2_j, z_j^2, h_j^2\},\\
    C_j^3:=&\{\ne \cee^3_j, z_j^3, h_j^3\},\\
    C_j^4:=&\{z_j^1,z_j^2,z_j^3\},\\
    C_j^5:=&\{z_j^1,h_j^2,g_j^1\},\\
    C_j^6:=&\{z_j^2,h_j^3,g_j^2\},\\
    C_j^7:=&\{z_j^1,h_j^3,g_j^3\}
\end{flalign*}
Now $L':=L\cup \bigcup_{j= 1}^{|\mc C|}\Big(\bigcup_{i=1}^3\{z_j^i,h_j^i,g_j^i\}\Big)$ and $\mc C':=\bigcup_{j=1}^{|\mc C|}\Big(\bigcup_{i=1}^7\{C_j^i\}\Big)$.
An example reduction is shown in \autoref*{fig:ESAT to OSAT}.
\begin{figure}[ht]
    \begin{center}
        \begin{flalign*}
            \ESAT:=&[L,\mc C],\text{ with}\\
            L&:=\{\ell_1,...,\ell_4,\ne\ell_1,...,\ne\ell_4\},\\
            \mc C&:=\big\{\{\ell_1,\ell_2,\ne\ell_3\},\{\ne\ell_1,\ell_2,\ell_4\}\big\}\\
            \mapsto\OSAT:=&[L',\mc C'],\text{ with}\\
            L'&:=\{\ell_1,...,\ell_4,\ne\ell_1,...,\ne\ell_4,\\
            &\hspace*{.8cm}z_1^1,z_1^2,z_1^3,h_1^1,h_1^2,h_1^3,g_1^1,g_1^2,g_1^3,\\
            &\hspace*{.8cm}z_2^1,\hspace*{1.65cm} ... \hspace*{1.65cm},g_2^3,\\
            &\hspace*{.8cm}\ne z_1^1,\ne z_1^2,\ne z_1^3,\ne h_1^1,\ne h_1^2,\ne h_1^3,\ne g_1^1,\ne g_1^2,\ne g_1^3,\\
            &\hspace*{.8cm}\ne z_2^1,\hspace*{1.65cm} ... \hspace*{1.65cm},\ne g_2^3\},\\
            \mc C'&:=\big\{\{\ne \ell_1, z_1^1, h_1^1\},\{\ne \ell_2, z_1^2, h_1^2\},\{\ell_3, z_1^3, h_1^3\},\\
            &\hspace*{.8cm}\{z_1^1,z_1^2,z_1^3\},\{z_1^1,h_1^2,g_1^1\},\{z_1^2,h_1^3,g_1^2\},\{z_1^1,h_1^3,g_1^3\}\\
            &\hspace*{.8cm}\{\ell_1, z_2^1, h_2^1\},\{\ne \ell_2, z_2^2, h_2^2\},\{\ne\ell_4, z_2^3, h_2^3\},\\
            &\hspace*{.8cm}\{z_2^1,z_2^2,z_2^3\},\{z_2^1,h_2^2,g_2^1\},\{z_2^2,h_2^3,g_2^2\},\{z_2^1,h_2^3,g_2^3\}\big\}
        \end{flalign*}
        \caption{\footnotesize{Example \textsc{Exact 3-Satisfiability} to \textsc{1-in-3-Satisfiability}}}
        \label{fig:ESAT to OSAT}
    \end{center}
\end{figure}
\begin{proof}
    Let $S$ be a satisfying assignment for the \ESAT instance. Then this assignment $S$ is equivalent to a partial assignment 
    $S'_{part}\subset S'$. However $S'_{part}$ implies a full assignment $S'$ in the \OSAT instance as follows. We can restrict this 
    proof to a single clause $C\in\mc C$, as each clause creates seven destinct new clauses with no new overlapping literals. 
    Therefore if this proof holds for instances with $|\mc C|=1$, it also holds for all $|\mc C|>1$. \\
    Let $\mc C$ be the \ESAT clause set with $\mc C:=\{C\}$ and let the clause $C$ be denoted as $C:=\{\cee^1, \cee^2, \cee^3\}$, with the literals $\cee^i$ 
    corresponding to literals in $L$. Then let the clauses in $\mc C'$, created from $C$, be denoted as $C^i$ and let $S'_{part}:=S$. 
    Now we have seven cases:
    \begin{enumerate}
        \item ($\cee^1\in S'_{part}$; $\cee^2, \cee^3\notin S'_{part}$)
        \begin{flalign*}
            \{\cee^1,\ne\cee^2,\ne\cee^3\}\subseteq S'& \implies C^2, C^3 \text{ satisfied by } \ne \cee^2, \ne \cee^3\\
             &\implies z^2,z^3,h^2,h^3\notin S' &\implies \{\ne z^2,\ne z^3,\ne h^2,\ne h^3\}\subseteq S'\\
             &\implies z^1 \in S' \text{ to satisfy } C^4 &\implies \{z^1\}\subseteq S'\\
             &\implies C^1, C^5, C^7\text{ satisfied by } z^1\\
             &\implies h^1 , g^1 ,g^3\notin S'&\implies \{\ne h^1,\ne g^1 ,\ne g^3\}\subseteq S'\\
             &\implies g^2 \in S' \text{ to satisfy } C^6 &\implies \{g^2\} \subseteq S'
        \end{flalign*}
        It follows that
        $$S=\{\cee^1, \ne \cee^2,\ne \cee^3\} \implies S'=\{\cee^1, \ne \cee^2,\ne \cee^3, z^1, \ne z^2, \ne z^3,\ne h^1, \ne h^2, \ne h^3, \ne g^1, g^2, \ne g^3\}$$
        
        \item ($\cee^2\in S'_{part}$; $\cee^1, \cee^3\notin S'_{part}$) \& 3. ($\cee^3\in S'_{part}$; $\cee^1, \cee^2\notin S'_{part}$)\\
        Analogous to 1.
        \item[4.] ($\cee^1,\cee^2\in S'_{part}$; $\cee^3\notin S'_{part} $)
        \begin{flalign*}
            \{\cee^1,\cee^2,\ne\cee^3\}\subseteq S'&\implies C_3\text{ satisfied by } \ne \cee^3 \\
            &\implies z^3,h^3\notin S' &\implies \{\ne z^3,\ne h^3\}\subseteq S'\\
            &\implies h_1, z_2 \in S' \text{ to satisfy } C_1,C_2\text{, as } &\implies \{h^1,z^2\}\subseteq S'\\
            &\hspace*{1.5cm} (C^1 \land \cee_1\in S') \implies (z^1\in S'\oplus h_1\in S') \text{ and } \\
            &\hspace*{1.5cm} (C^2 \land \cee_2\in S') \implies (z^2\in S'\oplus h_2\in S') \text{ and } \\
            &\hspace*{1.5cm} (C^4 \land z^3\notin S')\implies (z^1\in S'\oplus  z^2\in S' )\text{ and } \\
            &\hspace*{1.5cm} C^5 \implies \neg(z^1\in S'\land  h^2\in S')\\
            &\implies C^4,C^6 \text{ satisfied by }z^2\\
            &\implies z^1,h^2,g^2\notin S' &\implies \{\ne z^1,\ne h^2,\ne g^2\}\subseteq S'\\
            &\implies g^1,g^3\in S' \text{ to satisfy }C^5 , C^7 &\implies \{ g^1,g^3\}\subseteq S'
        \end{flalign*}
        It follows that
        $$S=\{\cee^1, \cee^2,\ne \cee^3\} \implies S'=\{\cee^1 \cee^2,\ne \cee^3, \ne z^1, z^2, \ne z^3, h^1, \ne h^2, \ne h^3, g^1, \ne g^2, g^3\}$$
        
        \item[5.] ($\cee^1,\cee^3\in S'_{part}$; $\cee^2\notin S'_{part}$) \& 6. ($\cee^2,\cee^3\in S'_{part}$; $\cee^1\notin S'_{part}$)\\
        Analogous to 1.
        \item[7.] ($\cee^1,\cee^2,\cee^3\in S'_{part}$)\\
        If $h^3$ were in $S'$ then the clauses $C^3, C^6$ and $C^7$ would imply, that neither $z^1, z^2$ nor $z^3$ could be in
        the solution $S'$. However clause $C^4$ then cannot be satisfied. Therefore $h^3\notin S'$.
        \begin{flalign*}
            \{\cee^1,\cee^2,\cee^3,\ne h^3\}\subseteq S' \implies& z^3 \in S' \text{ to satisfy } C^3 &\implies \{z^3\}\subseteq S'\\
            \implies& z^1,z^2 \notin S'&\implies \{\ne z^1, \ne z^2 \}\subseteq S'\\
            \implies& h^1, h^2 \in S'\text{ to satisfy } C^1,C^2 \\
            &\text{and }g^2, g^3 \in S'\text{ to satisfy } C^6,C^7&\implies \{h^1, h^2,g^2, g^3\}\subseteq S'\\
            \implies& g^1 \notin S' &\implies \{\ne g^1\}\subseteq S'
        \end{flalign*}
        It follows that 
        $$S=\{\cee^1, \cee^2, \cee^3\} \implies S'=\{\cee^1 \cee^2,\cee^3, \ne z^1, \ne z^2, z^3, h^1, h^2, \ne h^3, \ne g^1, g^2, g^3\}$$
     \end{enumerate}
     Therefore every satisfying assignment $\S\ESATu$ implies exactly one satisfying assingment $\Sr\OSATu$.\\
     Let now $S'$ be a satisfying assignment for $\OSAT$. For every set of $i\in [7]$ new clauses $C_j^i$, based of of clause $C_j\in \mc C$, we know that 
     $C_j^4=\{z_j^1,z_j^2,z_j^3\}$ forces at least one $z_j^i$ to be in $S'$. Therefore from the corresponding 
     \ESAT clause $C_j$  at least one of the literals $\cee_j^1,\cee_j^2$ or $\cee_j^3$ has to be in $S'$, as 
     $z_j^i\in S' \implies \ne \cee_j^i \notin S'$. It follows that the literals $\ell'\in S'$ (these are equivalent to the literals $\ell\in L$)
     form a solution $\S\ESATu$.
\end{proof}
\noindent The reduction function $g$ is computable in polynomial time and is dependent on the input size of the \ESAT instance $\big|[L,\mc C]\big|$,
as for each clause in $\mc C$ we create a fixed number of clauses and literals in the new instance. Therefore $\ESAT\leq \OSAT$ holds.

\paragraph{SSP \& Parsimonious }
As seen above there is a direct one-to-one correspondence between the literals in $L$ and the equivalent literals in $L'$, 
however the reduction function added further literals to $L'$. We can now use \lemref{theorem 1} to prove that 
$\ESAT\SSPR\OSAT$ holds as every set $\S\ESATu$ implies
exactly one set $\Sr\OSATu$.
\begin{proof}
    Let $\Snev:=\emptyset$ and let $\Srep$ be the set $L'\cap \hspace*{1pt}L$, aka the set of literals 
    directly corresponding to the literals in the \ESAT instance. Now let
    $$r_I:\Srep\rightarrow L,~r(\ell')=\ell,$$
    with $\cee'$ and $\cee$ being equivalent in their respective instances.
    As seen in the reduction proof above, every solution $\S\ESATu$ implies exactly one solution
    $\Sr\OSATu$ and therefore the representatives $\mf R_{S'}$ uniquely define a set of $z_j^i\hspace*{0pt}'s,h_j^i\hspace*{0pt}'s$ 
    and $g_j^i\hspace*{0pt}'s$. With this we get 
    $$\Slink:=\bigcup_{\Sr{\OSATu}} \mf L\big(\mf R_{S'}\big)=\bigcup_{j= 1}^{|\mc C|}\Big(\bigcup_{i=1}^3\{z_j^i,h_j^i,g_j^i\}\Big).$$
    As now $L'=\Srep \cupdot \Slink$ with the bijective functions $(r_I)_{I\in\mc I}$, \lemref{theorem 1} holds.
\end{proof}

\subsubsection[... to Maximum Cardinality Independent Set]{Exact 3-Satisfiability to Maximum Cardinality Independent Set}
\label{sec:ESAT to MIS}
\paragraph{Problem Definition}
\subfile{../../Problems/MIS.tex}

\paragraph{Reduction}
We base this reduction on the reduction presented in \cite{ssp}. Let $[L,C]$ be a \ESAT instance, then $[G:=(V,E),k]$ is a \MIS instance 
with the following mappings. \\
For every literal pair $(\ell,\ne \ell)\in L^2$ we map these two to the vertices $\ell\in V$ and $\ne \ell\in V$ and create the edge 
$\{\ell,\ne \ell\}\in E$. Now for every clause $C_j\in \mc C$ with $C_j = \{c_1,c_2,c_3\}$ we create a connected triangle of vertices $c_j^1,c_j^2,c_j^3\in V$. 
Every literal $c_i \in C_j$ corresponds to a literal $\ell \in L$ and we create the edge 
$\{c_j^i,\ell\}$, if $c^i\in L^-$ and the edge $\{c_j^i,\ne\ell\}$, if $c^i\in L^+$. This way every literal is 
connected to each of its occurrences in the clauses. Finally we set $k$ to $k:= |L^2| + |\mc C|$. \\
Let $\mc C = \{\{\ell_1, \ne \ell_2, \ell_3\}, \{\ne\ell_1, \ell_2,  \ell_3\}\}$ and $L=\{\ell_1, \ell_2, \ell_3, \ne\ell_1, \ne\ell_2, \ne\ell_3\}$ 
define a \textsc{E3S} instance, then \autoref*{fig:ESAT to MIS} shows the corresponding $\MIS$ instance.
\begin{figure}[ht]
    \begin{center}
    \scalebox{.85}{
        \begin{tikzpicture}[vertex/.style = {draw, circle, fill, inner sep = 1.5}, node distance = 1.5cm]
            \node[vertex, label = $\ell_1$](l1){};
            \node[vertex, right of = l1, label = $\ne\ell_1$](nl1){};
            \node[vertex, right of = nl1, label = $\ell_2$](l2){};
            \node[vertex, right of = l2, label = $\ne\ell_2$](nl2){};
            \node[vertex, right of = nl2, label = $\ell_3$](l3){};
            \node[vertex, right of = l3, label = $\ne\ell_3$](nl3){};

            \node[vertex, below of = nl1, label = below:$c_1^2$](c12){};
            \node[vertex, below left of = c12, label = below:$c_1^1$](c11){};
            \node[vertex, below right of = c12, label = below:$c_1^3$](c13){};

            \node[vertex, below of = l3, label = below:$c_2^2$](c22){};
            \node[vertex, below left of = c22, label = below:$c_2^1$](c21){};
            \node[vertex, below right of = c22, label = below:$c_2^3$](c23){};

            \node[right = 1cm of c23](k){$k=5$};

            \draw 
                (l1) -- (nl1)
                (l2) -- (nl2)
                (l3) -- (nl3)

                (c11) -- (c12)
                (c11) -- (c13)
                (c13) -- (c12)
                
                (c21) -- (c22)
                (c21) -- (c23)
                (c23) -- (c22)

                (c11) -- (nl1)
                (c12) -- (l2)
                (c13) -- (nl3)
                (c21) -- (l1)
                (c22) -- (nl2)
                (c23) -- (nl3)
            ;
        \end{tikzpicture} 
    }\\
    \caption{\footnotesize{Example \textsc{Exact 3-Satisfiability} to \textsc{Minimum Cardinality Independent Set}}}
    \label{fig:ESAT to MIS}
    \end{center}
\end{figure}
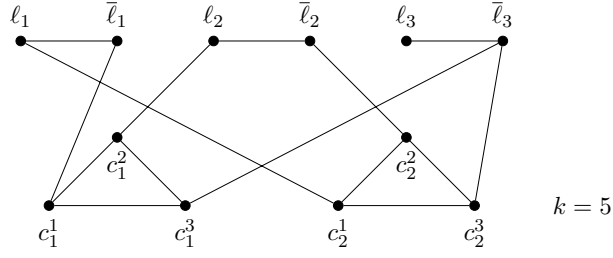
\begin{proof}
Now every \ESAT solution $S\in \mc S_\ESATu$ corresponds to a \MIS solution $S'\in \mc S_\MIS$, where $S'$ contains the vertices equivalent to literals
not in $S$, as well as one of the vertices $c_j^i$ for every clause $C_j$, satisfied by the solution $S$. This is an independent set of exactly size 
$|L^2| + |\mc C|$, as for each triangle pertaining to a clause $C_j\in \mc C$ exactly one vertex is in the solution $S'$ and also exactly one of each pair 
$(\ell,\ne\ell)$ is not in the solution $S$ and therefore the other one of the pair is in $S'$.\\
Also every \MIS solution $S'\in \mc S_\MIS$ corresponds to a \ESAT solution $S\in \mc S_\ESATu$, when we take the literals $\ell$ that are 
not in S'. This works,
because if there were three literal vertices in $S'$ that were connected to all three $c_j^i$ of a triangle, the independent set could not have size 
$|L^2| + |\mc C|$ anymore, as only one vertex of each literal pair and of each triangle can be in the independent set. Therefore each clause in $\mc C$ is 
satisfied by a literal not in $S'$.
\end{proof}
\noindent The reduction function $g$ is computable in polynomial time and is dependent on the input size of the \ESAT instance $\big|[L,\mc C]\big|$,
as for each clause in $\mc C$ we create a fixed number of vertices in the new instance. Therefore $\ESAT\leq \MIS$ holds.

\paragraph{SSP}
As for any solution $S\in \mc S_\ESATu$, the literal vertices corresponding to the literals in $S$ are also a solution in $S'\in \mc S_\MIS$ this reduction 
has the SSP property.
\begin{proof}
Let $f_I$ be a function set defined as follows $f_I:L\rightarrow V,~f(\cee) = \ell$, where $\cee$ denotes a literal $\ell\in L$ and $\ell$ the corresponding
vertex in $V$. 
With this injective function we get
\begin{flalign*}
    \{f(S):S\in \mc S_\ESATu\} &= \big\{\{f(\ell)\mid\cee\in S\}:S\in \mc S_\ESATu\big\}\\
    &= \big\{S'\backslash \big\{c_j^i\mid i\in [3], j\in\big[|\mc C|\big]\big\}:S' \in \mc S_\MIS\big\}\\
    &= \big\{S'\cap f(L):S'\in \mc S_\MIS\big\}
\end{flalign*}
and therefore \ESAT $\SSP$ \MIS holds.
\end{proof}

\paragraph{Parsimonious}
As we show in the following example, this reduction is not Parsimonious or strongly so. This necessitates the use of the transitive 
property of Parsimonious reductions. \\
Let $\mc C = \{\{\ell_1,\ell_2, \ell_3\}\}$ and $L=\{\ell_1, \ell_2, \ell_3, \ne\ell_1, \ne\ell_2, \ne\ell_3\}$ define a \textsc{E3S} 
instance, then \autoref*{fig:ESAT to MIS pars} shows the corresponding \MIS instance with $k=4$. We can easily see that $S=\{\ell_1,\ell_2,\ell_3\}$ is a solution in 
$\mc S_\ESATu$, however this solution has three possible equivalent solutions $S'\in \mc S_\MIS$, those being $S\cup \{c_1^i\}$ for every $i\in[3]$. 
Therefore $|\mc S_\ESATu|<|\mc S_\MIS|$ and it follows that there cannot be a bijective function between these solution sets.
\begin{figure}[ht]
    
    \begin{center}
    \scalebox{.85}{
        \begin{tikzpicture}[vertex/.style = {draw, circle, fill, inner sep = 1.5}, node distance = 1.5cm]
            \node[vertex, label = $\ell_1$](l1){};
            \node[vertex, right of = l1, label = $\ne\ell_1$](nl1){};
            \node[vertex, right of = nl1, label = $\ell_2$](l2){};
            \node[vertex, right of = l2, label = $\ne\ell_2$](nl2){};
            \node[vertex, right of = nl2, label = $\ell_3$](l3){};
            \node[vertex, right of = l3, label = $\ne\ell_3$](nl3){};

            \node[vertex, below right of = l2, label = below:$c_1^2$](c21){};
            \node[vertex, below left of = c21, label = below:$c_1^1$](c11){};
            \node[vertex, below right of = c21, label = below:$c_1^3$](c31){};

            \draw 
                (l1) -- (nl1)
                (l2) -- (nl2)
                (l3) -- (nl3)

                (c11) -- (c21)
                (c11) -- (c31)
                (c31) -- (c21)

                (c11) -- (nl1)
                (c21) -- (nl2)
                (c31) -- (nl3)
            ;
        \end{tikzpicture} 
    }\\
    \caption{\footnotesize{Example \ESAT $\SSP$\MIS} is not Parsimonious}
    \label{fig:ESAT to MIS pars}
    \end{center}
\end{figure}

\paragraph{Alternative transitive reduction with both SSP \& SPR properties}
To prove that $\ESAT \SSPR \MIS$ holds, we make use of the transitive property of Parsimonious reductions. In \autoref{sec:ESAT to OSAT} 
we show that $\ESAT \SSPR \OSAT$ and in \autoref{sec:OSAT to MIS} we see that $\OSAT \SSPR \MIS$. Therefore $\ESAT \SSPR \OSAT\SSPR\MIS$ 
holds and thus there is a strongly 
Parsimonious reduction between \textsc{Exact 3-Satisfiability} and \textsc{Maximum Cardinality Independent Set} that is also SSP.

\subsubsection[... to Minimum Cardinality Vertex Cover]{Exact 3-Satisfiability to Minimum Cardinality Vertex Cover} 
\label{sec:ESAT to MVC}
\paragraph{Problem Definitions}
\subfile{../../Problems/MVC.tex}

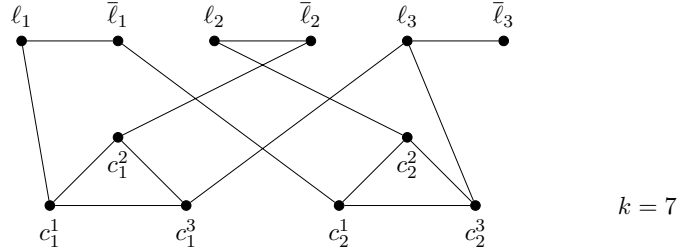
\begin{figure}[ht]
    \begin{center}
    \scalebox{.85}{
        \begin{tikzpicture}[vertex/.style = {draw, circle, fill, inner sep = 1.5}, node distance = 1.5cm]
            \node[vertex, label = $\ell_1$](l1){};
            \node[vertex, right of = l1, label = $\ne\ell_1$](nl1){};
            \node[vertex, right of = nl1, label = $\ell_2$](l2){};
            \node[vertex, right of = l2, label = $\ne\ell_2$](nl2){};
            \node[vertex, right of = nl2, label = $\ell_3$](l3){};
            \node[vertex, right of = l3, label = $\ne\ell_3$](nl3){};

            \node[vertex, below of = nl1, label = below:$c_1^2$](c12){};
            \node[vertex, below left of = c12, label = below:$c_1^1$](c11){};
            \node[vertex, below right of = c12, label = below:$c_1^3$](c13){};

            \node[vertex, below of = l3, label = below:$c_2^2$](c22){};
            \node[vertex, below left of = c22, label = below:$c_2^1$](c21){};
            \node[vertex, below right of = c22, label = below:$c_2^3$](c23){};

            \node[right = 2cm of c23](k){$k=7$};

            \draw 
                (l1) -- (nl1)
                (l2) -- (nl2)
                (l3) -- (nl3)

                (c11) -- (c12)
                (c11) -- (c13)
                (c13) -- (c12)
                
                (c21) -- (c22)
                (c21) -- (c23)
                (c23) -- (c22)

                (c11) -- (l1)
                (c12) -- (nl2)
                (c13) -- (l3)
                (c21) -- (nl1)
                (c22) -- (l2)
                (c23) -- (l3)
            ;
        \end{tikzpicture} 
    }\\
    \caption{\footnotesize{Example \textsc{Exact 3-Satisfiability} to \textsc{Minimum Cardinality Vertex Cover}}}
    \label{fig:ESAT to MVC}
    \end{center}
\end{figure}

\paragraph{Reduction}
We use the reduction we use in \autoref{sec:ESAT to MIS} with only a few minor changes. We do not connect the vertices $c_j^i$ with $\ne\ell$ if $\cee_j^i$ corresponds
to $ \ell$, but instead we connect it to $\ell$ directly and analogous for $\cee_j^i\cor \ne\ell$ and we also set $k$ to $k:= |L^2|+ 2\cdot|\mc C|$.\\
Let $\mc C = \{\{\ell_1, \ne \ell_2, \ell_3\}, \{\ne\ell_1, \ell_2,  \ell_3\}\}$ and 
$L=\{\ell_1, \ell_2, \ell_3, \ne\ell_1, \ne\ell_2, \ne\ell_3\}$ define a \ESAT instance, then \autoref*{fig:ESAT to MVC} shows
the corresponding \MVC instance.
\begin{proof}
Each \ESAT solution $S\in\mc S_\ESATu$ corresponds to a solution $S'\in\mc S_\MVC$, where if a literal in $(\ell_i,\ne\ell_i)$ is in $S$ then 
the corresponding vertex is in $S'$ and therefore all
the edges $\{\ell_i,\ne\ell_i\}$ are covered. For every triangle $c_j^1,c_j^2,c_j^3$ at least one edge connecting 
it to the literal pairs is covered as every clause is satified in $S$. The other two outgoing edges and the edges in every triangle can be covered by the $|\mc C|\cdot 2$ vertices we can 
still assign to the vertex cover, as $k=|L^2|+|\mc C|\cdot 2$.\\
Every solution $S'\in\mc S_\MVC$ also corresponds to a solution $S\in\mc S_\ESATu$, as vor every edge $\{\ell,\ne\ell\}$ one of the vertices needs to be
in the vertex cover and as we only have $|\mc C|\cdot 2$ vertices left in the cover, they need to be assigned to every triangle $c_j^1,c_j^2,c_j^3$, or else 
the edges inside it would not be covered. This leaves one uncovered edge per triangle left which connects it to the literal pairs and these edges are only 
covered if at least one literal would satisfy the corresponding clause. Therefore the literal vertices in $S'$ are a solution $S\in\mc S_\ESATu$.
\end{proof}
\noindent The reduction function $g$ is computable in polynomial time and is dependent on the input size of the \ESAT instance $\big|[L,\mc C]\big|$,
as for each clause in $\mc C$ we create a fixed number of vertices in the new instance. Therefore $\ESAT\leq \MVC$ holds.

\paragraph{SSP}
As for any solution $S\in \mc S_\ESATu$, the literal vertices corresponding to the literals in $S$ are also a solution in $S'\in \mc S_\MVC$ this reduction 
has the SSP property.
\begin{proof}
Let $f_I$ be a function set defined as follows $f_I:L\rightarrow V,~f(\cee) = \ell$, where $\cee$ denotes a literal $\ell\in L$ and $\ell$ the corresponding
vertex in $V$. 
With this injective function we get
\begin{flalign*}
    \{f(S):S\in \mc S_\ESATu\} &= \big\{\{f(\ell)\mid \ell\in S\}:S\in \mc S_\ESATu\big\}\\
    &= \big\{S'\backslash \big\{c_j^i\mid i\in [3], j\in\big[|\mc C|\big]\big\}:S' \in \mc S_\MVC\big\}\\
    &= \big\{S'\cap f(L):S'\in \mc S_\MVC\big\}
\end{flalign*}
and therefore \ESAT $\SSP$ \MVC holds.
\end{proof}

\paragraph{Parsimonious}
$\ESAT\SSP\MVC$ is not Parsimonious, as seen in the following example:
\begin{figure}[ht]
    Let $\mc C = \{\{\ell_1,\ell_2, \ell_3\}\}$ and 
    $L=\{\ell_1, \ell_2, \ell_3, \ne\ell_1, \ne\ell_2, \ne\ell_3\}$ define a \textsc{E3S} instance, then
    \begin{center}
    \scalebox{.85}{
        \begin{tikzpicture}[vertex/.style = {draw, circle, fill, inner sep = 1.5}, node distance = 1.5cm]
            \node[vertex, label = $\ell_1$](l1){};
            \node[vertex, right of = l1, label = $\ne\ell_1$](nl1){};
            \node[vertex, right of = nl1, label = $\ell_2$](l2){};
            \node[vertex, right of = l2, label = $\ne\ell_2$](nl2){};
            \node[vertex, right of = nl2, label = $\ell_3$](l3){};
            \node[vertex, right of = l3, label = $\ne\ell_3$](nl3){};

            \node[vertex, below right of = l2, label = $c_1^2$](c21){};
            \node[vertex, below left of = c21, label = below:$c_1^1$](c11){};
            \node[vertex, below right of = c21, label = below:$c_1^3$](c31){};

            \draw 
                (l1) -- (nl1)
                (l2) -- (nl2)
                (l3) -- (nl3)

                (c11) -- (c21)
                (c11) -- (c31)
                (c31) -- (c21)

                (c11) -- (l1)
                (c21) -- (l2)
                (c31) -- (l3)
            ;
        \end{tikzpicture} 
    }\\
    \caption{\footnotesize{Example \textsc{E3S} $\SSP$ \textsc{MVC}} is not Parsimonious}
    \end{center}
    is a \textsc{MVC} instance with $k = 5$.
\end{figure}\\
Now we look at $S = \{\ell_1,\ell_2,\ell_3\} \in \mc S_\ESATu$. For the reduction to be Parsimonious we need this solution to have exactly 
one equivalent in $\mc S_\MVC$. However all three of the solutions $S'_1= \{\ell_1,\ell_2,\ell_3, c_1^1, c_2^1\},~
S'_2= \{\ell_1,\ell_2,\ell_3, c_1^1, c_3^1\}$ and $S'_3= \{\ell_1,\ell_2,\ell_3, c_2^1, c_3^1\} $ are possible equivalents, while every other 
$S \in \mc S_\ESATu$ still has at least one.  Therefore $|\mc S_\ESATu|<|\mc S_\MVC|$ and 
it follows that there cannot be a bijective function between these solution sets.

\paragraph{Alternative transitive reduction with both SSP \& SPR properties}
To prove that $\ESAT \SSPR \MVC$ holds, we make use of the transitive property of Parsimonious reductions. In \autoref{sec:ESAT to OSAT} 
we show that $\ESAT \SSPR \OSAT$ and in \autoref{sec:OSAT to MVC} we see that $\OSAT \SSPR \MVC$. Therefore $\ESAT \SSPR \OSAT\SSPR\MVC$ 
holds and thus there is a strongly 
Parsimonious reduction between \textsc{Exact 3-Satisfiability} and \textsc{Minimum Cardinality Vertex Cover} that is also SSP.

\subsubsection[... to Minimum Cardinality Dominating Set]{Exact 3-Satisfiability to Minimum Cardinality Dominating Set}
\label{sec:ESAT to MDS}
\paragraph{Problem Definition}
\subfile{../../Problems/MDS.tex}

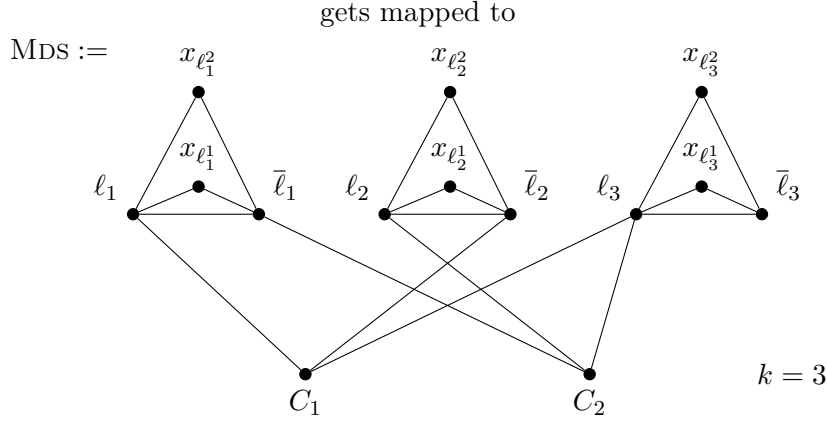
\begin{figure}[ht]
    \begin{center}
        $$\ESAT:=[\mc C = \{\{\ell_1, \ne \ell_2, \ell_3\}, \{\ne\ell_1, \ell_2,  \ell_3\}\}, 
        L=\{\ell_1, \ell_2, \ell_3, \ne\ell_1, \ne\ell_2, \ne\ell_3\}].$$\\
        gets mapped to\\
    \scalebox{1}{
        \begin{tikzpicture}[vertex/.style = {draw, circle, fill, inner sep = 1.5}, node distance = 1.5cm]
            \node[vertex, label = above left:$\ell_1$](l1){};
            \node[vertex, above right = .25cm and .75cm of l1, label = $x_{\ell_1^1}$](x11){};
            \node[vertex, above right = 1.5cm and .75cm of l1, label = $x_{\ell_1^2}$](x12){};
            \node[vertex, right = 1.5cm of l1, label = above right:$\ne\ell_1$](nl1){};
            \node[vertex, right = 1.5cm of nl1, label = above left:$\ell_2$](l2){};
            \node[vertex, above right = .25cm and .75cm of l2, label = $x_{\ell_2^1}$](x21){};
            \node[vertex, above right = 1.5cm and .75cm of l2, label = $x_{\ell_2^2}$](x22){};
            \node[vertex, right = 1.5cm of l2, label = above right:$\ne\ell_2$](nl2){};
            \node[vertex, right = 1.5cm of nl2, label = above left:$\ell_3$](l3){};
            \node[vertex, above right = .25cm and .75cm of l3, label = $x_{\ell_3^1}$](x31){};
            \node[vertex, above right = 1.5cm and .75cm of l3, label = $x_{\ell_3^2}$](x32){};
            \node[vertex, right = 1.5cm of l3, label = above right:$\ne\ell_3$](nl3){};

            \node[vertex, below right = 2cm and .5 cm of nl1, label = below:$C_1$](c1){};
            \node[vertex, below left = 2cm and .5 cm of l3, label = below:$C_2$](c2){};

            \node[above left = .25cm and 1 cm of x12](G){$\MDS:=$};
            \node[right = 2cm of c2](k){$k=3$};

            \draw 
                (l1) -- (nl1)
                (l1) -- (x11)
                (x11) -- (nl1)
                (l1) -- (x12)
                (x12) -- (nl1)

                (l2) -- (nl2)
                (l2) -- (x21)
                (x21) -- (nl2)
                (l2) -- (x22)
                (x22) -- (nl2)

                (l3) -- (nl3)
                (l3) -- (x31)
                (x31) -- (nl3)
                (l3) -- (x32)
                (x32) -- (nl3)

                (c1) -- (l1)
                (c1) -- (nl2)
                (c1) -- (l3)

                (c2) -- (nl1)
                (c2) -- (l2)
                (c2) -- (l3)
            ;
        \end{tikzpicture} 
    }\\
    \caption{\footnotesize{Example \textsc{Exact 3-Satisfiability} to \textsc{Minimum Cardinality Dominating set}}}
    \label{fig:ESAT to MDS}
    \end{center}
\end{figure}

\paragraph{Reduction}
For each Literal pair $(\ell_i,\ne \ell_i) \in L^2$ we create the three vertices $\ell_i$, $\ne\ell_i$, $x_{\ell_i^1}$ and $x_{\ell_i^2}$ as well as edges connecting the literal
pair to each other and to each of the $x_{\ell_i^y}$. For every clause $C_j \in \mc C$ we create the vertex $C_j$. Finally we connect each vertex $C_j$ with each of the three literals 
contained in it and set $k$ to $k:=|L^2|$. \autoref*{fig:ESAT to MDS} shows and example reduction.
\begin{proof}
A satisfying assignment $S\in \mc S_\ESATu$ is equivalent to a dominating set $S'\in \mc S_\MDS$ with every literal in $S$ being
mapped to the corresponding vertex $\ell_i$ or $\ne \ell_i$. As $|S|=|L^2|$, each vertex $C_i$ is dominated by the vertex $\ell_i$ or $\ne \ell_i$
 satisfying the corresponding clause and as every literal has an assignment, every $x_{\ell_i^j}$ and the 
negated literal are also dominated. \\
Every dominating set $S'\in \mc S_\MDS$ consists of only vertices pertaining to literals, as they are exactly $k$ literal gadgets and therefore no vertex $C_i$ 
can be in the dominating set. Also inside these gadgets we can only take either $\ell$ or $\ne \ell$ into the set, as otherwise one of the $x_{\ell^j}$ gadets 
cannot be dominated. However if we have a dominating set containing only literals, this is also a satisfying assignment for $\ESAT$, as each clause $C_i\in \mc C$
is satified by the vertex dominating $C_i\in V$.
\end{proof}
\noindent The reduction function $g$ is computable in polynomial time and is dependent on the input size of the \ESAT instance $\big|[L,\mc C]\big|$,
as for each clause in $\mc C$ we create a fixed number of vertices in the new instance. Therefore $\ESAT\leq \MDS$ holds.

\paragraph{SSP \& Parsimonious}
As shown in the reduction above, each dominating set in $\mc S_\MDS$ contains only literal vertices. This means there is a direct 
one-to-one correspondence between the two solution sets, which makes the reduction SSP and SPR according to \lemref{lem:SSP and SPR}.
\begin{proof} Let $\Sall = \Slink = \emptyset$. Also let 
    $$\Snev:= \big\{C_j\mid j\in \big[|\mc C|\big]\big\} \cup \big\{x_{\ell_i^y}\mid i\in \big[|L^2|\big], y\in \{1,2\}\big\},$$
    as neither the clause vertices, nor the new vertices $x_{\ell_i^y}$ can be in any dominating set, due to $k$ being restricted.
    With the functions $f_I:L\rightarrow V, f(\cee) = \ell$, where $\cee$ denotes a literal $\ell\in L$ and $\ell$ the corresponding
    vertex in $V$, we get
    \begin{flalign*}
        \{f(S):S\in \mc S_\ESATu\} &= \big\{\{f(\ell)\mid \ell\in S\} :S\in \mc S_\ESATu\big\} \\
        &= \big\{S'\backslash \Snev : S'\in \mc S_\MDS\big\} \\
        &= \{S'\cap f(L) : S'\in \mc S_\MDS\}
    \end{flalign*}
    Thus, as $f_I':L\rightarrow V\backslash \Snev$ is bijective, \ESAT $\SSPR$ \MDS holds.
\end{proof}

\subsubsection[... to Clique]{Exact 3-Satisfiability to Clique}
\label{sec:ESAT to CQ}
\paragraph{Problem Definitions}
\subfile{../../Problems/CQ.tex}

\paragraph{Reduction}
This reduction is based on an idea by Lance Fortnow and William Gasarch \cite{sharpclique}. For each clause $C_j=\{\ell_1,\ell_2,\ell_3\}$ 
in $\mc C$ we create $7\cdot 2^{|L^2|-3}$ vertices, each representing one of the satisfying assignments for this clause. This number stems
from first looking at the $2^3=8$ possible assignments for the three literals in the clause and removing the one assignment which does not
satisfy the clause ($8-1=7$). Now every literal not in this clause can be assigned to be either positive or negative and for these we can
assign every possible one of the $2^{|L^2|-3}$ combinations in addition to our previous satisfying assignments. Thus we get 
$7\cdot 2^{|L^2|-3}$ possible satisfying assingments for every \ESAT clause.
We now connect every pair of identical assignments for different clauses via an edge and finally set $k$ to $k:=|\mc C|$.\\
To showcase how the reduction works, but also to keep it legible, we use two examples. \autoref*{fig:ESAT to CQ} shows the reduction on
a \ESAT instance with two clauses, but four different literal pairs whereas \autoref*{fig:ESAT to CQ'} showcases this with three clauses, but
only three literal pairs.
\begin{enumerate}
    \item Let $\mc C = \{\{\ell_1,\ell_2, \ell_3\},\{\ne\ell_1,\ne\ell_2, \ell_4\}\}$ and 
    $L=\{\ell_1, \ell_2, \ell_3,\ell_4, \ne\ell_1, \ne\ell_2, \ne\ell_3,\ne\ell_4\}$ define a \textsc{E3S} 
    instance, then \autoref*{fig:ESAT to CQ} shows the corresponding graph $G$ of the \CQ instance with $k=2$.
    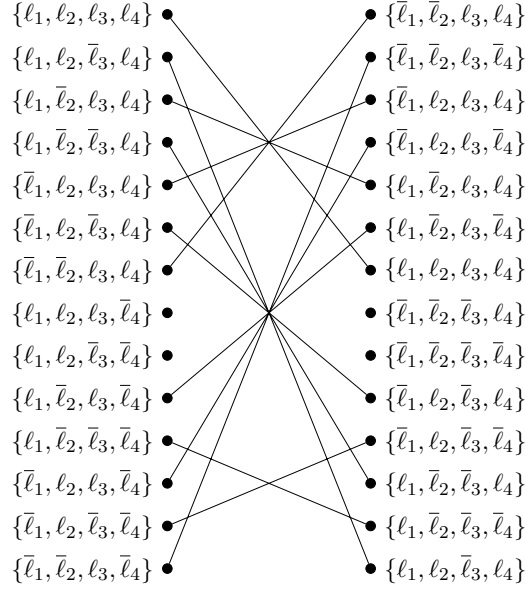
\begin{figure}[ht]
        \begin{center}
        \scalebox{.85}{
            \begin{tikzpicture}[vertex/.style = {draw, circle, fill, inner sep = 1.5}, node distance = 1.5cm]
                \node[vertex, label =left:$\{\ell_1\komma\ell_2\komma\ell_3\komma\ell_4\}$](c11){};
                \node[vertex, below = .5cm of c11, label =left:$\{\ell_1\komma\ell_2\komma\ne\ell_3\komma\ell_4\}$](c12){};
                \node[vertex, below = .5cm of c12, label =left:$\{\ell_1\komma\ne\ell_2\komma\ell_3\komma\ell_4\}$](c13){};
                \node[vertex, below = .5cm of c13, label =left:$\{\ell_1\komma\ne\ell_2\komma\ne\ell_3\komma\ell_4\}$](c14){};
                \node[vertex, below = .5cm of c14, label =left:$\{\ne\ell_1\komma\ell_2\komma\ell_3\komma\ell_4\}$](c15){};
                \node[vertex, below = .5cm of c15, label =left:$\{\ne\ell_1\komma\ell_2\komma\ne\ell_3\komma\ell_4\}$](c16){};
                \node[vertex, below = .5cm of c16, label =left:$\{\ne\ell_1\komma\ne\ell_2\komma\ell_3\komma\ell_4\}$](c17){};
                \node[vertex, below = .5cm of c17, label =left:$\{\ell_1\komma\ell_2\komma\ell_3\komma\ne\ell_4\}$](c18){};
                \node[vertex, below = .5cm of c18, label =left:$\{\ell_1\komma\ell_2\komma\ne\ell_3\komma\ne\ell_4\}$](c19){};
                \node[vertex, below = .5cm of c19, label =left:$\{\ell_1\komma\ne\ell_2\komma\ell_3\komma\ne\ell_4\}$](c110){};
                \node[vertex, below = .5cm of c110, label =left:$\{\ell_1\komma\ne\ell_2\komma\ne\ell_3\komma\ne\ell_4\}$](c111){};
                \node[vertex, below = .5cm of c111, label =left:$\{\ne\ell_1\komma\ell_2\komma\ell_3\komma\ne\ell_4\}$](c112){};
                \node[vertex, below = .5cm of c112, label =left:$\{\ne\ell_1\komma\ell_2\komma\ne\ell_3\komma\ne\ell_4\}$](c113){};
                \node[vertex, below = .5cm of c113, label =left:$\{\ne\ell_1\komma\ne\ell_2\komma\ell_3\komma\ne\ell_4\}$](c114){};
    
                \node[vertex, right = 3cm of c11, label =right:$\{\ne\ell_1\komma\ne\ell_2\komma\ell_3\komma\ell_4\}$](c21){};
                \node[vertex, below = .5cm of c21, label =right:$\{\ne\ell_1\komma\ne\ell_2\komma\ell_3\komma\ne\ell_4\}$](c22){};
                \node[vertex, below = .5cm of c22, label =right:$\{\ne\ell_1\komma\ell_2\komma\ell_3\komma\ell_4\}$](c23){};
                \node[vertex, below = .5cm of c23, label =right:$\{\ne\ell_1\komma\ell_2\komma\ell_3\komma\ne\ell_4\}$](c24){};
                \node[vertex, below = .5cm of c24, label =right:$\{\ell_1\komma\ne\ell_2\komma\ell_3\komma\ell_4\}$](c25){};
                \node[vertex, below = .5cm of c25, label =right:$\{\ell_1\komma\ne\ell_2\komma\ell_3\komma\ne\ell_4\}$](c26){};
                \node[vertex, below = .5cm of c26, label =right:$\{\ell_1\komma\ell_2\komma\ell_3\komma\ell_4\}$](c27){};
                \node[vertex, below = .5cm of c27, label =right:$\{\ne\ell_1\komma\ne\ell_2\komma\ne\ell_3\komma\ell_4\}$](c28){};
                \node[vertex, below = .5cm of c28, label =right:$\{\ne\ell_1\komma\ne\ell_2\komma\ne\ell_3\komma\ne\ell_4\}$](c29){};
                \node[vertex, below = .5cm of c29, label =right:$\{\ne\ell_1\komma\ell_2\komma\ne\ell_3\komma\ell_4\}$](c210){};
                \node[vertex, below = .5cm of c210, label =right:$\{\ne\ell_1\komma\ell_2\komma\ne\ell_3\komma\ne\ell_4\}$](c211){};
                \node[vertex, below = .5cm of c211, label =right:$\{\ell_1\komma\ne\ell_2\komma\ne\ell_3\komma\ell_4\}$](c212){};
                \node[vertex, below = .5cm of c212, label =right:$\{\ell_1\komma\ne\ell_2\komma\ne\ell_3\komma\ne\ell_4\}$](c213){};
                \node[vertex, below = .5cm of c213, label =right:$\{\ell_1\komma\ell_2\komma\ne\ell_3\komma\ell_4\}$](c214){};
    
                \draw   (c11) -- (c27)
                        (c12) -- (c214)
                        (c13) -- (c25)
                        (c14) -- (c212)
                        (c15) -- (c23)
                        (c16) -- (c210)
                        (c17) -- (c21)
                        (c110) -- (c26)
                        (c111) -- (c213)
                        (c112) -- (c24)
                        (c113) -- (c211)
                        (c114) -- (c22)
                ;
            \end{tikzpicture} 
        }\\
        \caption{\footnotesize{Example 1 \textsc{Exact 3-Satisfiability} to \textsc{Clique}}}
        \label{fig:ESAT to CQ}
        \end{center}
    \end{figure}

    \item Let $\mc C = \{\{\ell_1,\ell_2, \ell_3\},\{\ne\ell_1,\ne\ell_2, \ne\ell_3\},\{\ne\ell_1,\ne\ell_2, \ell_3\}\}$ and 
    $L=\{\ell_1, \ell_2, \ell_3, \ne\ell_1, \ne\ell_2, \ne\ell_3\}$ define a \textsc{E3S} 
    instance, then \autoref*{fig:ESAT to CQ'} shows the corresponding graph $G$ of the \CQ instance with $k=3$.
    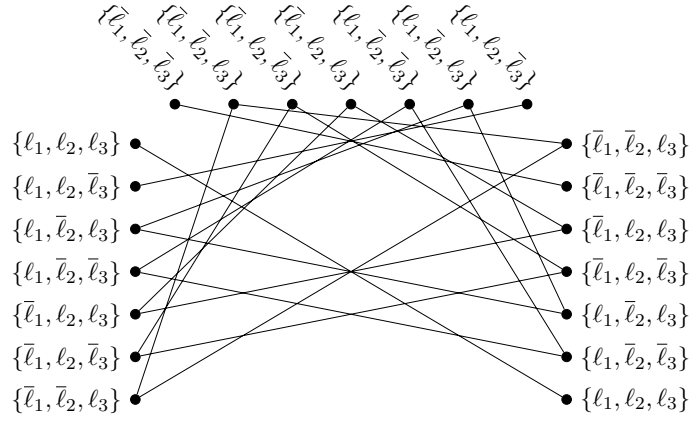
\begin{figure}[ht]
        \begin{center}
        \scalebox{.85}{
            \begin{tikzpicture}[vertex/.style = {draw, circle, fill, inner sep = 1.5}, node distance = 1.5cm]
                \node[vertex, label =left:$\{\ell_1\komma\ell_2\komma\ell_3\}$](c11){};
                \node[vertex, below = .5cm of c11, label =left:$\{\ell_1\komma\ell_2\komma\ne\ell_3\}$](c12){};
                \node[vertex, below = .5cm of c12, label =left:$\{\ell_1\komma\ne\ell_2\komma\ell_3\}$](c13){};
                \node[vertex, below = .5cm of c13, label =left:$\{\ell_1\komma\ne\ell_2\komma\ne\ell_3\}$](c14){};
                \node[vertex, below = .5cm of c14, label =left:$\{\ne\ell_1\komma\ell_2\komma\ell_3\}$](c15){};
                \node[vertex, below = .5cm of c15, label =left:$\{\ne\ell_1\komma\ell_2\komma\ne\ell_3\}$](c16){};
                \node[vertex, below = .5cm of c16, label =left:$\{\ne\ell_1\komma\ne\ell_2\komma\ell_3\}$](c17){};
    
                \node[vertex, above right = .5cm and .5cm of c11, label ={[label distance=-.1cm, rotate=-45]above left:$\{\ne\ell_1\komma\ne\ell_2\komma\ne\ell_3\}$}](c31){};
                \node[vertex, right = .75cm of c31, label ={[label distance=-.1cm, rotate=-45]above left:$\{\ne\ell_1\komma\ne\ell_2\komma\ell_3\}$}](c32){};
                \node[vertex, right = .75cm of c32, label ={[label distance=-.1cm, rotate=-45]above left:$\{\ne\ell_1\komma\ell_2\komma\ne\ell_3\}$}](c33){};
                \node[vertex, right = .75cm of c33, label ={[label distance=-.1cm, rotate=-45]above left:$\{\ne\ell_1\komma\ell_2\komma\ell_3\}$}](c34){};
                \node[vertex, right = .75cm of c34, label ={[label distance=-.1cm, rotate=-45]above left:$\{\ell_1\komma\ne\ell_2\komma\ne\ell_3\}$}](c35){};
                \node[vertex, right = .75cm of c35, label ={[label distance=-.1cm, rotate=-45]above left:$\{\ell_1\komma\ne\ell_2\komma\ell_3\}$}](c36){};
                \node[vertex, right = .75cm of c36, label ={[label distance=-.1cm, rotate=-45]above left:$\{\ell_1\komma\ell_2\komma\ne\ell_3\}$}](c37){};
                
                \node[vertex, below right = .5cm and .5cm of c37, label =right:$\{\ne\ell_1\komma\ne\ell_2\komma\ell_3\}$](c21){};
                \node[vertex, below = .5cm of c21, label =right:$\{\ne\ell_1\komma\ne\ell_2\komma\ne\ell_3\}$](c22){};
                \node[vertex, below = .5cm of c22, label =right:$\{\ne\ell_1\komma\ell_2\komma\ell_3\}$](c23){};
                \node[vertex, below = .5cm of c23, label =right:$\{\ne\ell_1\komma\ell_2\komma\ne\ell_3\}$](c24){};
                \node[vertex, below = .5cm of c24, label =right:$\{\ell_1\komma\ne\ell_2\komma\ell_3\}$](c25){};
                \node[vertex, below = .5cm of c25, label =right:$\{\ell_1\komma\ne\ell_2\komma\ne\ell_3\}$](c26){};
                \node[vertex, below = .5cm of c26, label =right:$\{\ell_1\komma\ell_2\komma\ell_3\}$](c27){};
    
                \draw   (c11) -- (c27)
                        (c13) -- (c25)
                        (c14) -- (c26)
                        (c15) -- (c23)
                        (c16) -- (c24)
                        (c17) -- (c21)

                        (c12) -- (c37)
                        (c13) -- (c36)
                        (c14) -- (c35)
                        (c15) -- (c34)
                        (c16) -- (c33)
                        (c17) -- (c32)

                        (c21) -- (c32)
                        (c22) -- (c31)
                        (c23) -- (c34)
                        (c24) -- (c33)
                        (c25) -- (c36)
                        (c26) -- (c35)
                ;
            \end{tikzpicture} 
        }\\
        \caption{\footnotesize{Example 2 \textsc{Exact 3-Satisfiability} to \textsc{Clique}}}
        \label{fig:ESAT to CQ'}
        \end{center}
    \end{figure}
\end{enumerate}
\begin{proof}
    Now every satisfying assignment $S\in \mc S_\ESATu$ is equivalent to a clique in $G$, consisting of $|\mc C|$ vertices labeled 
    with this assignment.\\
    Let $S\in \mc S_\ESATu$ be a satisfying assignment of a \ESAT instance. Then this assignment would satisfy every clause in $\mc C$ 
    and there would therefore be a vertex with this assignment in every vertex cluster corresponding to a clause. As this assignment 
    is the same for each vertex, they would all be interconnected via edges and would therefore be a clique.\\
    Let $S'\in \mc S_\CQ$ be a clique in the \CQ instance $G$. Then this clique would connect $k:=|\mc C|$ different vertices with the
    same label. As each label corresponds to an assignment satisfying one of the $|\mc C|$ clauses, this assignment satisfies all clauses
    and is therefore a satisfying assignment in the \ESAT instance.
\end{proof}
\noindent The reduction function $g$ is computable in polynomial time and is dependent on the input size of the \ESAT instance $\big|[L,\mc C]\big|$,
as for each clause in $\mc C$ we create a fixed number of vertices in the new instance. Therefore $\ESAT\leq \CQ$ holds.

\paragraph{SSP}
This reduction is not SSP, as the universe $L$ of the \ESAT instance cannot be clearly mapped to the vertices of the \CQ instance, as these
represent the possible solutions for each clause in which every literal appears multiple times each.

\paragraph{Parsimonious}
As seen above there is a direct one-to-one correspondence between the satisfying assignments and the cliques in which each 
clique is made up of $|\mc C|$ vertices labeled with the satisfying assignment. This means we can define the following 
functions $p_I$ to be a bijective function between the two solution sets and therefore we prove that this reduction is strongly Parsimonious. 
\begin{proof}
    Let 
    \begin{flalign*}
        p_I:&\mc S_\ESATu \rightarrow \mc S_\CQ,~p(S)=\bigcup\nolimits_{|\mc C|}\{S\}\\
        p_I^{-1}:&\mc S_\CQ \rightarrow \mc S_\ESATu,~p(S') = S'[1]
    \end{flalign*}
    where $S'[i]$ denotes the $i^{th}$ element in set $S'$, then
    \begin{flalign*}
        \{p(S):S\in \mc S_\ESATu\} &= \Big\{\bigcup\nolimits_{|\mc C|}\{S\} :S\in \mc S_\ESATu\Big\}= \{S':S'\in \mc S_\CQ\}\\
        \text{and }\{p_1(S'):\Sr\CQ\} &= \{S'[1] :\Sr\CQ\}= \{S:\S\ESATu\}
    \end{flalign*}
    and thus \ESAT $\SPR$ \CQ holds.
\end{proof}

\paragraph{Alternative transitive reduction with both SSP \& SPR properties}
We can use the transitive property of SSP reductions to show that a reduction between these problems that is both SSP and strongly 
Parsimonious does indeed exist. As we have shown in \autoref*{sec:ESAT to MIS} there is a transitive SSP and SPR reduction 
$\ESAT\SSPR\MIS$ via $\OSAT$. Using this and the reduction $\MIS\SSPR\CQ$ shown in \autoref*{sec:MIS to CQ}, we get 
$\ESAT\SSPR\MIS\SSPR\CQ$ and therefore $\ESAT\SSPR\CQ$.

\subsubsection[... to Subset Sum]{Exact 3-Satisfiability to Subset Sum}
\label{sec:ESAT to SS}
\paragraph{Problem Definition}
\subfile{../../Problems/SS.tex}

\paragraph{Reduction}
The reduction \ESAT $\leq$ \SS is based on a reduction by Sipser \cite{Sip97}, where the idea is to transform the 
literals and clauses into specific bits in a base-2 number. Each of these numbers has exactly $|L^2|\cdot 3|\mc C|$ bits. 
Now, each literal pair $(\ell_i, \ne\ell_i)\in L^2$ is transformed 
into a base-2 number ($a_i$ and $a_{\ne i}$), and we create numbers $a_{C_j^1}$ and $a_{C_j^2}$ for each clause $C_j\in \mc C$ as well as a number $M$ 
(the target value). The first $|L^2|$ bits of the numbers $a_i$ and $a_{\ne i}$ denote 
which literal pair $(\ell_i,\ne\ell_i)$ they correspond to.
This is done by setting the same bit to '1' for both of them. We also set this bit to '1' in the target value $M$, 
meaning that any subset of numbers adding up to $M$ can only include either $a_i$ or $a_{\ne i}$.
Each clause $C_j\in\mc C$ also has a dedicated set of three bits each in every number $a_x$ (with $x\in\{i,\ne i\}$), which get set to '001'
if the literal corresponding to $a_x$ is in the clause $C_j$ and to '000' otherwise. In the numbers $a_{C_j^1}$ and $a_{C_j^2}$ these bits get set 
to 001 and 010 respectively and in the target value $M$ the corresponding bits are set to 100. Now, to get a subset of numbers to add up to $100_2=4_{10}$,
at least one literal per clause needs to be in the subset, as the bits in $a_{C_j^1}$ and $a_{C_j^2}$ only add up to $011_2=3_{10}$.
This way we compartmentalize the target value $M$ into sets of bits, which need to be considered seperatly, as any bit overflows would 
result in the sum being larger than $M$. Formally the reduction works as follows.\\
Let $[L,\mc C]$ be a \ESAT instance, then 
$[A,M]$ is the corresponding \SS instance when we define the universe and target values as follows. 
First, let $|A|:=|L|+2\cdot|\mc C|$. Then according to \autoref{tab:ESAT to SS table} we transform 
each literal pair $(\ell_i,\ne\ell_i)$ into two binary numbers $a_i\in A$ and $a_{\ne i}\in A$ with $|L^2|+3\cdot |\mc C|$ digits each
and set $M:=1^{|L^2|}(100)^{(|\mc C|)}$. We also create the two numbers $a_{C_j^1}\in A$ and $a_{C_j^2}\in A$ for every clause $C_j$ and 
set the values in their column to $001$ and $010$ respectively. An example of this reduction is depicted in \autoref*{fig:ESAT to SS}.
\begin{table}[h!]
    \begin{center}
        \begin{adjustbox}{max width=\textwidth}
        \begin{tabular}{c|||c|c|c|c||c|c|c|c}
            & $\frac{\ell_1}{\ne\ell_1}$ & $\hdots$ &$\frac{\ell_{i}}{\ne\ell_{i}}$ & $\hdots$ & $C_1$ & $\hdots$ & $C_{j}$ & $\hdots$\\[5pt]
            \hline
            \hline
            \hline 
            \vdots & \vdots && \vdots & & \vdots & & \vdots & \\
            \hline 
            $a_i $ & 0 & $\hdots$ & 1 & $\hdots$ &
            $\begin{cases}001, & \ell_i\in C_1\\000, & \ell_i\notin C_1\end{cases}$& $\hdots$ &
            $\begin{cases}001, & \ell_i\in C_j\\000, & \ell_i\notin C_j\end{cases}$& $\hdots$ \\
            \hline 
            $a_{\ne i} $ & 0 & $\hdots$ & 1 & $\hdots$ & 
            $\begin{cases}001, & \ne\ell_i\in C_1\\000, & \ne\ell_i\notin C_1\end{cases}$& $\hdots$ &
            $\begin{cases}001, & \ne\ell_i\in C_j\\000, & \ne\ell_i\notin C_j\end{cases}$& $\hdots$ \\
            \hline
            \vdots & \vdots && \vdots & & \vdots & & \vdots & \\
            \hline
            \hline 
            \vdots & \vdots && \vdots & & \vdots & & \vdots & \\
            \hline
            $a_{C_j^1}$ & 0 & $\hdots$ & 0 & $\hdots$ & 000 & $\hdots$ & 001 & $\hdots$ \\[2pt]
            \hline
            $a_{C_j^2}$ & 0 & $\hdots$ & 0 & $\hdots$ & 000 & $\hdots$ & 010 & $\hdots$ \\[2pt]
            \hline
            \vdots & \vdots && \vdots & & \vdots & & \vdots & \\
            \hline
            \hline 
            \hline 
            $M$ & 1 & $\hdots$ & 1 & $\hdots$ & 100 & $\hdots$ & 100 & $\hdots$ \\
        \end{tabular}
        \end{adjustbox}
        \caption{\footnotesize{Converting \textsc{Exact 3-Satisfiability} to \textsc{Subset Sum}}}
        \label{tab:ESAT to SS table}
    \end{center}
\end{table}
\begin{figure}[ht]
    \begin{center}
        $$\ESAT:=\big[\{\{\ne\ell_1,\ne\ell_2,\ell_3\},\{\ell_1,\ne\ell_2,\ne\ell_3\}\},\{\ell_1,\ell_2,\ell_3,\ne\ell_1,\ne\ell_2,\ne\ell_3\}\big]$$\\
        gets mapped to
        \begin{flalign*}
            \SS:=\big[\{a_1=1~0~0~000~001,\hspace*{13.75pt}a_{\ne 1}=1~0~0~001~000&,\\
            a_2=0~1~0~000~000,\hspace*{13.75pt}a_{\ne 2}=0~1~0~001~001&,\\
            a_3=0~0~1~001~000,\hspace*{13.75pt}a_{\ne 3}=0~0~1~000~001&,\\
            a_{C_1^1}=0~0~0~001~000,~~a_{C_1^2}=0~0~0~010~000&,\\
            a_{C_2^1}=0~0~0~000~001,~~a_{C_2^2}=0~0~0~000~010&\},\\
            M=1~1~1~100~100&\big]
        \end{flalign*}
    \end{center}
    \caption{\footnotesize{Example \textsc{Exact 3-Satisfiability} to \textsc{Subset Sum}}}
    \label{fig:ESAT to SS}
\end{figure}
\begin{proof}
    Let $S\in \mc S_\ESATu$ be a solution for an \ESAT instance, then exactly one of each literal pair $(\ell_i, \ne\ell_i)$ is included
    in $S$. This means that either $a_i\in S'$ or $a_{\ne i}\in S'$ for an $S'\in \mc S_\SS$ and therefore the corresponding bits sum 
    up to 1. And as every clause $C_j$ is satisfied in $S$ the target value of $100$ can be reached in every clause column. 
    Either $a_{C_j^1},a_{C_j^2}$ or both are also in $S'$, depending on how many of the three literals in $C_j$ are included in $S$.
    This means that any satisfying assignment for a \ESAT instance has a corresponding satisfying subset in \SS.\\
    Let $S'$ be a solution for \SS, then each set of bits is satisfied independently, as otherwise there would be a bit overflow and
    the sum would be larger than the target value. As each set of bits adds up to the corresponding set of bits in the target value
    independently, exactly one of each pair $a_i$ or $a_{\ne i}$ is in $S'$. As the sets of bits in the target value corresponding
    to the clauses all sum up to 100, at least one of the numbers corresponding to the literals in the clause is in $S'$.
    The set of literals corresponding to the numbers in $S'$ is therefore a satisfying assignment in $\mc S_\ESATu$.
\end{proof}
\noindent The reduction function $g$ is computable in polynomial time and is dependent on the input size of the \ESAT instance $\big|[L,\mc C]\big|$,
as for each clause in $\mc C$ we create a fixed number of numbers $a\in \mc A$ in the new instance. Therefore $\ESAT\leq \SS$ holds.

\paragraph{SSP \& Parsimonious}
As seen in the proof above, any solution $S\in \mc S_\ESATu$ is equivalent to the solution $S'\in \mc S_\SS$, where for every $\ell_i\in
S$ we have $a_i\in S'$ and $S'$ also includes one or both of the numbers $a_{C_j}$ for each clause $C_j\in \mc C$. This means we can apply 
\lemref{theorem 1}.
\begin{proof}
    We define the functions $$f_I:L\rightarrow A,~f(\cee_i)=
    \begin{cases}
        a_i, & \cee_i \in L^+ \\
        a_{\ne i}, &\cee_i \in L^-
    \end{cases}$$ 
    Let now $\Snev:=\emptyset$ and $\Srep:=\bigcup_{\ell\in L}f(\ell)$. Each set $\mf R_{S'}$ of numbers $a_i$ and $a_{\ne i}$ 
    directly imply a solution $\Sr\SS$, as the bits in each of the clause columns add up to either one, two or three, depending
    on how many literals satisfy the clause in the corresponding \ESAT solution. If the numbers in column $C_j$ add up to three, 
    then $a_{C_j^1}$ needs to be added to $S'$, if they add up to two, $a_{C_j^2}$, and if they only add up to one, then both numbers need
    to be in the solution. Therefore let 
    $$\Slink:=\bigcup_{\Sr{\SS}} \mf L\big(\mf R_{S'}\big)=\big\{C_j^y\mid j\in \big[|\mc C|\big], y\in\{1,2\}\big\}$$
    And thus $A= \Srep\cupdot\Slink$ with the bijective functions $(r_I)_{I\in\mc I}$ being 
    $$r_I:=\Srep\rightarrow L, ~r(x)=f^{-1}(x)$$
    and so 
    $\ESAT \SSPR \SS$, according to \lemref{theorem 1}.
\end{proof}

\subsubsection[... to Exact Cover]{Exact 3-Satisfiability to Exact Cover}
\label{sec:ESAT to EC}
\paragraph{Problem Definition}
\subfile{../../Problems/EC.tex}

\paragraph{Reduction}
The reduction presented in \cite[p. 58-61]{Femke} is already both SSP and SPR and, as that thesis includes the whole reduction, 
proof and an example, we only briefly introduce the idea behind the reduction here.\\
Let $\ESAT:=[L, \mc C]$, then the \EC instance is created by making sets $\psi_{2i}$ and $\psi_{2i-1}$ for every pair of literals
$(\ell_i,\ne\ell_i)\in L^2$ that both consist of new variables $\ell_i'$ and $Z_{h,j}$, denoting which clauses 
$C_j$ the literals corresponding to the sets occur in at position $h\in[3]$. We also create seven new sets 
$\psi_{|L|+j\cdot k}$ with $1\leq k \leq 7$ for every clause $C_j\in\mc C$. These sets each correspond to exactly one satisfying 
assignment of the clause.

\paragraph{SSP \& Parsimonious}
This reduction is both SSP and Parsimonious, as each \ESAT solution corresponds to exactly one solution in $\mc S_\EC$, 
where the literals in the satisfying assignment are mapped to the sets representing them, and in addition to this each 
satisfied clause is represented by a set. Therefore we
can use \lemref{theorem 1} as follows.
\begin{proof}
    Let the functions $f_I$ be defined as
    $$f_I:L\rightarrow \{\psi_1,...,\psi_n\},~f(\cee_i)=\begin{cases}
        \psi_{2i}, & \cee_i\in L^+ \\
        \psi_{2i-1}, & \cee_i\in L^-
    \end{cases}$$
    and let $\Sall=\Snev=\emptyset$ and 
    $$\Slink :=\bigcup\nolimits\slink = \big\{\psi_i~\big|~ |L| < i \leq n\big\},$$
    with the satisfying assignments $\S\ESATu$ each being linked to the unique set 
    $$\slink\subseteq\{\psi_{|L|+1},...,\psi_{|L|+7|\mc C|}\}$$ representing that assignment.
    With this we get
    \begin{flalign*}
        \{f(S):\S\ESATu\}&=\big\{\{f(\ell)\mid\cee\in S\}:\S\ESATu\}\\
        &=\big\{\{\psi\in S'\}\backslash \Slink:\Sr\EC\big\}\\
        &=\{S'\backslash \Slink:\Sr\EC\}\\
        &=\{S'\cap f(L) :\Sr\EC\}
    \end{flalign*}
    and therefore $\ESAT\SSPR\EC$ holds.
\end{proof}

\subsubsection[... to Directed Hamiltonian Path]{Exact 3-Satisfiability to Directed Hamiltonian Path}
\label{sec:ESAT to DHP}
\paragraph{Problem Definition}
\subfile{../../Problems/DHP.tex}

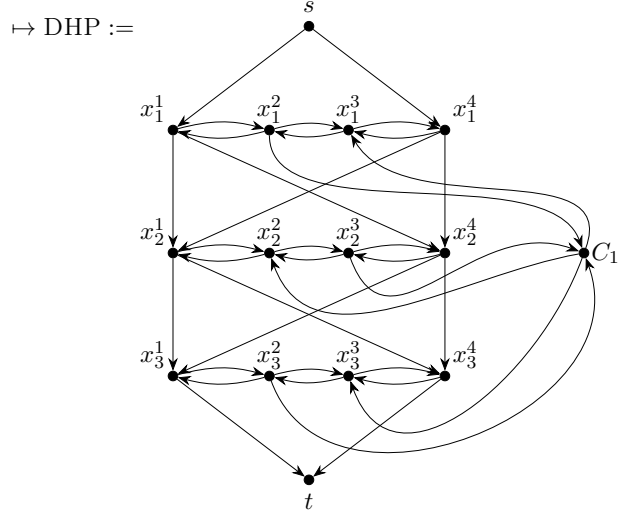
\begin{figure}[ht]
    \begin{center}
    \scalebox{.85}{
        \begin{tikzpicture}[vertex/.style = {draw, circle, fill, inner sep = 1.5}, node distance = 1.5cm, label distance=-.1cm]
            \node[vertex, label = {[label distance=0cm] above:$s$}](s){};
            \node[above = 1cm of s](SAT){$\ESAT:=\big(\mc C = \{\{\ell_1, \ne \ell_2, \ell_3\}\}, 
            L=\{\ell_1, \ell_2, \ell_3, \ne\ell_1, \ne\ell_2, \ne\ell_3\}\big)$};

            \node[vertex, below left = 1.5cm and 2cm of s, label = above left:$x_1^1$](x11){};
            \node[vertex, below left = 1.5cm and .5cm of s, label = above:$x_1^2$](x12){};
            \node[vertex, below right = 1.5cm and .5cm of s, label = above:$x_1^3$](x13){};
            \node[vertex, below right = 1.5cm and 2cm of s, label = above right:$x_1^4$](x14){};
            
            \node[vertex, below = 1.75cm of x11, label = above left:$x_2^1$](x21){};
            \node[vertex, below = 1.75cm of x12, label = above:$x_2^2$](x22){};
            \node[vertex, below = 1.75cm of x13, label = above:$x_2^3$](x23){};
            \node[vertex, below = 1.75cm of x14, label = above right:$x_2^4$](x24){};
            
            \node[vertex, below = 1.75cm of x21, label = above left:$x_3^1$](x31){};
            \node[vertex, below = 1.75cm of x22, label = above:$x_3^2$](x32){};
            \node[vertex, below = 1.75cm of x23, label = above:$x_3^3$](x33){};
            \node[vertex, below = 1.75cm of x24, label = above right:$x_3^4$](x34){};
            
            \node[vertex, below right = 1.5cm and 2cm of x31, label = {[label distance=0cm] below:$t$}](t){};

            \node[vertex, right = 2 cm of x24, label = right:$C_1$](C1){};

            \node[left = 2.5cm of s](GHP){$\mapsto\DHP:=$};

            \draw   (s) edge[\xarrow] (x11)
                    (s) edge[\xarrow] (x14)

                    (x11) edge[bend left = 15, \xarrow] (x12)
                    (x12) edge[bend left = 15, \xarrow] (x13)
                    (x13) edge[bend left = 15, \xarrow] (x14)
                    (x12) edge[bend left = 15, \xarrow] (x11)
                    (x13) edge[bend left = 15, \xarrow] (x12)
                    (x14) edge[bend left = 15, \xarrow] (x13)

                    (x11) edge[\xarrow] (x21)
                    (x14) edge[\xarrow] (x21)
                    (x11) edge[\xarrow] (x24)
                    (x14) edge[\xarrow] (x24)
                    
                    (x21) edge[bend left = 15, \xarrow] (x22)
                    (x22) edge[bend left = 15, \xarrow] (x23)
                    (x23) edge[bend left = 15, \xarrow] (x24)
                    (x22) edge[bend left = 15, \xarrow] (x21)
                    (x23) edge[bend left = 15, \xarrow] (x22)
                    (x24) edge[bend left = 15, \xarrow] (x23)

                    (x21) edge[\xarrow] (x31)
                    (x24) edge[\xarrow] (x31)
                    (x21) edge[\xarrow] (x34)
                    (x24) edge[\xarrow] (x34)
                    
                    (x31) edge[bend left = 15, \xarrow] (x32)
                    (x32) edge[bend left = 15, \xarrow] (x33)
                    (x33) edge[bend left = 15, \xarrow] (x34)
                    (x32) edge[bend left = 15, \xarrow] (x31)
                    (x33) edge[bend left = 15, \xarrow] (x32)
                    (x34) edge[bend left = 15, \xarrow] (x33)

                    (x31) edge[\xarrow] (t)
                    (x34) edge[\xarrow] (t)

                    (x12) edge[out=-90, in=110, looseness=.75, \xarrow] (C1)
                    (C1) edge[out=70, in=-60, \xarrow] (x13)
                    
                    (x23) edge[out=-70, in=160, looseness=1.25, \xarrow] (C1)
                    (C1) edge[out=190, in=-70, \xarrow] (x22)

                    (x32) edge[out=-70, in=-70, looseness=1.25, \xarrow] (C1)
                    (C1) edge[out=-110, in=-70, looseness=1.25, \xarrow] (x33)
            ;
        \end{tikzpicture} 
    }\\
    \caption{\footnotesize{Example \textsc{Exact 3-Satisfiability} to \textsc{Directed Hamiltonian Path}}}
    \label{fig:ESAT to DHP}
    \end{center}
\end{figure}

\paragraph{Reduction}
This reduction is based on a reduction by from Arora and Barak \cite{AB09}. Let $\ESAT:=[L,\mc C]$ then we transform this instance 
into a \DHP instance $[G=(V,A),s,t]$ with the following steps. First we introduce the two vertices $s,t\in V$. Next, for
each literal pair $(\ell_i,\ne\ell_i)\in L^2$ we create $m:=4\cdot |\mc C|$ new vertices and denote them as $x_i^1,...,x_i^{m}$. We also 
create arcs $(x_i^k,x_i^{k+1})$ and $(x_i^{k+1},x_i^k)$ for all $1\leq k <m$. This way there is a path from $x_i^1$ to $x_i^{m}$ 
and back again for every literal pair. In addition to this we connect $(s, x_1^1)$ and $(s, x_1^m),~(x_{|L^2|}^1, t)$ and 
$(x_{|L^2|}^{m}, t)$ and also $(x^1_i, x_{i+1}^1)$, $(x^m_i, x_{i+1}^m)$, $(x^1_i, x_{i+1}^m)$ and $(x_i^m , x_{i+1}^1)$ for 
all $1\leq i<|L^2|$. Finally we create a new 
vertex $C_j$ for every clause $C_j\in\mc C$ and connect these via the following arcs. If the literal $\ell_i$ appears in the 
clause $C_j$ we create the arcs $(x_i^{4j-2},C_j)$ and $(C_j , x_i^{4j-1})$ otherwise, if $\ne\ell_i\in C_j$ then we create 
the arcs $(x_i^{4j-1},C_j)$ and $(C_j , x_i^{4j-2})$.\\
The idea is that now, if a directed Hamiltonian path exist in this graph, it includes exactly one of the directions from 
$x_i^1$ to $x_i^{m}$ or back for every literal pair $(\ell_i,\ne\ell_i)$ and we interpret the direction $x_i^1$ to $x_i^{m}$
as $\ell_i$ and $x_i^m$ to $x_i^{1}$ as $\ne\ell_i$. \autoref*{fig:ESAT to DHP} shows an example reduction.

\begin{proof}
    Let there be an ordered \ESAT solution $\S\ESATu$. Then there is an equivalent \DHP solution $S'$, where for each
    literal pair $\ell_i$ or $\ne\ell_i$ in $S$ the path goes either from $x_i^1$ to $x_i^m$, if $\ell_i\in S$, or the other
    way, if $\ne\ell_i\in S$. For each clause $C_j$ in $\mc C$ every literal $\ell_i$ or $\ne\ell_i$ in $S$ that satisfies $C_j$ 
    extends the path over the clause vertex via the paths created over $x_i^{4j-2}$ and $x_i^{4j-1}$. As every clause is satisfied 
    in $S$ and also every literal pair is taken into consideration there is an equivalent path from $s$ to $t$ over every vertex.\\
    Now let $\Sr\DHP$ be a directed Hamiltonian path. Then every clause vertex is also considered. This means that there is an 
    equivalent solution $\S\ESATu$ with $\ell_i\in S$ if the path goes from $x_i^1$ to $x_i^m$ and $\ne\ell_i\in S$ if the path goes 
    from $x_i^m$ to $x_i^1$. Each literal can either appear positive or negated, but never both, as the path cannot go over the same
    vertex twice. As every set of vertices $x_i^k$ only has paths leading to either $x_{i+1}^1$, $x_{i+1}^m$ or $t$ each literal pair is 
    also considered in $S'$ and paths that go over a clause node cannot "jump down" this order, as previous vertices then cannot be 
    visited.
\end{proof}
\noindent The reduction function $g$ is computable in polynomial time and is dependent on the input size of the \ESAT instance $\big|[L,\mc C]\big|$,
as for each clause in $\mc C$ and each literal in $L$ we create a fixed number of vertices in the new instance. Therefore $\ESAT\leq \DHP$ holds.

\paragraph{SSP}
This reduction is SSP. The idea behind this is, that for each set of vertices $x_i^k$ the direction in which the path goes is indicated 
by checking if $(x_i^1,x_i^2)$ or $(x_i^2,x_i^1)$ is in the solution. This way we can define functions from the literals in $S$ to
the arcs in $S'$ as follows.
\begin{proof}
    Let
    $$f_I:L\rightarrow A,~f(\cee_i)=\begin{cases}
        (x_i^1,x_i^2), & \cee_i\in L^+ \\
        (x_i^2,x_i^1), & \cee_i\in L^-
    \end{cases}$$
    With this we get 
    \begin{flalign*}
        \{f(S):\S\ESATu\} & = \big\{\{f(\ell_i)\mid\cee_i\in S\}:\S\ESATu\big\}\\
        & = \Big\{\big\{\big((x_i^1,x_i^2)\xor(x_i^2,x_i^1)\big)\in S'\big\}:\Sr\DHP\Big\}\\
        &= \{S'\cap f(L):\Sr\DHP\}
    \end{flalign*}
    and so $\ESAT\SSP\DHP$ holds.
\end{proof}

\begin{figure}[ht]
    \begin{center}
    \scalebox{.85}{
        \begin{tikzpicture}[vertex/.style = {draw, circle, fill, inner sep = 1.5}, node distance = 1.5cm, label distance=-.1cm]
            \node[vertex, label = {[label distance=0cm] above:$s$}](s){};
            \node[above = 1cm of s](SAT){$\ESAT:=\big(\mc C = \{\{\ell_1, \ne \ell_2, \ell_3\}\}, 
            L=\{\ell_1, \ell_2, \ell_3, \ne\ell_1, \ne\ell_2, \ne\ell_3\}\big)$};

            \node[vertex, below left = 1.5cm and 2cm of s, label = above left:$x_1^1$](x11){};
            \node[vertex, below left = 1.5cm and .5cm of s, label = above:$x_1^2$](x12){};
            \node[vertex, below right = 1.5cm and .5cm of s, label = below:$x_1^3$](x13){};
            \node[vertex, below right = 1.5cm and 2cm of s, label = right:$x_1^4$](x14){};
            
            \node[vertex, below = 1.75cm of x11, label = above left:$x_2^1$](x21){};
            \node[vertex, below = 1.75cm of x12, label = above:$x_2^2$](x22){};
            \node[vertex, below = 1.75cm of x13, label = above:$x_2^3$](x23){};
            \node[vertex, below = 1.75cm of x14, label = above right:$x_2^4$](x24){};
            
            \node[vertex, below = 1.75cm of x21, label = above left:$x_3^1$](x31){};
            \node[vertex, below = 1.75cm of x22, label = above:$x_3^2$](x32){};
            \node[vertex, below = 1.75cm of x23, label = above:$x_3^3$](x33){};
            \node[vertex, below = 1.75cm of x24, label = above right:$x_3^4$](x34){};
            
            \node[vertex, below right = 1.5cm and 2cm of x31, label = {[label distance=0cm] below:$t$}](t){};

            \node[right = 3cm of x24](anchor){};

            \node[vertex, above = 0.25cm of anchor, label = above right:$c_1^1$](C1){};
            \node[vertex, below left = 0.35cm and 0.35cm of anchor, label = above right:$c_1^2$](C2){};
            \node[vertex, below right = 0.35cm and 0.35cm of anchor, label = right:$c_1^3$](C3){};

            \node[left = 2.5cm of s](GHP){$\mapsto\DHP:=$};

            \draw   (s) edge[\xarrow] (x11)
                    (s) edge[\xarrow] (x14)

                    (x11) edge[bend left = 15, \xarrow] (x12)
                    (x13) edge[bend left = 15, \xarrow] (x14)
                    (x12) edge[bend left = 15, \xarrow] (x11)
                    (x13) edge[bend left = 15, \xarrow] (x12)
                    (x14) edge[bend left = 15, \xarrow] (x13)

                    (x11) edge[\xarrow] (x21)
                    (x14) edge[\xarrow] (x21)
                    (x11) edge[\xarrow] (x24)
                    (x14) edge[\xarrow] (x24)
                    
                    (x21) edge[bend left = 15, \xarrow] (x22)
                    (x22) edge[bend left = 15, \xarrow] (x23)
                    (x23) edge[bend left = 15, \xarrow] (x24)
                    (x22) edge[bend left = 15, \xarrow] (x21)
                    (x24) edge[bend left = 15, \xarrow] (x23)

                    (x21) edge[\xarrow] (x31)
                    (x24) edge[\xarrow] (x31)
                    (x21) edge[\xarrow] (x34)
                    (x24) edge[\xarrow] (x34)
                    
                    (x31) edge[bend left = 15, \xarrow] (x32)
                    (x33) edge[bend left = 15, \xarrow] (x34)
                    (x32) edge[bend left = 15, \xarrow] (x31)
                    (x33) edge[bend left = 15, \xarrow] (x32)
                    (x34) edge[bend left = 15, \xarrow] (x33)

                    (x31) edge[\xarrow] (t)
                    (x34) edge[\xarrow] (t)

                    (C1) edge[bend right = 10, \xarrow] (C2)
                    (C2) edge[bend right = 10, \xarrow] (C3)
                    (C3) edge[bend right = 10, \xarrow] (C1)

                    (x12) edge[out=-90, in=180, looseness=.25, \xarrow, green, dashed] (C1)
                    (C1) edge[out=110, in=40, \xarrow, green, dashed] (x13)
                    (C2) edge[out=90, in=40, \xarrow, green, dashed] (x13)
                    (C3) edge[out=70, in=40, looseness=1.5, \xarrow, green, dashed] (x13)
                    
                    (x23) edge[out=-70, in=180, looseness=.5, \xarrow, blue, dashed] (C2)
                    (C1) edge[out=-160, in=-40, looseness=.75, \xarrow, blue, dashed] (x22)
                    (C2) edge[out=-110, in=-40, looseness=.5, \xarrow, blue, dashed] (x22)
                    (C3) edge[out=-110, in=-40, looseness=.5, \xarrow, blue, dashed] (x22)

                    (x32) edge[out=-70, in=-90, looseness=.5, \xarrow, red, dashed] (C3)
                    (C1) edge[out=-20, in=-50, looseness=1.75, \xarrow, red, dashed] (x33)
                    (C2) edge[out=-90, in=-50, looseness=1.25, \xarrow, red, dashed] (x33)
                    (C3) edge[out=-70, in=-50, looseness=1.4, \xarrow, red, dashed] (x33)
            ;
        \end{tikzpicture} 
    }\\
    \caption{\footnotesize{Example adjusted \textsc{Exact 3-Satisfiability} to \textsc{Directed Hamiltonian Path}}}
    \label{fig:adj ESAT to DHP}
    \end{center}
\end{figure}
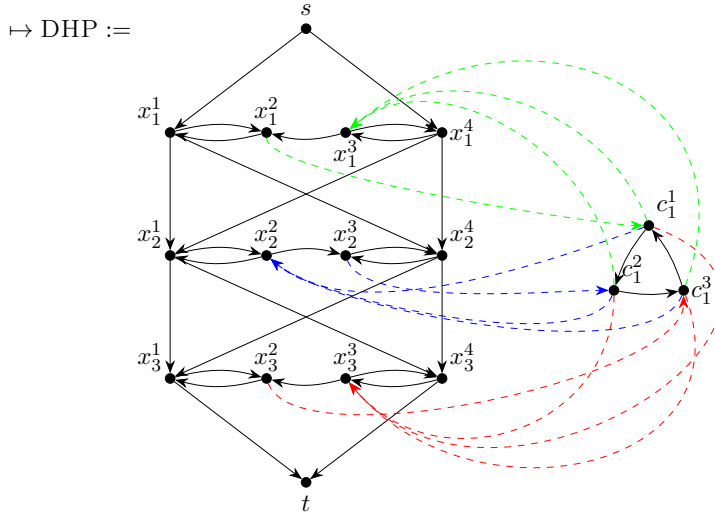

\paragraph{Parsimonious}
This reduction, however, is not Parsimonious. Observe the example shown in \autoref*{fig:ESAT to DHP}. Here $S=\{\ell_1\ell_2\ell_3\}$
is a satisfying assignment for the given \ESAT instance. Yet this solution has two equivalent solutions in $\Sr\DHP$, as either the
arcs $(x_1^2,C_1),(C_1,x_1^3)$ are in $S'$ or $(x_3^2,C_1)$ and $(C_1,x_3^3)$ are. Both cannot be true, as each vertex can only be visited 
once. Hence $S$ has two equivalent solutions in $\mc S_\DHP$ and therefore $|\mc S_\ESATu|<|\mc S_\DHP|$, which means there cannot be a 
bijective function between the solution spaces. \\
It follows, that we need to change the above reduction to fulfill the requirements needed for a strongly Parsimonious reduction. To do this
we need to ensure, that there is only one possible path for every given satisfying assignment. We do this by introducing three new vertices
$c_j^1, c_j^2$ and $c_j^3$ instead of the one vertex made for each clause $C_j$. These new vertices represent the three literals in $C_j$.
We connect them with the arcs $(c_j^1,c_j^2),(c_j^2,c_j^3)$ and $(c_j^3,c_j^1)$. This means there is only one direction the path can move along the 
triangle. W.l.o.g. we assume that every clause is ordered. Next we connect these new vertices 
with the paths representing the corresponding vertices as we do in the original reduction. This means if $\ell_i$ is the first literal
in the ordered clause $C_j$ we connect $(x_i^{4j-2},c_j^1)$ and $(c_j^1,x_i^{4j-1})$. We also create the arcs $(c_i^2,x_i^{4j-1})$ and 
$(c_i^3,x_i^{4j-1})$. Unlike in the original reduction we do not create both arcs $(x_i^{4j-2},x_i^{4j-1})$ and $(x_i^{4j-1},x_i^{4j-2})$,
but only $(x_i^{4j-1},x_i^{4j-2})$. Now a path going from $x_i^1$ to $x_i^m$ has to go over $c_j^1$. We do this for every literal in every 
clause. If the literal appears negated we switch $x_i^{4j-2}$ and $x_i^{4j-1}$ in every created arc. The rest of the reduction stays 
the same. \autoref*{fig:adj ESAT to DHP} shows the adjusted reduction on the above example, with colors added for increased legibility.

Now, if only one literal in a clause $C_j$ gets satisfied by a satisfying assignment, then the path can travel over all three clauses.
$c_j^1, c_j^2$ and $c_j^3$. However if all three literals in clause $C_j$ get satisfied by $S$, the paths visits all three vertices 
individually. And if two literals are satisfied, each combination corresponds to exactly one, unique path. 
\begin{itemize}
    \item The literals corresponding to $c_j^1$ and $c_j^2$ are in $S$, with $c_j^1\in\{\ell_i,\ne\ell_i\}$ and $c_j^2\in\{\ell_k,\ne\ell_k\}$,
    then the path travels from $c_j^2$ directly to $c_j^3$ and then back to the corresponding set of vertices $x_i$ and $c_j^1$ gets visited
    seperatly in the context of the vertices $x_k$.
    \item The literals corresponding to $c_j^1$ and $c_j^3$ are in $S$, with $c_j^1\in\{\ell_i,\ne\ell_i\}$ and $c_j^3\in\{\ell_k,\ne\ell_k\}$,
    then the path travels from $c_j^1$ directly to $c_j^2$ and then back to the corresponding set of vertices $x_i$ and $c_j^3$ gets visited
    seperatly in the context of the vertices $x_k$.
    \item The literals corresponding to $c_j^2$ and $c_j^3$ are in $S$, with $c_j^2\in\{\ell_i,\ne\ell_i\}$ and $c_j^3\in\{\ell_k,\ne\ell_k\}$,
    then the path travels from $c_j^3$ directly to $c_j^1$ and then back to the corresponding set of vertices $x_i$ and $c_j^2$ gets visited
    seperatly in the context of the vertices $x_k$.
\end{itemize}
This means that for each satisfying assignment $S$ we have exactly one corresponding directed Hamiltonian path $S'$. With this knowledge
we can now define a bijective function $$f:\mc S_\ESATu\rightarrow\mc S_\DHP,$$ which maps every literal in a solution $\S\ESATu$ to the 
corresponding path from $x_i^1$ to $x_i^m$ (or the other direction for negated literals), going over the necessary clause vertices for every 
clause that the literal is in. $S'$ also includes the arcs from $s$ to the first vertex of the set $x_1$, the arcs inbetween sets $x_i$ 
and the last arc from $x_{|L^2|}$ to $t$. This function is bijective as every path $\Sr\DHP$ encodes exactly one possible combination of 
literals satisfying $S$. The function defined in the previous section for SSP still applies and therefore for this adjusted reduction 
\ESAT$\SSPR$\DHP holds.

\subsubsection[... to Directed Hamiltonian Cycle]{Exact 3-Satisfiability to Directed Hamiltonian Cycle}
\paragraph{Problem Definition}
\label{sec:ESAT to DHC}
\subfile{../../Problems/DHC.tex}

\paragraph{Reduction}
We use the same adjusted reduction shown in the Parsimonious section  of the \ESAT $\SSPR$ \DHP reduction (\ref{sec:ESAT to DHP}). 
However, we add an arc from $t$ to $s$, forcing every Hamiltonian Cycle to go over this arc. \autoref*{fig:ESAT to DHC} shows an example
reduction.

\begin{figure}[ht]
    \begin{center}
    \scalebox{.85}{
        \begin{tikzpicture}[vertex/.style = {draw, circle, fill, inner sep = 1.5}, node distance = 1.5cm, label distance=-.1cm]
            \node[vertex, label = {[label distance=0cm] above:$s$}](s){};
            \node[above = 1cm of s](SAT){$\ESAT:=\big(\mc C = \{\{\ell_1, \ne \ell_2, \ell_3\}\}, 
            L=\{\ell_1, \ell_2, \ell_3, \ne\ell_1, \ne\ell_2, \ne\ell_3\}\big)$};

            \node[vertex, below left = 1.5cm and 2cm of s, label = above left:$x_1^1$](x11){};
            \node[vertex, below left = 1.5cm and .5cm of s, label = above:$x_1^2$](x12){};
            \node[vertex, below right = 1.5cm and .5cm of s, label = below:$x_1^3$](x13){};
            \node[vertex, below right = 1.5cm and 2cm of s, label = right:$x_1^4$](x14){};
            
            \node[vertex, below = 1.75cm of x11, label = above left:$x_2^1$](x21){};
            \node[vertex, below = 1.75cm of x12, label = above:$x_2^2$](x22){};
            \node[vertex, below = 1.75cm of x13, label = above:$x_2^3$](x23){};
            \node[vertex, below = 1.75cm of x14, label = above right:$x_2^4$](x24){};
            
            \node[vertex, below = 1.75cm of x21, label = above left:$x_3^1$](x31){};
            \node[vertex, below = 1.75cm of x22, label = above:$x_3^2$](x32){};
            \node[vertex, below = 1.75cm of x23, label = above:$x_3^3$](x33){};
            \node[vertex, below = 1.75cm of x24, label = above right:$x_3^4$](x34){};
            
            \node[vertex, below right = 1.5cm and 2cm of x31, label = {[label distance=0cm] below:$t$}](t){};

            \node[right = 3cm of x24](anchor){};

            \node[vertex, above = 0.25cm of anchor, label = above right:$c_1^1$](C1){};
            \node[vertex, below left = 0.35cm and 0.35cm of anchor, label = above right:$c_1^2$](C2){};
            \node[vertex, below right = 0.35cm and 0.35cm of anchor, label = right:$c_1^3$](C3){};

            \node[left = 2.5cm of s](GHP){$\mapsto\DHC:=$};

            \draw   (s) edge[\xarrow] (x11)
                    (s) edge[\xarrow] (x14)

                    (x11) edge[bend left = 15, \xarrow] (x12)
                    (x13) edge[bend left = 15, \xarrow] (x14)
                    (x12) edge[bend left = 15, \xarrow] (x11)
                    (x13) edge[bend left = 15, \xarrow] (x12)
                    (x14) edge[bend left = 15, \xarrow] (x13)

                    (x11) edge[\xarrow] (x21)
                    (x14) edge[\xarrow] (x21)
                    (x11) edge[\xarrow] (x24)
                    (x14) edge[\xarrow] (x24)
                    
                    (x21) edge[bend left = 15, \xarrow] (x22)
                    (x22) edge[bend left = 15, \xarrow] (x23)
                    (x23) edge[bend left = 15, \xarrow] (x24)
                    (x22) edge[bend left = 15, \xarrow] (x21)
                    (x24) edge[bend left = 15, \xarrow] (x23)

                    (x21) edge[\xarrow] (x31)
                    (x24) edge[\xarrow] (x31)
                    (x21) edge[\xarrow] (x34)
                    (x24) edge[\xarrow] (x34)
                    
                    (x31) edge[bend left = 15, \xarrow] (x32)
                    (x33) edge[bend left = 15, \xarrow] (x34)
                    (x32) edge[bend left = 15, \xarrow] (x31)
                    (x33) edge[bend left = 15, \xarrow] (x32)
                    (x34) edge[bend left = 15, \xarrow] (x33)

                    (x31) edge[\xarrow] (t)
                    (x34) edge[\xarrow] (t)

                    (C1) edge[bend right = 10, \xarrow] (C2)
                    (C2) edge[bend right = 10, \xarrow] (C3)
                    (C3) edge[bend right = 10, \xarrow] (C1)

                    (x12) edge[out=-90, in=180, looseness=.25, \xarrow, green, dashed] (C1)
                    (C1) edge[out=110, in=40, \xarrow, green, dashed] (x13)
                    (C2) edge[out=90, in=40, \xarrow, green, dashed] (x13)
                    (C3) edge[out=70, in=40, looseness=1.5, \xarrow, green, dashed] (x13)
                    
                    (x23) edge[out=-70, in=180, looseness=.5, \xarrow, blue, dashed] (C2)
                    (C1) edge[out=-160, in=-40, looseness=.75, \xarrow, blue, dashed] (x22)
                    (C2) edge[out=-110, in=-40, looseness=.5, \xarrow, blue, dashed] (x22)
                    (C3) edge[out=-110, in=-40, looseness=.5, \xarrow, blue, dashed] (x22)

                    (x32) edge[out=-70, in=-90, looseness=.5, \xarrow, red, dashed] (C3)
                    (C1) edge[out=-20, in=-50, looseness=1.75, \xarrow, red, dashed] (x33)
                    (C2) edge[out=-90, in=-50, looseness=1.25, \xarrow, red, dashed] (x33)
                    (C3) edge[out=-70, in=-50, looseness=1.4, \xarrow, red, dashed] (x33)

                    (t) edge[out=-170, in=170, looseness = 2, \xarrow] (s)
            ;
        \end{tikzpicture} 
    }\\
    \caption{\footnotesize{Example \textsc{Exact 3-Satisfiability} to \textsc{Directed Hamiltonian Cycle}}}
    \label{fig:ESAT to DHC}
    \end{center}
\end{figure}
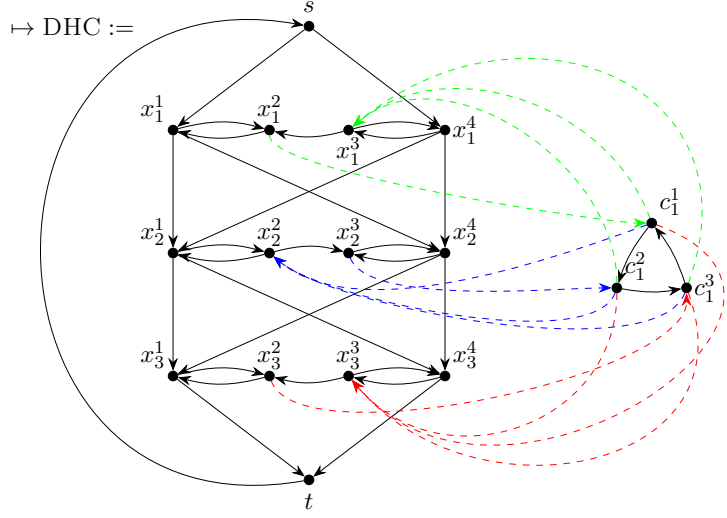
\begin{proof}
    As every Hamiltonian Cycle has to go over the arc $(t,s)$, every cycle includes a Hamiltonian path from $s$ to $t$. And this part of the 
    reduction holds according to section \ref{sec:ESAT to DHP}.
\end{proof}
\noindent The reduction function $g$ is computable in polynomial time and is dependent on the input size of the \ESAT instance $\big|[L,\mc C]\big|$,
as for each clause in $\mc C$ and each literal in $L$ we create a fixed number of vertices in the new instance. Therefore $\ESAT\leq \DHC$ holds.

\paragraph{SSP \& Parsimonious}
This reduction is SSP. The idea is that the direction of the cycle is set by the arc $(t,s)$ and for each set of vertices $x_i^m$ the 
direction in which the path from $s$ to $t$ goes is indicated by checking if $(x_i^1,x_i^2)$ or $(x_i^2,x_i^1)$ is in the solution. The
rest of the path, as well as the clause vertices $c_j^k$, are dependent on which literals are in the solution $S$. This means we can apply 
\lemref{lem:SSP and SPR} as follows.
\begin{proof} Let $\Sall:=(t,s)$, let $\Snev:=\emptyset$ and let 
 the functions $f_I$ be defined as
    $$f_I:L\rightarrow A,~f(\cee_i)=\begin{cases}
        (x_i^1,x_i^2), & \cee_i\in L^+ \\
        (x_i^2,x_i^1), & \cee_i\in L^-
    \end{cases}$$
    With these arcs being set, the rest of the path is set as well, as seen above. Therefore we get
    \begin{flalign*}
        \Slink:=&\big\{(s,x_1^m),(x_{|L^2|}^m,t) \mid m\in\big[4\cdot|\mc C|\big] \big\}\\
        &\cup\big\{(x_i^1,x_{i+1}^{4|\mc C|}),(x_{i}^{4|\mc C|}, x_{i+1}^1)\mid i\in \big[|L^2|\big]\big\}\\
        &\cup\Big(\big\{(x_i^m,x_i^{m+1}),(x_i^{m+1},x_i^m)\mid i\in\big[|L^2|\big], m\in \{2,..., 4\cdot|\mc C|\}\big\}\cap  A\Big)\\
        &\cup\big\{(c_j^1,c_j^2),(c_j^2,c_j^3),(c_j^3,c_j^1)\mid j\in\big[|\mc C|\big]\big\} \\
        &\cup \Big(\big\{(x_i^{4j-2},c_j^n), (c_j^n,x_i^{4j-1})\mid i\in\big[|L^2|\big], j\in\big[|\mc C|\big], n\in [3]\big\}\cap  A\Big)
    \end{flalign*}
for our linked elements.
Then the fnctions $f_I':L\rightarrow A\backslash(\Sall\cup\Slink)$ are bijective and 
    \begin{flalign*}
        \{f(S):\S\ESATu\} & = \big\{\{f(\ell_i)\mid\cee_i\in S\}:\S\ESATu\big\}\\
        & = \Big\{\big\{\big((x_i^1,x_i^2)\xor(x_i^2,x_i^1)\big)\in S'\big\}:\Sr\DHC\Big\}\\
        &= \{S'\cap f(L):\Sr\DHC\}
    \end{flalign*}
    and so $\ESAT\SSPR\DHC$ holds.
\end{proof}

\ifSubfilesClassLoaded{
  \bibliography{../../ref}
}{}
\end{document}

%% file: Problems/OSAT.tex
\noindent\fbox{
    \parbox{\textwidth}{
        \textsc{1-in-3-Satisfiability ($\OSAT$)}\\
        \begin{tabular}{cl}
        ~ & \textbf{Instances:} Literal Set $L = \{\ell_1, . . . , \ell_n\} \cup \{\ne\ell_1, . . . , \ne\ell_n\},$ Clauses $\mc C \subseteq 2^L$ \\
        ~ & s.t. $\forall C\in \mc C:|C| = 3$. \\
        ~ & \textbf{Universe:} The Literal set $\mathcal{U}:=L$.\\
        ~ & \textbf{Solution set:} The set of all sets $S\subseteq L$ such that for all $i \in\{1,...,n\}$ we have \\
        ~ &$|S\cap\{\ell_i,\ne\ell_i\}|=1$, and $|S\cap C|= 1$ for all $C\in \mc C$.
        \end{tabular}
    }
}

\ifSubfilesClassLoaded{
  \bibliography{../ref}
}{}
\end{document}

%% file: Problems/MIS.tex
\noindent\fbox{
    \parbox{\textwidth}{
        \textsc{Maximum Cardinality Independent Set ($\MIS$)}\\
        \begin{tabular}{cl}
        ~ & \textbf{Instances:} Graph $G=(V,E)$, number $k\in \mathbb{N}$\\
        ~ & \textbf{Universe:} Vertex set $\mathcal{U}:=V$.\\
        ~ & \textbf{Feasible solution set:} The set of all independent sets $\mc S_\textsc{IS}$.\\
        ~ & \textbf{Solution set:} The set of all independent sets of size at least $k$  \\
        ~ & s.t. $k = \max\{|S| \mid S \in \mc S_\textsc{IS}\}$.
        \end{tabular}
    }
}

\ifSubfilesClassLoaded{
  \bibliography{../ref}
}{}
\end{document}

%% file: Problems/MVC.tex
\noindent\fbox{
    \parbox{\textwidth}{
        \textsc{Minimum Cardinality Vertex Cover ($\MVC$)}\\
        \begin{tabular}{cl}
        ~ & \textbf{Instances:} Graph $G=(V,E)$, number $k\in \mathbb{N}$\\
        ~ & \textbf{Universe:} Vertex set $\mathcal{U}:=V$.\\
        ~ & \textbf{Feasible solution set:} The set of all vertex covers $\mc S_\textsc{VC}$.\\
        ~ & \textbf{Solution set:} The set of all vertex covers of size at most $k$, \\
        ~ & s.t. $k = \min\{|C| \mid C \in \mc S_\textsc{VC}\}$
        \end{tabular}
    }
}

\ifSubfilesClassLoaded{
  \bibliography{../ref}
}{}
\end{document}

%% file: Problems/MDS.tex
\noindent\fbox{
  \parbox{\textwidth}{
      \textsc{Minimum Cardinality Dominating Set ($\MDS$)}\\
      \begin{tabular}{cl}
      ~ & \textbf{Instances:} Graph $G=(V,E)$, number $k\in \mathbb{N}$\\
      ~ & \textbf{Universe:} Vertex set $\mathcal{U}:=V$.\\
      ~ & \textbf{Feasible solution set:} The set of all dominating sets $\mc S_\DS$.\\
      ~ & \textbf{Solution set:} The set of all dominating sets covers of size at most $k$ \\
      ~ & s.t. $k = \min\{|S| \mid S \in \mc S_\DS\}$.
      \end{tabular}
  }
}

\ifSubfilesClassLoaded{
  \bibliography{../ref}
}{}
\end{document}

%% file: Problems/CQ.tex
\noindent\fbox{
    \parbox{\textwidth}{
        \textsc{Clique ($\CQ$)}\\
        \begin{tabular}{cl}
        ~ & \textbf{Instances:} Graph $G=(V,E)$, number $k\in\mathbb N$\\
        ~ & \textbf{Universe:} Vertex Set $\mathcal{U}:=V$.\\
        ~ & \textbf{Feasible solution set:} The set of all cliques $\mc S_\CQ$.\\
        ~ & \textbf{Solution set:} The set of all cliques $S\in\mc S_\CQ$ of size at least $k$.
        \end{tabular}
    }
}

\ifSubfilesClassLoaded{
  \bibliography{../ref}
}{}
\end{document}

%% file: Problems/SS.tex
\noindent\fbox{
    \parbox{\textwidth}{
        \textsc{Subset Sum ($\SS$)}\\
        \begin{tabular}{cl}
        ~ & \textbf{Instances:} Numbers $A:=\{a_1,...,a_n\}\subseteq \mathbb{N}$, and target value $M\in\mathbb N$\\
        ~ & \textbf{Universe:} Set of numbers $\mathcal{U}:=A$.\\
        ~ & \textbf{Feasible solution set:} The set of all sets $S\subseteq \mathcal U$ with $\sum_{a_i\in S} a_i=M$.
        \end{tabular}
    }
}

\ifSubfilesClassLoaded{
  \bibliography{../ref}
}{}
\end{document}

%% file: Problems/EC.tex
\noindent\fbox{
    \parbox{\textwidth}{
        \textsc{Exact Cover ($\EC$)}\\
        \begin{tabular}{cl}
        ~ & \textbf{Instances:} Sets $\psi_i\subseteq \Psi = \{1,...,m\}$ for $i\in[n]$.\\
        ~ & \textbf{Universe:} Sets $\mathcal{U}:=\{\psi_1,...,\psi_n\}$.\\
        ~ & \textbf{Solution set:} The set of all $S\subseteq\{\psi_1,..,\psi_n\}$ s.t. \\
        ~ & $\forall \psi_i,\psi_j\in S, i\neq j:\psi_i\cap\psi_j=\emptyset$ and $\bigcup_{\psi_i\in S} \psi_i = \Psi$.
        \end{tabular}
    }
}

\ifSubfilesClassLoaded{
  \bibliography{../ref}
}{}
\end{document}

%% file: Problems/DHP.tex
\noindent\fbox{
    \parbox{\textwidth}{
        \textsc{Directed Hamiltonian Path ($\DHP$)}\\
        \begin{tabular}{cl}
        ~ & \textbf{Instances:} Directed Graph $G=(V,A)$, Vertices $s,t\in V$.\\
        ~ & \textbf{Universe:} Arc set $\mathcal{U}:=A$.\\
        ~ & \textbf{Feasible solution set:} The set of all sets $S\subseteq A$ forming a Hamiltonian path\\
        ~ & going from $s$ to $t$.
        \end{tabular}
    }
}

\ifSubfilesClassLoaded{
  \bibliography{../ref}
}{}
\end{document}

%% file: Problems/DHC.tex
\noindent\fbox{
    \parbox{\textwidth}{
        \textsc{Directed Hamiltonian Cycle ($\DHC$)}\\
        \begin{tabular}{cl}
        ~ & \textbf{Instances:} Directed Graph $G=(V,A)$.\\
        ~ & \textbf{Universe:} Arc set $\mathcal{U}:=A$.\\
        ~ & \textbf{Solution set:} The set of all sets $S\subseteq A$ forming a Hamiltonian cycle.
        \end{tabular}
    }
}

\ifSubfilesClassLoaded{
  \bibliography{../ref}
}{}
\end{document}

%% file: Reductions/1-in-3-Sat/1-in-3-Sat.tex
\subsection{Reductions from 1-in-3-Satisfiability}

\subfile{../../Problems/OSAT.tex}
\subsubsection[... to Steiner Tree]{1-in-3-Satisfiability to Steiner Tree}
\label{sec:OSAT to STT}
\paragraph{Problem Definition}
\subfile{../../Problems/STT.tex}

\paragraph{Reduction}
We use a modified version of the reduction presented in \cite{ssp}. Let $[L,\mc C]$ denote a \OSAT instance then we map this to a \STT instance $[G=(N\cup T,E),N,T,w,k]$ as follows.
Firstly, we create two terminal vertices $T:=\{s,t\}$. Next for every literal $\cee_i\in L$ we map this to the Steiner vertex $\cee_i\in N$. Now, for $1\leq i <|L^2|$
we create the vertices $v_i\in N$ and denote $s:=v_0$. Then we connect the created vertices into a so called 'diamond chain', where we begin by
connecting $s$ to both vertices $\ell_1$ and $\ne\ell_1$. These then are both connected to $v_1$, which is then connected to $\ell_2$ and $\ne\ell_2$
and so on until at last $\ell_{|L^2|}$ and $\ne\ell_{|L^2|}$ are connected to $t$.\\
For every clause $C_j\in \mc C$ we now create a corresponding terminal vertex $C_j$. 
The vertex $C_j$ is then connected to the three literal vertices corresponding
to the literals contained in $C_j$ via paths $1\leq i\leq 3$ over the Steiner vertices $c_{j_i}^1,...,c_{j_i}^{|L|}\in N$. 
Next, also for every clause $C_j$, we create the vertices $H_j^l,H_j^{l+}$ and $H_j^{l-}$ with $l\in [3]$, but only the $H_j^{l+}$ and $H_j^{l-}$
vertices are terminals.
Analogously to the terminal vertices $C_j$, the vertices $H_j^1$, $H_j^2$ and $H_j^3$ are connected to the three literals that correspond to the compliments of the vertices 
in the clause corresponding to $C_j$ (via the Steiner vertices $h_{j_i}^1,...,h_{j_i}^{|L|}\in N$). 
Now, let $M$ be a number $\geq |\mc C|\cdot|L|$. Via paths of length $M$, we connect each $H_j^l$ with both $H_j^{l+}$ and $H_j^{l-}$,
as well as $H_j^{1-}$ to $H_j^{2+}$, $H_j^{2-}$ to $H_j^{3+}$ and $H_j^{3-}$ to $H_j^{1+}$ through paths of length $M+1$.\\
Finally, the weight of every edge is set to 1 and 
the threshold is set to 
\begin{flalign*}
    k:=&|L|+|\mc C|\cdot (|L|+1)+|\mc C|\cdot \bigl(2\cdot (|L|+1) + 6M +2\bigr)\\
    =&|L|+|\mc C|\cdot \bigl(3\cdot (|L|+1) + 6M +2\bigr)
\end{flalign*}
An example reduction is shown in \autoref*{fig:1-3S to StT}.

\begin{figure}[ht]
    \begin{center}
        $$\OSAT:=\big[\mc C = \{\{\ne\ell_1, \ne \ell_2, \ell_3\}\}, L=\{\ell_1, \ell_2, \ell_3, \ne\ell_1, \ne\ell_2, \ne\ell_3\}\big]$$\\
        gets mapped to\\
        \begin{tikzpicture}[vertex/.style = {draw, circle, fill, inner sep = 1.5}, terminal/.style = {draw, rectangle, fill, inner sep = 2}, node distance = 1.5cm, label distance=0cm]
            \node[](SAT){};
            
            \node[below left = .1cm and -2.225cm of SAT](STT){$\STT:=$};

            \node[terminal, below right = 5cm and .5cm of STT, label = $s$](s){};
            \node[vertex, above right = .75cm and .75cm of s, label = below:$\ell_1$](l1){};
            \node[vertex, below right = .75cm and .75cm of s, label = $\ne\ell_1$](nl1){};
            \node[vertex, below right = .75cm and .75cm of l1, label = $v_1$](v1){};
            \node[vertex, above right = .75cm and .75cm of v1, label = below:$\ell_2$](l2){};
            \node[vertex, below right = .75cm and .75cm of v1, label = $\ne\ell_2$](nl2){};
            \node[vertex, below right = .75cm and .75cm of l2, label = $v_2$](v2){};
            \node[vertex, above right = .75cm and .75cm of v2, label = below:$\ne\ell_3$](nl3){};
            \node[vertex, below right = .75cm and .75cm of v2, label = $\ell_3$](l3){};
            \node[terminal, below right = .75cm and .75cm of nl3, label = $t$](t){};

            \node[vertex, below left = .75cm and .5cm of nl1, label = {[label distance= -.1cm]left:$c_{1_1}^1$}](c111){};
            \node[vertex, below = 1cm of c111, label = {[label distance= -.1cm]left:$c_{1_1}^6$}](c117){};
            \node[vertex, below = .75cm  of nl2, label = {[label distance= -.1cm]left:$c_{1_2}^1$}](c121){};
            \node[vertex, below = 1cm of c121, label = {[label distance= -.1cm]left:$c_{1_2}^6$}](c127){};
            \node[vertex, below right = .75cm and .5cm of l3, label = {[label distance= -.1cm]left:$c_{1_3}^1$}](c131){};
            \node[vertex, below = 1cm of c131, label = {[label distance= -.1cm]left:$c_{1_3}^6$}](c137){};
            
            \node[terminal, below = 1cm of c127, label= below:$C_1$](C1){};

            \node[vertex, above left = .75cm and .5cm of l1, label = {[label distance= -.1cm]left:$h_{1_1}^1$}](h111){};
            \node[vertex, above = 1cm of h111, label = {[label distance= -.1cm]left:$h_{1_1}^6$}](h117){};

            \node[vertex, above = .75cm of l2, label = {[label distance= -.1cm]left:$h_{1_2}^1$}](h121){};
            \node[vertex, above = 1cm of h121, label = {[label distance= -.1cm]left:$h_{1_2}^6$}](h127){};

            \node[vertex, above right = .75cm and .5cm of nl3, label = {[label distance= -.1cm]left:$h_{1_3}^1$}](h131){};
            \node[vertex, above = 1cm of h131, label = {[label distance= -.1cm]left:$h_{1_3}^6$}](h137){};

            \node[above = 2cm of h127](H12){};
            \node[vertex, below = 1 of H12, label= {[label distance= -.1cm]below right:$H_{1}^{2}$}](sH12){};
            \node[terminal, left = .15 of H12, label= {[label distance= .1cm, scale = .75]above:$H_{1}^{2+}$}](gH12){};
            \node[terminal, right = .15 of H12, label= {[label distance= .1cm, scale = .75]above:$H_{1}^{2-}$}](nH12){};
            \node[above left = 1 and .66 of H12](H11){};
            \node[vertex, above left = .75 and .75 of H11, label= {[label distance= -.1cm]above:$H_{1}^{1}$}](sH11){};
            \node[terminal, above right = .1 and .1 of H11, label= {[label distance= -.25cm, scale = .75]below right:$H_{1}^{1+}$}](gH11){};
            \node[terminal, below left = .1 and .1 of H11, label= {[label distance= .1cm, scale = .75]right:$H_{1}^{1-}$}](nH11){};
            \node[above right = 1 and .66 of H12](H13){};
            \node[vertex, above right = .75 and .75 of H13, label= {[label distance= -.1cm]above:$H_{1}^{3}$}](sH13){};
            \node[terminal, below right = .1 and .1 of H13, label= {[label distance= .1cm, scale = .75]left:$H_{1}^{3+}$}](gH13){};
            \node[terminal, above left = .1 and .1 of H13, label= {[label distance= -.25cm, scale = .75]below left:$H_{1}^{3-}$}](nH13){};

            \draw   (s) edge[thick, red] (l1)
                    (s) -- (nl1)
                    (v1) edge[thick, red] (l1)
                    (v1) -- (nl1)
                    (v1) edge[thick, red] (l2)
                    (v1) -- (nl2)
                    (v2) edge[thick, red] (l2)
                    (v2) -- (nl2)
                    (v2) edge[thick, red] (l3)
                    (v2) -- (nl3)
                    (t) edge[thick, red] (l3)
                    (t) -- (nl3)

                    (nl1) -- (c111)
                    (nl2) -- (c121)
                    (l3) edge[thick, red] (c131)
            
                    (c111) edge["$\small{c_{1_1}^{2...5}}$", dashed, inner sep=.35cm] (c117)
                    (c111) edge[decorate, decoration={brace,amplitude=5pt,raise=.1cm}] (c117)

                    (c121) edge["$\small{c_{1_2}^{2...5}}$", dashed, inner sep=.35cm] (c127)
                    (c121) edge[decorate, decoration={brace,amplitude=5pt,raise=.1cm}] (c127)

                    (c131) edge[dashed, thick, red, inner sep=.35cm] (c137)
                    (c131) edge[decorate, "$\small{c_{1_3}^{2...5}}$", inner sep=.35cm, decoration={brace,amplitude=5pt,raise=.1cm}] (c137)

                    (c117) -- (C1)
                    (c127) -- (C1)
                    (c137) edge[out = -90, in = 25, looseness = .5, thick, red] (C1)

                    (h117) edge[dashed, thick, red] (h111)
                    (h117) edge[decorate,"\small{$h_{1_1}^{2...5}$}", inner sep=.35cm, decoration={brace,amplitude=5pt,raise=.1cm}] (h111)

                    (h127) edge[dashed, thick, red] (h121)
                    (h127) edge[decorate,"\small{$h_{1_2}^{2...5}$}", inner sep=.35cm, decoration={brace,amplitude=5pt,raise=.1cm}] (h121)

                    (h137) edge["\small{$h_{1_3}^{2...5}$}", dashed, inner sep=.35cm] (h131)
                    (h137) edge[decorate, decoration={brace,amplitude=5pt,raise=.1cm}] (h131)

                    (l1) edge[thick, red] (h111)
                    (l2) edge[thick, red] (h121)
                    (nl3) edge[in = -100, looseness = .5] (h131)

                    (sH11) edge[dashed, thick, red] node [scale = .75, color = black] [yshift = .55cm, xshift = .3cm] {${M}$} (gH11)
                    (sH11) edge[decorate, decoration={brace,amplitude=5pt,raise=.1cm}] (gH11)
                    (nH11) edge[dashed, thick, red] node [scale = .75, color = black] [yshift = -.3cm, xshift = -.55cm] {${M}$} (sH11)
                    (nH11) edge[decorate, decoration={brace,amplitude=5pt,raise=.1cm}] (sH11)
                    (sH12) edge[dashed, thick, red] node [scale = .75, color = black] [yshift = -.1cm, xshift = 1.1cm] {${M}$} (gH12)
                    (sH12) edge[decorate, decoration={brace,amplitude=5pt,raise=.1cm}] (gH12)
                    (nH12) edge[dashed, thick, red] node [scale = .75, color = black] [yshift = -.1cm, xshift = -1.1cm] {${M}$} (sH12)
                    (nH12) edge[decorate, decoration={brace,amplitude=5pt,raise=.1cm}] (sH12)
                    (nH13) edge[dashed] node [scale = .75] [yshift = .55cm, xshift = -.3cm] {${M}$} (sH13)
                    (nH13) edge[decorate, decoration={brace,amplitude=5pt,raise=.1cm}] (sH13)
                    (sH13) edge[dashed] node [scale = .75] [yshift = -.3cm, xshift = .55cm] {${M}$} (gH13)
                    (sH13) edge[decorate, decoration={brace,amplitude=5pt,raise=.1cm}] (gH13)

                    (gH11) edge[dashed, thick, red] node [scale = .75, color = black] [above, yshift = .3cm] {${M+1}$} (nH13)
                    (gH11) edge[decorate, decoration={brace,amplitude=5pt,raise=.1cm}] (nH13)
                    (gH13) edge[dashed, thick, red] node [scale = .75, color = black] [below right, xshift = .2cm, yshift = -.2cm] {${M+1}$} (nH12)
                    (gH13) edge[decorate, decoration={brace,amplitude=5pt,raise=.1cm}] (nH12)
                    (gH12) edge[dashed] node [scale = .75] [below left, xshift = -.2cm, yshift = -.2cm] {${M+1}$} (nH11)
                    (gH12) edge[decorate, decoration={brace,amplitude=5pt,raise=.1cm}] (nH11)

                    (h117) edge[out = 90, in = -180, looseness = .25, thick, red] (sH11)
                    (h127) edge[thick, red] (sH12)
                    (h137) edge[out = 90, in = 0, looseness = .25]  (sH13)
            ;
        \end{tikzpicture} 
    \\The thick red edges are equivalent to the \OSAT solution $\{\ell_1,\ell_2,\ell_3\}$.

    \caption{\footnotesize{Example \textsc{1-in-3-Satisfiability} to \textsc{Steiner Tree}}}
    \label{fig:1-3S to StT}
    \end{center}
\end{figure}

\begin{proof}
Let $S$ be a solution for the \OSAT instance, then this means exactly one literal $\ell$ per clause $C_j$ is satisfied. 
It follows that for every clause 
$C_j$ in $\mc C$ the corresponding path in $G$ from the literal vertex $\ell$ to the clause vertex $C_j$ can be taken and adds up to 
$|\mc C|\cdot (|L|+1)$. Analogously, for the two literals not satisfied in $C_j$, we can include the paths to their respective $H_j^l$'s,
which add up to $2\cdot|\mc C|\cdot (|L|+1)$. The only possibility to connect the terminal vertices $H_j^{l+/-}$ is to connect the two already 
connected $H_j^l$'s to their respective counterparts and then to connect the terminal vertices of the missing $H_j^l$ to the two closest, connected
$H_j^{l+/-}$'s. In total this adds up to a path of weight $4M+2\cdot(M+1) = 6M+2$ for every clause.\\
Every
terminal vertex $C_j$ is now in the tree, however we still need to connect the two vertices $s$ and $t$. This can be done via a path 
from $s$ to $t$ over
all the literals in $S$. Any path from $s$ to $t$ over the literal vertices is at least of length $|L|$ and so, in total, 
we have a Steiner Tree of size 
\begin{flalign*}
    k &= |\mc C|\cdot (|L|+1) + 2 \cdot |\mc C|\cdot (|L|+1) + (6M+2)\cdot |\mc C| + |L| \\
    &= |L|+|\mc C|\cdot \bigl(3\cdot (|L|+1) + 6M+2\bigr)
\end{flalign*}

Now, let $S'$ be a solution for the \STT instance. Since $s$ and $t$ are connected in the tree, there has to be a path over the 
diamond chain connecting literal vertices, as the alternative paths either go through a clause vertex $C_j$ or through two of the helper vertices 
$H_j^1$, $H_j^2$ and $H_j^3$.
\begin{enumerate}
    \item If the path went over a clause vertex, then this would cost $|L|+1$ extra, instead of the $|L|$ it would cost to go via the chain.
    This would result in a tree of at least size
    $$(|L|+1)+|\mc C|\cdot \bigl(3\cdot (|L|+1) + 6M+2\bigr) > k$$
    which violates the constraints of the Steiner Tree.

    \item If the path went through the helper vertices, it would cost at least $M+1$ more to connect the two helpers.
    This would result in a path of at least size
    $$(M+1)+|\mc C|\cdot \bigl(3\cdot (|L|+1) + 6M+2\bigr) > k$$
    which violates the constraints of the Steiner Tree as $M \geq |\mc C| \cdot |L|$.

\end{enumerate}

Therefore every literal pair is visited once on the $|L|$ long path from $s$ to $t$ and each of the visited literals has a direct path 
(of exactly length $|L|+1$) 
to either a vertex corresponding to a satisfied clause, or to a helper vertex, as otherwise we would again violate the threshold. 
Each of the vertices $C_j$ correspond to a clause and if we take the literals visited along the 
$s$-$t$ path as a solution for $\OSAT$, every clause is visited, also known as satisfied, via exactly one literal.
\end{proof}
\noindent The reduction function $g$ is computable in polynomial time and is dependent on the input size of the \OSAT instance $\big|[L,\mc C]\big|$,
as for each clause in $\mc C$ and each literal in $L$ we create a fixed number of vertices in the new instance. Therefore $\OSAT\leq \DHP$ holds.

\paragraph{SSP \& Parsimonious}
As seen above, an $s$-$t$ path on the Steiner Tree encodes the literals in a satisfying assignment and vice versa. This means there exists a direct 
one-to-one correspondence between the solution spaces, where each solution $\S\OSATu$ gets encodes as part of an $s$-$t$ path, with the rest of 
the solution $\Sr\STT$ 
being dependent on that path and each of the terminal vertices
$C_j$ and $H_j^{l+}, H_j^{l-}$ for $l\in[3]$ being visited exactly once. The paths between the diamond chain and the aforementioned 
terminal vertices are therefore predetermined by the literals visited along the diamond chain. 
Therefore we can apply \lemref{lem:SSP and SPR} as follows.
\begin{proof}
    Let 
    $$f_I:L\rightarrow E,~f(\cee_i)= \{v_{i-1},\ell_i\}$$
    and let $\Sall=\Snev=\emptyset$. For every clause $C_j$, let $\ell_i(j)$ be the literal satisfying $C_j$ and $\ne{\ell_i^1}(j)$ 
    and $\ne{\ell_i^2}(j)$ the negation of the two other literals in $C_j$. Then
    \begin{itemize}
        \item[-] $\mf L\big(\{v_{i-1},\ell_i(j)\}\big)$ is the set of edges connecting $\ell_i(j)$ to $C_j$, as well as the edge $\{\ell_i,v_i\}$ and 
        \item[-] $\mf L\big(\{v_{i-1},\ne{\ell_i^1}(j)\}\big)$ and $\mf L\big(\{v_{i-1},\ne{\ell_i^2}(j)\}\big)$ are the sets of edges connecting
        these literals to their respective helper vertices $H_j^l$, the two edges on either side connecting the $H_j^l$'s to their respective 
        $H_j^{l+}$ and $H_j^{l-}$ and, finally, the two edges connecting these to the remaining two terminal helper vertices corresponding to the 
        negation of the literal satisfying the clause $C_j$ as well as the edges $\{\ne{\ell_i^1}(j), v_{i}\}$ and $\{\ne{\ell_i^2}(j), v_{i}\}$ 
        respectively.
    \end{itemize}
    With this we define
    $$\Slink:=\bigcup_{\S{\textsc{1in3SAT}}} \bigg(\bigcup_{x\in S} \mf L\big(f(x)\big)\bigg).$$
    Now $f_I':L\rightarrow E\backslash\Slink$ is bijective and 
    \begin{flalign*}
        \{f(S):\S\OSATu\} &= \big\{f\big(\{\cee_i\in S\}\big):\S\OSATu\}\\
        &=\Big\{\big\{\{v_{i-1},\ell_i\} \in S'\big\} :\Sr\STT\Big\}\\
        &=\{S'\backslash \Slink:\Sr\STT\}\\
        &=\{S'\cap f(L):\Sr\STT\}
    \end{flalign*}
    Thus, $\OSAT\SSPR\STT$ holds.
\end{proof}

\subsubsection[... to Minimum Cardinality Vertex Cover]{1-in-3-Satisfiability to Minimum Cardinality Vertex Cover} 
\label{sec:OSAT to MVC}
\paragraph{Problem Definitions}

\begin{figure}[ht]
    \begin{center}
    \scalebox{.85}{
        \begin{tikzpicture}[vertex/.style = {draw, circle, fill, inner sep = 1.5}, node distance = 1.5cm]
            \node[vertex, label = left:$\ell_1$](l1){};
            \node[vertex, right of = l1, label = right:$\ne\ell_1$](nl1){};
            \node[vertex, right of = nl1, label = left:$\ell_2$](l2){};
            \node[vertex, right of = l2, label = right:$\ne\ell_2$](nl2){};
            \node[vertex, right of = nl2, label = left:$\ell_3$](l3){};
            \node[vertex, right of = l3, label = right:$\ne\ell_3$](nl3){};

            \node[vertex, below right = 2 and .5 of l2, label = below:$c_1^2$](c12){};
            \node[vertex, below left of = c12, label = below:$c_1^1$](c11){};
            \node[vertex, below right of = c12, label = below:$c_1^3$](c13){};

            \node[vertex, above left = 2 and .5 of nl2, label = above:$c_2^2$](c22){};
            \node[vertex, above left of = c22, label = above:$c_2^1$](c21){};
            \node[vertex, above right of = c22, label = above:$c_2^3$](c23){};

            \node[below right = .1 and 2 of c13](k){$k=9$};

            \node[vertex, black!25!red, above = 1 of l1, label = {[black!25!red]left:$h_1^2$}](h12){};
            \node[vertex, black!25!red, below = 1 of nl1, label = {[black!25!red]right:$\ne h_1^1$}](nh11){};

            \node[vertex, black!25!red, below = 1 of l2, label = {[black!25!red]left:$h_2^1$}](h21){};
            \node[vertex, black!25!red, above = 1 of nl2, label = {[black!25!red]right:$\ne h_2^2$}](nh22){};

            \node[vertex, black!25!red, below = 1 of nl3, label = {[black!25!red]right:$\ne h_3^1$}](nh31){};
            \node[vertex, black!25!red, above = 1 of nl3, label = {[black!25!red]right:$\ne h_3^2$}](nh32){};

            \draw 
                (l1) -- (nl1)
                (l2) -- (nl2)
                (l3) -- (nl3)

                (c11) -- (c12)
                (c11) -- (c13)
                (c13) -- (c12)
                
                (c21) -- (c22)
                (c21) -- (c23)
                (c23) -- (c22)

                (c11) -- (l1)
                (c12) -- (nl2)
                (c13) -- (l3)
                (c21) -- (nl1)
                (c22) -- (l2)
                (c23) -- (l3)

                (l1) edge[black!25!red] (h12)
                (c21) edge[black!25!red] (h12)
                (nl1) edge[black!25!red] (nh11)
                (c11) edge[black!25!red] (nh11)
                
                (l2) edge[black!25!red] (h21)
                (c12) edge[black!25!red] (h21)
                (nl2) edge[black!25!red] (nh22)
                (c22) edge[black!25!red] (nh22)

                (nl3) edge[black!25!red] (nh31)
                (c13) edge[black!25!red] (nh31)
                (nl3) edge[black!25!red] (nh32)
                (c23) edge[black!25!red] (nh32)
            ;
        \end{tikzpicture} 
    }\\
    \caption{\footnotesize{Example \textsc{1-in-3-Satisfiability} to \textsc{Minimum Cardinality Vertex Cover}}}
    \label{fig:OSAT to MVC}
    \end{center}
\end{figure}
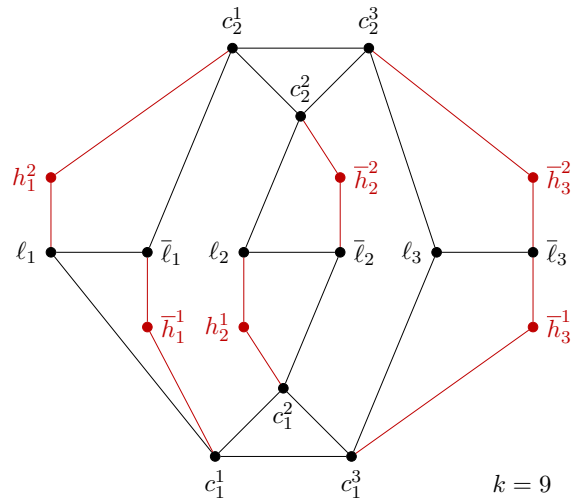

\subfile{../../Problems/MVC.tex}
\paragraph{Reduction}
The reduction presented here is a modified version of the reduction used in \autoref*{sec:ESAT to MVC} for $\ESAT\SSP\MVC$. Let $[L,\mc C]$ be 
a \textsc{1-in-3-Satisiability} instance. First we take the graph $G:=(V,E)$ created by the $\ESAT\SSP\MVC$ reduction. Next, for every edge 
$\{\ell_i,c_j^k\}\in E$ we create the vertex $\ne h_i^j$ and the edges $\{\ne\ell_i,\ne h_i^j\}$ and $\{\ne h_i^j,c_j^k\}$. 
Finally we set $k:=|L^2|+3|\mc C|$. Now $[G'=(V',E'),k]$ is the corresponding $\MVC$ instance,
where $V'$ and $E'$ are the modified sets as seen above.\\
Let $\mc C = \bigl\{\{\ell_1, \ne \ell_2, \ell_3\}, \{\ne\ell_1, \ell_2,  \ell_3\}\bigr\}$ and 
$L=\{\ell_1, \ell_2, \ell_3, \ne\ell_1, \ne\ell_2, \ne\ell_3\}$ define a \OSAT instance, then \autoref*{fig:OSAT to MVC} shows
the corresponding \MVC instance.

\begin{proof}
    Let $S$ be a solution for the \OSAT instance. Then exactly one literal is satisfied in every clause $C\in \mc C$ and one literal
    per pair is in $S$. Let $S'$ be the corresponding \MVC solution, where $S\subseteq S'$. This means all the edges between 
    literal pairs are covered, as well as exactly one of the six outgoing edges in every triangle. A vertex cover would also need to
    include at least two vertices in every triangle, to cover its interior edges. Now every edge is covered by the $|L^2|+2|\mc C|$ chosen vertices, 
    except for the edges leading from the triangles back to the compliment of the literal which satisfied the corresponding clause.
    However these can be covered by adding the $|\mc C|$ helper variables located on these paths to the vertex cover. Therefore every \OSAT solution
    $S$ has exactly one corresponding \MVC solution $S'$.\\
    Now, let $S'$ be a solution for the \MVC instance. This means every edge is covered by one of the $|L^2|+3|\mc C|$ vertices. Of these, $2|\mc C|$
    are needed to cover all the edges in the triangles, and $|L^2|$, to cover at least one literal vertex per literal pair. This leaves the 
    edges between these constructs. Four of the six edges going out from each triangle are covered by the two vertices that also cover the 
    interior edges. This means that, if for each literal pair, the vertex not connected directly to these two is in the cover, the edges going out 
    from the two helper variables connected to them are now covered from both ends. This leaves the last vertex in each triangle. The edge 
    connecting this triangle directly to the literal pairs now has to be covered by the literal vertex. This leaves the two edges connecting to the 
    helper variable of this literal's partner. As $S'$ is a \MVC solution, each of these helper variables is in the vertex cover, which means there 
    are exactly $|\mc C|$ of these. Therefore there are exactly $|\mc C|$ literals in $S'$, which connect directly to the clause triangles, one for
    each clause. Now let $S$ be the literals corresponding to the literal vertices in $S'$. As each clause triangle directly connects to 
    exactly one of these literal vertices, each clause in $\mc C$ is satisfied by exactly one literal. Therefore $S$ is a solution for the \OSAT instance.
\end{proof}
\noindent The reduction function $g$ is computable in polynomial time and is dependent on the input size of the \OSAT instance $\big|[L,\mc C]\big|$,
as for each clause in $\mc C$ and each literal in $L$ we create a fixed number of vertices in the new instance. Therefore $\OSAT\leq \MVC$ holds.

\paragraph{SSP \& Parsimonious}
As shown above, there is an one-to-one correspondence between the solutions $\S{\OSAT}$ and the solutions $\Sr{\MVC}$. We can therefore 
use \lemref{lem:SSP and SPR} as follows, to show $\OSAT\SSPR\MVC$.
\begin{proof}
    Let $\cee$ denote a literal in $L$ and let $\ell$ be the corresponding vertex in $V$ then let
    $$f_I:L\rightarrow V,~f(\cee)= \ell.$$
    Next, let $\Sall = \Snev =\emptyset$ and let $\cee_i(C_j)$ denote
    the $i^{th}$ literal in clause $C_j$. Then the linking functions can be defined as
    $$\mf L\big(\cee_i(C_j)\big) = \big\{c_j^{(i-1)\div  3},c_j^{(i-2)\div  3}, \ne h_i^j\big\} \cap V'$$
    where $\div$ denotes a modified modulo operator, with $i\div i =i$ and $i\div j = (i\mod j)$ for $i\neq j$. Note that $h_i^j$ is negated,
    if $\cee_i(C_j)$ corresponds to a negated literal. Now let
    $$\Slink:=\bigcup_{\S\OSATu}\mf L(S) = \big\{c_j^i\mid j\in \big[|\mc C|], i\in [3]\big\}$$
    and with this
    $$f_I':L\rightarrow V\backslash(\Slink)$$
    is a bijective function. Finally we get
    \begin{flalign*}
        \{f(S):\S\OSATu\}&=\big\{\{f(\ell)\mid\ell\in S\}:\S\OSATu\big\}\\
        &=\big\{\{\ell\in S'\}\backslash \Slink:\Sr\MVC\big\}\\
        &=\{S'\cap f(L):\Sr\MVC\}
    \end{flalign*}
    and therefore $\OSAT\SSPR\MVC$ holds according to \lemref{lem:SSP and SPR}.
\end{proof}

\subsubsection[... to Maximum Cardinality Independent Set]{1-in-3-Satisfiability to Maximum Cardinality Independent Set} 
\label{sec:OSAT to MIS}
\paragraph{Problem Definitions}

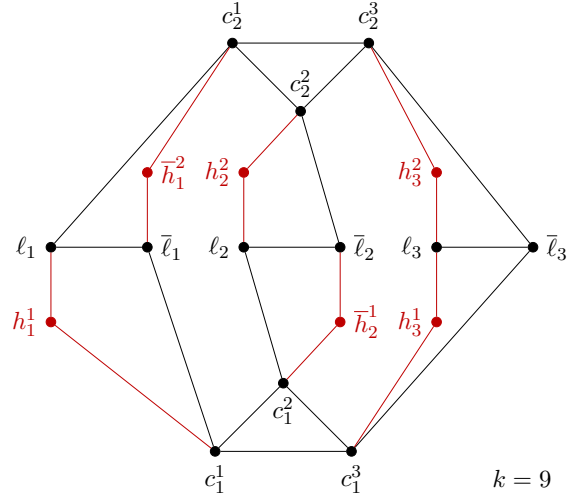
\begin{figure}[ht]
    \begin{center}
    \scalebox{.85}{
        \begin{tikzpicture}[vertex/.style = {draw, circle, fill, inner sep = 1.5}, node distance = 1.5cm]
            \node[vertex, label = left:$\ell_1$](l1){};
            \node[vertex, right of = l1, label = right:$\ne\ell_1$](nl1){};
            \node[vertex, right of = nl1, label = left:$\ell_2$](l2){};
            \node[vertex, right of = l2, label = right:$\ne\ell_2$](nl2){};
            \node[vertex, right of = nl2, label = left:$\ell_3$](l3){};
            \node[vertex, right of = l3, label = right:$\ne\ell_3$](nl3){};

            \node[vertex, below right = 2 and .5 of l2, label = below:$c_1^2$](c12){};
            \node[vertex, below left of = c12, label = below:$c_1^1$](c11){};
            \node[vertex, below right of = c12, label = below:$c_1^3$](c13){};

            \node[vertex, above left = 2 and .5 of nl2, label = above:$c_2^2$](c22){};
            \node[vertex, above left of = c22, label = above:$c_2^1$](c21){};
            \node[vertex, above right of = c22, label = above:$c_2^3$](c23){};

            \node[below right = .1 and 2 of c13](k){$k=9$};

            \node[vertex, black!25!red, below = 1 of l1, label = {[black!25!red]left:$h_1^1$}](nh11){};
            \node[vertex, black!25!red, above = 1 of nl1, label = {[black!25!red]right:$\ne h_1^2$}](h12){};

            \node[vertex, black!25!red, below = 1 of nl2, label = {[black!25!red]right:$\ne h_2^1$}](h21){};
            \node[vertex, black!25!red, above = 1 of l2, label = {[black!25!red]left:$h_2^2$}](nh22){};

            \node[vertex, black!25!red, below = 1 of l3, label = {[black!25!red]left:$h_3^1$}](nh31){};
            \node[vertex, black!25!red, above = 1 of l3, label = {[black!25!red]left:$h_3^2$}](nh32){};

            \draw 
                (l1) -- (nl1)
                (l2) -- (nl2)
                (l3) -- (nl3)

                (c11) -- (c12)
                (c11) -- (c13)
                (c13) -- (c12)
                
                (c21) -- (c22)
                (c21) -- (c23)
                (c23) -- (c22)

                (c11) -- (nl1)
                (c12) -- (l2)
                (c13) -- (nl3)
                (c21) -- (l1)
                (c22) -- (nl2)
                (c23) -- (nl3)

                (nl1) edge[black!25!red] (h12)
                (c21) edge[black!25!red] (h12)
                (l1) edge[black!25!red] (nh11)
                (c11) edge[black!25!red] (nh11)
                
                (nl2) edge[black!25!red] (h21)
                (c12) edge[black!25!red] (h21)
                (l2) edge[black!25!red] (nh22)
                (c22) edge[black!25!red] (nh22)

                (l3) edge[black!25!red] (nh31)
                (c13) edge[black!25!red] (nh31)
                (l3) edge[black!25!red] (nh32)
                (c23) edge[black!25!red] (nh32)
            ;
        \end{tikzpicture} 
    }\\
    \caption{\footnotesize{Example \textsc{1-in-3-Satisfiability} to \textsc{Minimum Cardinality Independent Set}}}
    \label{fig:OSAT to MIS}
    \end{center}
\end{figure}

\subfile{../../Problems/MIS.tex}
\paragraph{Reduction}
The reduction $\OSAT\leq\MIS$ is almost completely equivalent to the $\OSAT\leq\MVC$ reduction shown above in \autoref*{sec:OSAT to MVC}. The only change 
is that the graph $G$, which we take as a basis, comes from the $\ESAT\leq\MIS$ (\autoref*{sec:ESAT to MIS}) reduction, instead of the 
$\ESAT\leq\MVC$ reduction. \autoref*{fig:OSAT to MIS} shows an example reduction for the following instance, but we refrain from repeating the
proof here, as it is analogous to the proof presented above.\\
Let $\mc C = \{\{\ell_1, \ne \ell_2, \ell_3\}, \{\ne\ell_1, \ell_2,  \ell_3\}\}$ and 
$L=\{\ell_1, \ell_2, \ell_3, \ne\ell_1, \ne\ell_2, \ne\ell_3\}$ define a \OSAT instance, then the corresponding \MIS instance looks as follows.

\paragraph{SSP \& Parsimonious}
As with the $\OSAT\SSPR\MVC$ reduction, there exists a one-to-one correspondence between the solution spaces. 
We can therefore use \lemref{lem:SSP and SPR} as follows, to show $\OSAT\SSPR\MIS$.
\begin{proof}
    Let $\cee$ denote a literal in $L$ and let $\ell$ be the corresponding vertex in $V$ then let
    $$f_I:L\rightarrow V,~f(\cee)= \ell.$$
    Next, let $\Sall = \Snev =\emptyset$ and let 
    $$\mf L\big(\cee_i(C_j)\big) = \big\{c_j^i, \ne h_i^j\big\} \cap V'$$
    where $\cee_i(C_j)$ denotes the $i^{th}$ literal $\cee_i$ 
    in clause $C_j$. Note that $h_i^j$ is negated, if $\cee_i(C_j)$ corresponds to a negated literal. Now let
    $$\Slink:=\bigcup_{\S\OSATu}\mf L(S) = \big\{c_j^i\mid j\in \big[|\mc C|], i\in [3]\big\}$$
    and with this
    $$f_I':L\rightarrow V\backslash(\Slink)$$
    is a bijective function. Finally we get
    \begin{flalign*}
        \{f(S):\S\OSATu\}&=\big\{\{f(\ell)\mid\ell\in S\}:\S\OSATu\big\}\\
        &=\big\{\{\ell\in S'\}\backslash \Slink:\Sr\MIS\big\}\\
        &=\{S'\cap f(L):\Sr\MIS\}
    \end{flalign*}
    and therefore $\OSAT\SSPR\MIS$ holds according to \lemref{lem:SSP and SPR}.
\end{proof}

\subsubsection[... to 1+3-Dimensional Matching]{1-in-3-Satisfiability to 1+3-Dimensional Matching} 
\label{sec:OSAT to ODM}
\paragraph{Problem Definition}
\subfile{../../Problems/ODM.tex}

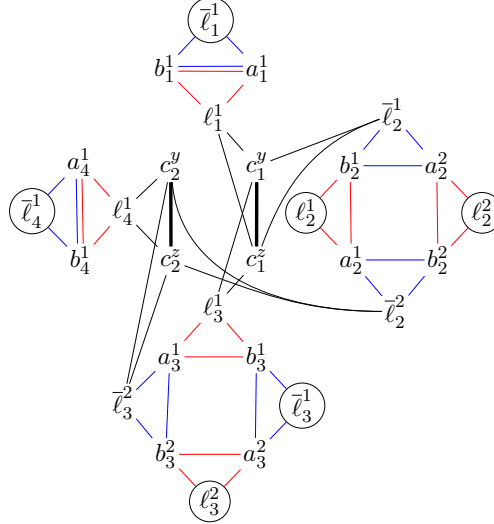
\begin{figure}[ht]
    \begin{center}
        $$\OSAT:=\big[\mc C = \{\{\ell_1, \ne \ell_2, \ell_3\},\{\ne \ell_2, \ne\ell_3,\ell_4\}\}, 
        L=\{\ell_1, \ell_2, \ell_3, \ell_4, \ne\ell_1, \ne\ell_2, \ne\ell_3,\ne \ell_4\}\big]$$\\
        gets mapped to\\
        \begin{flalign*}
            \ODM:=\big[X,&Y,Z,T,\Psi] ,\text{ with}\\
            X:= \{&\ell_1^1,\ne\ell_1^1,~\ell_2^1,\ell_2^2,\ne\ell_2^1,\ne\ell_2^2,\\
            &\ell_3^1,\ell_3^2,\ne\ell_3^1,\ne\ell_3^2,~\ell_4^1,\ne\ell_4^1\} \\
            Y:=\{&a_1^1,~a_2^1,a_2^2,~a_3^1,a_4^2,~a_4^1,\\
            &c_1^y,c_2^y\}\\
            Z:=\{&b_1^1,~b_2^1,b_2^2,~b_3^1,b_4^2,~b_4^1,\\
            &c_1^z,c_2^z\}
        \end{flalign*}
        and where $T$ and $\Psi$ are depicted as follows \\
        (a triangle indicates a triplet, a circled element a singleton)\\
        \scalebox{.85}{
        \begin{tikzpicture}[vertex/.style = {inner sep = 1},single/.style = {circle,draw,inner sep = 1}, node distance = 1.5cm, label distance=0cm]  
            \node[single](nl11){$\ne\ell_1^1$};
            \node[vertex, below right = .25 and .25 of nl11](a11){$a_1^1$};
            \node[vertex, below left = .25 and .25 of nl11](b11){$b_1^1$};
            \node[vertex, below left = .25 and .25 of a11](l11){$\ell_1^1$};

            \node[single, below right = 1 and 1cm of l11](l21){$\ell_2^1$};
            \node[vertex, below right = .25 and .25  of l21](a21){$a_2^1$};
            \node[vertex, above right = .25 and .25  of l21](b21){$b_2^1$};
            \node[vertex, above right = .25 and .25  of b21](nl21){$\ne\ell_2^1$};
            \node[vertex, below right = .25 and .25  of nl21](a22){$a_2^2$};
            \node[single, below right = .25 and .25 of a22](l22){$\ell_2^2$};
            \node[vertex, below left = .25 and .25  of l22](b22){$b_2^2$};
            \node[vertex, below left = .25 and .25  of b22](nl22){$\ne\ell_2^2$};
            
            \node[vertex, below left = 1 and 1 of l21](l31){$\ell_3^1$};
            \node[vertex, below left = .25 and .25  of l31](a31){$a_3^1$};
            \node[vertex, below right = .25 and .25  of l31](b31){$b_3^1$};
            \node[single, below right = .25 and .25  of b31](nl31){$\ne\ell_3^1$};
            \node[vertex, below left = .25 and .25  of nl31](a32){$a_3^2$};
            \node[single, below left = .25 and .25 of a32](l32){$\ell_3^2$};
            \node[vertex, above left = .25 and .25  of l32](b32){$b_3^2$};
            \node[vertex, above left = .25 and .25  of b32](nl32){$\ne\ell_3^2$};
            
            \node[vertex, above left = 1 and 1 of l31](l41){$\ell_4^1$};
            \node[vertex, above left = .25 and .25 of l41](a41){$a_4^1$};
            \node[vertex, below left = .25 and .25 of l41](b41){$b_4^1$};
            \node[single, below left = .25 and .25 of a41](nl41){$\ne\ell_4^1$};
            
            \node[vertex, below left = .25 and .25 of l11](c2y){$c_2^y$};
            \node[vertex, below left = 1.75 and .25 of l11](c2z){$c_2^z$};
            \node[vertex, below right = .25 and .25 of l11](c1y){$c_1^y$};
            \node[vertex, below right = 1.75 and .25 of l11](c1z){$c_1^z$};

            \draw   (l11) edge[] (c1y)
                    (l11) edge[] (c1z)
                    (c1y) edge[line width = 1.5pt] (c1z)
                    
                    (nl21) edge[] (c1y)
                    (nl21) edge[out = -160, in = 70] (c1z)
                    (c1y) edge[] (c1z)

                    (l31) edge[] (c1y)
                    (l31) edge[] (c1z)
                    (c1y) edge[] (c1z)

                    (l41) edge[] (c2y)
                    (l41) edge[] (c2z)
                    (c2y) edge[line width = 1.5pt] (c2z)

                    (nl32) edge[] (c2y)
                    (nl32) edge[] (c2z)
                    (c2y) edge[] (c2z)

                    (nl22) edge[out = 180, in = -85] (c2y)
                    (nl22) edge[out = 180, in = -20] (c2z)
                    (c2y) edge[] (c2z)

                    (l11) edge[red] (b11)
                    (l11) edge[red] (a11)
                    (a11) edge[red, in = -10, out = -170, looseness = 0] (b11)
                    (a11) edge[blue, in = 10, out = 170, looseness = 0] (b11)
                    (nl11) edge[blue] (b11)
                    (nl11) edge[blue] (a11)

                    (l21) edge[red] (b21)
                    (l21) edge[red] (a21)
                    (b21) edge[red] (a21)
                    (l22) edge[red] (b22)
                    (l22) edge[red] (a22)
                    (b22) edge[red] (a22)
                    (nl21) edge[blue] (b21)
                    (nl21) edge[blue] (a22)
                    (b21) edge[blue] (a22)
                    (nl22) edge[blue] (b22)
                    (nl22) edge[blue] (a21)
                    (b22) edge[blue] (a21)
                    
                    (l31) edge[red] (b31)
                    (l31) edge[red] (a31)
                    (b31) edge[red] (a31)
                    (l32) edge[red] (b32)
                    (l32) edge[red] (a32)
                    (b32) edge[red] (a32)
                    (nl31) edge[blue] (b31)
                    (nl31) edge[blue] (a32)
                    (b31) edge[blue] (a32)
                    (nl32) edge[blue] (b32)
                    (nl32) edge[blue] (a31)
                    (b32) edge[blue] (a31)

                    (l41) edge[red] (b41)
                    (l41) edge[red] (a41)
                    (a41) edge[red, in = 80, out = -80, looseness = 0] (b41)
                    (a41) edge[blue, in = 100, out = -100, looseness = 0] (b41)
                    (nl41) edge[blue] (b41)
                    (nl41) edge[blue] (a41)
            ;
        \end{tikzpicture} 
        }
    \caption{\footnotesize{Example \textsc{1-in-3-Satisfiability} to \textsc{1+3-Dimensional Matching}}}
    \label{fig:1-3S to ODM}
    \end{center}
\end{figure}

\paragraph{Reduction}
This reduction originated in \cite{Sch78} and is needed for the reduction $\OSAT\SSPR\DM$ (\autoref{sec:OSAT to DM}). 
Let $[L,\mc C]$ define a $\OSAT$ instance, then 
the corresponding $\ODM$ instance $[X,Y,Z,T,\Psi,\varphi]$ can be built as follows. For every literal pair $(\ell_i,\ne\ell_i)\in L^2$ let $o_{\max}(i)$
denote the number of occurrences of either one in $\mc C$, therefore if $\ell_i$ and $\ne\ell_i$ both occur once, then $o_{\max}(i)$ equals $2$. 
Now for every pair $(\ell_i,\ne\ell_i)$ we create $4\cdot o_{\max}(i)$ new elements: $\ell_i^1,...,\ell_i^{o_{\max}(i)} \in X$ and 
$\ne \ell_i^1,...,\ne \ell_i^{o_{\max}(i)} \in X$, $a_i^1,...,a_i^{o_{\max}(i)}\in Y$ and $b_i^1,...,b_i^{o_{\max}(i)}\in Z$. 
Next, for $1\leq k \leq o_{\max}(i)$ we add triplets $(\ell_i^k,a_i^k,b_i^k)$ and $(\ne \ell_i^k,a_i^k,b_i^{k-1})$, 
where $b_0 := b_{o_{\max}(i)}$. This way any solution includes either all triplets $(\ell_i^k,a_i^k,b_i^k)$ or all $(\ne \ell_i^k,a_i^k,b_i^{k-1})$ for $k\in [o_{\max}(i)]$
and therefore the entire collection of $\ell_i^k$'s are left uncovered, or all the $\ne \ell_i^k$'s.\\
Then we create two new elements $c_j^y\in Y$ and $c_j^z\in Z$ for every clause $C_j\in\mc C$. Let $o(i)$ denote the $k^{th}$ occurance of the literal 
$(\ell_i,\ne\ell_i)$ in $\mc C$ and connect $c_j^y$ and $c_j^z$ in triplets with the previously created elements in $X$: 
$$\forall \ell_i\in C_j:(\cee_i^{o(i)},c_j^y, c_j^z).$$
Let the set of singletons be defined as $\Psi:= \bigcup_{C\in\mc C}\Big(\bigcup_{\ell_i\in C}\ne\ell_i^{o(i)}\Big)$. Finally, we define the 
function $$\phi:\Psi\rightarrow T,~\phi(\ell_i^k)=\begin{cases}
    (\ne\ell_i^k,a_i^k,b_i^k), & \ell_i\in L^+\\
    (\ne\ell_i^k,a_i^k,b_i^{k-1}), & \ell_i\in L^-
\end{cases}$$
and with this, we define the binding function as 
$$\varphi:2^\Psi\rightarrow 2^T,~\varphi(S^\Psi) = \bigcup_{x\in S^\Psi} \phi(x).$$
\autoref*{fig:1-3S to ODM} shows an example reduction.
\begin{proof}
    Let $\S\OSATu$ be a satisfying assignment. Then there exists exactly one equivalent matching in the \ODM instance,
    where we include the triplet $(\cee_i^k,c_j^y,c_j^z)$ in $S'$ for every clause $C_j$ that is satisfied by the $k^{th}$ occurance
    of the literal $\cee_i$. Now the various $a$ and $b$ elements can be covered by including every triplet of the form 
    $(\ne\cee_i,a,b)$ for $\cee\in S$. The remaining elements are now only $\cee_i^k$'s, where $\ne\cee_i$ is the $k^{th}$ occurance of 
    the pair $(\cee_i,\ne\cee_i)$ as these are singletons they can all be taken into $S'$. Note that these singletons also directly imply 
    this specific solution.\\
    Now let $\Sr\ODMu$ be a valid matching. Then the clause elements $c_j^y$ and $c_j^z$ are contained in a triplet for every $j\in |\mc C|$.
    The last elements in each triplet are therefore a satisfying assignment in $\OSAT$, as the negated versions of these are contained
    in triplets with the various $a$ and $b$ elements.
\end{proof}
\noindent The reduction function $g$ is computable in polynomial time and is dependent on the input size of the \OSAT instance $\big|[L,\mc C]\big|$,
as for each clause in $\mc C$ and each literal in $L$ we create a fixed number of elements in the new instance. Therefore $\ESAT\leq \ODM$ holds.

\paragraph{SSP \& Parsimonious}
As we have seen above, there is a direct one-to-one correspondence between the literals in $S$ satisfying the clauses $C_j$ and the 
triplets containing the elements $c_j^y$ and $c_j^z$. Therefore we can use \lemref{lem:SSP and SPR} to prove $\OSAT\SSPR\ODM$.
\begin{proof}
    Let the functions $f_I$ be defined as follows
    $$f_I:L\rightarrow T\cup \Psi,~f(\cee_i)=(\cee_i^{o(i)},c_{C(i)}^y,c_{C(i)}^z)$$
    where $o(i)$ denotes the $k^{th}$ occurance of literal pair $(\cee_i,\ne\cee_i)$ in $|\mc C|$ and 
    $C(i)$ denotes the clause $C_j$ that is satisfied by $\cee_i$.\\
    Now let $\Sall=\Snev=\emptyset$ and 
    $$\Slink:=\Psi\cup \big\{(\ell_i^k,a_i^k,b_i^k),\big(\ne\ell_i^k,a_i^k,b_i^{k-1})\mid i\in |L^2|, k\in[o_{\max}(i)]\big\}$$
    then $f'_I:L\rightarrow (T\cup\Psi) \backslash \Slink$ is bijective and we get
    \begin{flalign*}
        \{f(S):\S\OSATu\} & = \big\{\{f(\ell)\mid\cee\in S\}:\S\OSATu\big\}\\
        &= \big\{\big\{(\cee_i^{o(i)},c_{C(i)}^y,c_{C(i)}^z)\in S'\big\}:\Sr\ODMu\big\}\\
        &= \{S'\backslash\Slink :\Sr\ODMu\}\\
        &=\{S'\cap f(L):\Sr\ODMu\}
    \end{flalign*}
    Therefore $\OSAT\SSPR\ODM$ holds.
\end{proof}

\subsubsection[... to 3-Dimensional Exact Matching]{1-in-3-Satisfiability to 3-Dimensional Exact Matching}
\label{sec:OSAT to DM}
\paragraph{Problem Definitions}

\subfile{../../Problems/ODM.tex}
\subfile{../../Problems/DM.tex}

\paragraph{Reduction}
This reduction uses the reduction $\OSAT\SSPR\ODM$, presented above in \autoref{sec:OSAT to ODM}. We therefore start with a \ODM instance.
We know that there is a binding $\varphi$ between triplets and singletons, such that a matching $\S\ODMu$ is destinct for a 
given set of singletons $\psi\subseteq S$. Let $\ODM:=[X,Y,Z,T,\Psi,\varphi]$, then we create the \DM instance $[X',Y',Z',T']$ as follows.
First, we create three destinctly labled copies each of $X$, $Y$, $Z$ and $T$. Now, let $X':= X_1\cup Y_2\cup Z_3$, 
$Y':= X_3\cup Y_1\cup Z_2$ and $Z':= X_2\cup Y_3\cup Z_1$ and let $T':=T_1\cup T_2\cup T_3\cup \{(\psi_1,\psi_2,\psi_3)\mid \psi\in \Psi\}$, 
where $\psi_i$ is a copy of an element $\psi\in \Psi$. This concludes the reduction, an example is shown in \autoref*{fig:ODM to DM}.
\begin{figure}[ht]
    \centering
    \begin{tabular}{RLRL}
        \ODM:=&\big[X,Y,Z,T,\Psi] ,\text{ with} \hspace*{1cm} &\hspace*{.5cm} \DM:= &\big[X',Y',Z',T'] ,\text{ with}\\
         X=& \{a,b,c\} & X'=& \{a_1,b_1,c_1,~p_2,~u_3,v_3\} \\
         Y=&\{p\} & Y=&\{a_2,b_2,c_2,~p_3,~u_1,v_1\} \\
         Z=&\{u,v\} & Z= &\{a_3,b_3,c_3,~p_1,~u_2,v_2\} 
    \end{tabular}\\
        \begin{subfigure}[t]{.3\textwidth}
            and $T, \Psi$ depicted as follows\\
            ~\\
        \begin{tikzpicture}[vertex/.style = {inner sep = 1},single/.style = {circle,draw,inner sep = 1}, node distance = 1.5cm, label distance=0cm]  
            \node[single](a){$a$};
            \node[single, right = .65 of a](b){$b$};
            \node[single, right = .65 of b](c){$c$};
            \node[vertex, below = .5 of b](p){$p$};
            \node[single, below left = .5 and .25 of p](u){$u$};
            \node[single, below right = .5 and .25 of p](v){$v$};
            \node[below = 4.55cm of b](){};

            \draw   (a) edge[red!80!black!100!] (p)
                    (a) edge[red!80!black!100!] (u)
                    (u) edge[red!80!black!100!] (p)

                    (c) edge[blue!80!black!100!] (p)
                    (c) edge[blue!80!black!100!] (v)
                    (v) edge[blue!80!black!100!] (p)
            ;
        \end{tikzpicture} 
        \end{subfigure} 
        \begin{subfigure}[t]{.4\textwidth}
            \hspace{2cm}and $T'$ depicted as follows\\
            ~\\
        \begin{tikzpicture}[vertex/.style = {inner sep = 1},single/.style = {circle,draw,inner sep = 1}, node distance = 1.5cm, label distance=0cm]  
            \node[vertex](a){$a_1$};
            \node[vertex, right = .35 of a](b){$b_1$};
            \node[vertex, right = .35 of b](c){$c_1$};
            \node[vertex, below = .25 of b](p){$p_1$};
            \node[vertex, below left = .25 and .15 of p](u){$u_1$};
            \node[vertex, below right = .25 and .15 of p](v){$v_1$};

            \node[vertex, black!75!, above left = .1 and .5 of a](a12){$a_2$};
            \node[vertex, black!75!, above left = .5 and .1 of a](a13){$a_3$};
            \node[vertex, black!75!, above left = .5 and .1 of b](b12){$b_2$};
            \node[vertex, black!75!, above right = .5 and .1 of b](b13){$b_3$};
            \node[vertex, black!75!, above right = .5 and .1 of c](c12){$c_2$};
            \node[vertex, black!75!, above right = .1 and .5 of c](c13){$c_3$};

            \node[vertex, below left = .5 and .5 of u](v2){$v_2$};
            \node[vertex, below right = .5 and .5 of v2](u2){$u_2$};
            \node[vertex, below = .5 of v2](p2){$p_2$};
            \node[vertex, below = .5 of p2](a2){$a_2$};
            \node[vertex, above left = .1 and .1 of a2](b2){$b_2$};
            \node[vertex, left = .5 of p2](c2){$c_2$};

            \node[vertex, black!75!, below right = .5 and .0 of a2](a23){$a_3$};
            \node[vertex, black!75!, below left = .5 and .0 of a2](a21){$a_1$};
            \node[vertex, black!75!, below left = .5 and .1 of b2](b23){$b_3$};
            \node[vertex, black!75!, below left = .1 and .5 of b2](b21){$b_1$};
            \node[vertex, black!75!, below left = .0 and .5 of c2](c23){$c_3$};
            \node[vertex, black!75!, above left = .0 and .5 of c2](c21){$c_1$};

            \node[vertex, below right = .5 and .5 of v](u3){$u_3$};
            \node[vertex, below left = .5 and .5 of u3](v3){$v_3$};
            \node[vertex, below = .5 of u3](p3){$p_3$};
            \node[vertex, right = .5 of p3](a3){$a_3$};
            \node[vertex, below left = .1 and .1 of a3](b3){$b_3$};
            \node[vertex, below = .5 of p3](c3){$c_3$};

            \node[vertex, black!75!, below right = .0 and .5 of a3](a32){$a_3$};
            \node[vertex, black!75!, above right = .0 and .5 of a3](a31){$a_1$};
            \node[vertex, black!75!, below right = .5 and .1 of b3](b32){$b_2$};
            \node[vertex, black!75!, below right = .1 and .5 of b3](b31){$b_1$};
            \node[vertex, black!75!, below right = .5 and .0 of c3](c31){$c_1$};
            \node[vertex, black!75!, below left = .5 and .0 of c3](c32){$c_2$};

            \draw   (a) edge[red!80!black!100!] (p)
                    (a) edge[red!80!black!100!] (u)
                    (u) edge[red!80!black!100!] (p)

                    (c) edge[blue!80!black!100!] (p)
                    (c) edge[blue!80!black!100!] (v)
                    (v) edge[blue!80!black!100!] (p)

                    (a2) edge[red!80!black!100!] (p2)
                    (a2) edge[red!80!black!100!] (u2)
                    (u2) edge[red!80!black!100!] (p2)

                    (c2) edge[blue!80!black!100!] (p2)
                    (c2) edge[blue!80!black!100!] (v2)
                    (v2) edge[blue!80!black!100!] (p2)

                    (a3) edge[red!80!black!100!] (p3)
                    (a3) edge[red!80!black!100!] (u3)
                    (u3) edge[red!80!black!100!] (p3)

                    (c3) edge[blue!80!black!100!] (p3)
                    (c3) edge[blue!80!black!100!] (v3)
                    (v3) edge[blue!80!black!100!] (p3)

                    (u) -- (u2)
                    (u) -- (u3)
                    (u2) -- (u3)
                    (v) -- (v2)
                    (v) -- (v3)
                    (v2) -- (v3)

                    (a) edge[dashed, black!75!] (a12)
                    (a) edge[dashed, black!75!] (a13)
                    (a2) edge[dashed, black!75!] (a21)
                    (a2) edge[dashed, black!75!] (a23)
                    (a3) edge[dashed, black!75!] (a31)
                    (a3) edge[dashed, black!75!] (a32)

                    (b) edge[dashed, black!75!] (b12)
                    (b) edge[dashed, black!75!] (b13)
                    (b2) edge[dashed, black!75!] (b21)
                    (b2) edge[dashed, black!75!] (b23)
                    (b3) edge[dashed, black!75!] (b31)
                    (b3) edge[dashed, black!75!] (b32)

                    (c) edge[dashed, black!75!] (c12)
                    (c) edge[dashed, black!75!] (c13)
                    (c2) edge[dashed, black!75!] (c21)
                    (c2) edge[dashed, black!75!] (c23)
                    (c3) edge[dashed, black!75!] (c31)
                    (c3) edge[dashed, black!75!] (c32)
            ;
        \end{tikzpicture} 
    \end{subfigure}\\~\\
    (a triangle indicates a triplet and a circled element a singleton)
    \caption{\footnotesize{Example \textsc{1+3-Dimensional Matching} to \textsc{3-Dimensional Exact Matching}}}
    \label{fig:ODM to DM}
\end{figure}
\begin{proof}
    Now every matching $\S\ODMu$ is equivalent to a matching $\Sr\DMu$, where, for all $(x,y,z)\in S$ we choose
    $(x_1,y_1,z_1)$, $(y_2,z_2,x_2)$ and $(z_3,x_3,y_3)$ for $S'$, as well as the triples $(\psi_1,\psi_2,\psi_3)$ for the
    $\psi\in S$. Also, every matching $\Sr\DMu$ consits only of triplets that either consist of three copies of the same element
    or of three elements of the same copied instance. This triplet must then exist for every copy, as in our assumption, we stated that
    a choice of singletons must indicate exactly one solution. Hence, this is equivalent to a $\ODM$ matching.
\end{proof}
\noindent The reduction function $g$ is computable in polynomial time and is dependent on the input size of the \ODM 
instance $\big|[X,Y,Z,T,\Psi]\big|$, as for each set in the instance, we create a fixed number of elements in the new instance.
Therefore $\ODM\leq \DM$ holds.

\paragraph{SSP \& Parsimonious}
As seen above, exists a one-to-one correspondence between the triplets and singletons in $\ODM$ and the triplets in $\DM$.
Therefore we can apply \lemref{lem:SSP and SPR} as follows to prove $\ODM\SSPR\DM$.
\begin{proof}
    Let the functions $f_I$ be defined as
    $$f_I:T\cup\Psi \rightarrow T', ~f(\pi) = \begin{cases}
        (x_1,y_1,z_1) &\mid \pi = (x,y,z) \in T \\
        (\psi_1,\psi_2,\psi_3) &\mid \pi=\psi \in \Psi
    \end{cases}$$
    and let $\Sall=\Snev=\emptyset$ and let $\Slink$ be defined by
    $$\mf L:T'\rightarrow 2^{T'},~\mf L\big((x_1,y_1,z_1)\big) = \big\{(y_2,z_2,x_2),(z_3,x_3,y_3)\big\}$$
    then the functions $f'_I:T\cup\Psi\rightarrow T'\backslash \Slink$ are bijective and we get
    \begin{flalign*}
        \{f(S):\S\ODMu\} &=\Big\{\big\{f\big((x,y,z)\big)\mid (x,y,z)\in S\}\cup\{f(\psi)\mid\psi\in S\}:\S\ODMu\Big\}\\
        &= \big\{\{(x_1,y_1,z_1)\in S'\}\cup\{(\psi_1,\psi_2,\psi_3)\in S'\}:\Sr\DMu\big\}\\
        &= \{S'\backslash \Slink :\Sr\DMu\}\\
        &= \{S'\cap f(T\cap \Psi) :\Sr\DMu\}
    \end{flalign*}
    and therefore $\ODM\SSPR\DM$ holds.
\end{proof}

\ifSubfilesClassLoaded{
  \bibliography{../../ref}
}{}
\end{document}

%% file: Problems/STT.tex
\noindent\fbox{
  \parbox{\textwidth}{
      \textsc{Steiner Tree ($\STT$)}\\
      \begin{tabular}{cl}
      ~ & \textbf{Instances:} Graph $G=(N\cup T,E)$, set of Steiner (nonterminal) vertices $N\subseteq V$, \\
      ~ & set of terminal vertices $T\subseteq V$, edge weight function $w:E\rightarrow\mathbb N$ and number $k\in \mathbb N$.\\
      ~ & \textbf{Universe:} Edge set $\mathcal{U}:=E$.\\
      ~ & \textbf{Feasible solution set:} The set of all sets $S\subseteq E$, such that $S$ is a tree connecting all\\
      ~ & terminal vertices in $T$.\\
      ~ & \textbf{Solution set:} The set of feasible solutions $S$ with $\sum_{e\in S}w(e)\leq k$.
      \end{tabular}
  }
}

\ifSubfilesClassLoaded{
  \bibliography{../ref}
}{}
\end{document}

%% file: Problems/ODM.tex
\noindent\fbox{
    \parbox{\textwidth}{
        \textsc{1+3-Dimensional Matching ($\ODM$)}\\
        \begin{tabular}{cl}
        ~ & \textbf{Instances:} Sets $X,Y,Z$, such that $\exists n\in \mathbb N: n=|X|=|Y|=|Z|$, \\
        ~ & triplets $T\subseteq X\times Y\times Z$, singleton set $\Psi\subseteq X\cup Y\cup Z$ and binding function $\varphi:2^\Psi\rightarrow 2^T$.\\
        ~ & \textbf{Universe:} Set $\mathcal{U}:=T\cup \Psi$.\\
        ~ & \textbf{Solution set:} The set $S = S^T\cup S^\Psi$ with $S^T \subseteq T$ and $S^\Psi\subseteq \Psi$, such that\\
        ~ & $\big(\bigcup S^T\big) \cup  S^\Psi = X\cup Y\cup Z$; $\big(\bigcup S^T\big) \cap  S^\Psi = \emptyset$, \\
        ~ & $\forall~t_i,t_j\in S^T, i\neq j: t_i\cap t_j = \emptyset$ and $\varphi(S^\Psi)\in S \Leftrightarrow S^T\in S$.
        \end{tabular}
    }
}

\ifSubfilesClassLoaded{
  \bibliography{../ref}
}{}
\end{document}

%% file: Problems/DM.tex
\noindent\fbox{
    \parbox{\textwidth}{
        \textsc{3-Dimensional Exact Matching ($\DM$)}\\
        \begin{tabular}{cl}
        ~ & \textbf{Instances:} Sets $X,Y,Z$, such that $\exists n\in \mathbb N: n=|X|=|Y|=|Z|$, triplets $T\subseteq X\times Y\times Z$.\\
        ~ & \textbf{Universe:} Set $\mathcal{U}:=T$.\\
        ~ & \textbf{Solution set:} The set $S\subseteq T$, such that for any\\
        ~ & $(x_i,y_i,z_i),(x_j,y_j,z_j)\in S, i\neq j: x_i\neq x_j\land y_i\neq y_j\land z_i\neq z_j\land |S|=m$.
        \end{tabular}
    }
}

\ifSubfilesClassLoaded{
  \bibliography{../ref}
}{}
\end{document}

%% file: Reductions/Vertex_Cover/Vertex_Cover.tex
\subsection{Reductions from Minimum Cardinality Vertex Cover}

\subfile{../../Problems/MVC.tex}
\subsubsection[... to Minimum Cardinality Dominating Set]{Minimum Cardinality Vertex Cover to Minimum Cardinality Dominating Set}
\label{sec:MVC to MDS}
\paragraph{Problem Definition}

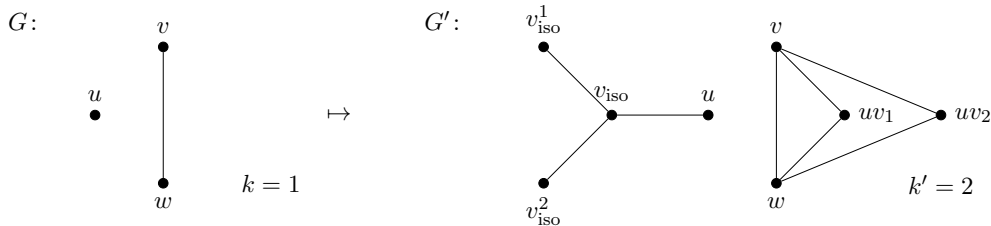
\begin{figure}[h]
    \begin{center}
    \scalebox{.85}{
        \begin{tikzpicture}[vertex/.style = {draw, circle, fill, inner sep = 1.5}, node distance = 1.5cm]
            \node[label = above:$G\colon$](G){};
            \node[vertex, below right of = G, label = $u$](u){};
            \node[vertex, above right of = u, label = $v$](v){};
            \node[vertex, below right of = u, label = below:$w$](w){};

            \node[right = 1cm of w](k){$k=1$};
            
            \node[above right of = k](a){$\mapsto$};

            \node[above right of = a, label = above right:$G'\colon$](G'){};
            \node[below right of = G'](x){};
            \node[vertex, above right of = x, label = $v_\text{iso}^1$](iso1){};
            \node[vertex, below right of = x, label = below:$v_\text{iso}^2$](iso2){};
            \node[vertex, below right of = iso1, label = $v_\text{iso}$](iso){};
            \node[vertex, right of = iso, label = $u$](u'){};
            \node[vertex, above right of = u', label = $v$](v'){};
            \node[vertex, below right of = u', label = below:$w$](w'){};
            \node[vertex, below right of = v', label = right:$uv_1$](1){};
            \node[vertex, right of = 1, label = right:$uv_2$](2){};
            \node[below of = 2, label = $k'\equals 2$](){};

            \draw   (v) -- (w)
                    (iso) -- (iso1)
                    (iso) -- (iso2)
                    (iso) -- (u')
                    (v') -- (w')
                    (v') -- (1)
                    (v') -- (2)
                    (w') -- (1)
                    (w') -- (2)
            ;
        \end{tikzpicture} 
    }
    \caption{\footnotesize{Example \textsc{Minimum Cardinality Vertex Cover} to \textsc{Minimum Cardinality Dominating Set}}}
    \label{fig:VC to DS}
    \end{center}
\end{figure}

\subfile{../../Problems/MDS.tex}
\paragraph{Reduction}
The reduction \textsc{Minimum Cardinality Vertex Cover} $\SSP$ \textsc{Minimum Cardinality Dominating Set} on a graph $G=(V,E)$ works as follows. 
First we identify any isolated vertices in $G$ and connect these to a new vertex $v_\text{iso}$ not in $V$. We then increase $k$ by one for the 
bounds of the Dominating Set instance and add $k+2$ new vertices connected to $v_\text{iso}$. We also add $k+1$ new vertices for every edge found 
in the original graph $G$ and connect each of them to the two endpoints of the edge. Formally the reduction is as follows:\\
Let $G:=(V,E)$ and $k\in \mathbb N$ be a \textsc{Minimum Cardinality Vertex Cover} instance. Then
\begin{flalign*}
    G'&:=(V',E')\text{, where}\\
    &V':=V\cup \{v_\text{iso}\}\cup \big\{v_\text{iso}^i \bigm| i \in [k+2]\big\} \cup \big\{uv_i\bigm|(u,v)\in E,i \in [k+1]\big\}\\
    &E':=E\cup \big\{(v,v_\text{iso}) \bigm| v\in V \text{ is isolated}\big\}\cup \big\{(v_\text{iso}, v_\text{iso}^i) \bigm| i\in [k+2]\big\}\\
    & \hspace*{1.33cm}\cup\big\{(u, uv_i),(v,uv_i)\bigm|(u,v)\in E, i \in [k+1]\big\}\\
    k' & := k + 1\
\end{flalign*}
is a \textsc{Minimum Cardinality Dominating Set} instance. An example of this for\\ $\MVC:=[G=(\{u,v,w\},\{(v,w)\}),~ k= 1]$ is shown in 
\autoref*{fig:VC to DS}. \\
\\
The solution sets of this example would be $\mc S_\MVC = \{\{v\}, \{w\}\}$ and $\mc S_\MDS = \{\{v_\text{iso}, v\}, \{v_\text{iso}, w\}\}$.
\begin{proof}
    For every vertex cover $S\in \mc S_\MVC$ in $G$ the set $S\cup \{v_\text{iso}\}$ is a dominating set in $G'$, as $v_\text{iso}$ dominates all 
previously isolated vertices, as well as the new nodes connected to $v_\text{iso}$, and the vertices in $S$ dominate all other vertices via the 
edge they cover for the \textsc{Minimum Cardinality Vertex Cover}. \\
For every dominating set $S'\in \mc S_\MDS$ in $G'$ the set $S'\backslash \{v_\text{iso}\}$ is a vertex cover in $G$. This also holds as $v_\text{iso}$ 
has to be in $S'$, as otherwise the vertices connected to $v_\text{iso}$ cannot be dominated and therefore all other vertices in $S'$ also have to be a 
vertex cover in $G$, as not all $k'$ vertices added for each edge in $E$ can be dominated by the $k'-1$ vertices in $S'\backslash \{v_\text{iso}\}$, but 
since we added these vertices for every edge, at least one vertex of every edge needs to be in $S'$. \\
Therefore the reduction holds.
\end{proof}
\noindent The reduction function $g$ is computable in polynomial time and is dependent on the input size of the \MVC instance $\big|[G=(V,E),k]\big|$,
as for each vertex in $V$ we create a fixed number of vertices and edges in the new instance. Therefore $\MVC\leq \SC$ holds.

\paragraph{SSP \& Parsimonious}
As for any solution $S$ of a \MVC instance, $S\cup\{v_{iso}\}$ is a solution for the corresponding \MDS instance, we can use \lemref{lem:SSP and SPR}
to prove that the reduction is both SSP and strongly Parsimonious. 
\begin{proof}
    Let $\Snev=\Slink=\emptyset$ and let $\Sall:=\{v_{iso}\}$. Let the functions be defined as 
    $$f_I:V\rightarrow V',~f(v) = v',$$ where $v$ and $v'$ are the 
    same vertex in their respective instances. Now $f_I':V\rightarrow V'\backslash\Sall$ are bijective functions and
    \begin{flalign*}
        \{f(S) :S \in \mc S_\MVC\} & = \big\{\{f(v)\mid v\in S\} : S\in \mc S_\MVC\big\}\\
        & = \big\{\{v'\in S'\}\backslash\{v_{iso}\}: S'\in \mc S_\MDS\big\}\\
        &= \{S' \cap f(V) : S' \in \mc S_\MDS\}
    \end{flalign*}
    Therefore $\MVC\SSPR\MDS$ holds according to \lemref{lem:SSP and SPR}.
\end{proof}

\subsubsection[... to Set Cover]{Minimum Cardinality Vertex Cover to Set Cover}
\paragraph{Problem Definition}
\subfile{../../Problems/SC.tex}

\paragraph{Reduction}
The reduction $\MVC \leq \SC$ as presented by Karp \cite{karp} works as follows. Let $\MVC:=[G=(V,E),k]$ and
let the vertices $v\in V$ and edges $e\in E$ be denoted by their indices. By mapping every vertex $v\in V$ to a set $\psi_v$, 
which is made up of the indices of its incident edges, we get $n:=|V|$ sets $\psi$ consiting of subsets of $\Psi:=\{1,...,|E|\}$.\\
An example reduction looks as follows.
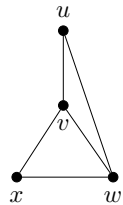
\begin{center}
    \scalebox{.85}{
        \begin{tikzpicture}[vertex/.style = {draw, circle, fill, inner sep = 1.5}, node distance = 1.5cm]
            \node[vertex, label = above:$u$](u){};
            \node[vertex, below = 1cm of u, label = below:$v$](v){};
            \node[vertex, below right = 1cm and .66cm of v, label = below:$w$](w){};
            \node[vertex, left of = w, label = below:$x$](x){};

            \draw   (u) -- (v)
                    (v) -- (w)
                    (w) -- (x)
                    (x) -- (v)
                    (u) -- (w)
            ;
        \end{tikzpicture} 
    }
    \begin{flalign*}
        &\MVC:=\Big[G = \big(\{u, v, w, x\},\big\{\{u, v\},\{u, w\},\{v, w\}, \{w, x\},\{x, v\}\big\}\big),k\Big]\\
        \mapsto~&\SC:=\big[\Psi = \{ 1, 2, 3, 4, 5\},~\psi_1 = \{ 1, 2\}, ~\psi_2= \{1, 3, 5\}, ~\psi_3= \{2, 3, 4\},~\psi_4= \{4, 5\},k\big]
    \end{flalign*}
\end{center}
\begin{proof}
    Now, let $S\in \mc S_\MVC$ be a solution for $\MVC$. Then the set $\{\psi_{\ind(v)}\mid v\in S\}$ (Note that $\ind(v)$ denotes the index of $v$ in $V$) 
    is a solution $S'\in \mc S_\SC$, as a vertex cover
    includes at least one vertex incident to every edge and therefore the corresponding set cover covers every edge in $\{1,...,|E|\}$.
\end{proof}
\noindent The reduction function $g$ is computable in polynomial time and is dependent on the input size of the \MVC instance $\big|[G=(V,E),k]\big|$,
as for each vertex in $V$ we create a fixed number of sets the new instance. Therefore $\MVC\leq \SC$ holds.

\paragraph{SSP \& Parsimonious}
The reduction is also SSP, as every vertex $v$ in a solution $S$ is directly mapped to a set $S_v$ in $S'$. As no other sets are created by the
reduction, we have a direct one-to-one correspondence between the universes and can therefore use \lemref{lem:SSP and SPR 2} to prove that this reduction
is both SSP and SPR.
\begin{proof}
    Let the mapping be defined by the functions 
    $$f_I: V \rightarrow \{\psi_i\mid i\in[n]\}, ~f(v) = \psi_{\ind(v)},$$ 
    then we also get 
    $$f^{-1}_I:\{\psi_i\mid i\in[n]\} \rightarrow V,~f^{-1}(\psi_i) = v_i$$
    and now
    \begin{flalign*}
        \{f(S): S\in \mc S_\MVC\} & = \big\{\{f(v)\mid v\in S\} :S\in \mc S_\MVC \big\}\\
        & = \big\{\{\psi_{\ind(v)}\mid v\in S\} :S\in \mc S_\MVC \big\}\\
        &= \big\{\{\psi_{i}\in S'\} :S'\in \mc S_\SC \big\}\\
        &= \{S'\cap f(V) : S'\in \mc S_\SC\}\\
        &= \{\Sr\SC\}
    \end{flalign*}
    and therefore it holds that \MVC $\SSPR$ \textsc{SC}.
\end{proof}

\subsubsection[... to Hitting Set]{Minimum Cardinality Vertex Cover to Hitting Set}
\paragraph{Problem Definition}
\subfile{../../Problems/HS.tex}

\paragraph{Reduction}
This reduction is also very simple as Hitting Set is basically the same problem as Vertex Cover. Each vertex $v\in V$ gets mapped to
its index in the set $\Psi = \{1,...,|V|\}$ and each edge $\{u,v\}\in E$ gets mapped to the set $\psi_{\ind(e)} = \{\ind(u),\ind(v)\}$, where $\ind$ is 
the index function. \\
An example of this would be:
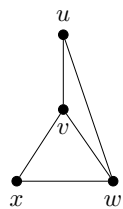
\begin{center}
    \scalebox{.85}{
        \begin{tikzpicture}[vertex/.style = {draw, circle, fill, inner sep = 1.5}, node distance = 1.5cm]
            \node[vertex, label = above:$u$](u){};
            \node[vertex, below = 1cm of u, label = below:$v$](v){};
            \node[vertex, below right = 1cm and .66cm of v, label = below:$w$](w){};
            \node[vertex, left of = w, label = below:$x$](x){};

            \draw   (u) -- (v)
                    (v) -- (w)
                    (w) -- (x)
                    (x) -- (v)
                    (u) -- (w)
            ;
        \end{tikzpicture} 
    }
    \end{center}
\begin{flalign*}
    &\MVC:=\Big[G = \big(\{u, v, w, x\}, \big\{\{u, v\},\{u, w\},\{v, w\}, \{w, x\},\{x, v\}\big\}\big),k\Big]\\
    \mapsto~&\HS:=\big[\Psi = \{ 1, 2, 3, 4\},~\psi_1 = \{ 1, 2\}, ~\psi_2 = \{1, 3\}, ~\psi_3 = \{2, 3\},~\psi_4 = \{3, 4\},~\psi_5 = \{4,2\},k\big]
\end{flalign*}
\begin{proof}
    This reduction holds, as every solution $S\in \mc S_\MVC$ is equivalent in a solution in $\mc S_\HS$ when we map every 
    $v\in S$ to its index.
\end{proof}
\noindent The reduction function $g$ is computable in polynomial time and is dependent on the input size of the \MVC instance $\big|[G=(V,E),k]\big|$,
as for each vertex in $V$ we create a fixed number of sets the new instance. Therefore $\MVC\leq \HS$ holds.

\paragraph{SSP \& Parsimonious}
As for every solution set $S\in \mc S_\MVC$ the set of indices of the vertices in $S$ is a solution $S'\in \mc S_\HS$ we have a direct one-to-one correspondence 
between the universes and can therefore use \lemref{lem:SSP and SPR 2} to prove that this reductionis both SSP and SPR.
\begin{proof}
    Let the functions $f_I$ be defined as
    $$f_I: V\rightarrow \Psi,~f(v) = \ind(v),$$ 
    then we get
    $$f^{-1}_I:\Psi \rightarrow V,~f^{-1}(i) = v_i$$
    and now
    \begin{flalign*}
        \{f(S): S\in \mc S_\MVC \} &= \big\{\{f(v)\mid v\in S\}: S\in \mc S_\MVC \big\}\\
        &= \big\{\{\ind(v) \mid v\in S\}: S\in \mc S_\MVC \big\}\\
        &= \big\{\{i\in S'\}: S'\in S_\HS\big\}\\
        &= \{S' \cap f(V): S'\in S_\HS \}\\
        &= \{\Sr\HS\}
    \end{flalign*}
    and therefore \MVC $\SSPR$ \HS.
\end{proof}

\subsubsection[... to Feedback Vertex Set]{Minimum Cardinality Vertex Cover to Feedback Vertex Set}
\paragraph{Problem Definition}
\subfile{../../Problems/FVS.tex}

\paragraph{Reduction}
Karp's reduction \cite{karp} works as follows: Let $G=(V,E), k$ be a \MVC instance, then, if we map every vertex $v\in V$ to itself
($v\in V \Leftrightarrow v\in V'$) and every edge $\{u,v\}\in E$ is mapped to the two arcs $(u,v), (v,u) \in A$ in either direction,
we get the \FVS instance $G'=(V',A), k$. \\
\begin{figure}[h]
    \begin{center}
    \scalebox{.85}{
        \begin{tikzpicture}[vertex/.style = {draw, circle, fill, inner sep = 1.5}, node distance = 1.5cm]
            \node[label = above:$G\colon$](G){};
            \node[vertex, below right =  .25cm and 1cm of G, label = $u$](u){};
            \node[vertex, right of = u, label = $v$](v){};

            \node[right = 1cm of v, label = right:$\mapsto$](to){};
            
            \node[above right = .25 and 2cm of to, label = above:$G'\colon$](G'){};
            \node[vertex, below right =  .25cm and 1cm of G', label = $u$](u'){};
            \node[vertex, right of = u', label = $v$](v'){};

            \draw   (u) -- (v)
                    (u') edge[bend right = 30, ->] (v')
                    (v') edge[bend right = 30, ->] (u');
        \end{tikzpicture} 
    }
    \caption{\footnotesize{Example \textsc{Minimum Cardinality Vertex Cover} to \textsc{Feedback Vertex Set}}}
    \end{center}
\end{figure}
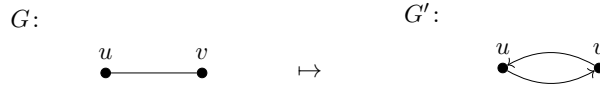
\begin{proof}
    This reduction holds, as now every edge is mapped to a cycle and therefore any vertex cover $S\in \mc S_\MVC$ also covers every cycle in $G'$ and every
    solution $S'\in \mc S_\FVS$ also covers every edge in $G$, as every edge was mapped to its own arc cycle.
\end{proof}
\noindent The reduction function $g$ is computable in polynomial time and is dependent on the input size of the \MVC instance $\big|[G=(V,E),k]\big|$,
as for each vertex in $V$ we create a fixed number of vertices the new instance. Therefore $\MVC\leq \FVS$ holds.

\paragraph{SSP \& Parsimonious}
This reduction also upholds the SSP and strongly Parsimonious properties, as the solutions are equivalent in both instances. We can use 
\lemref{lem:SSP and SPR 2} as no new vertices are created.
\begin{proof}
    We define 
    $$f_I: V\rightarrow V',~f(v)=v',$$ 
    where $v$ and $v'$ are equivalent in their respective instances. As the functions $f_I$ are basically the identity functions,
    they are bijective and the following holds
    \begin{flalign*}
        \{f(S):S\in\mc S_\MVC\} &= \big\{\{f(v)\mid v\in S\} : S\in\mc S_\MVC\big\}\\
        &=\big\{\{v'\in S'\} : S'\in\mc S_\FVS\big\}\\
        &=\{S'\cap f(V):S'\in \mc S_\FVS\}\\
        &=\{\Sr\FVS\}
    \end{flalign*}
    Therefore we get \MVC $\SSPR$ \FVS.
\end{proof}

\subsubsection[... to Feedback Arc Set]{Minimum Cardinality Vertex Cover to Feedback Arc Set}
\paragraph{Problem Definition}
\subfile{../../Problems/FAS.tex}

\paragraph{Reduction}
This reduction is based on the reduction by Karp \cite{karp}, but is modified to fit the SSP and Parsimonious frameworks. Let 
$\MVC:=[G=(V,E),k]$ be a \textsc{Minimum Vertex Cover} instance. We now create a new vertex set $V'$ consiting of, among others, vertices $v_0$ and $v_1$ for each 
vertex $v\in V$. We then connect these vertices via the arc $(v_0,v_1)\in A$. For every edge $\{u,v\}\in E$ we create $|V|+1$ paths over 
new vertices $uv_1,...,uv_{|V|+1}$ leading from $u_1$ to $v_0$ and also $|V|+1$ paths over new vertices $vu_1,...,vu_{|V|+1}$ 
leading from $v_1$ to $u_0$. Finally we leave $k$ as is and get $\FAS:=[G'=(V',A),k]$ as the corresponding \textsc{Feedback Arc Set} 
instance. An example reduction is shown in \autoref*{fig:MVC to FAS}. 
\begin{figure}[h]
    \begin{center}
    \scalebox{.85}{
        \begin{tikzpicture}[vertex/.style = {draw, circle, fill, inner sep = 1.5}, node distance = 1.5cm]
            \node[label = above:$G\colon$](G){};
            \node[vertex, below right =  .25cm and 1cm of G, label = $u$](u){};
            \node[vertex, right of = u, label = $v$](v){};
            \node[below left =  .25cm and 1cm of u, label = below:$k\equals 1$](k){};

            \node[right = 1cm of v, label = right:$\mapsto$](to){};
            
            \node[above right = .25 and 2cm of to, label = above:$G'\colon$](G'){};
            \node[vertex, below right =  .25cm and 1cm of G', label = left:$u_0$](u0){};
            \node[vertex, below = 2cm of u0, label = left:$u_1$](u1){};
            \node[below left =  .25cm and 1cm of u1, label = below:$k\equals 1$](k){};

            \node[vertex, right of = u0, label = $vu_2$](vu2){};
            \node[vertex, above = .5cm of vu2, label = $vu_1$](vu1){};
            \node[vertex, below = .5cm of vu2, label = $vu_3$](vu3){};

            \node[vertex, right of = u1, label = below:$uv_2$](uv2){};
            \node[vertex, above = .5cm of uv2, label = below:$uv_1$](uv1){};
            \node[vertex, below = .5cm of uv2, label = below:$uv_3$](uv3){};

            \node[vertex, right of = vu2, label = right:$v_1$](v1){};
            \node[vertex, below = 2cm of v1, label = right:$v_0$](v0){};

            \draw   (u) -- (v)
                    (u0) edge[bend right = 30, ->] (u1)
                    (v0) edge[bend right = 30, ->] (v1)
                    
                    (u1) edge[->] (uv1)
                    (u1) edge[->] (uv2)
                    (u1) edge[->] (uv3)
                    (uv1) edge[->] (v0)
                    (uv2) edge[->] (v0)
                    (uv3) edge[->] (v0)
                    
                    (v1) edge[->] (vu1)
                    (v1) edge[->] (vu2)
                    (v1) edge[->] (vu3)
                    (vu1) edge[->] (u0)
                    (vu2) edge[->] (u0)
                    (vu3) edge[->] (u0)
                ;
        \end{tikzpicture} 
    }
    \caption{\footnotesize{Example \textsc{Minimum Cardinality Vertex Cover} to \textsc{Feedback Arc Set}}}
    \label{fig:MVC to FAS}
    \end{center}
\end{figure}
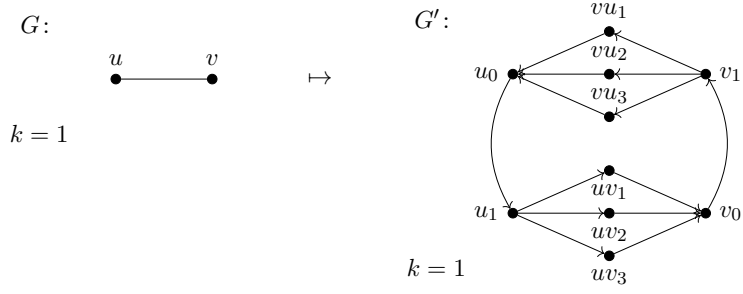
\begin{proof}
    Let $S$ be a solution for the \MVC instance, then the corresponding arc set $\{(v_0,v_1)\mid v\in S\}$ is a solution $S'$
    in the \FAS instance as due to the multiple arcs addes in for every edge, we have 9 different cycles in $G'$ for every 
    edge $\{u,v\}$ in $G$. However every one of these cycles flows over the arcs $(u_0,u_1)$ and $(v_0,v_1)$ and so if one of 
    these arcs is removed in $G'$ for every edge in $G$, we get a cycle-free Graph.\\
    Let $S'$ now be a solution for the \FAS instance, then, as $k$ was not changed, this solution needs to include at least one of the 
    arcs $(u_0,u_1)$ or $(v_0,v_1)$ for every edge $\{u,v\}$ in $G$. Otherwise not every cycle could be covered, as every edge adds 9 cycles.
    These arcs correspond to a set of vertices covering all the edges in $G$ and thus $S:=\{v\mid(v_0,v_1)\in S'\}$ is a solution for the
    \MVC instance.
\end{proof}
\noindent The reduction function $g$ is computable in polynomial time and is dependent on the input size of the \MVC instance $\big|[G=(V,E),k]\big|$,
as for each vertex in $V$ we create a fixed number of vertices the new instance. Therefore $\MVC\leq \FAS$ holds.

\paragraph{SSP \& Parsimonious}
We have shown that there is a correspondence between the vertices in $S\in \mc S_\MVC$ and the arcs created between them in the 
\FAS instance. Due to the abundance of extra cycles added and $k$ being minimal, we know that these solutions are actually equivalent
in their respective instances as, with these restrictions, no extra solutions are possible in the \FAS instance. We can use \lemref{lem:SSP and SPR} to prove
that this reduction is both SSP and SPR.
\begin{proof}
    Let $\Sall=\Slink=\emptyset$ and let 
    $$\Snev:=\big\{(u_1,uv_i),(uv_i,v_0),(v_1,vu_i),(vu_i,u_0)\mid \{u,v\}\in E, i\in\big[|V|+1]\big\}$$
    Now let 
    $$f_I:V\rightarrow A,~f(v)=(v_0,v_1),$$
    then $f_I':V\rightarrow A\backslash\Snev$ are bijective functions and
    \begin{flalign*}
        \{f(S):\S\MVC\} &= \big\{\{f(v)\mid v\in S\}:\S\MVC\big\}\\
        &= \big\{\{(v_0,v_1)\in S'\}:\Sr\FAS\big\}\\
        &= \{S'\cap f(V):\Sr\FAS\}
    \end{flalign*}
    and thus $\MVC\SSPR\FAS$ holds.
\end{proof}

\subsubsection[... to Uncapacitated Facility Location]{Minimum Cardinality Vertex Cover to Uncapacitated Facility Location}
\label{sec:MVC to UFL}
\paragraph{Problem Definition}
\subfile{../../Problems/UFL.tex}

\paragraph{Reduction}
Let $[G=(V,E),k]$ be a \MVC instance, then $[I,J,c_o,c_s,k]$ is a \UFL instance when we define the client set $I$ as $I:=\{\ind(e)\mid e\in E\}$, with $\ind$
being the index function,
and the set of potential facilities as $J$ as $J:= \{\ind(v)\mid v\in V\}$
and then let the costs for opening each facility be defined by the function $c_o:V\rightarrow \mathbb Z,~c_o(v)=1$ and the service costs by the function 
$$c_s:I\times J\rightarrow \mathbb Z,~c_s(i,j)=
\begin{cases}
    0, &v_j\in e_i\\
    |V|+1, &\text{otherwise}
\end{cases}.$$
We also leave $k$ unchanged. An example of this would be:
$$ V = \{u, v, w\},~E= \big\{\{u, v\},\{v, w\}\big\} $$
\begin{center}
        \begin{tikzpicture}[vertex/.style = {draw, circle, fill, inner sep = 1.5}, node distance = 1.5cm]
            \node[vertex, label = below:$u$](u){};
            \node[vertex, right = 1cm of u, label = below:$v$](v){};
            \node[vertex, right = 1cm of v, label = below:$w$](w){};

            \draw   (u) -- (v)
                    (v) -- (w)
            ;
        \end{tikzpicture} \\
    which gets mapped to
$$J = \{1, 2, 3\},~I= \{1,2\},~\forall j\in J:c_o(j)=1,$$
    \begin{tabular}{ccc}
        $c_s(1,1)=0$ & $c_s(2,1)=0$ & $c_s(3,1)=4$\\
        $c_s(1,2)=4$ & $c_s(2,2)=0$ & $c_s(3,2)=0$
    \end{tabular}
\end{center}
Where $[G=(V,E),1]$ is a \MVC instance and $[I,J, c_o,c_s,1]$ is the corresponding \UFL instance.
\begin{proof}
    A minimum cardinality vertex cover $S\in \mc S_\MVC$ is also a solution to the Uncapacitated Facility Location problem, as every vertex $v\in S$ represents an
    opened facility, so there are at most $k$ opened facilities, and all clients $i\in I$ are served, as each client corresponds to one edge and all edges are covered 
    by the vertex cover. \\
    On the other hand a solution for the \UFL problem $S'\in \mc S_\UFL$ has at most cost $k<|V|+1$, which means it includes at most $k$ facilities
    $j\in J$ and serves all clients $i\in I$. As the facilities correspond to vertices in the vertex cover and the clients can only correspond to edges in $G$ (due to the
    high costs $c(e,v)=|V|+1$ for $v\notin e$) this solution is also a vertex cover in $G$. 
\end{proof}
\noindent The reduction function $g$ is computable in polynomial time and is dependent on the input size of the \MVC instance $\big|[G=(V,E),k]\big|$,
as for each vertex in $V$ we create a fixed number of facilities the new instance. Therefore $\MVC\leq \UFL$ holds.

\paragraph{SSP \& Parsimonious}
As every solution $S\in\mc S_\MVC$ is equivalent to a solution $S'\in\mc S_\UFL$ and vice versa and $|V|=|J|$, we can use \lemref{lem:SSP and SPR 2}
to prove that the reduction is both SSP and SPR.
\begin{proof}
    Let the functions $f_I$ be defined as 
    $$f_I:V\rightarrow J, ~f(v)=\ind(v)$$ and 
    $$f_I^{-1}:J\rightarrow V, ~f^{-1}(j) = v_j$$ then we get 
    \begin{flalign*}
        \{f(S) :S\in \mc S_\MVC\}&=\big\{\{f(v)\mid v\in S\}:S\in \mc S_\MVC\big\}\\
        &=\big\{\{\ind(v)\mid v\in S\}:S\in \mc S_\MVC\big\}\\
        &=\big\{\{j\mid j\in S'\}:S'\in \mc S_\UFL\big\}\\
        &=\{S'\cap f(V):S'\in \mc S_\UFL\}\\
        &=\{\Sr\UFL\}
    \end{flalign*}
    and therefore it follows that $\MVC \SSPR \UFL$.
\end{proof}

\subsubsection[... to p-Center]{Minimum Cardinality Vertex Cover to p-Center}
\paragraph{Problem Definition}
\subfile{../../Problems/PCEN.tex}

\paragraph{Reduction}
This reduction is a slighty modified version of the \MVC $\leq$ \UFL reduction. Let $[G=(V,E),k]$ be a \MVC instance, then $[I,J,c,p,k']$ is a \PCen instance when we define the client set $I$ as $I:=\{\ind(e)\mid e\in E\}$
and the set of potential facilities as $J$ as $J:= \{\ind(v)\mid v\in V\}$
and then let the service costs be defined by the function 
$$c:I\times J\rightarrow \mathbb Z,~c(i,j)=
\begin{cases}
    0, &v_j\in e_i\\
    |V|+1, &\text{otherwise}
\end{cases}.$$
Finally we set $p$ to the size of the vertex cover $p=k$ and let $k'=0$. An example of this would be:
$$ V = \{u, v, w\},~E= \big\{\{u, v\},\{v, w\}\big\} 
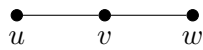
$$
\begin{center}
        \begin{tikzpicture}[vertex/.style = {draw, circle, fill, inner sep = 1.5}, node distance = 1.5cm]
            \node[vertex, label = below:$u$](u){};
            \node[vertex, right = 1cm of u, label = below:$v$](v){};
            \node[vertex, right = 1cm of v, label = below:$w$](w){};

            \draw   (u) -- (v)
                    (v) -- (w)
            ;
        \end{tikzpicture} \\
    which gets mapped to
    $$J = \{1, 2, 3\},~I= \{1,2\}$$
    \begin{tabular}{ccc}
        $c(1,1)=0$ & $c(2,1)=0$ & $c(3,1)=4$\\
        $c(1,2)=4$ & $c(2,2)=0$ & $c(3,2)=0$
    \end{tabular}
\end{center}
Where $[G=(V,E),1]$ is a \MVC instance and $[I,J, c,1,0]$ is the corresponding \PCen instance.\\
The proof that this reduction holds works analogously to the proof showing $\MVC \leq \UFL$ shown in section \ref{sec:MVC to UFL}, as the same one-to-one correspondence 
exists here as well.\\
\noindent The reduction function $g$ is computable in polynomial time and is dependent on the input size of the \MVC instance $\big|[G=(V,E),k]\big|$,
as for each vertex in $V$ we create a fixed number of facilities the new instance. Therefore $\MVC\leq \PCen$ holds.

\paragraph{SSP \& Parsimonious}
As every solution $S\in\mc S_\MVC$ is equivalent to a solution $S'\in\mc S_\PCen$ and vice versa and $|V|=|J|$, we can use \lemref{lem:SSP and SPR 2}
to prove that the reduction is both SSP and SPR.
\begin{proof}
    Let the functions $f_I$ be defined as 
    $$f_I:V\rightarrow J, ~f(v)=\ind(v)$$ and 
    $$f_I^{-1}:J\rightarrow V, ~f^{-1}(j) = v_j$$ then we get 
    \begin{flalign*}
        \{f(S) :S\in \mc S_\MVC\}&=\big\{\{f(v)\mid v\in S\}:S\in \mc S_\MVC\big\}\\
        &=\big\{\{\ind(v)\mid v\in S\}:S\in \mc S_\MVC\big\}\\
        &=\big\{\{j\mid j\in S'\}:S'\in \mc S_\PCen\big\}\\
        &=\{S'\cap f(V):S'\in \mc S_\PCen\}\\
        &=\{\Sr\PCen\}
    \end{flalign*}
    and therefore it follows that $\MVC \SSPR \PCen$.
\end{proof}

\subsubsection[... to p-Median]{Minimum Cardinality Vertex Cover to p-Median}
\paragraph{Problem Definition}
\subfile{../../Problems/PMED.tex}

\paragraph{Reduction}
This reduction works analogously to the reduction $\MVC\leq\PCen$. Let $[G=(V,E),k]$ be a \MVC instance, then $[I,J,c,p,k']$ is a \PMed instance where
 we define the client set $I$ as $I:=\{\ind(e)\mid e\in E\}$
and the set of potential facilities as $J$ as $J:= \{\ind(v)\mid v\in V\}$
and then let the service costs be defined by the function 
$$c:I\times J\rightarrow \mathbb Z,~c(i,j)= 
\begin{cases}
    0, &v_j\in e_i\\
    |V|+1, &\text{otherwise}
\end{cases}.$$
Finally we set $p$ to the size of the vertex cover $p=k$ and let $k'=0$. An example of this would be:
$$ V = \{u, v, w\},~E= \big\{\{u, v\},\{v, w\}\big\} $$
\begin{center}
        \begin{tikzpicture}[vertex/.style = {draw, circle, fill, inner sep = 1.5}, node distance = 1.5cm]
            \node[vertex, label = below:$u$](u){};
            \node[vertex, right = 1cm of u, label = below:$v$](v){};
            \node[vertex, right = 1cm of v, label = below:$w$](w){};

            \draw   (u) -- (v)
                    (v) -- (w)
            ;
        \end{tikzpicture} 
    \\
    which gets mapped to
    $$ J = \{1, 2, 3\},~I= \{1,2\}$$
    \begin{tabular}{ccc}
        $c(1,1)=0$ & $c(2,1)=0$ & $c(3,1)=4$\\
        $c(1,2)=4$ & $c(2,2)=0$ & $c(3,2)=0$
    \end{tabular}
\end{center}
Where $[G=(V,E),1]$ is a \MVC instance and $[I,J, c,1,0]$ is the corresponding \PMed instance.\\
The proof that this reduction holds works analogously to the proof showing $\MVC \leq \UFL$ shown in section \ref{sec:MVC to UFL}, as the same one-to-one correspondence 
exists here as well.\\
\noindent The reduction function $g$ is computable in polynomial time and is dependent on the input size of the \MVC instance $\big|[G=(V,E),k]\big|$,
as for each vertex in $V$ we create a fixed number of facilities in the new instance. Therefore $\MVC\leq \PMed$ holds.

\paragraph{SSP \& Parsimonious}
As every solution $S\in\mc S_\MVC$ is equivalent to a solution $S'\in\mc S_\PMed$ and vice versa and $|V|=|J|$, we can use \lemref{lem:SSP and SPR 2}
to prove that the reduction is both SSP and SPR.
\begin{proof}
    Let the functions $f_I$ be defined as 
    $$f_I:V\rightarrow J, ~f(v)=\ind(v)$$ and 
    $$f_I^{-1}:J\rightarrow V, ~f^{-1}(j) = v_j$$ then we get 
    \begin{flalign*}
        \{f(S) :S\in \mc S_\MVC\}&=\big\{\{f(v)\mid v\in S\}:S\in \mc S_\MVC\big\}\\
        &=\big\{\{\ind(v)\mid v\in S\}:S\in \mc S_\MVC\big\}\\
        &=\big\{\{j\mid j\in S'\}:S'\in \mc S_\PMed\big\}\\
        &=\{S'\cap f(V):S'\in \mc S_\PMed\}\\
        &=\{\Sr\PMed\}
    \end{flalign*}
    and therefore it follows that $\MVC \SSPR \PMed$.
\end{proof}

\subsubsection[... to Vertex Cover with one fixed Vertex]{Minimum Cardinality Vertex Cover to Vertex Cover with one fixed Vertex}
\label{sec:MVC to VCV}
\subfile{../../Problems/VCV.tex}

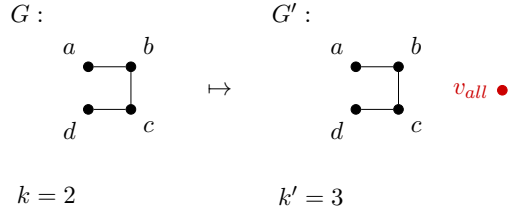
\begin{figure}[h]
    \begin{center}
    \scalebox{.85}{
        \begin{tikzpicture}[vertex/.style = {draw, circle, fill, inner sep = 1.5}, node distance = 1.5cm]
            \node[](G){$G:$};
            \node[vertex, below right = .5 and .5 of G, label = above left:$a$](a){};
            \node[vertex, right = .5 of a, label = above right:$b$](b){};
            \node[vertex, below = .5 of b, label = below right:$c$](c){};
            \node[vertex, below = .5 of a, label = below left:$d$](d){};
            \node[below left = 1 and 0 of d](k){$k=2$};

            \node[below right = .1 and 1 of b](to){$\mapsto$};

            \node[above right = .5 and 2 of b](G'){$G':$};
            \node[vertex, below right = .5 and .5 of G', label = above left:$a$](a'){};
            \node[vertex, right = .5 of a', label = above right:$b$](b'){};
            \node[vertex, below = .5 of b', label = below right:$c$](c'){};
            \node[vertex, below = .5 of a', label = below left:$d$](d'){};
            \node[below left = 1 and 0 of d'](k'){$k'=3$};
            \node[vertex, below right = .25 and 1.5 of b', label ={[red!80!black] left:$v_{all}$}, red!80!black](v){};

            \draw   (a) -- (b)
                    (b) -- (c)
                    (c) -- (d)

                    (a') -- (b')
                    (b') -- (c')
                    (c') -- (d')
            ;
        \end{tikzpicture}
    }
    \caption{\footnotesize{Example \textsc{Maximum Cardinality Independent Set} to \textsc{Vertex Cover with one fixed Vertex}}}
    \label{fig:MVC to VCV}
    \end{center}
\end{figure}

\paragraph{Reduction}
This reduction is included, as, if the \textsc{Vertex Cover with one fixed Vertex} problem is SSP-NP-hard, then the reductions shown in
the following sections (\autoref{sec:MVC to DHC} \& \autoref{sec:MVC to UHC}) are not easily modified to be Parsimonious. Now,
let $\MVC:= \big[G=(V,E),k\big]$, then we create a \VCV instance by a new vertex $v_{all}$ to $V$. 
Now $V':=V\cup\{v_{all}\}$, $E':=E$ and $k':=k+1$ and we get the $\VCV$ instance $\big[G'=(V',E'),k',v_{all}\big]$.
An example reduction is shown in \autoref*{fig:MVC to VCV}.
\begin{proof} Now, every solution in the \MVC cover instance is also a solution for $\VCV$, as the graphs are equivalent, except for the
    fixed vertex $v_{all}$. But as $v_{all}$ is not connected to the rest of the graph, and $k'-1=k$, any vertex cover $S'$ 
    in $G'$ is a vertex cover $S\backslash \{v_{all}\}$ in $G$ and vice versa.
\end{proof}
\noindent The reduction function $g$ is computable in polynomial time and is dependent on the input size of the \MVC instance 
$\big|[G=(V,E),k]\big|$, as we create exactly one extra vertex in the new instance. Therefore $\MVC\leq \VCV$ holds.

\paragraph{SSP \& Parsimonious}
As seen above, there exists a one-to-one correspondence between the vertices of the vertex cover in $G$, and the 
vertices in $G'$, when we ignore $v_{all}$. We can therefore apply \lemref{lem:SSP and SPR} as follows, to prove
$\MVC\SSPR\VCV$.
\begin{proof}
    Let $\Snev=\Slink=\emptyset$, $\Sall=\{v_{all}\}$ and let the functions $f_I$ be defined as
    $$f_I: V\rightarrow V',~f(v)=v.$$
    Now $f(V)=V'\backslash\Sall$ and 
    \begin{flalign*}
        \{f(S):\S\MVC\}&=\big\{\{f(v)\mid v\in S\}:\S\MVC\big\}\\
        &=\big\{\{v\in S'\}\backslash \Sall :\Sr\VCV\big\}\\
        &= \{S'\cap f(V):\Sr\VCV\}
    \end{flalign*}
    and therefore $\MVC\SSPR\VCV$ holds.
\end{proof}

\subsubsection[... to Directed Hamiltonian Cycle]{Minimum Cardinality Vertex Cover to Directed Hamiltonian Cycle}
\label{sec:MVC to DHC}

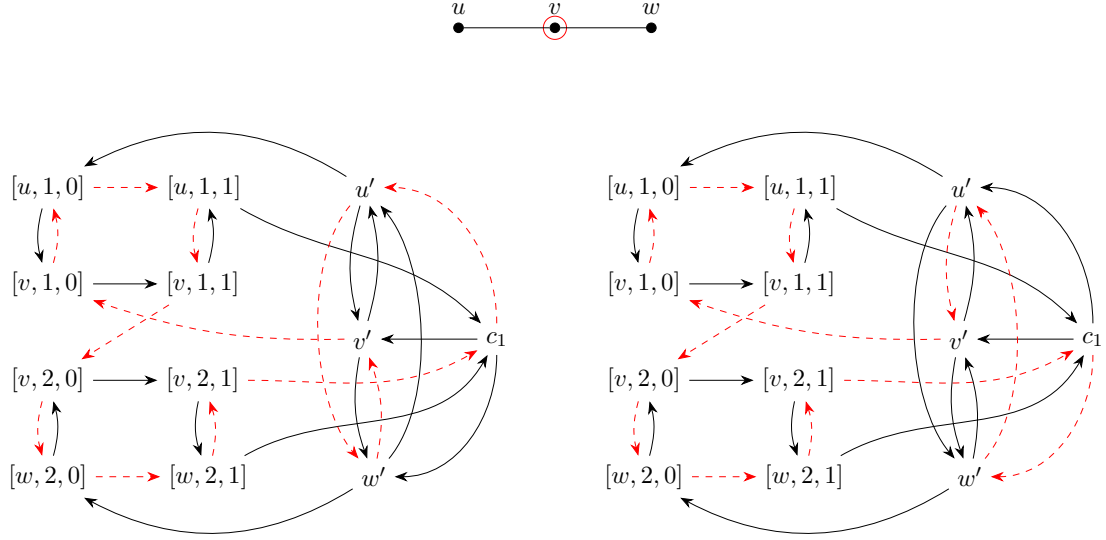
\begin{figure}[h]
    \begin{center}
    \scalebox{.85}{
        \begin{tikzpicture}[vertex/.style = {draw, circle, fill, inner sep = 1.5}, node distance = 1.5cm]
            \node[vertex, label = $u$](u){};
            \node[vertex, right of = u, label = {$v$}](v){};
            \node[circle, draw, red, right of = u](){};
            \node[vertex, right of = v, label = $w$](w){};

            \draw   (u) -- (v)
                    (v) -- (w)
            ;
        \end{tikzpicture} 
    }\\\vspace*{1cm}
    \scalebox{.85}{
        \begin{tikzpicture}[vertex/.style = {draw, circle, fill, inner sep = 1.5}, node distance = 1.5cm]
            \node[](u10){$[u,1,0]$};
            \node[right = 1cm of u10](u11){$[u,1,1]$};
            \node[below of = u10](v10){$[v,1,0]$};
            \node[right = 1cm of v10](v11){$[v,1,1]$};

            \node[below of = v10](v20){$[v,2,0]$};
            \node[right = 1cm of v20](v21){$[v,2,1]$};
            \node[below of = v20](w20){$[w,2,0]$};
            \node[right = 1cm of w20](w21){$[w,2,1]$};

            \node[right = 1.5cm of u11](u'){$u'$};
            \node[below right = .25cm and 1.5cm of v11](v'){$v'$};
            \node[right = 1.5cm of w21](w'){$w'$};
            \node[right = 1.5cm of v'](c1){$c_1$};

            \draw   (u10) edge[\xarrow, red, dashed] (u11)
                    (u10) edge[\xarrow, bend right = 15] (v10)
                    (v10) edge[\xarrow] (v11)
                    (v10) edge[\xarrow, bend right = 15, red, dashed] (u10)
                    (u11) edge[\xarrow, bend right = 15, red, dashed] (v11)
                    (v11) edge[\xarrow, bend right = 15] (u11)

                    (v11)edge[\xarrow, red, dashed] (v20)
                    
                    (v20) edge[\xarrow] (v21)
                    (v20) edge[\xarrow, bend right = 15, red, dashed] (w20)
                    (w20) edge[\xarrow, red, dashed] (w21)
                    (w20) edge[\xarrow, bend right = 15] (v20)
                    (v21) edge[\xarrow, bend right = 15] (w21)
                    (w21) edge[\xarrow, bend right = 15, red, dashed] (v21)

                    (u') edge[\xarrow, bend right = 30] (u10)
                    (v') edge[\xarrow, bend left = 10, red, dashed] (v10)
                    (w') edge[\xarrow, bend left = 30] (w20)

                    (u11) edge[\xarrow, out = -30, in = 135](c1)
                    (v21) edge[\xarrow, out = 0, in = -150, red, dashed](c1)
                    (w21) edge[\xarrow, out = 30, in = -120](c1)

                    (c1) edge[\xarrow, out = 90, in = 0, red, dashed] (u')
                    (c1) edge[\xarrow] (v')
                    (c1) edge[\xarrow, out = -90, in = 0] (w')

                    (u') edge[\xarrow, bend right = 15] (v')
                    (u') edge[\xarrow, out = -130, in = 130, looseness = .75, red, dashed] (w')
                    (v') edge[\xarrow, bend right = 15] (w')
                    (w') edge[\xarrow, out = 50, in = -50, looseness = .75] (u')
                    (w') edge[\xarrow, bend right = 15, red, dashed] (v')
                    (v') edge[\xarrow, bend right = 15] (u')
            ;
        \end{tikzpicture} \hspace*{1cm}
        \begin{tikzpicture}[vertex/.style = {draw, circle, fill, inner sep = 1.5}, node distance = 1.5cm]
            \node[](u10){$[u,1,0]$};
            \node[right = 1cm of u10](u11){$[u,1,1]$};
            \node[below of = u10](v10){$[v,1,0]$};
            \node[right = 1cm of v10](v11){$[v,1,1]$};

            \node[below of = v10](v20){$[v,2,0]$};
            \node[right = 1cm of v20](v21){$[v,2,1]$};
            \node[below of = v20](w20){$[w,2,0]$};
            \node[right = 1cm of w20](w21){$[w,2,1]$};

            \node[right = 1.5cm of u11](u'){$u'$};
            \node[below right = .25cm and 1.5cm of v11](v'){$v'$};
            \node[right = 1.5cm of w21](w'){$w'$};
            \node[right = 1.5cm of v'](c1){$c_1$};

            \draw   (u10) edge[\xarrow, red, dashed] (u11)
                    (u10) edge[\xarrow, bend right = 15] (v10)
                    (v10) edge[\xarrow] (v11)
                    (v10) edge[\xarrow, bend right = 15, red, dashed] (u10)
                    (u11) edge[\xarrow, bend right = 15, red, dashed] (v11)
                    (v11) edge[\xarrow, bend right = 15] (u11)

                    (v11)edge[\xarrow, red, dashed] (v20)
                    
                    (v20) edge[\xarrow] (v21)
                    (v20) edge[\xarrow, bend right = 15, red, dashed] (w20)
                    (w20) edge[\xarrow, red, dashed] (w21)
                    (w20) edge[\xarrow, bend right = 15] (v20)
                    (v21) edge[\xarrow, bend right = 15] (w21)
                    (w21) edge[\xarrow, bend right = 15, red, dashed] (v21)

                    (u') edge[\xarrow, bend right = 30] (u10)
                    (v') edge[\xarrow, bend left = 10, red, dashed] (v10)
                    (w') edge[\xarrow, bend left = 30] (w20)

                    (u11) edge[\xarrow, out = -30, in = 135](c1)
                    (v21) edge[\xarrow, out = 0, in = -150, red, dashed](c1)
                    (w21) edge[\xarrow, out = 30, in = -120](c1)

                    (c1) edge[\xarrow, out = 90, in = 0] (u')
                    (c1) edge[\xarrow] (v')
                    (c1) edge[\xarrow, out = -90, in = 0, red, dashed] (w')

                    (u') edge[\xarrow, bend right = 15, red, dashed] (v')
                    (u') edge[\xarrow, out = -130, in = 130, looseness = .75] (w')
                    (v') edge[\xarrow, bend right = 15] (w')
                    (w') edge[\xarrow, out = 50, in = -50, looseness = .75, red, dashed] (u')
                    (w') edge[\xarrow, bend right = 15] (v')
                    (v') edge[\xarrow, bend right = 15] (u')
            ;
        \end{tikzpicture} 
    }\\
    \caption{\footnotesize{\textsc{Maximum Cardinality Independent Set} to \textsc{Directed Hamiltonian Cycle} is not SPR}}
    \label{fig:VC to DHC not SPR}
    \end{center}
  \end{figure}

\subfile{../../Problems/DHC.tex}
\paragraph{Reduction, SSP \& Parsimonious}
The reduction presented in \cite[p. 18-21]{Femke} is unfortunately only SSP and not Parsimonious, as seen in \autoref{fig:VC to DHC not SPR}, where each of the Directed
Hamiltonian Cycles, indicated in red, are equivalent to the Vertex Cover. This reduction also can not be easily modified to be Parsimonious,
as we would need to know which arcs $(c,v')$, for all $v\in V$ and with $c\in \{c_1,...,c_k\}$, are included in the cycle. This is equivalent to
knowing whether or not a specific vertex is always in the vertex cover. As we can see in \autoref{sec:MVC to VCV} this is an
SSP-NP-hard problem and therefore there exists no simple solution computable in polynomial time.

\paragraph{Alternative transitive reduction with both SSP \& SPR properties}
To prove the existence of a reduction that is both SSP and SPR, we utilize \autoref{sec:Cook}, 
where we prove that Cook's reduction is both SSP and SPR. We also need $\SAT\SSPR\ESAT$ (\autoref{sec:SAT to 3SAT} \& \autoref{sec:3SAT to ESAT}) 
and $\ESAT\SSPR\DHC$ (\autoref{sec:ESAT to DHC}). With all of these we get
$$\MVC\SSPR\SAT\SSPR\TSAT\SSPR\ESAT\SSPR\DHC$$
and therefore $\MVC\SSPR\DHC$ holds.

\subsubsection[... to Undirected Hamiltonian Cycle]{Minimum Cardinality Vertex Cover to Undirected Hamiltonian Cycle}
\label{sec:MVC to UHC}
\paragraph{Problem Definition}
\subfile{../../Problems/UHC.tex}

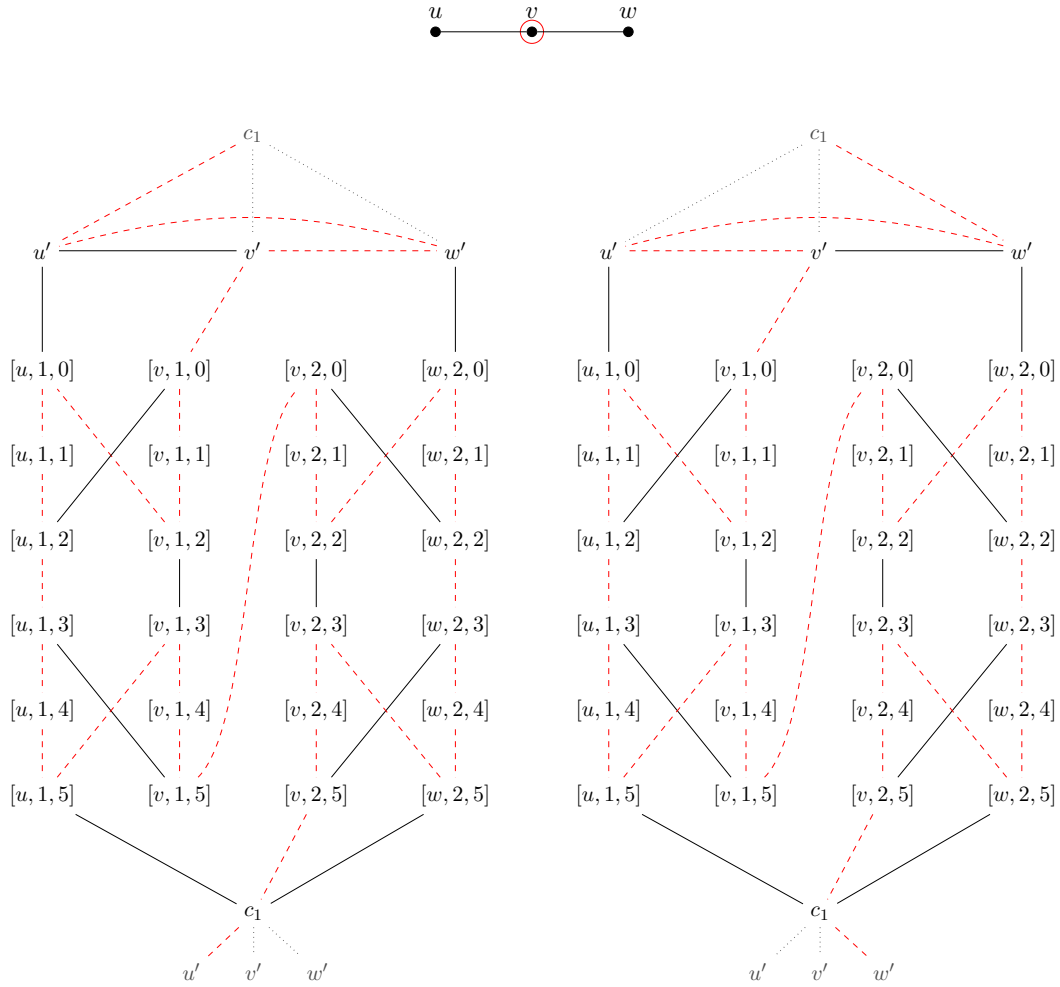
\begin{figure}[h]
    \begin{center}
    \scalebox{.85}{
        \begin{tikzpicture}[vertex/.style = {draw, circle, fill, inner sep = 1.5}, node distance = 1.5cm]
            \node[vertex, label = $u$](u){};
            \node[vertex, right of = u, label = {$v$}](v){};
            \node[circle, draw, red, right of = u](){};
            \node[vertex, right of = v, label = $w$](w){};

            \draw   (u) -- (v)
                    (v) -- (w)
            ;
        \end{tikzpicture} 
    }\\\vspace*{1cm}
    \scalebox{.75}{
        \begin{tikzpicture}[vertex/.style = {draw, circle, fill, inner sep = 1.5}, node distance = 1.5cm]
            \node[](u10){$[u,1,0]$};
            \node[below of = u10](u11){$[u,1,1]$};
            \node[below of = u11](u12){$[u,1,2]$};
            \node[below of = u12](u13){$[u,1,3]$};
            \node[below of = u13](u14){$[u,1,4]$};
            \node[below of = u14](u15){$[u,1,5]$};
            \node[right = 1cm of u10](v10){$[v,1,0]$};
            \node[below of = v10](v11){$[v,1,1]$};
            \node[below of = v11](v12){$[v,1,2]$};
            \node[below of = v12](v13){$[v,1,3]$};
            \node[below of = v13](v14){$[v,1,4]$};
            \node[below of = v14](v15){$[v,1,5]$};

            \node[right = 1cm of v10](v20){$[v,2,0]$};
            \node[below of = v20](v21){$[v,2,1]$};
            \node[below of = v21](v22){$[v,2,2]$};
            \node[below of = v22](v23){$[v,2,3]$};
            \node[below of = v23](v24){$[v,2,4]$};
            \node[below of = v24](v25){$[v,2,5]$};
            \node[right = 1 of v20](w20){$[w,2,0]$};
            \node[below of = w20](w21){$[w,2,1]$};
            \node[below of = w21](w22){$[w,2,2]$};
            \node[below of = w22](w23){$[w,2,3]$};
            \node[below of = w23](w24){$[w,2,4]$};
            \node[below of = w24](w25){$[w,2,5]$};

            \node[above = 1.5cm of u10](u'){$u'$};
            \node[above right = 1.5 and .3 of v10](v'){$v'$};
            \node[above = 1.5cm of w20](w'){$w'$};
            \node[below right = 1.5 and .30 of v15](c1){$c_1$};

            \node[below left = .5 and .5 of c1, black!70!](u'1){$u'$};
            \node[below = .5 of c1, black!70!](v'1){$v'$};
            \node[below right = .5 and .5 of c1, black!70!](w'1){$w'$};
            
            \node[above = 1.5 of v', black!70!](c11){$c_1$};

            \draw   (v') edge[red, dashed] (w')
                    (v') edge[] (u')
                    (w') edge[bend right = 15, red, dashed] (u')
            
                    (v') edge[red, dashed] (v10)
                    (v10) edge[red, dashed] (v11)
                    (v11) edge[red, dashed] (v12)
                    (v12) edge[] (v13)
                    (v13) edge[red, dashed] (v14)
                    (v14) edge[red, dashed] (v15)

                    (u') edge[] (u10)
                    (u10) edge[red, dashed] (u11)
                    (u11) edge[red, dashed] (u12)
                    (u12) edge[red, dashed] (u13)
                    (u13) edge[red, dashed] (u14)
                    (u14) edge[red, dashed] (u15)

                    (v12) edge[red, dashed] (u10)
                    (u12) edge[] (v10)
                    (v15) edge[] (u13)
                    (u15) edge[red, dashed] (v13)

                    (v15) edge[red, dashed, out = 45, in = 225, looseness = .5] (v20)
                    
                    (v20) edge[red, dashed] (v21)
                    (v21) edge[red, dashed] (v22)
                    (v22) edge[] (v23)
                    (v23) edge[red, dashed] (v24)
                    (v24) edge[red, dashed] (v25)

                    (w') edge[] (w20)
                    (w20) edge[red, dashed] (w21)
                    (w21) edge[red, dashed] (w22)
                    (w22) edge[red, dashed] (w23)
                    (w23) edge[red, dashed] (w24)
                    (w24) edge[red, dashed] (w25)

                    (v22) edge[red, dashed] (w20)
                    (w22) edge[] (v20)
                    (v25) edge[] (w23)
                    (w25) edge[red, dashed] (v23)

                    (u15) edge[] (c1)
                    (v25) edge[red, dashed] (c1)
                    (w25) edge[] (c1)

                    (c11) edge[red, dashed] (u')
                    (c11) edge[black!70!, dotted] (v')
                    (c11) edge[black!70!, dotted] (w')
                    
                    (c1) edge[red, dashed] (u'1)
                    (c1) edge[black!70!, dotted] (v'1)
                    (c1) edge[black!70!, dotted] (w'1)
            ;
        \end{tikzpicture} \hspace*{1cm}
        \begin{tikzpicture}[vertex/.style = {draw, circle, fill, inner sep = 1.5}, node distance = 1.5cm]
            \node[](u10){$[u,1,0]$};
            \node[below of = u10](u11){$[u,1,1]$};
            \node[below of = u11](u12){$[u,1,2]$};
            \node[below of = u12](u13){$[u,1,3]$};
            \node[below of = u13](u14){$[u,1,4]$};
            \node[below of = u14](u15){$[u,1,5]$};
            \node[right = 1cm of u10](v10){$[v,1,0]$};
            \node[below of = v10](v11){$[v,1,1]$};
            \node[below of = v11](v12){$[v,1,2]$};
            \node[below of = v12](v13){$[v,1,3]$};
            \node[below of = v13](v14){$[v,1,4]$};
            \node[below of = v14](v15){$[v,1,5]$};

            \node[right = 1cm of v10](v20){$[v,2,0]$};
            \node[below of = v20](v21){$[v,2,1]$};
            \node[below of = v21](v22){$[v,2,2]$};
            \node[below of = v22](v23){$[v,2,3]$};
            \node[below of = v23](v24){$[v,2,4]$};
            \node[below of = v24](v25){$[v,2,5]$};
            \node[right = 1 of v20](w20){$[w,2,0]$};
            \node[below of = w20](w21){$[w,2,1]$};
            \node[below of = w21](w22){$[w,2,2]$};
            \node[below of = w22](w23){$[w,2,3]$};
            \node[below of = w23](w24){$[w,2,4]$};
            \node[below of = w24](w25){$[w,2,5]$};

            \node[above = 1.5cm of u10](u'){$u'$};
            \node[above right = 1.5 and .3 of v10](v'){$v'$};
            \node[above = 1.5cm of w20](w'){$w'$};
            \node[below right = 1.5 and .30 of v15](c1){$c_1$};

            \node[below left = .5 and .5 of c1, black!70!](u'1){$u'$};
            \node[below = .5 of c1, black!70!](v'1){$v'$};
            \node[below right = .5 and .5 of c1, black!70!](w'1){$w'$};
            
            \node[above = 1.5 of v', black!70!](c11){$c_1$};

            \draw   (v') edge[] (w')
                    (v') edge[red, dashed] (u')
                    (w') edge[bend right = 15, red, dashed] (u')
            
                    (v') edge[red, dashed] (v10)
                    (v10) edge[red, dashed] (v11)
                    (v11) edge[red, dashed] (v12)
                    (v12) edge[] (v13)
                    (v13) edge[red, dashed] (v14)
                    (v14) edge[red, dashed] (v15)

                    (u') edge[] (u10)
                    (u10) edge[red, dashed] (u11)
                    (u11) edge[red, dashed] (u12)
                    (u12) edge[red, dashed] (u13)
                    (u13) edge[red, dashed] (u14)
                    (u14) edge[red, dashed] (u15)

                    (v12) edge[red, dashed] (u10)
                    (u12) edge[] (v10)
                    (v15) edge[] (u13)
                    (u15) edge[red, dashed] (v13)

                    (v15) edge[red, dashed, out = 45, in = 225, looseness = .5] (v20)
                    
                    (v20) edge[red, dashed] (v21)
                    (v21) edge[red, dashed] (v22)
                    (v22) edge[] (v23)
                    (v23) edge[red, dashed] (v24)
                    (v24) edge[red, dashed] (v25)

                    (w') edge[] (w20)
                    (w20) edge[red, dashed] (w21)
                    (w21) edge[red, dashed] (w22)
                    (w22) edge[red, dashed] (w23)
                    (w23) edge[red, dashed] (w24)
                    (w24) edge[red, dashed] (w25)

                    (v22) edge[red, dashed] (w20)
                    (w22) edge[] (v20)
                    (v25) edge[] (w23)
                    (w25) edge[red, dashed] (v23)

                    (u15) edge[] (c1)
                    (v25) edge[red, dashed] (c1)
                    (w25) edge[] (c1)

                    (c11) edge[black!70!, dotted] (u')
                    (c11) edge[black!70!, dotted] (v')
                    (c11) edge[red, dashed] (w')
                    
                    (c1) edge[black!70!, dotted] (u'1)
                    (c1) edge[black!70!, dotted] (v'1)
                    (c1) edge[red, dashed] (w'1)
            ;
        \end{tikzpicture} 
    }\\
    \caption{\footnotesize{\textsc{Maximum Cardinality Independent Set} to \textsc{Undirected Hamiltonian Cycle} is not SPR}}
    \label{fig:VC to UHC not SPR}
    \end{center}
  \end{figure}

\paragraph{Reduction, SSP \& Parsimonious}
Similarly to what we see in \autoref{sec:MVC to DHC} the reduction presented in \cite[p. 22-25]{Femke} is also only SSP and not Parsimonious, 
which can be seen in \autoref{fig:VC to UHC not SPR}, where each of the Undirected
Hamiltonian Cycles, indicated in red, are equivalent to the given Vertex Cover. To be able to modify this reduction, we also have the
same problem of needing to know whether or not a certain vertex is in the vertex cover. As this is shown to be NP-hard 
(\autoref{sec:MVC to VCV}) there exists no simple solution computable in polynomial time.

\paragraph{Alternative transitive reduction with both SSP \& SPR properties}
To prove the existence of a reduction that is both SSP and SPR, we utilize \autoref{sec:Cook}, 
where we prove that Cook's reduction is both SSP and SPR. We also need $\SAT\SSPR\ESAT$ (\autoref{sec:SAT to 3SAT} \& \autoref{sec:3SAT to ESAT}) 
and $\ESAT\SSPR\UHC$ (\autoref{sec:ESAT to DHC} \& \autoref{sec:DHC to UHC}). With all of these we get
$$\MVC\SSPR\SAT\SSPR\TSAT\SSPR\ESAT\SSPR\DHC\SSPR\UHC$$
and therefore $\MVC\SSPR\UHC$ holds.

\ifSubfilesClassLoaded{
  \bibliography{../../ref}
}{}
\end{document}

%% file: Problems/SC.tex
\noindent\fbox{
    \parbox{\textwidth}{
        \textsc{Set Cover ($\SC$)}\\
        \begin{tabular}{cl}
        ~ & \textbf{Instances:} Sets $\psi_i \subseteq \Psi = \{1,...,m\}$ for $i\in [n]$, number $k\in\mathbb{N}$\\
        ~ & \textbf{Universe:} Sets $\mathcal{U}:=\{\psi_1,...,\psi_n\}$.\\
        ~ & \textbf{Feasible solution set:} The set of all $S\subseteq\{\psi_1,...,\psi_n\}$ s.t. $\bigcup_{s\in S} s = \{1,...,m\}$\\
        ~ & \textbf{Solution set:} The set of all feasible solutions with $|S|\leq k$.
        \end{tabular}
    }
}

\ifSubfilesClassLoaded{
  \bibliography{../ref}
}{}
\end{document}

%% file: Problems/HS.tex
\noindent\fbox{
    \parbox{\textwidth}{
        \textsc{Hitting Set ($\HS$)}\\
        \begin{tabular}{cl}
        ~ & \textbf{Instances:} Sets $\psi_i \subseteq \Psi = \{1,...,m\}$ for $i\in [n]$, number $k\in\mathbb{N}$\\
        ~ & \textbf{Universe:} The set $\mathcal{U}:=\{1,...,m\}$.\\
        ~ & \textbf{Feasible solution set:} The set of all $S\subseteq\{1,...,m\}$ such that $S\cap \psi_i \neq \emptyset$ \\ 
        ~ & for all $i \in \{1,...,n\}$\\
        ~ & \textbf{Solution set:} The set of all feasible solutions with $|S|\leq k$.
        \end{tabular}
    }
}

\ifSubfilesClassLoaded{
  \bibliography{../ref}
}{}
\end{document}

%% file: Problems/FVS.tex
\noindent\fbox{
    \parbox{\textwidth}{
        \textsc{Feedback Vertex Set ($\FVS$)}\\
        \begin{tabular}{cl}
        ~ & \textbf{Instances:} Directed Graph $G=(V,A)$, number $k\in \mathbb N$\\
        ~ & \textbf{Universe:} Vertex set $V$, $\mathcal{U}:=V$.\\
        ~ & \textbf{Feasible solution set:} The set of all vertex sets $S\subseteq V$ such that\\ 
        ~ & after deleting $S$ from $G$, the resuting graph is cycle-free.\\
        ~ & \textbf{Solution set:} The set of all feasible solutions with $|S|\leq k$.
        \end{tabular}
    }
}

\ifSubfilesClassLoaded{
  \bibliography{../ref}
}{}
\end{document}

%% file: Problems/FAS.tex
\noindent\fbox{
    \parbox{\textwidth}{
        \textsc{Feedback Arc Set ($\FAS$)}\\
        \begin{tabular}{cl}
        ~ & \textbf{Instances:} Directed Graph $G=(V,A)$, number $k\in \mathbb N$\\
        ~ & \textbf{Universe:} Arc set $\mathcal{U}:=A$.\\
        ~ & \textbf{Feasible solution set:} The set of all arc sets $S\subseteq A$ such that\\ 
        ~ & after deleting $S$ from $G$, the resuting graph is cycle-free.\\
        ~ & \textbf{Solution set:} The set of all feasible solutions with $|S|\leq k$.
        \end{tabular}
    }
}

\ifSubfilesClassLoaded{
  \bibliography{../ref}
}{}
\end{document}

%% file: Problems/UFL.tex
\noindent\fbox{
    \parbox{\textwidth}{
        \textsc{Uncapacitated Facility Location ($\UFL$)}\\
        \begin{tabular}{cl}
        ~ & \textbf{Instances:} Set of clients $I=\{1,...,m\}$, set of potential facilities $J=\{1,...,n\}$,\\
        ~ & fixed costs of opening a facility function $c_o:J\rightarrow \mathbb Z$, service cost function $c_s:I\times J\rightarrow \mathbb Z$,\\
        ~ & cost threshold $k\in \mathbb Z$\\
        ~ & \textbf{Universe:} Facility set $\mathcal{U}:=J$.\\
        ~ & \textbf{Solution set:} The set of sets $J'\subseteq J$ s.t. $\sum_{j\in J'} f(j)+\sum_{i\in I} \min_{j\in J'}c(i,j)\leq k$.
        \end{tabular}
    }
}

\ifSubfilesClassLoaded{
  \bibliography{../ref}
}{}
\end{document}

%% file: Problems/PCEN.tex
\noindent\fbox{
    \parbox{\textwidth}{
        \textsc{p-Center ($\PCen$)}\\
        \begin{tabular}{cl}
        ~ & \textbf{Instances:} Set of clients $I=\{1,...,m\}$, set of potential facilities $J=\{1,...,n\}$,\\
        ~ & service cost function $c_s:I\times J\rightarrow \mathbb Z$, facility threshold $p\in \mathbb N$, cost threshold $k\in \mathbb Z$\\
        ~ & \textbf{Universe:} Facility set $\mathcal{U}:=J$.\\
        ~ & \textbf{Solution set:} The set of sets $J'\subseteq J$ s.t. $|J'|\leq p$ and $\max_{i\in I}\min_{j\in J'}c(i,j)\leq k$.
        \end{tabular}
    }
}

\ifSubfilesClassLoaded{
  \bibliography{../ref}
}{}
\end{document}

%% file: Problems/PMED.tex
\noindent\fbox{
    \parbox{\textwidth}{
        \textsc{p-Median ($\PMed$)}\\
        \begin{tabular}{cl}
        ~ & \textbf{Instances:} Set of clients $I=\{1,...,m\}$, set of potential facilities $J=\{1,...,n\}$,\\
        ~ & service cost function $c_s:I\times J\rightarrow \mathbb Z$, facility threshold $p\in \mathbb N$, cost threshold $k\in \mathbb Z$\\
        ~ & \textbf{Universe:} Facility set $\mathcal{U}:=J$.\\
        ~ & \textbf{Solution set:} The set of sets $J'\subseteq J$ s.t. $|J'|\leq p$ and $\sum_{i\in I}\min_{j\in J'}c(i,j)\leq k$.
        \end{tabular}
    }
}

\ifSubfilesClassLoaded{
  \bibliography{../ref}
}{}
\end{document}

%% file: Problems/VCV.tex
\noindent\fbox{
    \parbox{\textwidth}{
        \textsc{Vertex Cover with one fixed Vertex ($\VCV$)}\\
        \begin{tabular}{cl}
        ~ & \textbf{Instances:} Graph $G=(V,E)$, number $k\in \mathbb{N}$, vertex $v\in V$.\\
        ~ & \textbf{Universe:} Vertex set $\mathcal{U}:=V$.\\
        ~ & \textbf{Feasible solution set:} The set of all vertex covers $S\in\mc S_\textsc{VC}$, where $v\in S$.\\
        ~ & \textbf{Solution set:} The set of all vertex covers $S$ of size at most $k$, with $v\in S$.
        \end{tabular}
    }
}

\ifSubfilesClassLoaded{
  \bibliography{../ref}
}{}
\end{document}

%% file: Problems/UHC.tex
\noindent\fbox{
    \parbox{\textwidth}{
        \textsc{Undirected Hamiltonian Cycle ($\UHC$)}\\
        \begin{tabular}{cl}
        ~ & \textbf{Instances:} Graph $G=(V,E)$.\\
        ~ & \textbf{Universe:} Edge set $\mathcal{U}:=E$.\\
        ~ & \textbf{Solution set:} The set of all sets $S\subseteq E$ forming a Hamiltonian cycle.
        \end{tabular}
    }
}

\ifSubfilesClassLoaded{
  \bibliography{../ref}
}{}
\end{document}

%% file: Reductions/Independent_Set/Independent_Set.tex
\subsection{Reductions from Maximum  Cardinality Independent Set}

\subfile{../../Problems/MIS.tex}
\subsubsection[... to Minimum Cardinality Vertex Cover]{Maximum Cardinality Independent Set to Minimum Cardinality Vertex Cover}
\label{sec:MIS to MVC}
\paragraph{Problem Definition}

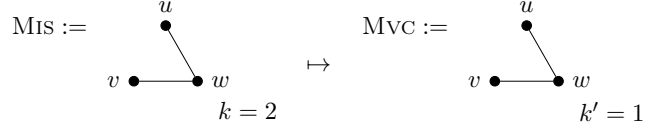
\begin{figure}[h]
    \begin{center}
    \scalebox{.85}{
        \begin{tikzpicture}[vertex/.style = {draw, circle, fill, inner sep = 1.5}, node distance = 1.5cm]
            \node[](G){$\MIS:=$};
            \node[vertex, right = 1cm of G, label = $u$](u){};
            \node[vertex, below left = 0.75cm and  .38cm of u, label = left:$v$](v){};
            \node[vertex, below right= 0.75cm and  .38cm of u, label = right:$w$](w){};
  
            \node[below right = .13cm and .13cm of w](k){$k=2$};
  
            \node[below right = .38cm and 2cm of u](to){$\mapsto$};

            \node[right = 4cm of G](G'){$\MVC:=$};

            \node[vertex, right = 1cm of G', label = $u$](u'){};
            \node[vertex, below left = 0.75cm and  .38cm of u', label = left:$v$](v'){};
            \node[vertex, below right= 0.75cm and  .38cm of u', label = right:$w$](w'){};
            
            \node[below right = .13cm and .13cm of w'](k'){$k'=1$};
  
            \draw   (u)--(w)
                    
                    (v)--(w)
                    (u')--(w')
                    
                    (v')--(w')
            ;
        \end{tikzpicture} 
    }
    \caption{\footnotesize{Example \textsc{Maximum Cardinality Independent Set to Minimum Cardinality Vertex Cover}}}
    \label{fig:Mis to Mvc}
    \end{center}
\end{figure}

\subfile{../../Problems/MVC.tex}
\paragraph{Reduction}
Let $\MIS:=[G=(V,E),k]$. Now for any independent set found on a graph, there also exists a vertex cover on the same graph consiting of
the vertices not in the set. It follows, that for any \MIS instance, $\MVC:=[G=(V,E),k']$ is the corresponding \textsc{Minimum Cardinality Vertex Cover} 
instance with $k':=|V|-k$. \autoref*{fig:Mis to Mvc} shows an example reduction.

\begin{proof}
    This holds as any independent Set 
    $S\in \mc S_\MIS$ has no edge $e=\{u,v\}$ for which both $u$ and $v$ are in $S$, hence every edge $e\in E$ has at least one
    endpoint not in $S$. It follows that $S':=V\backslash S$ is a vertex cover of size $|V|-k$.\\
    The other way around, if $S'$ is a vertex cover of size $k'$ then no edge $e=\{u,v\}$ exists, for which both $u\notin S'$
    and $v\notin S'$. Thus $S:=V\backslash S'$ is an independent set of size $k=|V|-k'$.
\end{proof}
\noindent The reduction function $g$ is computable in polynomial time and is dependent on the input size of the \MIS instance $\big|[G=(V,E),k]\big|$,
as we only change the number $k$. Therefore $\MIS\leq \MVC$ holds.

\paragraph{SSP}
This reduction is not SSP. We look at the example shown in \autoref*{fig:Mis to Mvc} and see, that for the solution $\S\MIS$ with 
$S=\{v,w\}$ the equivalent solution in $\mc S_\MVC$ is $S'=\{u\}$. Now the SSP functions could not map the vertices to themselves, as
with this 
$$f(\{v,w\})=\{v,w\} \neq \{u\} \cap f(V) = \emptyset ,$$
but the functions could also not map the vertices to other vertices as an injective function $f:\{v,w\}\rightarrow \{u\}$ cannot exist.
Therefore $\MIS\leq_{\cancel{\text{\tiny SSP}}}\MVC$.

\paragraph{Parsimonious}
This reduction, albeit not SSP, is still strongly Parsimonious, as shown with the following bijective function.
\begin{flalign*}
    p_I&:\mc S_\MIS \rightarrow \mc S_\MVC,~p(S)=V\backslash S\\
    \text{and }p_I^{-1}&:\mc S_\MVC \rightarrow \mc S_\MIS,~p^{-1}(S')=V\backslash S'
\end{flalign*}
Therefore $\MIS\SPR\MVC$ holds. 

\paragraph{Alternative transitive reduction with both SSP \& SPR properties}
To find a transitive reduction $\MIS\SSPR\MVC$, we utilize \autoref{sec:Cook}, 
where we prove that Cook's reduction is both SSP and SPR, as well as \autoref{sec:SAT to 3SAT}, \autoref{sec:3SAT to ESAT} and \autoref{sec:ESAT to MVC}. With all of these we get
$$\MIS\SSPR\SAT\SSPR\TSAT\SSPR\ESAT\SSPR\MVC$$
and therefore $\MIS\SSPR\MVC$ holds.

\subsubsection[... to Clique]{Maximum  Cardinality Independent Set to Clique}
\label{sec:MIS to CQ}
\paragraph{Problem Definition}

\subfile{../../Problems/CQ.tex}
\paragraph{Reduction}
Based on a reduction presented by Garey and Johnson \cite{garey} this reduction also works by using the complement of the graph. \\
Let $[G=(V,E),k]$ be a \MIS instance, then $[G',k']$ is a \CQ instance where $G':=\ne G$ and $k':=k$.
\begin{figure}[h]
  \begin{center}
  \scalebox{.85}{
      \begin{tikzpicture}[vertex/.style = {draw, circle, fill, inner sep = 1.5}, node distance = 1.5cm]
          \node[vertex, label = $u$](u){};
          \node[vertex, right= 1cm of u, label = $v$](v){};
          \node[vertex, below right= 1cm and .5cm of v, label = right:$w$](w){};
          \node[vertex, below left= 1cm and .5cm of w, label = below:$x$](x){};
          \node[vertex, left = 1cm of x, label = below:$y$](y){};
          \node[vertex, above left= 1cm and .5cm of y, label = left:$z$](z){};
          \node[above left = 1cm and .5cm of z, label ={[label distance=-.45cm]left:$G:$}](G){};

          \node[right = 1.5cm of x](k){$k=4$};

          \node[right = 3cm of w, label ={[label distance=-.45cm]left:$\mapsto$}](to){};

          \node[vertex, right= 2cm of to, label = left:$z$](z'){};
          \node[vertex, above right= 1cm and .5cm of z', label = $u$](u'){};
          \node[vertex, right= 1cm of u', label = $v$](v'){};
          \node[vertex, below right= 1cm and .5cm of v', label = right:$w$](w'){};
          \node[vertex, below left= 1cm and .5cm of w', label = below:$x$](x'){};
          \node[vertex, left = 1cm of x', label = below:$y$](y'){};
          \node[above left = 1cm and .5cm of z', label ={[label distance=-.45cm]left:$G':$}](G'){};
          
          \node[right = 1.5cm of x'](k){$k'=4$};

          \draw (u)--(w)
                (z)--(w)
                (z)--(x)
                (z)--(u)

                (u')--(v')
                (u')--(x')
                (u')--(y')
                (v')--(w')
                (v')--(x')
                (v')--(y')
                (v')--(z')
                (w')--(x')
                (w')--(y')
                (x')--(y')
                (y')--(z')
          ;
      \end{tikzpicture} 
  }\\
  \caption{\footnotesize{Example \textsc{Maximum Cardinality Independent Set to \textsc{Clique}}}}
  \label{fig:Mis to CQ}
  \end{center}
\end{figure}
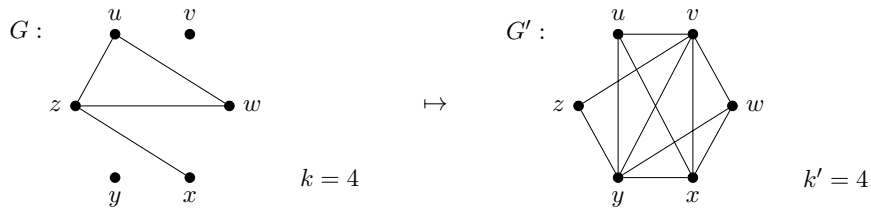
\begin{proof}
    Now, any $S\in\mc S_\MIS$ is also a solution in $\mc S_\CQ$, as a set of $k$ not connected vertices in $G$ is a clique of size 
    $k$ in $G'$.\\
    Likewise, for an $S'\in \mc S_\CQ$ this solution has an equivalent solution in $\mc S_\MIS$, as all vertices in a clique are not 
    connected at all in $G$, making this an independent set.
\end{proof}
\noindent The reduction function $g$ is computable in polynomial time and is dependent on the input size of the \MIS instance $\big|[G=(V,E),k]\big|$,
as for each vertex in $V$ we create one vertex the new instance. Therefore $\MIS\leq \CQ$ holds.

\paragraph{SSP \& Parsimonious}
Due to the complete equivalence between the two universes it is easy to construct functions $f_I$ $f$ between the two and therefore we use
\lemref{lem:SSP and SPR 2}. 
\begin{proof}
    Let 
    $$f_I:V\rightarrow V', f(v) = v',$$ where $v$ and $v'$ are the same vertex in the two instance respectively, and 
    $$f_I^{-1}:V'\rightarrow V, f(v') = v$$ then
    \begin{flalign*}
      \{f(S):S\in \mc S_\MIS\} &= \big\{\{f(v):v\in S\}:S\in \mc S_\MIS\big\}\\
      &= \big\{\{v'\in S'\}:S'\in \mc S_\CQ\big\}\\
      &= \{S'\cap f(V): S'\in \mc S_\CQ\}\\
      &= \{S'\in \mc S_\CQ\}
    \end{flalign*}
    and therefore \MIS $\SSPR$ \CQ.
\end{proof}

\subsubsection[... to Set Packing]{Maximum  Cardinality Independent Set to Set Packing}
\label{sec:MIS to SP}
\paragraph{Problem Definition}
\subfile{../../Problems/SP.tex}

\paragraph{Reduction}
The reduction presented in \cite[p. 13-14]{Femke} is already both SSP and SPR and, as that thesis includes the whole reduction, 
proof and an example, we only briefly introduce the idea behind the reduction here.\\
Let $\MIS:=[G=(V,E),k]$, then the corresponding \SP instance is created by mapping every vertex $v\in V$ to the set 
$\psi_{v}$, containing the edges $e=\{v,w\}$, which connect to the vertex $v$, as well as the vertex $v$ itself. The number $k$ stays the same in both instances. 
Any independent set $\S\MIS$ is now equivalent to a Set Packing solution $\Sr\SP$ by stating if $v\in S$, then $\psi_v\in S'$
 and vice versa.

\paragraph{SSP \& Parsimonious}
As seen above, as well as in the proof provided in \cite{Femke}, there exists a one-to-one correspondence between the sets $\psi_v$ and the vertices $v$ in 
the solutions. This means we can use \lemref{lem:SSP and SPR 2} as follows.
\begin{proof}
    Let $f$ be defined as follows.
    $$f_I:V\rightarrow \{\psi_1,...,\psi_{|V|}\}, ~f(v)=\psi_v$$
    Then $f$ is a bijective function, as for every vertex $v$ in $V$, we create exactly one corresponding set $\psi_v$. With this
    bijective function we get 
    \begin{flalign*}
        \{f(S):\S\MIS\}&=\big\{\{f(v):v\in S\}:\S\MIS\big\}\\
        &=\big\{\{\psi_v\in S'\}:\Sr\SP\big\}\\
        &=\{S'\cap f(V):\Sr\SP\}\\
        &=\mc S_{\SP}
    \end{flalign*}
    and therefore $\MIS\SSPR\SP$ holds.
\end{proof}

\ifSubfilesClassLoaded{
  \bibliography{../../ref}
}{}
\end{document}

%% file: Problems/SP.tex
\noindent\fbox{
    \parbox{\textwidth}{
        \textsc{Set Packing ($\SP$)}\\
        \begin{tabular}{cl}
        ~ & \textbf{Instances:} Sets $\psi_i\subseteq \Psi = \{1,...,m\}$ for $i\in[n]$, number $k\in \mathbb N$.\\
        ~ & \textbf{Universe:} Sets $\mathcal{U}:=\{\psi_1,...,\psi_n\}$.\\
        ~ & \textbf{Feasible solution set:} The set of all $S\subseteq\{\psi_1,..,\psi_n\}$ s.t. $\forall \psi_i,\psi_j\in S, i\neq j:\psi_i\cap\psi_j=\emptyset$.\\
        ~ & \textbf{Solution set:} All feasible solutions with $|S|\geq k$.
        \end{tabular}
    }
}

\ifSubfilesClassLoaded{
  \bibliography{../ref}
}{}
\end{document}

%% file: Reductions/Clique/Clique.tex
\subsection{Reductions from Clique}

\subfile{../../Problems/CQ.tex}
\subsubsection[... to Maximum Cardinality Independent Set]{Clique to Maximum Cardinality Independent Set}
\label{sec:CQ to MIS}

\subfile{../../Problems/MIS.tex}
\paragraph{Reduction}
Based on the reduction \textsc{Independent Set} $\leq$ \textsc{Clique} by Garey and Johnson \cite{garey} this reduction also works
by using the complement of the graph. \\Let $[G=(V,E),k]$ be a \CQ instance, then $[G',k']$ is a \MIS instance where $G':=\ne G$ and $k':=k$.
\begin{figure}[h]
  \begin{center}
  \scalebox{.85}{
      \begin{tikzpicture}[vertex/.style = {draw, circle, fill, inner sep = 1.5}, node distance = 1.5cm]
          \node[vertex, label = $u$](u){};
          \node[vertex, right= 1cm of u, label = $v$](v){};
          \node[vertex, below right= 1cm and .5cm of v, label = right:$w$](w){};
          \node[vertex, below left= 1cm and .5cm of w, label = below:$x$](x){};
          \node[vertex, left = 1cm of x, label = below:$y$](y){};
          \node[vertex, above left= 1cm and .5cm of y, label = left:$z$](z){};
          \node[above left = 1cm and .5cm of z, label ={[label distance=-.45cm]left:$G:$}](G){};

          \node[right = 1.5cm of x](k){$k=3$};

          \node[right = 3cm of w, label ={[label distance=-.45cm]left:$\mapsto$}](to){};

          \node[vertex, right= 2cm of to, label = left:$z$](z'){};
          \node[vertex, above right= 1cm and .5cm of z', label = $u$](u'){};
          \node[vertex, right= 1cm of u', label = $v$](v'){};
          \node[vertex, below right= 1cm and .5cm of v', label = right:$w$](w'){};
          \node[vertex, below left= 1cm and .5cm of w', label = below:$x$](x'){};
          \node[vertex, left = 1cm of x', label = below:$y$](y'){};
          \node[above left = 1cm and .5cm of z', label ={[label distance=-.45cm]left:$G':$}](G'){};
          
          \node[right = 1.5cm of x'](k){$k'=3$};

          \draw (u)--(w)
                (z)--(w)
                (z)--(x)
                (z)--(u)

                (u')--(v')
                (u')--(x')
                (u')--(y')
                (v')--(w')
                (v')--(x')
                (v')--(y')
                (v')--(z')
                (w')--(x')
                (w')--(y')
                (x')--(y')
                (y')--(z')
          ;
      \end{tikzpicture} 
  }\\
  \caption{\footnotesize{Example \textsc{Clique} to \textsc{Maximum Cardinality Independent Set}}}
  \label{fig:CQ to MIS}
  \end{center}
\end{figure}
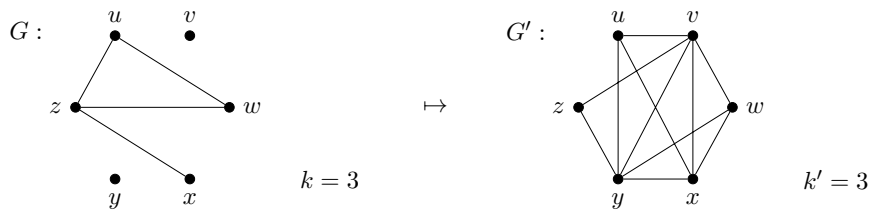
\begin{proof}
  Now, for an $S\in \mc S_\CQ$ this solution is also a solution in $\mc S_\MIS$, as all vertices in a clique are not connected at
  all in $G'$, making this an independent set.\\
  Likewise an $S'\in\mc S_\MIS$ is also a solution in $\mc S_\CQ$, as a set of $k$ not connected vertices in $G'$ is a clique of size 
  $k$ in $G$.
\end{proof}
\noindent The reduction function $g$ is computable in polynomial time and is dependent on the input size of the \CQ instance $\big|[G=(V,E),k]\big|$,
as for each vertex in $V$ we create one vertex the new instance. Therefore $\CQ\leq \MIS$ holds.

\paragraph{SSP \& Parsimonious}
Due to the complete equivalence between the two universes it is easy to construct bijective functions $f_I$ between the two and therefore we use
\lemref{lem:SSP and SPR 2}. 
\begin{proof}
    Let 
    $$f_I:V\rightarrow V', f(v) = v',$$ where $v$ and $v'$ are the same vertex in the two instance respectively, and 
    $$f_I^{-1}:V'\rightarrow V, f(v') = v$$ then
    \begin{flalign*}
      \{f(S):S\in \mc S_\CQ\} &= \big\{\{f(v): v\in S\}:S\in \mc S_\CQ\big\}\\
      &= \big\{\{v'\in S'\}:S'\in \mc S_\MIS\big\}\\
      &= \{S'\cap f(V): S'\in \mc S_\MIS\}\\
      &= \{S'\in \mc S_\MIS\}
    \end{flalign*}
    and therefore \CQ $\SSPR$ \MIS.
\end{proof}

\subsubsection[... to Minimum Cardinality Vertex Cover]{Clique to Minimum Cardinality Vertex Cover}
\label{sec:CQ to MVC}
\paragraph{Problem Definition}

\subfile{../../Problems/MVC.tex}
\paragraph{Reduction}
This reduction works analogously to the reduction \CQ $\leq$ \MIS shown in section \ref{sec:CQ to MIS}.\\
Let $[G=(V,E),k]$ be a \CQ instance, then $[G',k']$ is a \MVC instance where $G':=\ne G$ and $k':=|V|-k$.
\begin{figure}[h]
  \begin{center}
  \scalebox{.85}{
      \begin{tikzpicture}[vertex/.style = {draw, circle, fill, inner sep = 1.5}, node distance = 1.5cm]
          \node[vertex, label = $u$](u){};
          \node[vertex, right= 1cm of u, label = $v$](v){};
          \node[vertex, below right= 1cm and .5cm of v, label = right:$w$](w){};
          \node[vertex, below left= 1cm and .5cm of w, label = below:$x$](x){};
          \node[vertex, left = 1cm of x, label = below:$y$](y){};
          \node[vertex, above left= 1cm and .5cm of y, label = left:$z$](z){};
          \node[above left = 1cm and .5cm of z, label ={[label distance=-.45cm]left:$G:$}](G){};
          \node[below left = 2cm and 0cm of z, label ={[label distance=-.45cm]left:$k=4$}](k){};

          \node[right = 2cm of w, label ={[label distance=-.45cm]left:$\mapsto$}](to){};

          \node[vertex, right= 2cm of to, label = left:$z$](z'){};
          \node[vertex, above right= 1cm and .5cm of z', label = $u$](u'){};
          \node[vertex, right= 1cm of u', label = $v$](v'){};
          \node[vertex, below right= 1cm and .5cm of v', label = right:$w$](w'){};
          \node[vertex, below left= 1cm and .5cm of w', label = below:$x$](x'){};
          \node[vertex, left = 1cm of x', label = below:$y$](y'){};
          \node[above left = 1cm and .5cm of z', label ={[label distance=-.45cm]left:$G':$}](G'){};
          \node[below = 2cm of z', label ={[label distance=-.45cm]left:$k'=2$}](k){};

          \draw (u)--(w)
                (u)--(v)
                (v)--(w)
                (v)--(z)
                (z)--(w)
                (z)--(x)
                (z)--(u)
                
                (u')--(x')
                (u')--(y')
                (v')--(x')
                (v')--(y')
                (w')--(x')
                (w')--(y')
                (x')--(y')
                (y')--(z')
          ;
      \end{tikzpicture} 
  }\\
  \caption{\footnotesize{Example \textsc{Clique} to \textsc{Minimum Cardinality Vertex Cover}}}
  \label{fig:CQ to MVC}
  \end{center}
\end{figure}
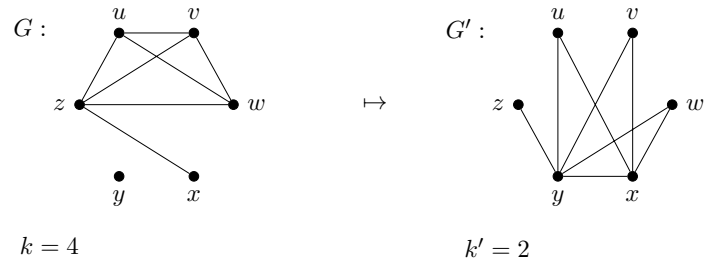
\begin{proof}
  Now, for a clique $S\in \mc S_\CQ$ of size $k$, $V\backslash S$ is a vertex cover of size $|V|-k$ in $G'$, as none of the vertices
  connected in the clique in $G$ have any edges between them in $G'$.\\
  Likewise a vertex cover $S'\in\mc S_\MVC$ of size $k'$ in $G'$ is also clique in $G'$, as the vertices not in the vertex cover do not 
  have edges between them in $G$, meaning they are a clique of size $|V|-k'$ in $G$.
\end{proof}
\noindent The reduction function $g$ is computable in polynomial time and is dependent on the input size of the \CQ instance $\big|[G=(V,E),k]\big|$,
as for each vertex in $V$ we create one vertex the new instance. Therefore $\CQ\leq \MIS$ holds.

\paragraph{SSP}
Even though we show in the following subsection that this reduction is strongly Parsimonious, it is not SSP. Due to the fact that there cannot 
be an injective function between the universes. See the example shown in \autoref*{fig:CQ to MVC}. In this there is one clique of size $k=4$ in 
$G$, which is $\{u,v,w,z\}$. In $G'$ there is the corresponding vertex cover $\{x,y\}$ of size 2. However there cannot be an injective function 
between these two sets, as $|\{u,v,w,z\}|=4>2=|\{x,y\}|$. And since the solution sets do not overlap at all, a function mapping each vertex to itself
would violate the stipulation 
$$\{f(S):S\in \mc S_\CQ\} = \{S'\cap f(V)\in \mc S_\MVC\}$$
as $S'\cap f(V) = \emptyset$. Thus this reduction is not SSP.

\paragraph{Parsimonious}
Due to the correspondence of the two solution sets it is easy to construct a bijective function $f$ between the two 
universes of the problems. Let 
$$p_I:V\rightarrow V',~p(C) = V\backslash C,~C\subseteq V$$ 
$$\text{and }p_I^{-1}:V'\rightarrow V,~p^{-1}(C) = V'\backslash C,~C\subseteq V'$$
then
\begin{flalign*}
  \{f(S):S\in \mc S_\CQ\} &= \big\{V\backslash S:S\in \mc S_\CQ\big\}\\
  &= \{S'\in \mc S_\MVC\}\\
\end{flalign*}
Thus $\CQ \SPR\MVC$ holds.

\paragraph{Alternative transitive reduction with both SSP \& SPR properties}
To find a transitive reduction $\CQ\SSPR\MVC$, we utilize \autoref{sec:Cook}, 
where we prove that Cook's reduction is both SSP and SPR, as well as \autoref{sec:SAT to 3SAT}, \autoref{sec:3SAT to ESAT} and \autoref{sec:ESAT to CQ}. With all of these we get
$$\CQ\SSPR\SAT\SSPR\TSAT\SSPR\ESAT\SSPR\MVC $$
and therefore $\CQ\SSPR\MVC$ holds.

\ifSubfilesClassLoaded{
  \bibliography{../../ref}
}{}
\end{document}

%% file: Reductions/Set_Packing/Set_Packing.tex
\subsection{Reductions from Set Packing}

\subfile{../../Problems/SP.tex}
\subsubsection[... to  Maximum Cardinality Independent Set]{Set Packing to Maximum Cardinality Independent Set}
\label{sec:SP to MIS}
\paragraph{Problem definition}

\subfile{../../Problems/MIS.tex}
\paragraph{Reduction}
The reduction works analogously to the reduction $\MIS\leq\SP$ (\autoref{sec:MIS to SP}) and is also shown in \cite[p. 14-15]{Femke}, therefore
we only briefly introduce the idea behind the reduction here.\\
Let $\SP:=[\{\psi_1,...,\psi_n\},k]$, then the corresponding \MIS instance is created by mapping every set $\psi_i$ to a vertex 
$i\in V$ and create edges between the vertices $i$ and $j$, if an element $x\in\Psi$ is in both sets $\psi_i$ and $\psi_j$. 
The number $k$ stays the same in both instances. Any Set Packing solution $\S\SP$ is now equivalent to an independent set $\Sr\MIS$ by 
stating if $\psi_i\in S$, then $i\in S'$ and vice versa.

\paragraph{SSP \& Parsimonious}
As seen above, as well as in the proof provided in \cite{Femke}, there exists a one-to-one correspondence between the sets 
$\psi_i$ and the vertices $i$ in the solutions. This means we can use \lemref{lem:SSP and SPR 2} as follows.
\begin{proof}
    Let $f$ be defined as follows.
    $$f_I:\{\psi_1,...,\psi_n\}\rightarrow V, ~f(\psi_i)= i$$
    Then $f$ is a bijective function, as for every set $\psi_i$, we create exactly one corresponding vertex $i$ in $V$. With this
    bijective function we get 
    \begin{flalign*}
        \{f(S):\S\SP\}&=\big\{\{f(\psi_i)\mid\psi_i\in S\}:\S\SP\big\}\\
        &=\big\{\{i\in S'\}:\Sr\MIS\big\}\\
        &=\{S'\cap f(\{\psi_1,...,\psi_n\}):\Sr\MIS\}\\
        &=\mc S_{\MIS}
    \end{flalign*}
    and therefore $\SP\SSPR\MIS$ holds.
\end{proof}

\ifSubfilesClassLoaded{
  \bibliography{../../ref}
}{}
\end{document}

%% file: Reductions/Subset_Sum/Subset_Sum.tex
\subsection{Reductions from Subset Sum}

\subfile{../../Problems/SS.tex}
\subsubsection[... to Knapsack]{Subset Sum to Knapsack}
\label{sec:SS to KS}
\subfile{../../Problems/KS.tex}

\paragraph{Reduction}
Let $[\{a_1,...,a_n\},M]$ be a \SS instance. We transform the numbers $\{a_1,...,a_n\}$ into objects with the same price and weigth 
$\{(a_1,a_1),...,(a_n,a_n)\}$ and also we map the target value $M$ to both limits ($P:=M,~W:=M$). Now both constraints 
$\sum_{(a_i,a_i)\in S} a_i\leq M$ and $\sum_{(a_i,a_i)\in S} a_i\geq M$ together are equivalent to $\sum_{(a_i,a_i)\in S} a_i = M$,
which is the feasible solution set of \SS. Thus we have a one-to-one correspondence between the solution sets. The following shows an
example reduction.
$$\SS:=[\{1,2,3,4\},5] \mapsto \KS:=[\{(1,1),(2,2),(3,3),(4,4)\},5,5]$$
\begin{proof}
  Due to the one-to-one correspondence between the two solution sets and the fact that 
  $$\Big(\sum_{(a_i,a_i)\in S} a_i\leq M~\land \sum_{(a_i,a_i)\in S} a_i\geq M \Big)\Leftrightarrow \sum_{(a_i,a_i)\in S} a_i = M$$
  holds, the reduction also holds.
\end{proof}
\noindent The reduction function $g$ is computable in polynomial time and is dependent on the input size of the \SS instance $\big|[\mc A,M]\big|$,
as for each number in $\mc A$ we create an object in the new instance. Therefore $\SS\leq \KS$ holds.

\paragraph{SSP \& Parsimonious}
As seen above, there is a direct correspondence between the two universes. This means there also exist bijective functions $f_I$ between
the universes and therefore we can use \lemref{lem:SSP and SPR 2}.
\begin{proof}
  Let 
  $$f: \mc A\rightarrow \mc O,~f(a)=(a,a)$$
  $$\text{and }f^{-1}:\mc O\rightarrow \mc A,~f\big((a,a)\big)=a,$$then
  \begin{flalign*}
    \{f(S):S\in \mc S_\SS\} &= \big\{\{f(a)\mid a\in S\}:S\in\mc S_\SS\big\}\\
    &= \big\{\{(a,a)\in S'\}:S'\in\mc S_\KS\big\}\\
    &= \big\{S'\cap f(\mc A):S'\in\mc S_\KS\big\}\\
    &= \big\{S'\in\mc S_\KS\big\}
  \end{flalign*}
  and thus \SS $\SSPR$ \KS holds.
\end{proof}

\subsubsection[... to Partition]{Subset Sum to Partition}
\label{sec:SS to P}
\subfile{../../Problems/P.tex}

\paragraph{Reduction}
This reduction is based on the reduction presented in \cite{ssp} and has the same constraint used there, which is that
w.l.o.g. the last element has to be in the solution. This prevents the symmetry of solutions 
$(S\in \mc S_\P \implies \mc U \backslash S\in \mc S_\P)$ which would destroy the solution preserving property. Let 
$[\mc A,M]$ be a \SS instance, then we define a new number 
$$a_f:=\Big(\sum_{a_i\in \mc A}a_i\Big)+1-M$$ 
and let the new instance be defined by
$\P:=[\mc A\cup\{M+1,a_f\}]$, with $a_f$ being the final number of the new set. The following shows an
example reduction.
$$\SS:=[\{1,2,3,4\},5] \mapsto \P:=[\{1,2,3,4,6,6\}]$$
\begin{proof}
  Let $S$ be a solution for \textsc{Subset Sum}, then
  $$\sum_{a\in S} a = M \text{ and also that } \sum_{a_j\notin S} a_j= \big(\sum_{a_i\in \mc A} a_i\big) -M$$ 
  and so
  we get a solution $S'$ for \textsc{Partition} with 
  $S'= S\cup \{a_f\}$ and $$\mc U_\P\backslash S' = (\mc U_\SS\backslash S)\cup \{M+1\}$$
  for which the following holds.
  \begin{flalign*}
    \sum_{a_i\in S'}a_i &= a_f + \sum_{a\in S} a \\
    &= \Big[\Big(\sum_{a_i\in \mc A}a_i\Big)+1-M\Big] + M \\
    & = \Big[\Big(\sum_{a_i\in \mc A}a_i\Big)-M\Big]+(M+1) \\
    &=\sum_{a_j\notin S} a_j + (M+1)\\
    &= \sum_{a_j\notin S'} a_j
  \end{flalign*}
  Let $S'$ now be a solution for \textsc{Partition}, then 
  $$\sum_{a_i\in S'}a_i = \Big[\Big(\sum_{a_i\in \mc A}a_i\Big)+1-M\Big] + M = a_f+M $$
  which means that $S\backslash\{a_f\}$ is a solution $S$ for $ $.
\end{proof}
\noindent The reduction function $g$ is computable in polynomial time and is dependent on the input size of the \SS instance $\big|[\mc A,M]\big|$,
as we create exactly one new number in the new instance. Therefore $\SS\leq \P$ holds.

\paragraph{SSP \& Parsimonious}
As seen above, any solution $S\in\mc S_\SS$ is equivalent to the solution $S'\in\mc S_\P$, in which $S'=S\cup\{a_f\}$. We can now use 
\lemref{lem:SSP and SPR} to prove that this reduction is both SSP and SPR.
\begin{proof}
  Let $f$ be defined as 
  $$f_I:\mc A_\SS\rightarrow \mc A_\P, ~f(a)=a'$$ where $a$ and $a'$ are the same numbers in the $\SS$ and $\P$ instances
  respectively. Let $\Snev=\Slink=\emptyset$ and let $\Sall := \{a_f\}$. Now $f_I:\mc A_\SS\rightarrow \mc A_\P\backslash \{a_f\}$ is bijective. We get
  \begin{flalign*}
    \{f(S):S\in\mc S_\SS\} &= \big\{\{f(a)\mid a\in S\}:S\in\mc S_\SS\big\}\\
    &= \big\{\big(\{a'\in S'\}\backslash \{a_f\}\big) :S'\in\mc S_\P\big\}\\
    &= \{S'\cap f(\mc A_\SS):S'\in\mc S_\P\}
  \end{flalign*}
  and therefore $\SS\SSPR\P$ holds
\end{proof}

\ifSubfilesClassLoaded{
  \bibliography{../../ref}
}{}
\end{document}

%% file: Problems/KS.tex
\noindent\fbox{
    \parbox{\textwidth}{
        \textsc{Knapsack ($\KS$)}\\
        \begin{tabular}{cl}
        ~ & \textbf{Instances:} Objects with prices and weights $\mc O:=\{(p_1,w_1),...,(p_n,w_n)\}\subseteq \mathbb N^2$, and \\
        ~ & limits $P,W\in \mathbb N$\\
        ~ & \textbf{Universe:} The set of objects $\mathcal{U}:=\mc O$.\\
        ~ & \textbf{Feasible solution set:} The set of all $S\in \mathcal U$ with $\sum_{(p_i,w_i)\in S} w_i\leq W$.\\
        ~ & \textbf{Solution set:} The set of all feasible $S$ with $\sum_{(p_i,w_i)\in S} p_i\geq P$.
        \end{tabular}
    }
}

\ifSubfilesClassLoaded{
  \bibliography{../ref}
}{}
\end{document}

%% file: Problems/P.tex
\noindent\fbox{
    \parbox{\textwidth}{
        \textsc{Partition ($\P$)}\\
        \begin{tabular}{cl}
        ~ & \textbf{Instances:} Numbers $\mc A:=\{a_1,...,a_n\}\subseteq\mathbb N$\\
        ~ & \textbf{Universe:} The set of numbers $\mathcal{U}:=\mc A$.\\
        ~ & \textbf{Feasible solution set:} The set of all sets $S\subseteq \mathcal U$ with $a_n\in S$ and 
        $\sum_{a_i\in S}a_i = \sum_{a_j\notin S}a_j$.
        \end{tabular}
    }
}

\ifSubfilesClassLoaded{
  \bibliography{../ref}
}{}
\end{document}

%% file: Reductions/Partition/Partition.tex
\subsection{Reductions from Partition}

\subfile{../../Problems/P.tex}
\subsubsection[... to Two-Machine-Scheduling]{Partition to Two-Machine-Scheduling}
\label{sec:P to TMS}
\paragraph{Problem definition}
\subfile{../../Problems/TMS.tex}

\paragraph{Reduction}
As in section \ref*{sec:SS to P} we demand w.l.o.g. that the last element $t_n$ be in the first machine to prevent symmetric solutions. 
Let $\P:=[\mc A=\{a_1,...,a_n\}]$, then we map each number to a job with processing time $a_i$ and get $\mc J:=\{a_1,...,a_n\}$ and finally 
set the time threshold at $T:=\frac 12\sum_{a\in \mc A} a$. This necessitates that the jobs get split into two equal sets, which is equivalent
to the \textsc{Partition} problem. An example of this looks as follows.
$$\P:=[\{1,2,3,4\}] \mapsto \TMS:=[\{1,2,3,4\},5]$$
\begin{proof}
    Let $S$ be a solution for the \P instance. Then the numbers can be split into two equal parts, each with a sum of $\sum_{a\in S} a$. This means that
    $S$ is equivalent to a solution $\Sr\TMS$ with $S':= S$.\\
    Now let $S'$ be a solution for the \TMS instance. Then $\sum_{t_i\in S'} t_i\leq T$ and $\sum_{t_j\notin S'} t_i\leq T$ and as $\mc J:=\{a_1,...,a_n\}$
    and $T:=\frac 12\sum_{a\in \mc A} a$ we get that
    $$\sum_{a_i\in S'} t_i\leq \frac 12\sum_{a\in \mc A} a \geq \sum_{a_j\notin S'} t_i$$
    and so, as $S'\cup (\mc J\backslash S')=\mc A$
    $$\sum_{a_i\in S'} t_i = \frac 12\sum_{a\in \mc A} a = \sum_{a_j\notin S'} t_i$$
    holds. Therefore $S'$ is equivalent to a solution $\S\P$ with $S:= S'$.
\end{proof}
\noindent The reduction function $g$ is computable in polynomial time and is dependent on the input size of the \P instance $\big|[\mc A,M]\big|$,
as for every number in $\mc A$ we create one job in the new instance. Therefore $\P\leq \TMS$ holds.

\paragraph{SSP \& Parsimonious}
As seen above the two solutions are equivalent, meaning we can define a bijective function $f$ between the two universes and use \lemref{lem:SSP and SPR 2} as follows.
\begin{proof} 
    Let
    $$f_I:\mc A\rightarrow \mc J,~f(a)=a'$$
    $$\text{and }f_I^{-1}:\mc J\rightarrow \mc A,~f(a')=a,$$
    where $a$ corresponds to $a'$ in their respective instances.
    And with $f_I$ we get 
    \begin{flalign*}
        \{f(S):\S\P\} &= \big\{\{f(a)\mid a\in S\}:\S\P\big\}\\
        &= \big\{\{a'\in S'\}:\Sr\TMS\big\}\\
        &= \{S'\cap f(\mc A):\Sr\TMS\}\\
        &= \{\Sr\TMS\}
    \end{flalign*}
    which means that $\P\SSPR\TMS$.
\end{proof}

\ifSubfilesClassLoaded{
  \bibliography{../../ref}
}{}
\end{document}

%% file: Problems/TMS.tex
\noindent\fbox{
    \parbox{\textwidth}{
        \textsc{Two-Machine-Scheduling ($\TMS$)}\\
        \begin{tabular}{cl}
        ~ & \textbf{Instances:} Jobs with processing time $\mathcal J:=\{t_1,...,t_n\}\subseteq\mathbb N$, threshold $T\in\mathbb N$ \\
        ~ & \textbf{Universe:} The set of jobs $\mathcal{U}:=\mc J$.\\
        ~ & \textbf{Solution set:} The set of all $J_1\in\mc J$ such that $t_n\in J_1$ and $\sum_{t_i\in J_1} t_i\leq T$ and \\
        ~ & $\sum_{t_j\in J_2} t_j\leq T$ with $J_2=\mc J\backslash J_1$, i.e. both machines finish in time $T$.
        \end{tabular}
    }
}

\ifSubfilesClassLoaded{
  \bibliography{../ref}
}{}
\end{document}

%% file: Reductions/Exact_Cover/Exact_Cover.tex
\subsection{Reductions from Exact Cover}

\subfile{../../Problems/EC.tex}
\subsubsection[... to Steiner Tree]{Exact Cover to Steiner Tree}
\label{sec:EC to STT}
\paragraph{Problem Definition}

\subfile{../../Problems/STT}
\paragraph{Reduction}
For this reduction we refer to the reduction presented in \cite[p. 25-29]{Femke}, as it already upholds both the SSP and Parsimonious properties.
The idea behind the reduction is to map every element in the \textsc{Exact Cover} instance to a terminal vertex in the \textsc{Steiner Tree}.
The rest of the tree is constructed from non-terminal vertices representing the sets $\psi\subseteq \Psi$ and one extra terminal vertex $v_0$ 
and the weights on the edges and the limit $k$ are set so that every \textsc{Steiner Tree} found in $G$ corresponds to exacty one \textsc{Exact Cover}, identifiable
over the non-terminals $\psi_i$ included in the tree. As \cite{Femke} includes a detailed reduction, proof and an example, we refrain from 
repeating them here.

\paragraph{SSP \& Parsimonious}
Every \textsc{Exact Cover} corresponds to exacty one \textsc{Steiner Tree} and vice versa, and every tree can be identified by the edges, connecting to
$v_0$, included in it. Therefore we can apply \autoref{lem:SSP and SPR} as follows
to prove that the reduction presented in \cite{Femke} is both SSP and SPR.
\begin{proof}
    Let the functions $f_I$ be defined as
    $$f_I:\{\psi_1,...,\psi_n\}\rightarrow E,~f(\psi_i)=\{v_0,\psi_i\}$$
    and let $\Sall=\Snev=\emptyset$. Also let
    $$\mf L\big(\{v_0,\psi_i\}\big):=\big\{\{\psi_i,x\}\mid \{\psi_i,x\}\in E, x\in \Psi\big\}$$
    and therefore let $\Slink$ be defined by
    $$\Slink:=\bigcup_{\S\EC}\Big(\bigcup_{x\in f(S)}\mf L(x)\Big).$$
    Now we get that $\mc U_\STT = E = f\big(\{\psi_1,...,\psi_n\}\big)\cupdot \Slink = \mc U_\EC\cupdot\Slink$ and with this
    \begin{flalign*}
        \{f(S):\S\EC\} &= \big\{\{f(\psi_i)\mid\psi_i\in S\}:\S\EC\big\}\\
        &=\Big\{\big\{\{v_0,\psi_i\}\in S'\big\}:\Sr\STT\Big\}\\
        &= \{S'\backslash\Slink:\Sr\STT\}\\
        &= \big\{S'\cap f\big(\{\psi_1,...,\psi_n\}\big):\Sr\STT\big\}
    \end{flalign*}
    holds, and therefore $\EC\SSPR\STT$ also holds.
\end{proof}

\subsubsection[... to Subset Sum]{Exact Cover to Subset Sum}
\label{sec:EC to SS}
\paragraph{Problem Definition}

\subfile{../../Problems/SS}
\paragraph{Reduction}
For this reduction we refer to the reduction presented in \cite[p. 33-35]{Femke}, as it already upholds both the SSP and Parsimonious properties.
The reduction works by mapping every set $\psi_i\subseteq \Psi$ to a corresponding number $a_i$. The values of the numbers $a\in A$, 
as well as the target value $M$, are then constructed in a way, so that only the numbers corresponding to an \textsc{Exact Cover} add up to $M$. 
As \cite{Femke} includes a detailed reduction, proof and an example, we refrain from 
repeating them here.

\paragraph{SSP \& Parsimonious}
There exists an exact one-to-one correspondence between the sets $\psi_i$ and the numbers $a_i$. We can therefore easily define a bijective
function between the universes of the two instances and use \autoref{lem:SSP and SPR 2} to prove that this reduction is both SSP and Parsimonious.
\begin{proof}
  Let the functions $f_I$ be defined as
  \begin{flalign*}
    f_I&:\{\psi_1,...,\psi_n\}\rightarrow A,~f(\psi_i)= a_i\\
    f_I^{-1}&: A \rightarrow \{\psi_1,...,\psi_n\}, ~f^{-1}(a_i)=\psi_i 
  \end{flalign*}
  then $f(\mc U_\EC)=\mc U_\SS$ and $f^{-1}(\mc U_\SS)=\mc U_\EC$. With this 
  \begin{flalign*}
    \{f(S):\S\EC\} &= \big\{\{f(\psi_i)\mid\psi\in S\}:\S\EC\big\}\\
    &=\big\{\{a_i\in S'\}:\Sr\SS\big\}\\
    &=\{\Sr\SS\}\\ 
    &=\big\{S'\cap f\big( \{\psi_1,...,\psi_n\}\big):\Sr\SS\big\}
  \end{flalign*}
  and therefore $\EC\SSPR\SS$ holds, according to \autoref{lem:SSP and SPR 2}.
\end{proof}

\subsubsection[... to 3-Dimensional Exact Matching]{Exact Cover to 3-Dimensional Exact Matching}
\label{sec:EC to DM}
\paragraph{Problem Definition}

\subfile{../../Problems/DM}
\paragraph{Reduction}
For this reduction we first examine the reduction presented in \cite[p. 29-32]{Femke}. The reduction presented there is SSP,
but does not fulfill the Parsimonious property. \autoref*{fig:EC to DM} shows an example, where this property is breached. The
sets surrounded by a box are in every equivalent solution, with the two different styles of boxes being alternative additions 
to the solutions in the \DM instance. As the shown \EC solution therefore has two equivalent solutions in the \DM instance,
this reduction is not Parsimonious.
\begin{figure}[ht]
  \centering
      \begin{subfigure}[t]{.3\textwidth}
        \begin{flalign*}
          \EC&:=\big[\{\psi_1,\psi_2,\psi_3\},\{1,2,3\}\big]&~\\
          &\psi_1=\fbox{\{1,2\}}\\
          &\psi_2=\fbox{\{3\}}\\
          &\psi_3=\{2,3\}
        \end{flalign*}
      \end{subfigure} 
      \begin{subfigure}[t]{.3\textwidth}
        \begin{flalign*}
          \DM:=\big[&X,Y,Z,T\big]\\
          X=&~\big\{(1, 1), (2, 1), (3, 2), (2, 3),(3,3)\}\\
          Y=&~X\\
          Z=&~X\\
          T=&\big\{\fbox{\big((1, 1), (1, 1), (1, 1)\big)},\\
          &~\fbox{\big((2, 1), (2, 1), (2, 1)\big)},\\
          &~\fbox{\big((3, 2), (3, 2), (3, 2)\big)},\\
          &~\big((2, 1), (2, 3), (2, 3)\big),\\
          &~\big((3, 2), (3, 3), (3, 3)\big),\\
          &~\big((2,3),(1,1),(2,1)\big),\\
          &~\big((2,3),(2,1),(1,1)\big),\\
          &~\big((2,3),(3,2),(3,2)\big),\\
          &~\doublebox{\big((2,3),(2,3),(3,3)\big)},\\
          &~\ovalbox{\big((2,3),(3,3),(2,3)\big)},\\
          &~\big((3,3),(1,1),(2,1)\big),\\
          &~\big((3,3),(2,1),(1,1)\big),\\
          &~\big((3,3),(3,2),(3,2)\big),\\
          &~\ovalbox{\big((3,3),(2,3),(3,3)\big)},\\
          &~\doublebox{\big((3,3),(3,3),(2,3)\big)}\big\}\\
        \end{flalign*}
    \end{subfigure}
  \caption{\footnotesize{Example \textsc{1+3-Dimensional Matching} to \textsc{3-Dimensional Exact Matching}}}
  \label{fig:EC to DM}
\end{figure}

\paragraph{Alternative transitive reduction with both SSP \& SPR properties}
To prove the existence of a reduction that is both SSP and Parsimonious, we utilize \autoref{sec:Cook}, 
where we prove that Cook's reduction is both SSP and SPR. We also use $\SAT\SSPR\ESAT$ (\autoref{sec:SAT to 3SAT} \& \autoref{sec:3SAT to ESAT}) 
and $\ESAT\SSPR\DM$ (\autoref{sec:ESAT to OSAT} \& \autoref{sec:OSAT to DM}). With all of these we get
$$\EC\SSPR\SAT\SSPR\TSAT\SSPR\ESAT\SSPR\OSAT\SSPR\DM$$
and therefore $\EC\SSPR\DM$ holds.

\ifSubfilesClassLoaded{
  \bibliography{../../ref}
}{}
\end{document}

%% file: Reductions/3-Dimensional_Matching/3-Dimensional_Matching.tex
\subsection{Reductions from 3-Dimensional Exact Matching}

\subfile{../../Problems/DM.tex}
\subsubsection[... to Partition]{3-Dimensional Exact Matching to Partition}
\label{sec:DM to P}
\paragraph{Problem Definition}

\subfile{../../Problems/P.tex}
\paragraph{Reduction}
For this reduction we refer to \cite[p. 50-54]{Femke}, as the reduction presented there is already both SSP and Parsimonious. The idea behind the reduction
is to create a number $a_i\in A$ for every triple $u_i\in T$. The reduction also adds two numbers $a_m~(a_{\text{\tiny matching}})$ and $a_g~(a_{\text{\tiny garbage}})$ for which the
numbers corresponding to the matching set plus $a_m$ then add up to the same value as the numbers not in the matching set plus $a_g$. 
Since the \textsc{Partition} problem, as presented in this paper, uses the constraint that the last element in $A$ has to be part of the solution we order 
the numbers, so that the last element is $a_m$. This prevents symmetric solutions, which would destroy the solution preserving
property. As \cite{Femke} includes a detailed reduction, proof and an example, we refrain from repeating them here.

\paragraph{SSP \& Parsimonious}
Every partition found in the \P instance has to include $a_m$ and therefore there exist a direct one-to-one correspondence between
the triplets in a matching and the numbers in a \textsc{Partition} solution. We can therefore apply \autoref{lem:SSP and SPR} as follows to 
prove that this reduction is both SSP and SPR.
\begin{proof}
    Let the functions $f_I$ be defined as 
    $$f_I:T\rightarrow A,~f(u_i)=a_i$$
    and let $\Slink=\emptyset$, $\Sall:=\{a_m\}$ and $\Snev:=\{a_g\}$. With this 
    $$\mc U_\P = A = f(T) \cup \Sall\cup\Snev = f(\mc U_\DMu) \cup \Sall\cup\Snev$$
    and we get
    \begin{flalign*}
        \{f(S):\S\DMu\}&=\big\{\{f(u)\mid u\in S\}:\S\DMu\big\}\\
        &=\big\{\{a\mid a\in S'\}\backslash \Sall :\Sr\P\big\}\\
        &=\{S'\backslash \Sall :\Sr\P\}\\
        &=\{S'\cap f(T) :\Sr\P\}
    \end{flalign*}
    Therefore $\DM\SSPR\P$ holds.
\end{proof}

\ifSubfilesClassLoaded{
  \bibliography{../../ref}
}{}
\end{document}

%% file: Reductions/Directed_Hamiltonian_Cycle/Directed_Hamiltonian_Cycle.tex
\subsection{Reductions from Directed Hamiltonian Cycle}

\subfile{../../Problems/DHC.tex}
\subsubsection[... to Undirected Hamiltonian Cycle]{Directed Hamiltonian Cycle to Undirected Hamiltonian Cycle}
\label{sec:DHC to UHC}
\paragraph{Problem Definition}

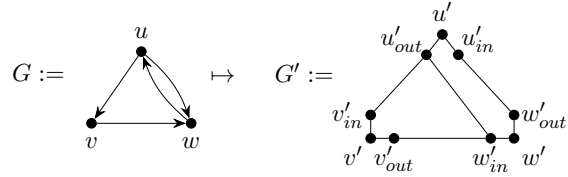
\begin{figure}[h]
    \begin{center}
        \scalebox{.85}{
            \begin{tikzpicture}[vertex/.style = {draw, circle, fill, inner sep = 1.5}, node distance = 1.5cm, label distance=0cm]
                \node[](G){$G:=$};
                \node[vertex, above right = 0cm and 1cm of G, label = $u$](u){};
                \node[vertex, below left = 1cm and .66cm of u, label = below:$v$](v){};
                \node[vertex, below right = 1cm and .66cm of u, label = below:$w$](w){};
                
                \node[below right = -.4cm and 2cm of G](to){$\mapsto$};

                \node[right = 3cm of G](G'){$G':=$};
                \node[vertex, above right = .25cm and 1.5cm of G', label = $u'$](u'){};
                \node[vertex, below right = .2cm and .133cm of u', label ={[label distance=-.2cm]above right:$u_{in}'$}](uin){};
                \node[vertex, below left = .2cm and .133cm of u', label ={[label distance=-.2cm]above left:$u_{out}'$}](uout){};
                \node[vertex, below left = 1.5cm and 1cm of u', label = {[label distance=-.1cm]below left:$v'$}](v'){};
                \node[vertex, right = .2cm of v', label ={[label distance=-.1cm]below:$v_{out}'$}](vout){};
                \node[vertex, above = .2cm of v', label ={[label distance=-.1cm]left:$v_{in}'$}](vin){};
                \node[vertex, below right = 1.5cm and 1cm of u', label = {[label distance=-.1cm]below right:$w'$}](w'){};
                \node[vertex, left = .2cm of w', label ={[label distance=-.1cm]below:$w_{in}'$}](win){};
                \node[vertex, above = .2cm of w', label ={[label distance=-.1cm]right:$w_{out}'$}](wout){};

                \draw   (u) edge[\xarrow] (v)
                        (u) edge[bend left = 15, \xarrow] (w)
                        
                        (w) edge[bend left = 15, \xarrow] (u)
                        (v) edge[bend left = 0, \xarrow] (w)

                        (uin) -- (u')
                        (uout) -- (u')
                        (uout) -- (vin)
                        (uout) -- (win)
                        
                        (vin) -- (v')
                        (vout) -- (v')
                        (vout) -- (win)
                        
                        (win) -- (w')
                        (wout) -- (w')
                        (wout) -- (uin)
                ;
            \end{tikzpicture}
        }
        \caption{\footnotesize{Example \textsc{Directed Hamiltonian Cycle} to \textsc{Undirected Hamiltonian Cycle}}}
        \label{fig:DHC to UHC}
    \end{center}
\end{figure}

\subfile{../../Problems/UHC.tex}
\paragraph{Reduction}
We use the reduction presented by Karp \cite{karp}. Let $\DHC:=[G=(V,A)]$ and $\UHC:=[G'=(V',E)]$. Now, to get $V'$, we replace
each vertex $v\in V$ with three vertices $v_{in}', v', v'_{out}\in V'$. We then add edges $\{v_{in}',v'\},\{v',v_{out}'\}$ and, for
every arc $(v,w)\in A$ we add the edge $\{v_{out}',w'_{in}\}\in E$. This preserves the one-to-one correspondence between the arcs and 
edges. \autoref*{fig:DHC to UHC} shows an example reduction.
\begin{proof}
    Let $S$ be a directed Hamiltonian cycle in $G$. Then every vertex in $G$ has at least one incoming and one outgoing arc. 
    It follows that the corresponding edges in $G'$ form a undirected Hamiltonian cycle when combined with the edges $\{v'_{in}, v'\}$
    and $\{v'_{out}, v'\}$ vor every vertex set.\\
    Now, let $S'$ be an undirected Hamiltonian cycle in $G'$. Then the edges $\{v'_{in}, v'\}$ and $\{v'_{out}, v'\}$ are included for
    every vertex set in $V'$. Due to the nature of our reduction and of Hamiltonian cycles every vertex $v'_{out}$ has to be connected
    to a vertex $u'_{in}$. It follows that the arc set $(v,u)$ for every edge $\{v'_{out}, u'_{in}\}$ is a directed Hamiltonian cycle in 
    $G$.
\end{proof}
\noindent The reduction function $g$ is computable in polynomial time and is dependent on the input size of the \DHC instance $\big|[G=(V,A)]\big|$,
as for each vertex in $V$ we create a fixed number of vertices the new instance. Therefore $\DHC\leq \UHC$ holds.

\paragraph{SSP \& Parsimonious}
There exists an exact one-to-one correspondence between the arcs in $G$ and the edges in $G'$ and therefore we can define a 
bijective function $f$. There also exists a fixed set of vertices that occur in every solution $\Sr\UHC$. We can therefore use
\lemref{lem:SSP and SPR 2} to prove that this reduction is both SSP and SPR.
\begin{proof}
    Let 
    \begin{flalign*}
        f_I&:A\rightarrow E,~f \big( (u,v)\big) = \{u_{out}', v_{in}'\}
    \end{flalign*}
    and let $\Snev=\Slink=\emptyset$,
    $$\Sall:= \bigcup_{v\in V}\big\{\{v'_{in},v'\},\{v'_{out},v'\}\big\}$$
    then $f_I':A\rightarrow E\backslash \Sall$ is bijective and
    \begin{flalign*}
        \{f(S):\S\DHC\} &= \Big\{\big\{f\big((u,v)\big)\mid (u,v)\in S\big\}:\S\DHC\Big\}\\
        &= \Big\{\big\{\{u'_{out},v'_{in}\}\in S'\big\}:\Sr\UHC\Big\}\\
        &= \{S'\cap f(A):\Sr\UHC\}
    \end{flalign*}
    Therefore $\DHC\SSPR\UHC$ holds.
\end{proof}

\ifSubfilesClassLoaded{
  \bibliography{../../ref}
}{}
\end{document}

%% file: Reductions/UHC/UHC.tex
\subsection{Reductions from Undirected Hamiltonian Cycle}

\subfile{../../Problems/UHC.tex}
\subsubsection[... to Traveling Salesman Problem]{Undirected Hamiltonian Cycle to Traveling Salesman Problem}
\label{sec:UHC to TSP}
\paragraph{Problem Definition}
\subfile{../../Problems/TSP.tex}

\paragraph{Reduction}
Let $\UHC:=[G=(V,E)]$, then the following is an easy folklore reduction mapping this instance to the $\TSP$ instance $[G'=(V',E'),w,k]$. First we map
every vertex to itself ($v\in V\implies v\in V'$) and create the edges $E'$ needed for a complete graph. Then we define the weight function as follows:
$$w:E'\rightarrow \mathbb Z,~w(e)=\begin{cases}
    0 &\mid e\in E\\
    1 &\mid e\notin E
\end{cases}$$
and finally set $k$ to $0$ resulting in only edges from $E$ being usable. An example reduction is shown in \autoref*{fig:UHC to TSP}.
\begin{figure}[h]
    \begin{center}
        \scalebox{.85}{
            \begin{tikzpicture}[vertex/.style = {draw, circle, fill, inner sep = 1.5}, node distance = 1.5cm, label distance=0cm, 
                                                every edge quotes/.style = {font=\footnotesize}, tight/.style={inner sep=1pt}]
                \node[](G){$G:=$};
                \node[vertex, right = 1cm of G, label = $u$](u){};
                \node[vertex, below left = .75cm and 1cm of u, label = left:$v$](v){};
                \node[vertex, below right = .75cm and 1cm of u, label = right:$y$](y){};
                \node[vertex, below right = 1cm and .33cm of v, label = below:$w$](w){};
                \node[vertex, below left = 1cm and .33cm of y, label = below:$x$](x){};
                
                \node[below right = -.4cm and 2cm of G](to){$\mapsto$};

                \node[right = 3cm of G](G'){$G':=$};
                \node[vertex, right = 1cm of G', label = $u$](u'){};
                \node[vertex, below left = .75cm and 1cm of u', label = left:$v$](v'){};
                \node[vertex, below right = .75cm and 1cm of u', label = right:$y$](y'){};
                \node[vertex, below right = 1cm and .33cm of v', label = below:$w$](w'){};
                \node[vertex, below left = 1cm and .33cm of y', label = below:$x$](x'){};
                
                \node[below right = .33cm and .33cm of x'](k){$k=0$};

                \draw   (u) -- (v)
                        (u) -- (w)
                        (u) -- (y)
                        (v) -- (w)
                        (x) -- (w)
                        (x) -- (y)

                        (u') edge["$0$" tight, above left] (v')
                        (u') edge["$0$" tight, above left] (w')
                        (u') edge["$0$" tight, above right] (y')
                        (v') edge["$0$" tight, below left] (w')
                        (x') edge["$0$" tight, below] (w')
                        (x') edge["$0$" tight, below right] (y')
                        
                        (u') edge["$1$" tight, above right, red, dashed] (x')
                        (v') edge["$1$" tight, below left, red, dashed] (x')
                        (y') edge["$1$" tight, below right, red, dashed] (w')
                        (y') edge["$1$" tight, above, red, dashed, in = -100, out = -800, looseness = 3] (v')
                ;
            \end{tikzpicture}
        }
        \caption{\footnotesize{Example \textsc{Undirected Hamiltonian Cycle} to \textsc{Traveling Salesman Problem}}}
        \label{fig:UHC to TSP}
    \end{center}
\end{figure}
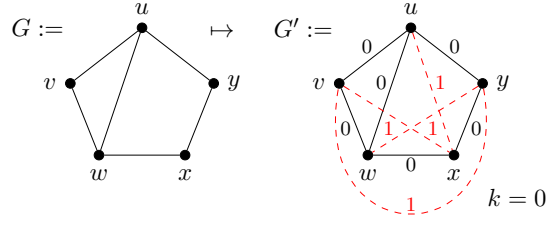
\begin{proof}
    This reduction holds, as there is a one-to-one correspondence between the edges of the \UHC instance and the usable edges of the 
    \TSP instance. Any cycle found in $G$ 
    is now also a TSP route and vice versa.
\end{proof}
\noindent The reduction function $g$ is computable in polynomial time and is dependent on the input size of the \UHC instance $\big|[G=(V,E)]\big|$,
as we only create a maximum of $\sum_{i=1}^{|V|-1} i$ edges in the new instance. Therefore $\UHC\leq \TSP$ holds.

\paragraph{SSP \& Parsimonious}
As we have a direct one-to-one correspondence between the instances, this reduction is both SSP and strongly Parsimonious.
\begin{proof}
    The following injective functions $f_I:E\rightarrow E',~f(e)=e'$,
    where $e$ and $e'$ are the same edge in the \UHC and \TSP instances respectively, can be defined. Due to the nature of our reduction
    there is a set of edges in $G'$ that will never be part of any solution 
    $$(\forall e\in E', \Sr\TSP: w(e)=1\implies e\notin S').$$
    We can therefore apply \lemref{lem:SSP and SPR} as with $\Snev:=\{e\in E'\mid w(e)=1\}$ and $\Sall=\Slink=\emptyset$ we get that 
    $$f_I:E\rightarrow E'\backslash\Snev,~f(e)=e'$$
    is bijective. Thus
    \begin{flalign*}
        \{f(S):\S\UHC\}&=\big\{\{f(e)\mid e\in S\}:\S\UHC\big\}\\
        &= \big\{\{e'\mid e'\in S'\}:\Sr\TSP\big\}\\
        &=\{S'\cap f(E):\Sr\TSP\}
    \end{flalign*}
    holds and so $\UHC\SSPR\TSP$.
    
\end{proof}

\ifSubfilesClassLoaded{
  \bibliography{../../ref}
}{}
\end{document}

%% file: Problems/TSP.tex
\noindent\fbox{
    \parbox{\textwidth}{
        \textsc{Traveling Salesman Problem ($\TSP$)}\\
        \begin{tabular}{cl}
        ~ & \textbf{Instances:} Complete Graph $G=(V,E)$, weight function $w:E\rightarrow \mathbb Z$, number $k\in\mathbb N$.\\
        ~ & \textbf{Universe:} Edge set $\mathcal{U}:=E$.\\
        ~ & \textbf{Feasible solution set:} The set of all TSP tours $S\subseteq E$.\\
        ~ & \textbf{Solution set:} The set of feasible $S$ with $w(S)\leq k$.
        \end{tabular}
    }
}

\ifSubfilesClassLoaded{
  \bibliography{../ref}
}{}
\end{document}

%% file: Reductions/Directed_Hamiltonian_Path/Directed_Hamiltonian_Path.tex
\subsection{Reductions from Directed Hamiltonian Path}

\subfile{../../Problems/DHP.tex}
\subsubsection[... to Undirected Hamiltonian Path]{Directed Hamiltonian Path to Undirected Hamiltonian Path}
\label{sec:DHP to UHP}
\paragraph{Problem Definition}
\subfile{../../Problems/UHP.tex}

\paragraph{Reduction}
For this reduction we refer to \cite[p. 42-45]{Femke}, as the reduction presented there is already both SSP and Parsimonious, where every
vertex $v$ is replaced by three vertices $v_{in}'$, $v'$ and $v'_{out}$ with edges $\{v_{in}',v'\}$ and $\{v',v'_{out}\}$. The arcs $(u,v)$ get 
mapped to the edges $\{u_{out}', v'_{in}\}$ and now the path starts at $s'_{in}$ and ends at $t'_{out}$. As \cite{Femke} includes a detailed
reduction, proof and an example, we refrain from repeating them here.

\paragraph{SSP \& Parsimonious}
The reduction maps every arc $(u,v)$ to the edges $\{u_{out}', v'_{in}\}$. This mean every solution $\S\DHP$ is equivalent to a solution $\Sr\UHP$,
where we include the edges from inside the vertex gadgets. It follows that we can apply \lemref{lem:SSP and SPR} as follows, to prove that 
the reduction is both SSP and SPR.
\begin{proof}
    Let the functions $f_I$ be defined as
    $$f_I:A\rightarrow E,~f\big((u,v)\big) = \{u_{out}', v'_{in}\}$$
    and let $\Snev=\Slink = \emptyset$ and
    $$\Sall :=\big\{ \{v_{in}',v'\},\{v',v'_{out}\}\mid v\in V\big\}$$
    then $f(E)\cup \Sall = A$ and we get 
    \begin{flalign*}
        \{f(S):\S\DHP\}&=\Big\{\big\{f\big((u,v)\big)\in S\big\}:\S\DHP\Big\}\\
        &=\Big\{\big\{\{u_{out}', v'_{in}\}\in S'\big\}:\Sr\UHP\Big\}\\
        &= \{S'\backslash\Sall:\Sr\UHP\}\\
        &=\{S'\cap f(E):\Sr\UHP\}
    \end{flalign*}
    and therefore $\DHP\SSPR\UHP$ holds.
\end{proof}

\ifSubfilesClassLoaded{
  \bibliography{../../ref}
}{}
\end{document}

%% file: Problems/UHP.tex
\noindent\fbox{
    \parbox{\textwidth}{
        \textsc{Undirected Hamiltonian Path ($\UHP$)}\\
        \begin{tabular}{cl}
        ~ & \textbf{Instances:} Graph $G=(V,E)$, Vertices $s,t\in V$.\\
        ~ & \textbf{Universe:} Edge set $\mathcal{U}:=E$.\\
        ~ & \textbf{Feasible solution set:} The set of all sets $S\subseteq E$ forming a Hamiltonian path\\
        ~ & going from $s$ to $t$.
        \end{tabular}
    }
}

\ifSubfilesClassLoaded{
  \bibliography{../ref}
}{}
\end{document}

%% file: Reductions/UHP/UHP.tex
\subsection{Reductions from Undirected Hamiltonian Path}

\subfile{../../Problems/UHP.tex}
\subsubsection[... to Undirected Hamiltonian Cycle]{Undirected Hamiltonian Path to Undirected Hamiltonian Cycle}
\label{sec:UHP to UHC}
\paragraph{Problem Definition}

\subfile{../../Problems/UHC.tex}
\paragraph{Reduction}
For this reduction we refer to \cite[p. 46-47]{Femke}, as the reduction presented there is already both SSP and Parsimonious. The idea is simple,
as the reduction only adds a new vertex $v_{new}$ and two edges connecting it with the $s$ and $t$ vertices of the $\UHP:=[G,s,t]$ instance. This 
way any Hamiltonian cycle has to include this new vertex and therefore goes over both new edges. The rest of the cycle is the equivalent
to an $s$-$t$ path in $G$.
As \cite{Femke} includes a detailed
reduction, proof and an example, we refrain from repeating them here.

\paragraph{SSP \& Parsimonious}
As we see above, every $s$-$t$ path in $G$ is part of a Hamiltonian cycle in $G'$, whith the edges $\{v_{new},s\}$ and $\{v_{new},t\}$ being part 
of every solution. We can therefore apply \autoref{lem:SSP and SPR} as follows to prove that the reduction is both SSP and Parsimonious.
\begin{proof}
  Let the functions $f_I$ be the identity functions defined as 
  $$f_I:E\rightarrow E', ~f\big(\{u,v\}\big)=\{u,v\}$$
  and let $\Snev=\Slink=\emptyset$ and $\Sall:=\big\{\{v_{new},s\},\{v_{new},t\}\big\}$. Now the functions
  $f_I:E\rightarrow E'\backslash\Sall$
  are bijective and we get
  \begin{flalign*}
    \{f(S):\S\UHP\}&=\Big\{\big\{f\big(\{u,v\}\big)\mid \{u,v\}\in S\big\}:\S\UHP\Big\}\\
    &=\Big\{\big\{\{u,v\}\mid \{u,v\}\in S'\big\}\backslash\Sall:\Sr\UHC\Big\}\\
    &=\{S'\backslash\Sall:\Sr\UHC\}\\
    &=\{S'\cap f(E):\Sr\UHC\}
  \end{flalign*}
  Therefore $\UHP\SSPR\UHC$ holds.
\end{proof}

\subsubsection[... to Traveling Salesman Problem]{Undirected Hamiltonian Path to Traveling Salesman Problem}
\label{sec:UHP to TSP}
\paragraph{Problem Definition}

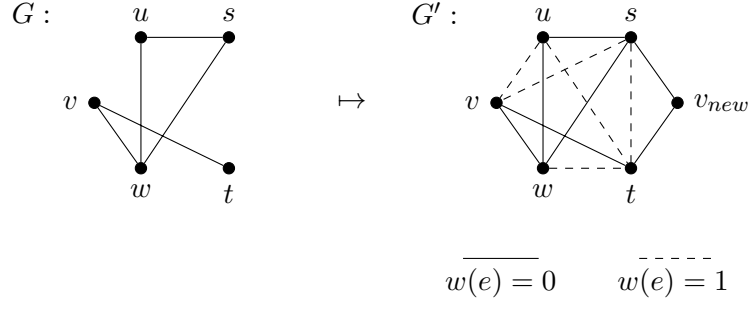
\begin{figure}[ht]
  \centering
    \begin{tikzpicture}[vertex/.style = {inner sep = 1.5, circle, draw, fill}, node distance = 1.5cm, label distance=0cm]  
          \node[](G){$G:$};
          \node[vertex, below right = 0 and 1 of G, label=$u$](u){};
          \node[vertex, right = 1 of u, label=$s$](s){};
          \node[vertex, below left = .75 and .5 of u, label=left:$v$](v){};
          \node[vertex, below right = .75 and .5 of v, label=below:$w$](w){};
          \node[vertex, right = 1 of w, label=below:$t$](t){};

          \node[right = 3 of v](to){$\mapsto$};

          \node[right = 4.5cm of G](G'){$G':$};
          \node[vertex, right = 1.5 of to, label=left:$v$](v'){};
          \node[vertex, below right = .75 and .5 of v', label=below:$w$](w'){};
          \node[vertex, above right = .75 and .5 of v', label=$u$](u'){};
          \node[vertex, right = 1 of u', label=$s$](s'){};
          \node[vertex, right = 1 of w', label=below:$t$](t'){};
          \node[vertex, below right = .75 and .5 of s', label=right:$v_{new}$](new){};

          \node[below left = 1 and 1 of w'](n1){};
          \node[right = 1 of n1](n2){};
          \node[below right = 1 and 1 of t'](d1){};
          \node[left = 1 of d1](d2){};

          \draw   (s) -- (u)
                  (s) -- (w)
                  (u) -- (w)
                  (v) -- (w)
                  (v) -- (t)

                  (s') -- (u')
                  (s') -- (w')
                  (u') -- (w')
                  (v') -- (w')
                  (v') -- (t')
                  (new) -- (t')
                  (new) -- (s')
                  (s') edge[dashed] (t')
                  (s') edge[dashed] (v')
                  (u') edge[dashed] (t')
                  (u') edge[dashed] (v')
                  (w') edge[dashed] (t')

                  (n2) edge["$w(e)=0$"] (n1)
                  (d1) edge["$w(e)=1$", dashed] (d2)
          ;
      \end{tikzpicture}
  \caption{\footnotesize{Example \textsc{Undirected Hamiltonian Path} to \textsc{Traveling Salesman Problem}}}
  \label{fig:UHP to TSP}
\end{figure}

\subfile{../../Problems/TSP.tex}
\paragraph{Reduction}
This reduction is based on the reduction presented in \cite[p. 44-46]{Femke}, but is slightly modified, to make proving the
Parsimonious property easier. Let $\UHP:=[G=(V,E),s,t]$, then we 
create the $\TSP$ instance as follows. First we create graph $G'=(V',E')$ by turning $G$ into a complete graph. Next we add a new vertex $v_{new}\in V'$
and the two edges $\{v_{new},s\}\in E'$ and $\{v_{new},t\}\in E'$. Finally we define the weight function as 
$$w:E'\rightarrow \mathbb Z,~w(e)=\begin{cases}
  0, & e\in E \text{ or } e\in \big\{\{v_{new},s\},\{v_{new},t\}\big\}\\
  1, &\text{otherwise}
\end{cases}$$
and set $k:=0$. Now $\TSP:=[G'=(V',E'), w, k]$ is the equivalent $\TSP$ instance. \autoref{fig:UHP to TSP} shows an example reduction.

\begin{proof}
  Let $\S\UHP$ be a Hamiltonian path. Then every vertex is visited once on the $s$-$t$ path. It follows that the same path is a path from $s$ to $t$
  in $G'$ of weight $k=0$. The only vertex not visited on this path is $v_{new}$, which connects $s$ and $t$ with edges of weight $0$. Therefore 
  $S':=S\cup\{\{v_{new},s\},\{v_{new},t\}\big\}$ is a valid solution of the \TSP instance.\\
  Now let $\Sr\TSP$ be a solution for the TSP instance. Then every vertex is visited once. As $k=0$, the solution only includes edges from $G$, 
  as well as the edges $\{v_{new},s\}$ and $\{v_{new},t\}$. Therefore the path from $s$ to $t$ that does not visit $v_{new}$ is a valid Hamiltonian path.
\end{proof}
\noindent The reduction function $g$ is computable in polynomial time and is dependent on the input size of the \UHP instance $\big|[G=(V,E),s,t]\big|$,
as we create only create a maximum of $2+\sum_{i=1}^{|V|-1}\leq |V|^2$ new edges and one vertex. Therefore $\UHP\leq \TSP$ holds.

\paragraph{SSP}
Every $s$-$t$ path is part of a \TSP solution, where only the edges $\{v_{new},s\}$ and $\{v_{new},t\}$ are added to the solution.
We can therefore use \autoref{lem:SSP and SPR} to prove that this reduction is both SSP and Parsimonious.
\begin{proof}
  Let the functions $f_I$ be the identity functions defined as 
  $$f_I:E\rightarrow E', ~f\big(\{u,v\}\big)=\{u,v\}$$
  and let $\Snev=\Slink=\emptyset$ and $\Sall:=\big\{\{v_{new},s\},\{v_{new},t\}\big\}$. Now the functions
  $f_I:E\rightarrow E'\backslash\Sall$
  are bijective and we get
  \begin{flalign*}
    \{f(S):\S\UHP\}&=\Big\{\big\{f\big(\{u,v\}\big)\mid \{u,v\}\in S\big\}:\S\UHP\Big\}\\
    &=\Big\{\big\{\{u,v\}\mid \{u,v\}\in S'\big\}\backslash\Sall:\Sr\TSP\Big\}\\
    &=\{S'\backslash\Sall:\Sr\TSP\}\\
    &=\{S'\cap f(E):\Sr\TSP\}
  \end{flalign*}
  Therefore $\UHP\SSPR\TSP$ holds.
\end{proof}

\ifSubfilesClassLoaded{
  \bibliography{../../ref}
}{}
\end{document}

%% file: Sections/Conclusion/Conclusion.tex
    \chapter{Conclusion}
    \label{chap:5}
    This thesis explores reductions upholding the SSP property 
    and how they relate to Parsimonious reductions, 
    offering new insights into computational complexity. By exploring 46 reductions between 30 SSP problems, 
    we lay out when and how these two concepts overlap and where they diverge. \\
    A big takeaway is that SSP and Parsimonious reductions are not always interchangeable and only co-exists under 
    certain conditions, which we formalize in \autoref*{theorem 1}. 
    We prove that Cook's theorem can be applied to the complexity class \SSPP, confirming that \textsc{Satisfiability} 
    is the canonical \SSPP-complete problem. 
    One unexpected challenge during this work was finding that
    \textsc{(Exact-)3-Satisfiability} is not a good basis for proving the 
    \SSPP-completeness of many problems, while the \textsc{1-in-3-Satisfiability} variant appears to be more suited.
    Another big hurdle was provided by the fact that we need to use the maximum or minimum cardinality versions 
    of certain problems, to ensure that the Parsimonious property would not be breached. Although we use 
    \textsc{Minimum Cardinality Vertex Cover} to prove the \SSPP-hardness of seven known problems, we nevertheless
    noticed that it has its own limitations, especially where reductions to graph problems are concerned, and therefore 
    does not always lend itself as a basis for proving that something is in \SSPP. On the topic 
    of known problems, we also had to introduce two more unknown, and slightly forced, variants of classic problems. 
    We use the \textsc{Vertex Cover with one fixed Vertex} problem, to prove that certain given SSP reductions
    can not be easily modified to also have the Parsimonious property, and we use \textsc{1+3-Dimensional Matching}
    as an intermediate problem, to prove the \SSPP-hardness of \textsc{3-Dimensional Exact Matching}.
    In the cases in which a single reduction presented in this paper does not meet the conditions necessary for proving \SSPP-hardness,
    we can use the transitive nature of the SSP and Parsimonious properties, 
    to prove that a reduction upholding both properties does indeed exist.
    This suggests that these types of challenges can generally be addressed with the right approach. \\
    ~\\
    Looking forward, future research could explore further SSP variants of known problems and identify further 
    key variants of classical problems, similar to how
    \textsc{1-in-3-Satisfiability} seems to be better suited for these reduction structures than \textsc{(Exact-)3-Satisfiability}.\\
    Finally, there is real potential to grow "The Reduction Network" into a central hub for researchers. 
    By adding more reductions, creating visualization tools and offering interactive 
    features it could become an invaluable resource for anyone working in this field.\\
    Ultimately, this thesis lays the groundwork for deeper exploration of \SSPP~and its related concepts. 
    These findings could drive progress in both theoretical research and practical applications, 
    particularly in optimization and other fields that rely on understanding computational complexity.
\ifSubfilesClassLoaded{
  \bibliography{../../ref}
}{}
\end{document}

%% file: Sections/Appendix/Appendix.tex
\begin{appendices}
\section{Problem Definitions}
\label{appendix}
Here we list all of the computational problem presented in this paper, sorted into three types.
\enquote{Satisfiability Problems} contain variants of \textsc{Satisfiability}, 
\enquote{Graph Problems} contain problems over a graph, divided up into directed and undirected graphs, 
and \enquote{Set Problems} contain the problems over a set of numbers, tuples or objects.

\subsection{Satisfiability Problems}
\label{sec:Sat Problems}
\subfile{../../Problems/SAT.tex}
\label{prob:SAT}

\subfile{../../Problems/TSAT.tex}
\label{prob:TSAT}

\subfile{../../Problems/ESAT.tex}
\label{prob:ESAT}

\subfile{../../Problems/OSAT.tex}
\label{prob:OSAT}

\subsection{Graph Problems}
\label{sec:Graph Problems}
\subsubsection{Undirected Graphs}
\subfile{../../Problems/CQ.tex}
\label{prob:CQ}

\subfile{../../Problems/DS.tex}
\label{prob:DS}

\subfile{../../Problems/MDS.tex}
\label{prob:MDS}

\subfile{../../Problems/MIS.tex}
\label{prob:MIS}

\subfile{../../Problems/STT.tex}
\label{prob:STT}

\subfile{../../Problems/TSP.tex}
\label{prob:TSP}

\subfile{../../Problems/UHC.tex}
\label{prob:UHC}

\subfile{../../Problems/UHP.tex}
\label{prob:UHP}

\subfile{../../Problems/VC.tex}
\label{prob:VC}

\subfile{../../Problems/MVC.tex}
\label{prob:MVC}

\subfile{../../Problems/VCV.tex}
\label{prob:VCV}

\subsubsection{Directed Graphs}
\subfile{../../Problems/DHC.tex}
\label{prob:DHC}

\subfile{../../Problems/DHP.tex}
\label{prob:DHP}

\subfile{../../Problems/FAS.tex}
\label{prob:FAS}

\subfile{../../Problems/FVS.tex}
\label{prob:FVS}

\subsection{Set Problems}
\label{sec:Set Problems}
\subfile{../../Problems/IP.tex}
\label{prob:IP}

\subfile{../../Problems/DM.tex}
\label{prob:DM}

\subfile{../../Problems/ODM.tex}
\label{prob:ODM}

\subfile{../../Problems/EC.tex}
\label{prob:EC}

\subfile{../../Problems/HS.tex}
\label{prob:HS}

\subfile{../../Problems/KS.tex}
\label{prob:KS}

\subfile{../../Problems/P.tex}
\label{prob:P}

\subfile{../../Problems/PCEN.tex}
\label{prob:PCEN}

\subfile{../../Problems/PMED.tex}
\label{prob:PMED}

\subfile{../../Problems/SC.tex}
\label{prob:SC}

\subfile{../../Problems/SP.tex}
\label{prob:SP}

\subfile{../../Problems/SS.tex}
\label{prob:SS}

\subfile{../../Problems/TMS.tex}
\label{prob:TMS}

\subfile{../../Problems/UFL.tex}
\label{prob:UFL}
\end{appendices}

\ifSubfilesClassLoaded{
  \bibliography{../../ref}
}{}
\end{document}

%% file: Problems/DS.tex
\noindent\fbox{
    \parbox{\textwidth}{
        \textsc{Dominating Set (DS)}\\
        \begin{tabular}{cl}
        ~ & \textbf{Instances:} Graph $G=(V,E)$, number $k\in \mathbb{N}$\\
        ~ & \textbf{Universe:} Vertex set $\mathcal{U}:=V$.\\
        ~ & \textbf{Feasible solution set:} The set of all dominating sets.\\
        ~ & \textbf{Solution set:} The set of all vertex covers of size at most $k$.
        \end{tabular}
    }
}

\ifSubfilesClassLoaded{
  \bibliography{../ref}
}{}
\end{document}

%% file: Problems/VC.tex
\noindent\fbox{
    \parbox{\textwidth}{
        \textsc{Vertex Cover (VC)}\\
        \begin{tabular}{cl}
        ~ & \textbf{Instances:} Graph $G=(V,E)$, number $k\in \mathbb{N}$\\
        ~ & \textbf{Universe:} Vertex set $\mathcal{U}:=V$.\\
        ~ & \textbf{Feasible solution set:} The set of all vertex covers.\\
        ~ & \textbf{Solution set:} The set of all vertex covers of size at most $k$.
        \end{tabular}
    }
}

\ifSubfilesClassLoaded{
  \bibliography{../ref}
}{}
\end{document}